%% file: long_mnras.tex
\PassOptionsToPackage{pdfpagelabels=false}{hyperref} 
\documentclass[fleqn,usenatbib]{mnras}
\usepackage{multirow}
\usepackage{mathptmx}
\usepackage{graphicx}   
\usepackage{amssymb}    
\usepackage{amsmath}    
\usepackage{hhline}

\title{Twenty-year monitoring of the surface magnetic fields of chemically peculiar stars}

\author[Giarruso, M. et al.]{
M. Giarrusso $^{1,2}$,
M. Cecconi$^{3}$,
R. Cosentino$^{3}$,
M. Munari$^{2}$, 
A. Ghedina$^{3}$,
F. Ambrosino$^{4}$,\and
W. Boschin$^{3,5,6}$,
F. Leone$^{7,2}$\\
$^{1}$Department of Physics and Astronomy, University of Florence, Largo Enrico Fermi 2, 50125, Firenze, Italy\\
$^{2}$ INAF - Osservatorio Astrofisico di Catania, Via S. Sofia 78, I--95123 Catania, Italy\\
$^{3}$INAF - Fund. Galileo Galilei, Rambla Jos\'e Ana Fern\'andez Perez 7, 38712 Bre\~na  Baja (La Palma), Canary Islands, Spain\\
$^{4}$INAF - Osservatorio Astrofisico di Roma, Via Frascati 33, I--00078, Monteporzio Catone (Roma) Italy\\
$^{5}$Instituto de Astrof\'isica de Canarias (IAC), Calle V\'ia L\'actea s/n, 38205 La Laguna, TF - Spain\\
$^{6}$Departamento de Astrof\'isica, Universidad de La Laguna (ULL), 38206 La Laguna, TF - Spain\\
$^{7}$ Universit\'a di Catania, Dipartimento di Fisica e Astronomia, Sezione Astrofisica, Via S. Sofia 78, I-95123 Catania, Italy}

\begin{document}

\date{Received To be inserted later, accepted To be inserted later}

\pagerange{\pageref{firstpage}--\pageref{lastpage}} \pubyear{2021}

\maketitle

\label{firstpage}

\begin{abstract}
Magnetic chemically peculiar stars of the main sequence can present rotational periods as
long as many decades. Here we report the results of an observational campaign started in
2001 aimed at establishing these very long periods from the variability of the integrated
magnetic field modulus, the so-called surface magnetic field $B_s$, as measured from
the Zeeman splitting of the Fe{\sc ii}\,6149.258\,{\AA} spectral line. Thirty-six stars
have been monitored with various high-resolution spectrographs at different telescopes,
totalling 412 newly collected spectra. To improve the phase coverage, we have also
exploited all public archives containing high-resolution spectra, many not yet published.
On the basis of these new $B_s$ variability curves, we
1) confirm or revisit the periods of 24 stars,
2)  extend the lower limits to the periods of HD\,55719 ($P > 38$\,yr), HD\,165474 ($P > 27$\,yr), HD\,177765 ($P > 37$\,yr),
3) establish for the first time the periods of HD\,29578 ($P = 10.95$\,yr), HD\,47103  ($P = 17.683$\,d), HD\,150562 ($P = 5.7$\,yr), HD\,216018 ($P = 34.044$\,d), and
4) set lower limits to the periods of  HD\,75445 ($P >> 14$\,yr), HD\,110066  ($P >> 29$\,yr), HD\,116114 ($P >  48$\,yr), and HD\,137949 ($P > 27$\,yr). 
As to $\gamma$\,Equ, whose period must exceed 90 years, we point out
a clear decrease in the field modulus, the maximum of which coincides within the uncertainties
with the minimum of the variation in the integrated longitudinal field.
\end{abstract}


\section{Introduction}

Following initial measurements of the surface magnetic field $B_s$ (average of the
magnetic field modulus over the visible stellar disk) of the star HD\,215441 by
\cite{Babcock1960}, large numbers of $B_s$ measurements in Magnetic Chemically
Peculiar (MCP) stars have been reported by \cite{Preston1971}, \cite{Mathys1997}
(=M97) and \cite{Mathys2017}(=M17). These days, the most straightforward and
accurate method to determine $B_s$ is based on the Zeeman splitting of the
Fe{\sc ii}\,6149.258\,{\AA} spectral line which features two Zeeman
$\pi$ subcomponents coinciding with two $\sigma$ subcomponents \citep{Mathys1990}.

Within the framework of the oblique rotator model first proposed by \cite{Babcock1949}
and further developed by \cite{Stibbs1950}, MCP stars present photometric, spectroscopic,
and magnetic variability with a single period as a consequence of stellar rotation.
Measuring the rotational period from the modulation of the surface magnetic field
is probably the most reliable approach to the study of very long-term variations. All
the necessary information is embedded in a single spectrum and there is no need for
standard stars or zero point adjustments as in the case of photometry.

This paper reports the results of an observational campaign started in 2001 to measure
and monitor the surface fields of 36 MCP stars whose rotational periods were expected
to be very long, up to decades, judging from the sharpness of their spectral lines.
As indicated above, field measurements rely on high-resolution spectra of the
Fe{\sc ii}\,6149.258\,{\AA} line. In order to extend as much as possible the time
frame, we have in addition explored all public astronomical archives and mined the
literature. We have also analysed some of our spectra dating back to 1995.

In Section 2, we present the: 1) high-resolution spectrographs we have operated,
2) reduction methods, 3) procedure to measure $B_s$ from the Fe{\sc ii}\,6149.258\,{\AA}
spectral line, and 4) method to establish the variability period. In Section 3, we
present the rotational periods star-by-star. Whenever possible, these periods have
been checked with the effective magnetic field $B_e$ (average over the visible stellar 
disk of the magnetic field component along the line of sight) and/or photometric
results, or determined from a combination of these data. Finally, we present an
empirical relation between rotational period and magnetic field strength.

\begin{table}   
\caption{List of spectrographs used to measure stellar surface magnetic fields
together with the spectral resolution R\,[k] = $\frac{\lambda}{\Delta\lambda}$/1000.
For every instrument, $N$ represents the number of $B_s$ measurements from spectra
acquired in this study. A two-letter identification is used for the instruments.}
\begin{tabular}{llrrl}\multicolumn2c{ \rm Spectrograph@Telescope}  & {\rm R [k]} &{N}&  {\rm Reference} \\\hline
CAOS@OAC               & CS &    55 & 123 & \cite{Leone2016a}      \\
HARPS@ESO\,3.6m   & HS &  115 & 20 & \cite{Mayor2003}         \\
HARPS-North@TNG         & HN &  115 &   81 & \cite{Cosentino2013}   \\
UCLES@AAT              & UC &  120 &  22 & \cite{Horton2012}       \\
SARG@TNG               & SG &  $\le$ 164 &   53 & \cite{Gratton2001}       \\
CES@ESO\,3.6m        & CE &  220 &     2  & \cite{Enard1982}        \\
 \hline
         \end{tabular}\label{Tab_Spectrographs}
         \end{table}

\begin{table}
\caption{Archives hosting high-resolution spectra of long period magnetic
chemically peculiar stars. For every instrument, $N$ represents the number
of $B_s$ measurements from archive spectra.}   
\begin{tabular}{llrrl}
 \multicolumn{2}{c}{\rm Spectrograph@Telescope}  & \rm R [k]& N & \rm Reference \\\hline
ELODIE@OHP-1.9m & EE  &   40 & 6 &\cite{Baranne1996} \\
FEROS@ESO-2.2m  & FS  &   48 & 6 &\cite{Kaufer1999} \\
EMMI@ESO-NTT      & EM   &   60 &  2 &\cite{Dodorico1990} \\
NES@BTA                  & NS & 60  & 6 &\cite{Panchuk09}\\
ESPaDOns@CFHT  & ES  &   65 & 85 &\cite{Silvester2012} \\ 
NARVAL@TBL           & NL  &   65 & 16 &\cite{Silvester2012} \\
UCLES@AAT             & UC &   90 & 10 &\citet{Horton2012} \\
UVES@ESO-UT2      & US  &  100 & 72 &\cite{Dekker2000}\\
HARPS@ESO-3.6m   & HS &  115 & 350 &\cite{Mayor2003}\\
GECKO@CFHT         & GO  &  120 & 21 & \cite{Glaspey1993} \\  
CES@ESO-3.6m       & CE  &  220 & 7 &\cite{Enard1982} \\
 \hline
         \end{tabular}\label{Tab_Archive}
         \end{table}

\begin{table*}
\scriptsize
\caption{Observed stars. Literature and adopted ephemerides of $B_s$ variability.
}
\begin{tabular}{ccclcllcc}
\hline
\label{Tab_Periods}
{HD  / HDE}& Other id.&$T_{\rm eff}[K]\,  /\, R[R_\odot]$  & \multicolumn2c{Literature Ephemeris}  & {Reference} & \multicolumn2c{ Here determined or adopted Ephemeris  }     \\
                    &              &   &   {JD = 2400000+}                      &  {Period (days)}     &                                             &  {$HJD_0$ = 240000+}                      &  Period (days) \\\hline
         965                   &    BD\,$-$00\,21  & 7027 / 2.46    & $B_e^{\rm min}$ = 51000.0     &   6030$\pm$200   & \cite{Mathys2019_HD965}   &   $B_e^{\rm min}$ = 51000.0   &    6030 \\
 \hline
\multirow{2}{*}{2453} & \multirow{2}{*}{BD\,+31\,59}  & \multirow{2}{*}{8355}  & $B_e^{\rm min}$ = 42213.0       &     521$\pm$2       & \cite{Mathys2017} &  \multirow{2}{*}{ $B_e^{\rm min}$ = 48440.911 }  &  \multirow{2}{*}{518.2} \\
                                     & &  & c$_{1}^{\rm max}$ = 48440.911 & 518.2$\pm$0.5    & \cite{Pyper2017}   &   &    \\   \hline
\multirow{2}{*}{9996}  & \multirow{2}{*}{HR\,465}  &   \multirow{2}{*}{9200}  & y$^{\rm max}$ = 53016.610                &7850$\pm$100     &  \cite{Pyper2017}                & \multirow{2}{*}{$B_s^{\rm max}$ = 49200.0  }         &  \multirow{2}{*}{7850} \\ 
                                        &     &  & $B_e^{\rm min}$ = 33301.360  &7936.522            &     \cite{Bychkov2019}                               &                                             &  &        \\                       \hline
\multirow{3}{*}{12288}  & \multirow{3}{*}{BD +68\,144}  & \multirow{3}{*}{8415}        &    $B_e^{\rm max}$ = 48499.87  &  34.9$\pm$0.2     &  \cite{Wade2000}        &      \multirow{3}{*}{$B_e^{\rm max}$ = 51131.9} & \multirow{3}{*}{34.993$\pm$0.003} \\  
                                     &  &   &  v$^{\rm max}$ = 51131.772       & 34.99$\pm$0.01  &  \cite{Pyper2017}        &                                                                      &         &         \\ 
                                     & &   &  V$^{\rm max}$ = 57218.6         &  35.73$\pm$ 0.2   &  \cite{Bernhard2020}  &                                                                       &          &          \\ \hline
\multirow{2}{*}{\rm \,14437 } & \multirow{2}{*}{BD\,+42\,502 } & \multirow{2}{*}{9660}& V$^{\rm max}$ = 57077.7                  &  26.78$\pm$0.1     &  \cite{Bernhard2020}  &   \multirow{2}{*}{v$^{\rm max} = 49228.820$} & \multirow{2}{*}{26.734$\pm$0.007}         \\
					&  & &  $B_e^{\rm max}$ = 48473.846  &  26.87$\pm$0.02  &  \cite{Wade2000}       &                 &            \\ \hline
\multirow{2}{*}{\rm \,18078 } & \multirow{2}{*}{BD +55 726 } & \multirow{2}{*}{7718 / 4.43 }&  \multirow{2}{*}{$B_{s}^{\rm max} = 49930.0$}  &  \multirow{2}{*}{$1358\pm$12} &  \cite{Pyper2017}  &    \multirow{2}{*}{$B_{s}^{\rm max} = 49916$}  &  \multirow{2}{*}{$1352\pm$6}       \\
                                        &    &           &                                                               &     &   \cite{Mathys2016}     &       &        \\\hline
29578                        &    CPD\,$-$54 685 &       7340 / 2.88            &                &    $>>$ 1800          &  \cite{Mathys2017}      &$B_{\rm s}^{\rm max}$ = 51950.0 & 4000\,/\,9370       \\\hline
47103                        &    BD\,+20\,1508 &            8108                &                    &       $>$ 10              &  \cite{Wraight2012}     & $B_{\rm s}^{\rm max}$ = 50098.99 &    17.683$\pm$0.004      \\\hline
50169                        &    BD\,$-$1\,1414   &         8901              &                       & 10600$\pm$300  &   \cite{Mathys2019_HD50169}     &      $B_{\rm s}^{\rm max}$ = 41600.0  &10600                       \\\hline
51684                        &   CoD\,$-$40\,2796 & 7546 / 2.99 &$B_{\rm s}^{\rm max}$ = 49947.0        & 371$\pm$ 6 &   \cite{Mathys2019_HD50169}     &      $B_{\rm s}^{\rm max}$ = 53617  & 366$\pm$1                       \\\hline
55719                        &    HR\,2727 &    9131 &                                               &  $ >>$ 3650    &    \cite{Mathys2017}        &  \,\,\,\,\,\,\,\,48500        &  $\ge$ 14000    \\\hline
61468                        &   CoD\,$-$27\,4341 & 9057 & $B_e^{\rm max}$ = 50058.5 &   322$\pm$3  &  \cite{Mathys2017}     &  $B_e^{\rm max}$ = 50058.5 &   321$\pm$1 \\\hline
75445                        &   CoD\,$-$38\,4907 &   9057     &        &     6.291$\pm$ 0.002\,\, ?    & \cite{Mathys2017}   &   &   $>$ 5000  \\\hline
\multirow{2}{*}{81009 } & \multirow{2}{*}{HR\,3724 } & \multirow{2}{*}{8430 } &v$^{\rm min}$  =  48646.878  & 33.987$\pm$0.002   & \cite{Pyper2017}  &  \multirow{2}{*}{$B_{\rm s}^{\rm max} =  48645.9$}    &  \multirow{2}{*}{33.987}\\
                                      &    &  & v$^{\rm max}$ =  44483.420  & 33.984$\pm$0.055  &  \cite{Wade2000_HD81009}  &                                                                        &          \\  \hline
93507      &    CoD\,$-$67\,1494    & 8999             &   $B_e^{\rm min}$ =   49800.0      &  556$\pm$ 22       &    \cite{Mathys1997}                   & $B_{\rm s}^{\rm max}$ = 48965.0 &  562$\pm$5   \\ \hline
94660      & HR\,4263      & 9571               &    $B_{\rm s}^{\rm min}$ =  47000.0  &  2800$\pm$200       &   \cite{Mathys2017}                   & $B_{\rm s}^{\rm max}$ = 48284.0 &  2830$\pm$140   \\ \hline
\multirow{2}{*}{110066} & \multirow{2}{*}{HR\,4816} &   \multirow{2}{*}{8878} &   \multicolumn2c{phot.\,not\,variable}  &    \cite{Pyper2017}      &   & \multirow{2}{*}{ $>$ 10\,500}    \\ 
                                     & &   &  $B_{\rm s}^{\rm min}$ =  49826.738  &  6.4769$\pm$0.0011    &    \cite{Bychkov2021}      &      &     \\ \hline
\multirow{2}{*}{116114} &\multirow{2}{*}{BD\,$-$17\,3829} &\multirow{2}{*}{7424 / 2.84} & m$^{\rm max}$ = 54352.057 & 5.3832 & \cite{Wraight2012} & \multirow{2}{*}{$B_{\rm s}^{\rm max} =  40350.0$}        &   \multirow{2}{*}{ $>$ 17700 }   \\
                                  &  &  &    $B_e^{\rm max}$ = 47539.000 &27.61 &     \cite{Mathys2017}   & &           \\                       \hline
\multirow{2}{*}{126515} &  \multirow{2}{*}{BD\,+1\,2927} &  \multirow{2}{*}{9422 } &  $B_e^{\rm max}$ = 37015.000 & 129.95                    &  \cite{Mathys2017}   &  \multirow{2}{*}{$B_e^{\rm max} =   37015.0$}    &  \multirow{2}{*}{129.95}              \\
                                      & &    &     v$^{\rm min}$ = 52031.708  & 129.95$\pm$0.02  & \cite{Pyper2017}     &    & \\ \hline    
\multirow{3}{*}{137949} &\multirow{3}{*}{33 Lib} &\multirow{3}{*}{7406 / 1.97} &  \multicolumn2c{many decades}           &  \cite{Landstreet2014}       &     & \multirow{3}{*}{ $>$ 10\,000}   \\
                                      &&&  \multicolumn2c{phot. not variable } &  \cite{Pyper2017}     &      & \\
                                      &&&  $B_e^{\rm max}$ = 38166         & 5195             &  \cite{Mathys2017}    &      &          \\                       \hline                                    
\multirow{2}{*}{142070} & \multirow{2}{*}{BD\,$-$0\,3026} & \multirow{2}{*}{8130} &   v$^{\rm min}$ = 50837.499  & 3.37189$\pm$0.00007  & \cite{Adelman2001}   & \multirow{2}{*}{$B_e^{\rm max} =  49878.2$} & \multirow{2}{*}{$B_e^{\rm max} =  3.3721\pm$0.0002}   \\
                                     &  &  &          $B_e^{\rm max}$ = 49878.2    &    3.3718$\pm$0.0011       &  \cite{Mathys2017}  &&  \\ \hline    
144897        & CoD\,$-$40\,10236  & 7398 / 2.89        &  $B_e^{\rm max} $ = 49133.7      &   48.57$\pm$0.15       & \cite{Mathys2017}        & $B_e^{\rm max}$ = 491157.1  & 48.60$\pm$0.02    \\ \hline
150562        & CoD\,$-$48\,11127   & 6390 / 2.13       &                                                      &   $>$ 1600                  & \cite{Mathys2017}       & $B_e^{\rm max}$ =  54317.0  &  2100$\pm$200  \\\hline
154708        & CD\,$-$57\,6753   & 6812 / 1.66        &  $B_e^{\rm max}$  = 54257.740  &   5.363$\pm$0.003     & \cite{Landstreet2014}  & $B_e^{\rm max}$ = 53662.57  & 5.367$\pm$0.001    \\ \hline
318107        & CoD\,$-$32 13074   & 9050   &  $B_e^{\rm max}$  = 48800.000  &   9.7088$\pm$0.0007 & \cite{Bailey2011_HDE318107} &  $B_e^{\rm max}$  = 48800.0  &   9.7089$\pm$0.0002   \\ \hline
165474      & BD\,+12\,3382  & 7692 / 2.82                   &                                                      &  $>>$3300                 & \cite{Mathys2017}  &  $B_e^{\rm max} $ = 52150.0  &   $\ge$ 9900    \\ \hline
166473      & CoD\,$-$37\,12303  & 7451 / 1.88        &  $B_{\rm s}^{\rm max}$ = 48660.0 & 3836$\pm$30          &  \cite{Mathys2020_HD166473}    &   $B_{\rm s}^{\rm max}$ = 48660.0 & 3836 \\ \hline
177765      & CoD\,$-$26\,13816  & 7002 / 3.52                  &                                                        &   $>> $1800             & \cite{Mathys2017} & & $\ge$13500\\ \hline
178892      & BD\,+14\,3811   & 7582 / 2.06           & V$^{\rm max}$    = 52708.562    &   8.2549          & \cite{Semenko_HD178892}     & $B_s^{\rm max}$ = 52696.850 & 8.2572$\pm$0.0016   \\ \hline
187474       & HR\,7552      & 9438             & $B_e^{\rm min}$ = 46766.000    & 2345  & \cite{Mathys2017}                                 & $B_{\rm s}^{\rm max}$ = 57176    &   2324$\pm$40            \\ \hline
\multirow{2}{*}{188041} & \multirow{2}{*}{HR\,7575} & \multirow{2}{*}{8400} &    v$^{\rm min}$     = 49904.860 & 223.826$\pm$0.040  &  \cite{Pyper2017}    &   \multirow{2}{*}{$B_e^{\rm max}$ =   49797.921} & \multirow{2}{*}{223.826}       \\
                                      & & &  $B_e^{\rm max}$ = 46319.5     &  223.78$\pm$0.10     &  \cite{Mathys2017} &            &                 \\ \hline
192678     & BD\,+53\,2368    & 9276            &  $B_e^{\rm max}$ = 44890.170  & 6.4193$\pm$0.003  &  \cite{Pyper2017}   &  $B_s^{\rm max}$ =   49112.76  &6.4199$\pm$0.0001\\ \hline
335238      & BD\,+29\,4202    & 9045                & \hspace{0.5cm} 47000                &    48.7$\pm$0.1                  &\cite{Mathys2017}  & $B_e^{\rm max}$ = 57222.7 & 48.985$\pm$0.007 \\ \hline
201601     & $\gamma$\,Equ   & 7457 / 2.16                  &   $B_e^{\rm min}$ = 52457.1      & 35462.5$\pm$1149            &     \cite{Bychkov_HD201601}              &      $B_e^{\rm max}$ =  52200.0         &  35462.5         \\\hline
\multirow{2}{*}{208217} &  \multirow{2}{*}{CPD\,$-$62\,6281}  &  \multirow{2}{*}{8140} & \multirow{2}{*}{$B_e^{\rm max}$ = 47028.0}     &  8.44475$\pm$0.00011      &  \cite{Mathys2017}       &   \multirow{2}{*}{$B_s^{\rm max}$ = 47027.094}  &  \multirow{2}{*}{8.445$\pm$0.005} \\
			             &   &          &                                        &  8.317$\pm$0.001             & \cite{David-Uraz2019} &    & \\\hline
\multirow{2}{*}{216018} &   \multirow{2}{*}{BD\,$-$12\,6357}  &   \multirow{2}{*}{7719 / 2.08}      &                                          &  $>$ 10                              & \cite{Wraight2012}  & \multirow{2}{*}{$B_e^{\rm max}$  =  49531.870}    &  \multirow{2}{*}{$34.044\pm$0.007}        \\
			             &    &       &                                          &  $>>$2000               & \cite{Mathys2017}  & & \\
\hline
\end{tabular}
\end{table*}

\section{Observations, archives, data reduction and period search}

We have carried out high-resolution spectroscopy of 36~MCP stars with the instruments
listed in Table\,\ref{Tab_Spectrographs}, for a total of 412 newly acquired spectra. The
data have been reduced using IRAF routines as described in \cite{Leone2017}. To extend as
much as possible the time frame for determining the variability periods we have exploited
all public archives storing high-resolution spectra (Table\,\ref{Tab_Archive}) with the
Fe{\sc ii}\,6149.258\,{\AA} line. A total of 581 spectra have been retrieved.

Following \cite{Mathys1990, Mathys2017} and \cite{ Mathys1997}, measurements of the
surface magnetic field $B_s$ have been derived (assuming a linear Zeeman effect regime)
from the distance $\Delta\lambda$ between the Zeeman subcomponents of the
Fe{\sc ii}\,6149.258\,{\AA} line:
\begin{eqnarray}
B_s\,[G] = 20974\, \Delta\lambda\,[{\AA}]
\label{ZWF}
\end{eqnarray}

\begin{figure}\center
\includegraphics[trim={1cm 0cm 0cm 0cm}, width=0.495\textwidth]{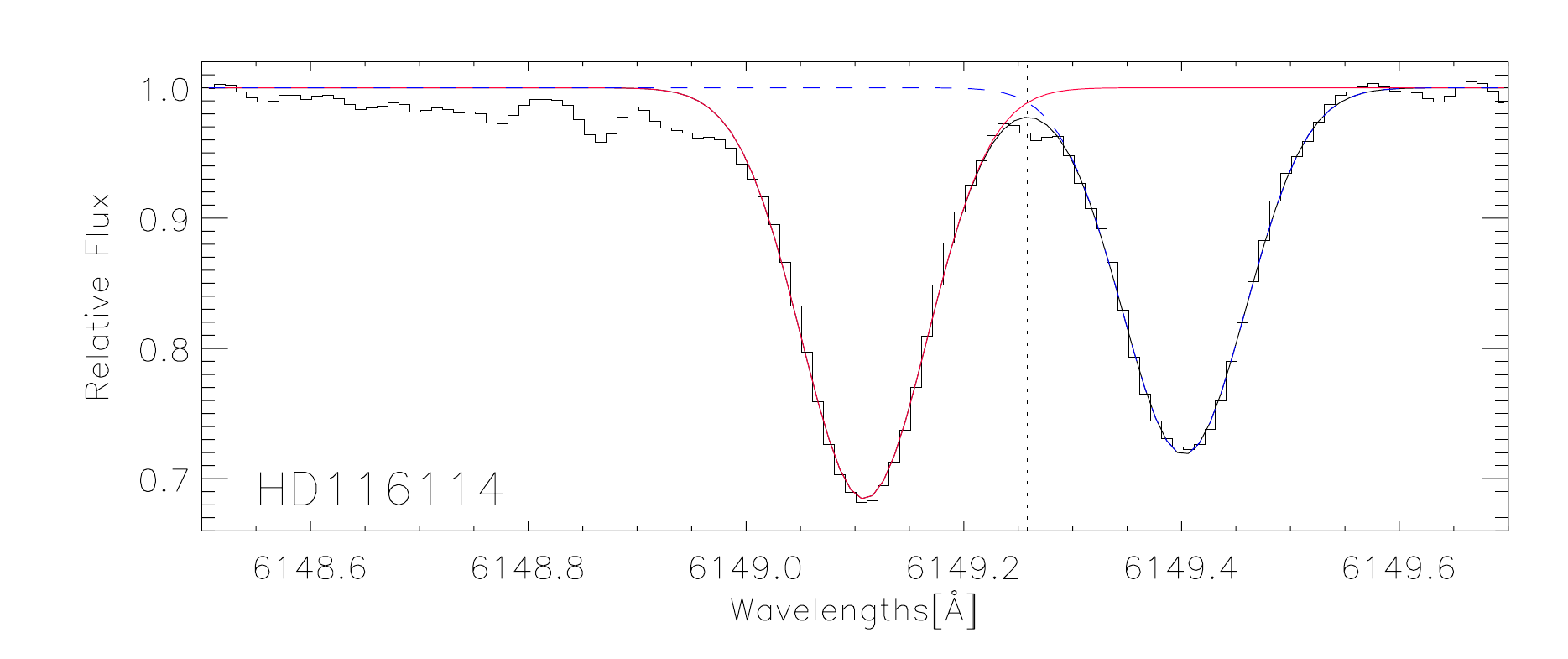}
\includegraphics[trim={1cm 0cm 0cm 0cm}, width=0.495\textwidth]{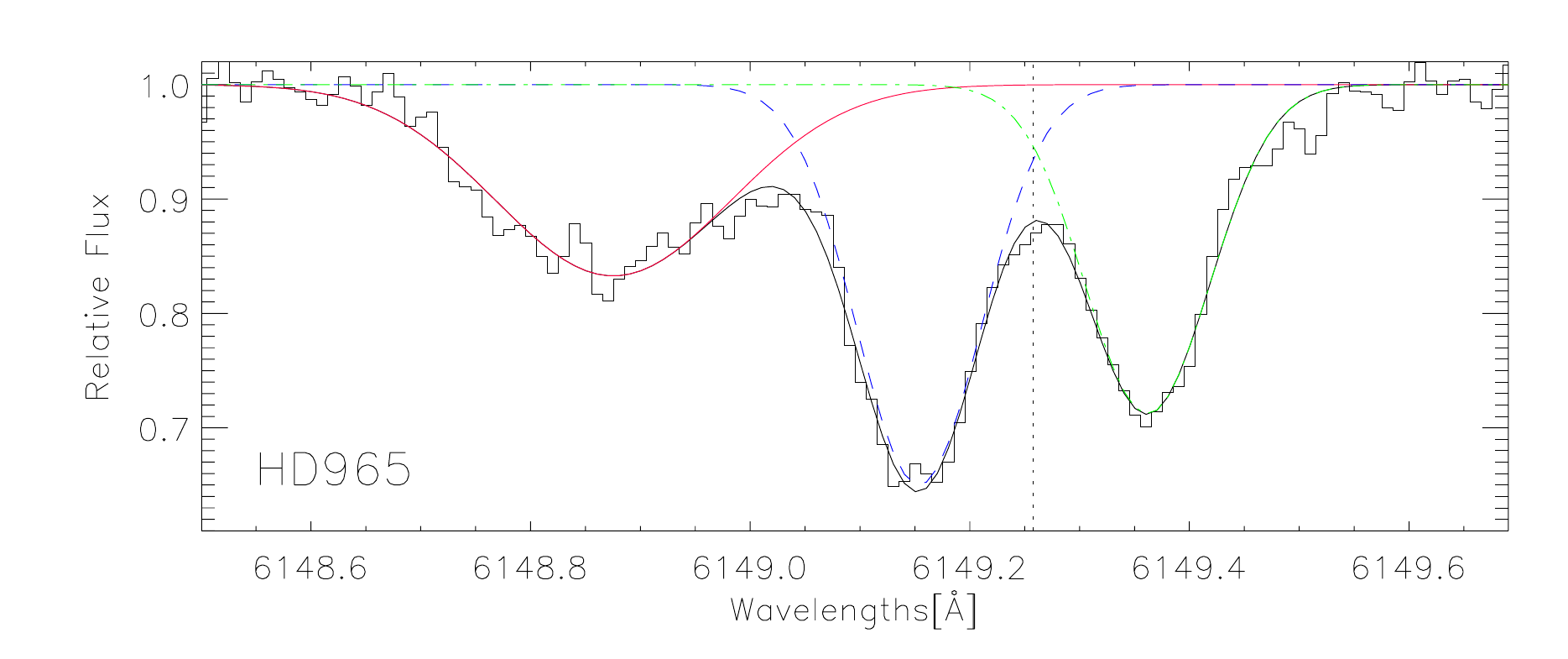}
\caption{The distance between the Zeeman subcomponents of the Fe{\sc ii}\,6149.258\,{\AA}
line is determined via (a) a double Gaussian fit or (b) a triple Gaussian fit including
the yet unidentified spectral line at $\sim $6148.84\,{\AA}.}
\label{Fig_gaussians}
\end{figure}

\begin{figure}\center
\includegraphics[trim={0.0cm 0cm 0cm 0cm},width=0.5\textwidth]{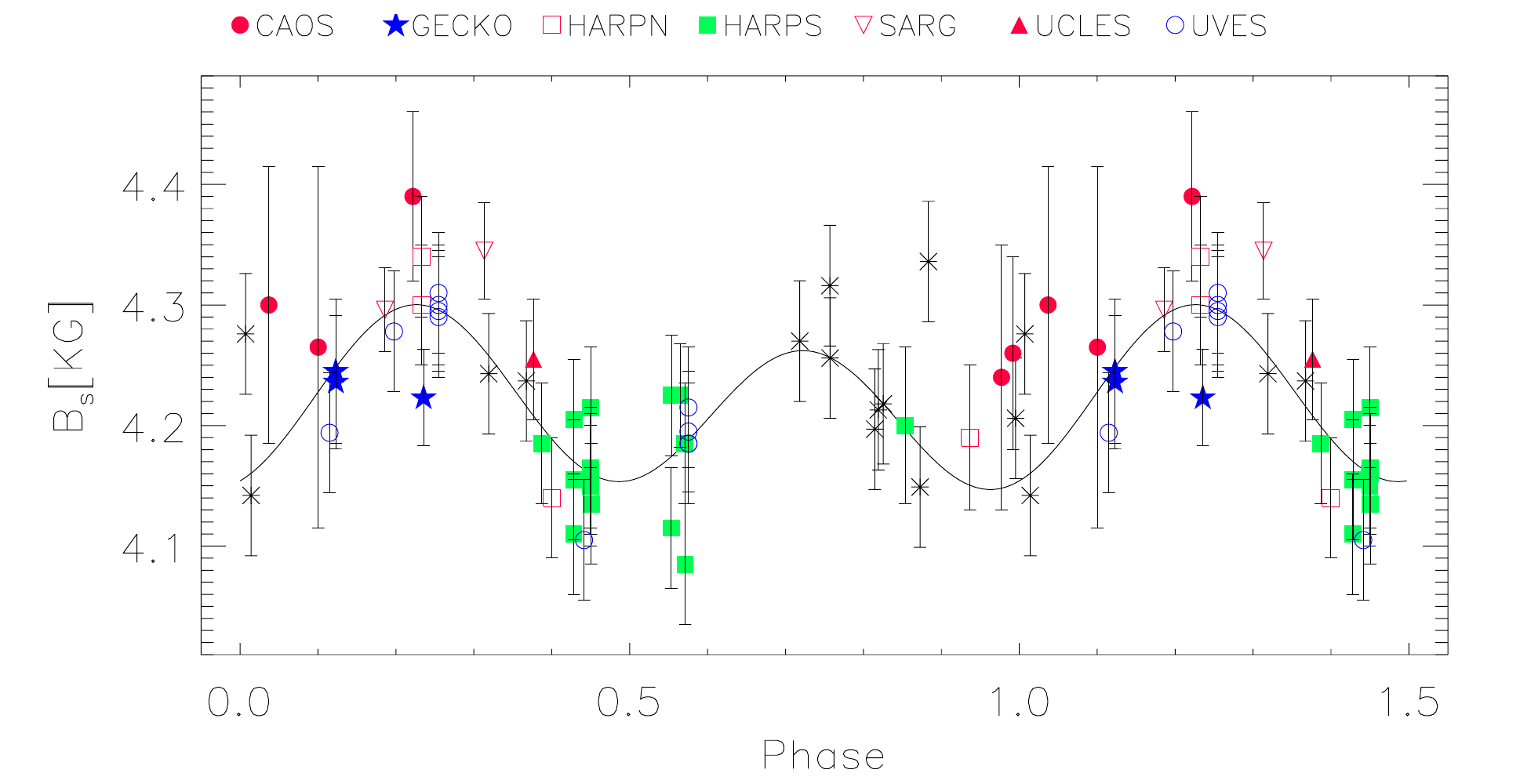}
\includegraphics[trim={0.0cm 0cm 0cm 0cm},width=0.5\textwidth]{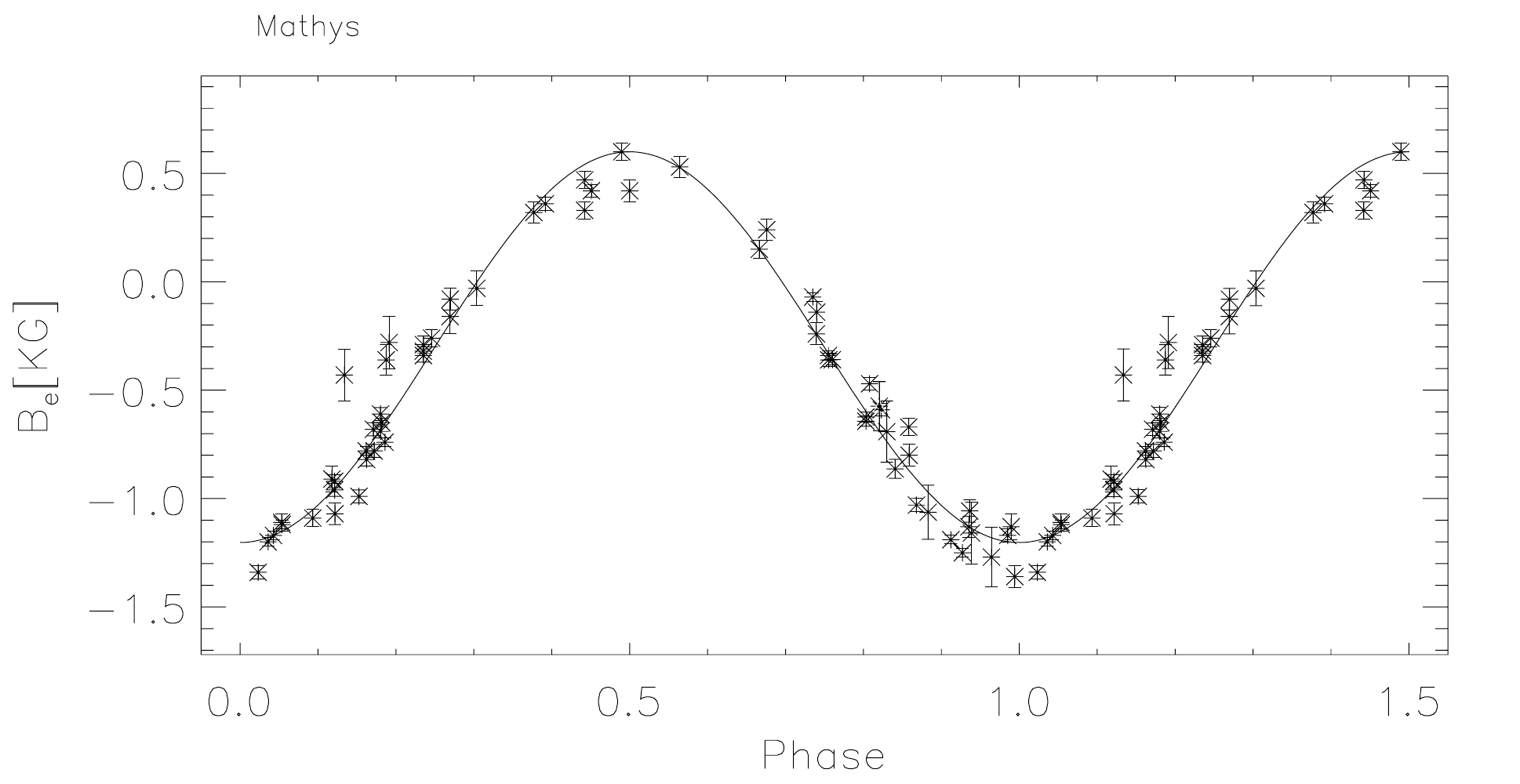}
\caption{HD\,965. $B_s$ and $B_e$ variations, folded with the 6030\,d period.}
\label{Fig_HD965}
\end{figure}

\noindent The wavelengths of Zeeman subcomponents have been obtained with a double Gaussian
fit (top panel of Figure\,\ref{Fig_gaussians}). The error in a surface magnetic
field measurement results from the propagation of errors in the determination of
subcomponent positions. 
In the case of stars, e.g. HD\,9996, with a blend of the Fe{\sc ii}\,6149.258\,{\AA} and
not yet identified, $\sim$6148.84\,{\AA} spectral lines, the previous 2 Gaussians have been
assumed identical in width, and  a 3rd Gaussian component has been added
(bottom panel of Figure\,\ref{Fig_gaussians}).

In this paper, the measurements of the surface magnetic field by \cite{Mathys1997}
(=M97) and \cite{Mathys2017} (=M17) are fundamental. The huge observational effort
of these authors represents a significant enlargement of the time base for many stars.
Whenever it was possible, we retrieved the original spectroscopic data of these
authors and derived the surface magnetic field. On average, our measurements are
50\,G smaller than Mathys' values. For this reason, when original spectra of Mathys
and coworkers were not available, we have combined $B_s$ values published by M97
and M17 with our values after the application of this shift.

Variability periods are determined here from the Lomb-Scargle \citep{Press1989}
periodogram of the surface magnetic field measurements: $LS(B_s)$. If necessary, the
periodograms of other observables ($LS(O^i))$ have also been computed and the final
period determined as the position of the highest peak in the product:
$LS(B_s, O^1, O^2,....)=\frac{LS(B_s)}{LS^{max}(B_s)} \times\Pi_i\frac{LS(O^i)}{LS^{max}(O^i)}$.
We assume that spurious peaks in the respective periodograms -- due to data sampling
and noise -- of these observables do not coincide and that they cancel or at least are 
greatly reduced in the product. Normalisation of any periodogram to its maximum value
is equivalent to assuming equal weight for all datasets.

Estimating the uncertainty in the determination of periods has always proved a
controversial subject; the literature abounds with proposed methods that we might choose
for this purpose. \cite{Pyper2017}(=P17) -- many of their stars are in common with our
sample -- wrote: {\it ``we used the practical method of comparing the two good data sets
most widely separated in time and determined how much the period had to be changed to see
a definite shift in phase''}. Whereas \cite{Mathys2019_HD965} state that {\it ``by
plotting a phase diagram of the measurements for a series of tentative values of the
period around the one suggested by the periodogram, one can visually identify the
period value that minimizes the phase shifts between field determinations from different
rotation cycles, and constrain the range around that value for which those phase shifts
remain reasonably small''}. We prefer to adopt a ``clinical'' decision based on the
$\sigma$ value of a Gaussian fit to the highest peak in the periodogram as a measure
of the uncertainty in the period determination. If our period coincides within errors
with the value quoted in the literature and if the latter has been determined to within
a smaller error, we have adopted the period taken from the literature.

\section{Individual stars}

Table\,\ref{Tab_Periods} lists the 36 MCP stars discussed in this paper, listing the
ephemerides taken from the literature and the ones adopted here. For every star, from
Table\,\ref{Tab_HD965} to \ref{Tab_HD216018}, measured surface field values are given
together with estimated errors, Heliocentric Julian Date (HJD) and the spectrographs
used -- as coded in Tables\,\ref{Tab_Spectrographs} and \ref{Tab_Archive}. Field
measurements by Mathys and coworkers are indicated with asterisks in the figures shown
hereafter. When data taken from the literature are used, sources and symbols are detailed
in the corresponding subsection and figures, respectively. If different photometric data
sets are available for a star, these are overplotted after the application of an ad hoc shift.

In a similar way to M17, $B_s$ measurements folded with the variability period are fitted
(by the CURVEFIT routine of Interactive Data Language Version 8)  with the trigonometric function
$$
B_s = B_0 + B_1 \sin\left(2\pi (\phi+\phi_1) \right)$$ or the function 
$$B_s = B_0 + B_1 \sin\left(2\pi (\phi+\phi_1) \right) + B_2 \sin \left(2\pi (2\phi+\phi_2) \right)$$
where $\phi = \frac{HJD-HJD_0}{P}$ is the phase,  the zero phase time $HJD_0$ and period $P$
are from Table\,\ref{Tab_Periods}. Measurements are weighted by the inverse of the square of uncertainty.
Fit coefficients are listed in Table\,\ref{Coef_fits}.

\subsection{HD\,965}

\cite{Mathys2019_HD965} noted that $B_s$ measurements of HD\,965, collected between 1993
and 2008, do not show significant variations. In contrast, these authors found $B_e$
measurements, collected between 1995 and 2017, to be variable with a period of
6030$\pm$200\,d. We have obtained high-resolution spectra of HD\,965 between 2001 and
2021, and have retrieved spectra from the CFHT and ESO archives. The list of our
$B_s$ measurements of HD\,965 is given in Table\,\ref{Tab_HD965}. There is no clear
evidence for a periodically variable surface magnetic field: the average value is
$<B_s> = 4240\pm$70\,G, where the r.m.s. is comparable to the error of single field
measurement. However, $B_s$ measurements folded with the period applicable to $B_e$
(Table\,\ref{Tab_Periods}) present a double wave variation suggesting minima in
coincidence with the extrema of the effective magnetic field, and maxima corresponding
to zero $B_e$ values (Fig.\,\ref{Fig_HD965}). The phase relations hint at a magnetic
field that is certainly not purely dipolar.

\begin{figure}\center
\includegraphics[trim={0.0cm 0cm 0cm 0cm},width=0.50\textwidth]{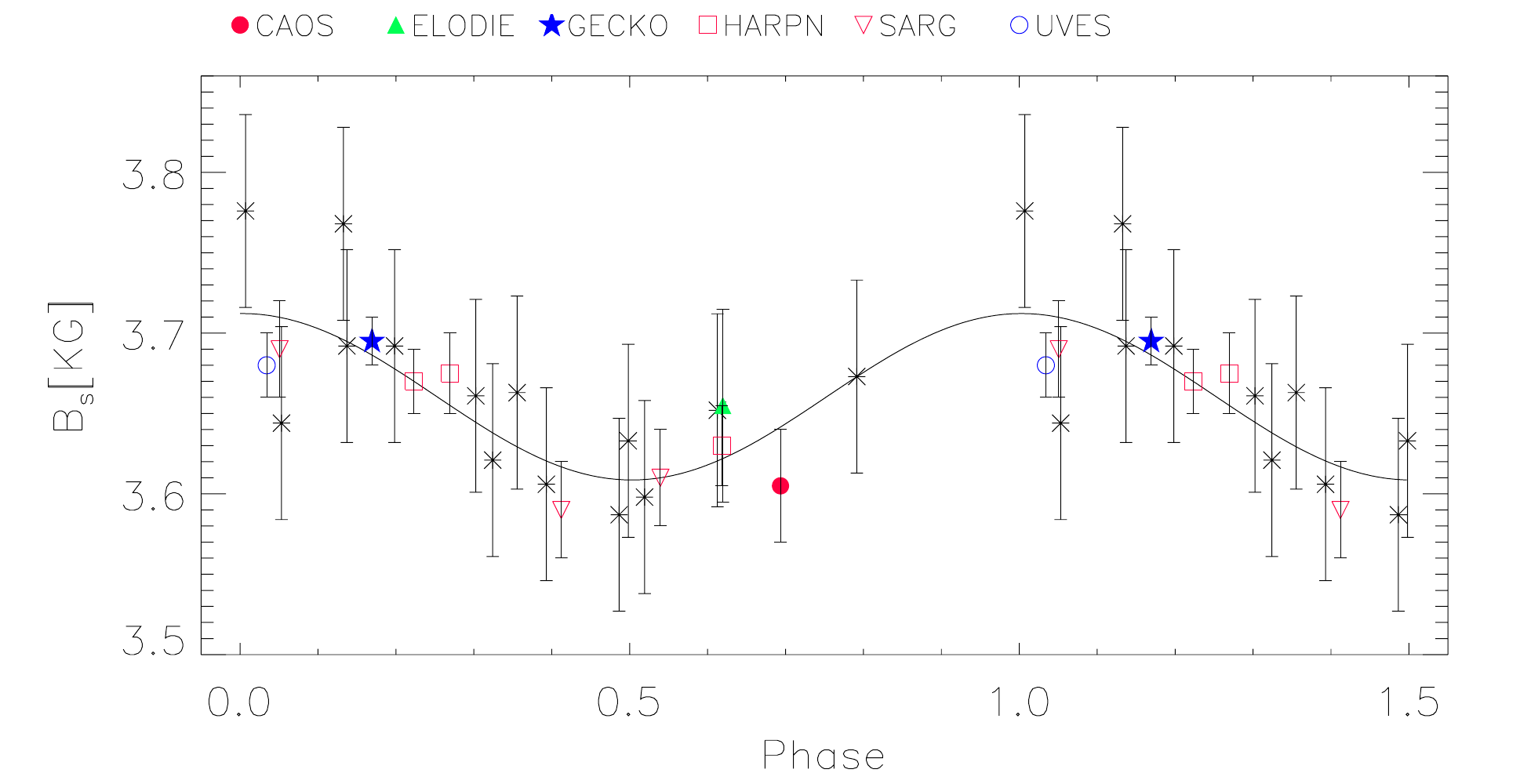}
\includegraphics[trim={0.0cm 0cm 0cm 0cm},width=0.50\textwidth]{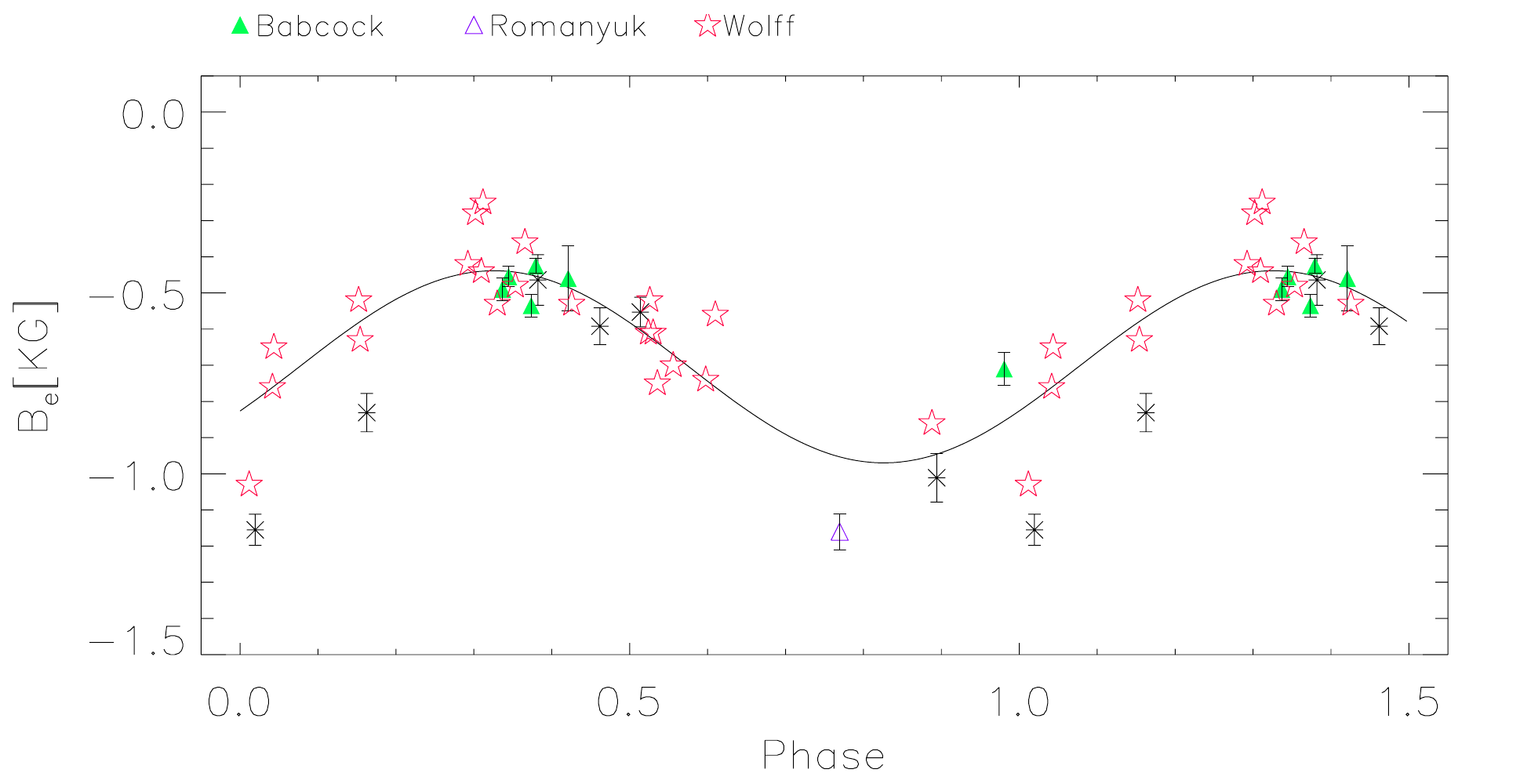}
\includegraphics[trim={0.0cm 0cm 0cm 0cm},width=0.50\textwidth]{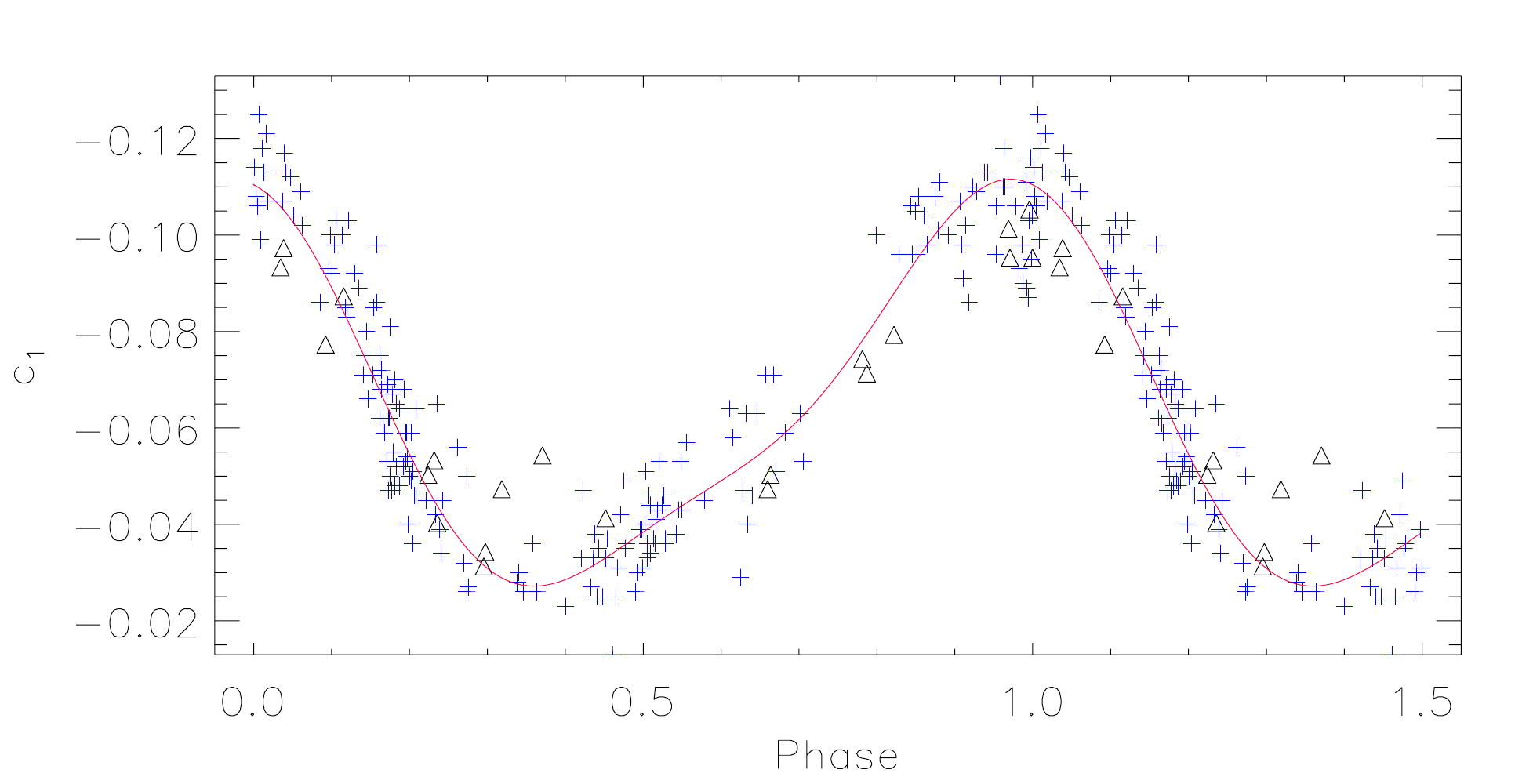} 
\caption{HD\,2453. $B_s$, $B_e$ and $c_1$ index variability, folded with the 518.2\,d
period. $c_1$ values are from \citet{Wolff1975} (+) and \citet{Pyper2017} ($\triangle$).}
\label{Fig_HD2453}
\end{figure}

\begin{table*}
\caption{Coefficients   of a weighted fit of $B_s$ (G) measurements with the function 
$B_0 + B_1 sin\left(2\pi(\frac{HJD-HJD_0}{P}+\phi_1)\right) + B_2 sin\left(2\pi(2\frac{HJD-HJD_0}{P}+\phi_2) \right)$.
Period ($P$) and zero phase Heliocentric Julian Date ($HJD_0$) are from Table\,\ref{Tab_Periods}.
$\chi^2 =\frac{1}{D_f} \sum_{i=1}^N{w_i(B_i - B_{fit})^2}$, where $D_f = N - M$ is the degrees of freedom (N is the number of measurements, and M is the number of coefficients) and the weight is equal to the inverse of the square of measurement uncertainty ($w_i = 1/\sigma_i^2$).
}
\begin{tabular}{rccrcrrc}
\multicolumn{1}{c}{HD} & \multicolumn{1}{c}{$B_0$} &  \multicolumn{1}{c}{$B_1$} &  \multicolumn{1}{c}{$\phi_1$} &\multicolumn{1}{c}{$B_2$} &  \multicolumn{1}{c}{$\phi_2$} & $ N\, $ & $\chi^2$ \\\hline
965 &     4229 $\pm$        7 &       31 $\pm$       11 &      0.161 $\pm$     0.049 &       51 $\pm$       10 &      0.792 $\pm$     0.031 &          50&     0.016\\
2453 &     3666$\pm$       7 &       32$\pm$       8 &      0.132 $\pm$    0.067 &  &  &          21&     0.056 \\
9996 &     2759 $\pm$        8 &     1680 $\pm$        9 &      0.244 $\pm$    0.001 &      537 $\pm$       11 &      0.199 $\pm$    0.002 &          24&      0.875\\
12288 &     8065 $\pm$       19 &      457 $\pm$       29 &      0.497 $\pm$    0.008 &      147 $\pm$       25 &      0.283 $\pm$     0.031 &          33&     0.045\\
14437 &     7379$\pm$      30 &      388$\pm$      29 &      0.856 $\pm$    0.014 &  &  &          40&      0.146 \\
18078 &     3395$\pm$       8 &      809$\pm$      12 &      0.253 $\pm$   0.001 &  &  &          20&       1.122 \\
47103 &    17211 $\pm$       20 &      397 $\pm$       26 &      0.215 $\pm$     0.012 &      241 $\pm$       30 &      0.167 $\pm$     0.018 &          11&      0.539\\
50169 &     5059 $\pm$        6 &      896 $\pm$       10 &      0.239 $\pm$    0.001 &      101 $\pm$        7 &      0.271 $\pm$     0.015 &          52&     0.022\\
51684 &     6013$\pm$       5 &      271$\pm$       8 &      0.242 $\pm$   0.004 &  &  &          13&      0.222 \\
61468 &     6720$\pm$       9 &     1122$\pm$      13 &      0.247 $\pm$   0.001 &  &  &          13&      0.559 \\
81009 &     8373$\pm$       9 &      997$\pm$      11 &      0.245 $\pm$   0.002 &  &  &          84&     0.072 \\
93507 &     7067$\pm$      13 &      317$\pm$      10 &      0.230 $\pm$    0.010 &  &  &          59&     0.052 \\
94660 &     6218$\pm$       7 &      165$\pm$       8 &      0.247 $\pm$   0.007 &  &  &          43&     0.028 \\
126515 &    12704$\pm$      11 &     3240$\pm$      12 &      0.240 $\pm$      0.000 &  &  &          80&      0.146 \\
166473 &     7084$\pm$       9 &     1420$\pm$      12 &      0.247 $\pm$   0.001 &  &  &          61&     0.036 \\
177765 &     3432$\pm$       6 &      107$\pm$      11 &      0.250 $\pm$    0.015 &  &  &          11&      0.175 \\
178892 &    18889$\pm$      96 &      911$\pm$      92 &      0.649 $\pm$    0.021 &  &  &          18&     0.046 \\
187474 &     5372 $\pm$        3 &      632 $\pm$        5 &       1.258 $\pm$    0.001 &      231 $\pm$        5 &      0.227 $\pm$    0.003 &          45&      0.111\\
188041 &     3596$\pm$       3 &       37$\pm$       5 &      0.155 $\pm$    0.022 &  &  &          47&     0.030 \\
192678 &     4624 $\pm$        6 &      103 $\pm$        9 &      0.236 $\pm$     0.014 &       39 $\pm$        8 &      0.122 $\pm$     0.035 &          42&     0.060\\
335238 &     9357 $\pm$       27 &     2064 $\pm$       22 &      0.232 $\pm$    0.003 &      977 $\pm$       37 &      0.242 $\pm$    0.004 &          36&      0.140\\
201601 &     3444$\pm$       6 &      492$\pm$       8 &      0.258 $\pm$   0.001 &  &  &         141&     0.055 \\
208217 &     7668 $\pm$       48 &      587 $\pm$       69 &      0.294 $\pm$     0.018 &      417 $\pm$       56 &      0.188 $\pm$     0.020 &          44&     0.053\\
216018 &     5570$\pm$       6 &       62$\pm$      10 &      0.233 $\pm$    0.022 &  &  &          38&     0.024 \\
\hline
\end{tabular}\label{Coef_fits}
\end{table*}

\subsection{HD\,2453}\label{Sec_HD2453}

Two periods have almost simultaneously been published for the variability of HD\,2453:
1) 521$\pm$\,2\,d by M17 from $B_e$ measurements and 2) 518.2$\pm$\,0.5\,d by P17 from
Str{\"o}mgren photometry. Our $B_s$ measurements of HD\,2453 (Table\,\ref{Tab_HD2453})
extend the 2700 day coverage by M17 to 11000 days. However, with an average value of
$B_s = 3690\pm$60\,G and a scatter comparable to the error of a single field measurement,
it is not possible to ascertain a clear variability of the surface magnetic field. The
single $B_e$ = $-$1160$\pm$50\,G measurement obtained by \cite{Romanyuk2016} on
JD = 2\,455\,075.417 makes the 518.2\,d period the most probable. The Lomb-Scargle
analysis of $B_e$ (from M17, \cite{Wolff1975}, \cite{Romanyuk2016}) and $c_1$ (from
\cite{Wolff1975}, \cite{Pyper2017}) produces a $LS(B_s, B_e, c_1)$ peaking at
517.7$\pm$1.8\,d. We have adopted the P17 period to fold the $B_s$, $B_e$ and $c_1$ data
(Fig.\,\ref{Fig_HD2453}). With this period, $B_s$ presents a single wave variation that
is in phase with the light curve and with a maximum coincident with the negative $B_e$
extremum. This is what one can expect from a dominant dipole component of the magnetic
field.

\subsection{HD\,9996}

From $B_e$ measurements, \cite{Metlova2014} and \cite{Bychkov2019} concluded at a
variability period of HD\,9996 of 7936.522\,d ($\sim$21.7 yr). \cite{Pyper2017}
determined a photometric period of 7850\,d. M17 adopted the 7936.522 day period,
indicating an incomplete phase coverage. This author always observed well split
Fe{\sc ii}\,6149.258\,{\AA} Zeeman components, consistent with $B_s > 4000$\,G,
an exception (unsplit line) occurring on JD = 2\,450\,797.312.

\begin{figure}
\includegraphics[width=0.50\textwidth]{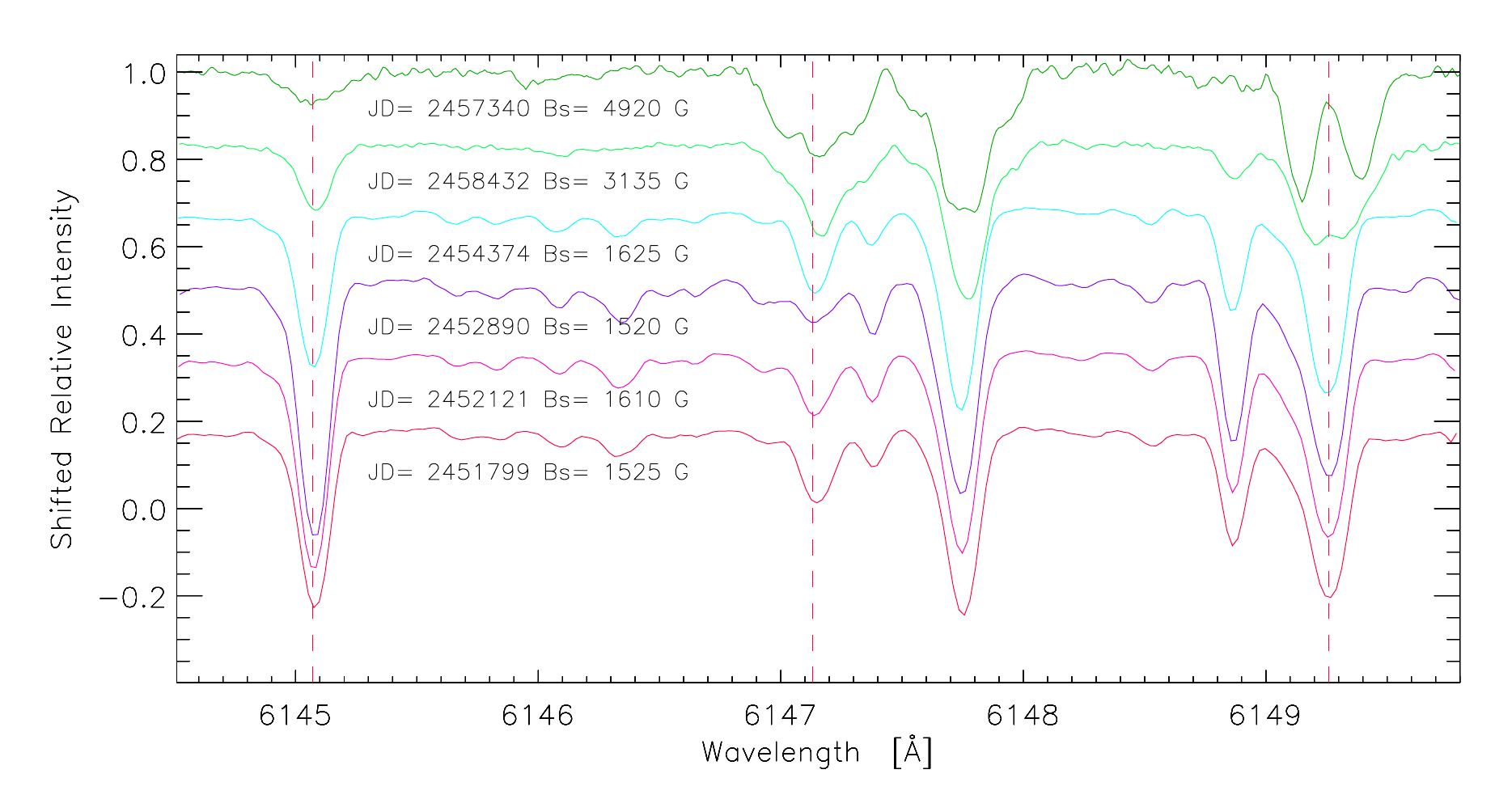}
 \includegraphics[trim={0.3cm 0cm 0cm 0cm},width=0.50\textwidth]{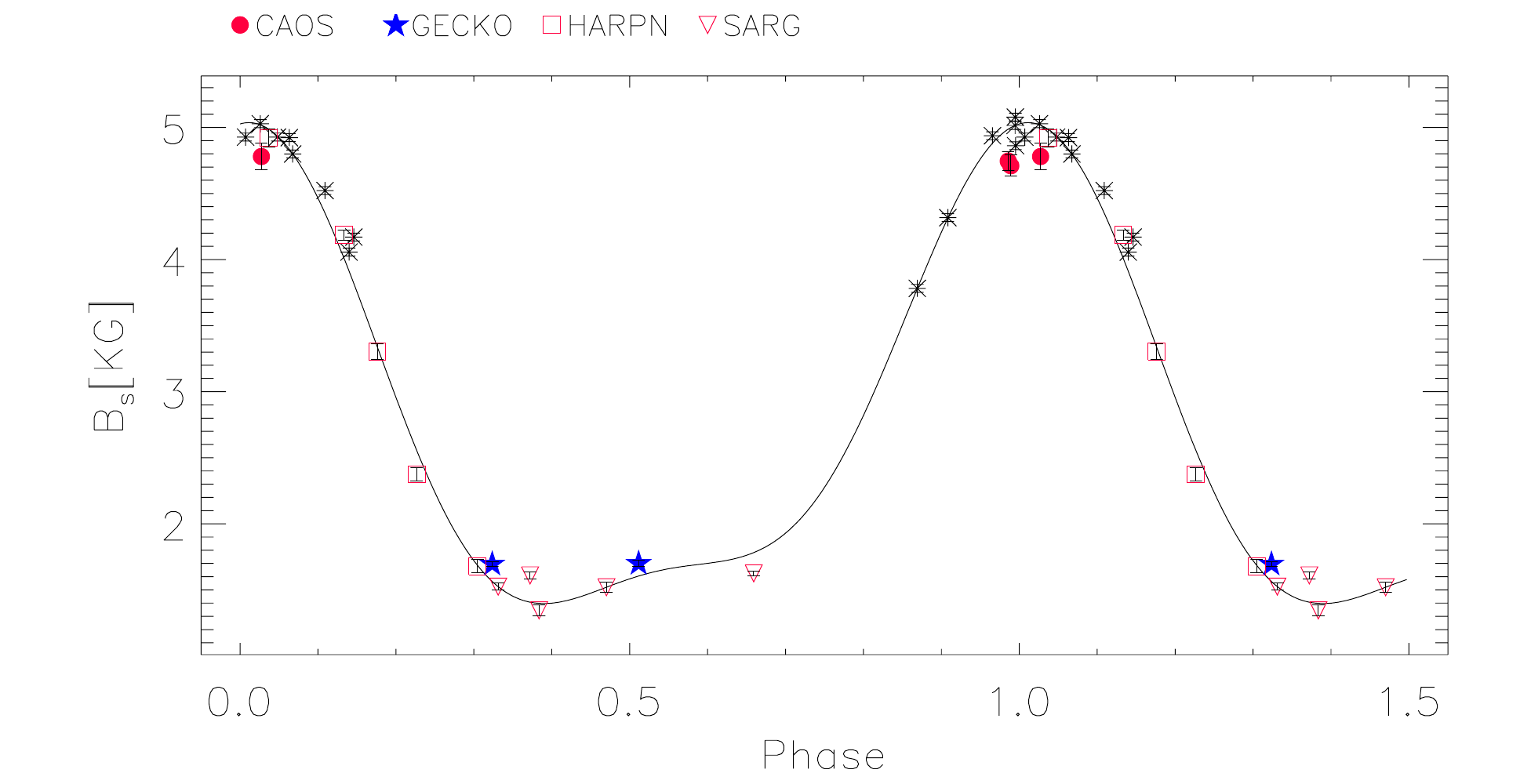}
 \includegraphics[trim={0.3cm 0cm 0cm 0cm},width=0.50\textwidth]{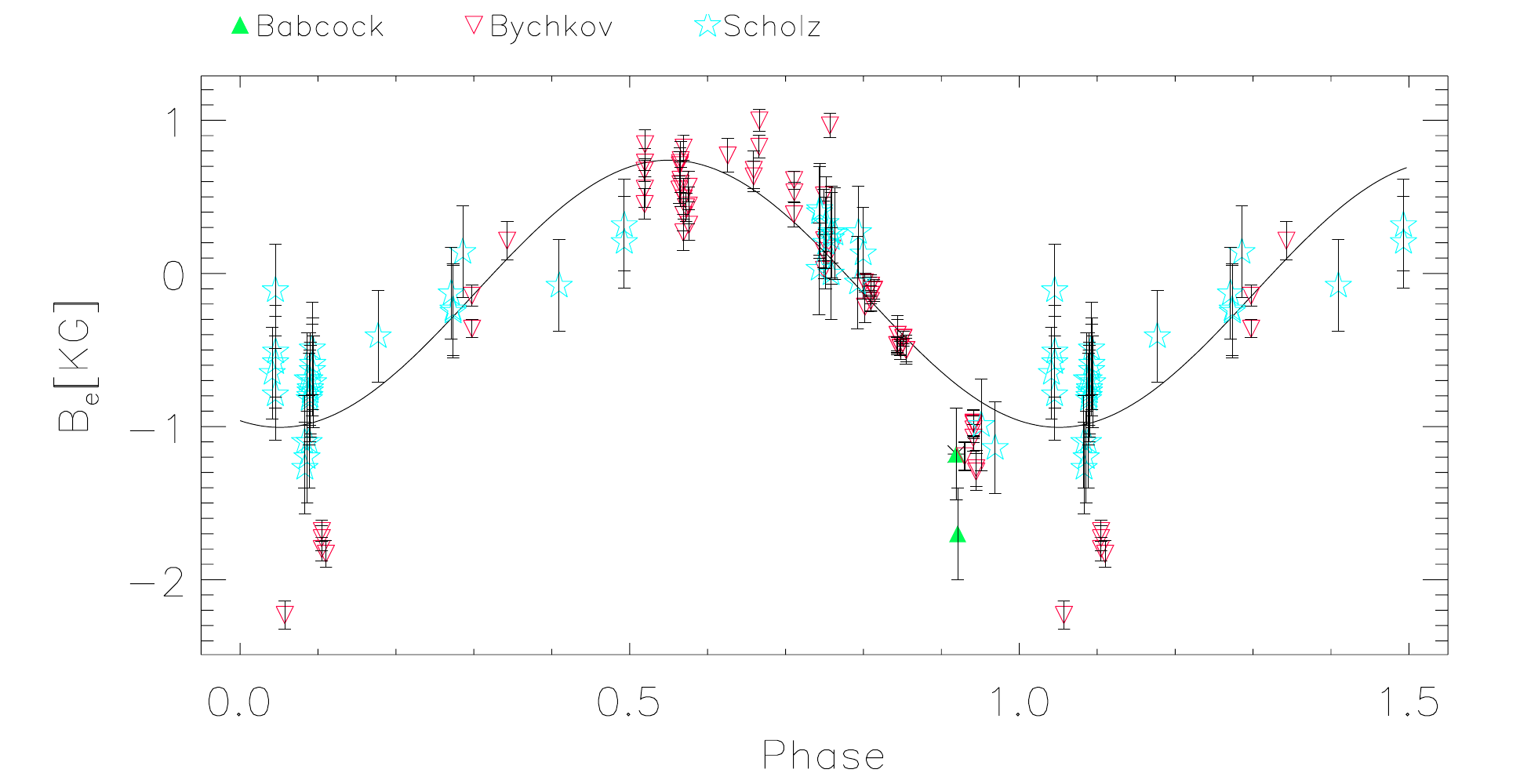}
\caption{HD\,9996. Top, chunks of spectra with the Fe{\sc ii}\,6149.258\,{\AA},
 Cr{\sc ii}\,6147.154\,{\AA} and Nd{\sc iii} 6145.070\,{\AA} lines marked.
 The normalised spectra are shifted vertically and ordered with
 decreasing values of $B_s$. Cr ($\lambda$\,6147) and  Nd ($\lambda$\,6145)
 abundances change out of phase. Central and lower  panels show the $B_s$
 and $B_e$ variations folded with the 7850\,d period. Errorbars (tens of G) for $B_s$ measurements are smaller than symbols and are barely visible.}
\label{Fig_HD9996}
\end{figure}

We have observed HD\,9996 between 2001 and 2021, in addition we have retrieved
2 GECKO spectra from the CFHT archive, one of those obtained in 2000. Our $B_s$
measurements of HD\,9996 are listed in Table\,\ref{Tab_HD9996}. We find that the
variability period given by P17 is representative of the $B_s$ and $B_e$ variations
of HD\,9996 (Figure\,\ref{Fig_HD9996}) with a well-defined maximum and a rather flat
minimum. This is not rare for this class of stars (see HD\,318107 or HD\,335238 later
in the text), although in the case of HD\,9996 it could be that the observed plateau is a consequence of the
difficulty in measuring such weak $B_s$ fields because of the merging of the Zeeman
subcomponents of the Fe{\sc ii}\,6149.258\,{\AA} line (see Figure\,\ref{Fig_HD9996}).
Near-infrared lines -- more sensitive because of the $\lambda^2$ dependence of
Zeeman-splitting -- should preferably be used to determine the real minimum in the
$B_s$ variations of HD\,9996. \cite{Leone2003} have shown an application of this
technique to magnetic stars by selecting suitable spectral lines with large effective
Land{\'e} factors \citep{Solanki1994}. 

Figure\,\ref{Fig_HD9996} also shows some of these spectra ordered with time; it confirms
the finding by \cite{Preston1970_HD9996} that the respective equivalent widths of chromium
and rare-earth spectral lines change out of phase. The $B_s$ maximum coincides with the
negative extremum of $B_e$, while from phases 0.3 through 0.7 $B_e$ presents a maximum,
$B_s$ staying constant at its minimum value. If the flat minimum is real, HD\,9996 presents
a magnetic field that is not purely dipolar. With a value $q = B_s^{max}/B_s^{min} \ge 3$,
HD\,9996 appears a rather extreme MCP star, since there is no other known case with
$q$ larger than 2 (see figure\,2 in M17).

\begin{figure}
\includegraphics[trim={0.3cm 0cm 0cm 0cm},width=0.50\textwidth]{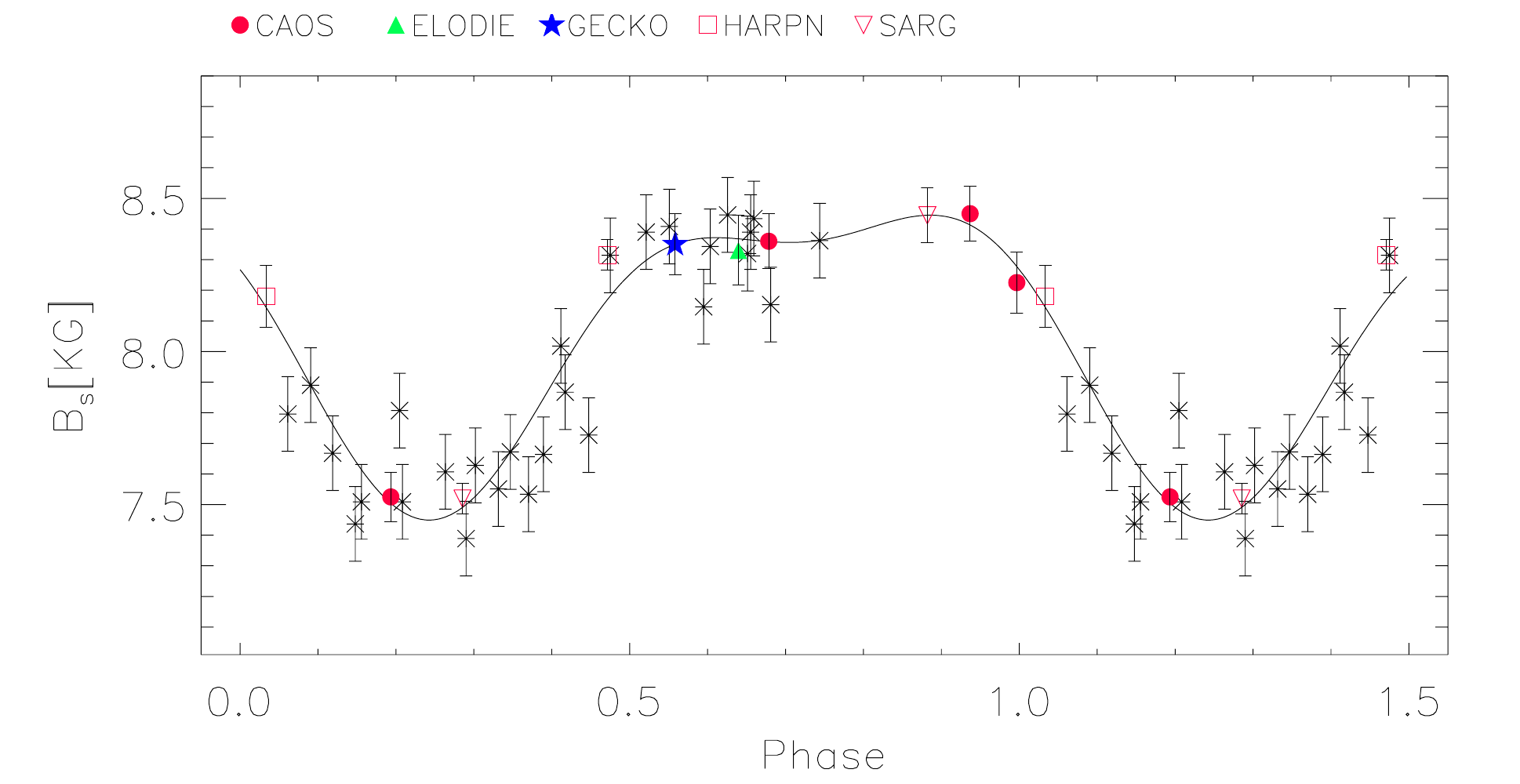}
\includegraphics[trim={0.3cm 0cm 0cm 0cm},width=0.50\textwidth]{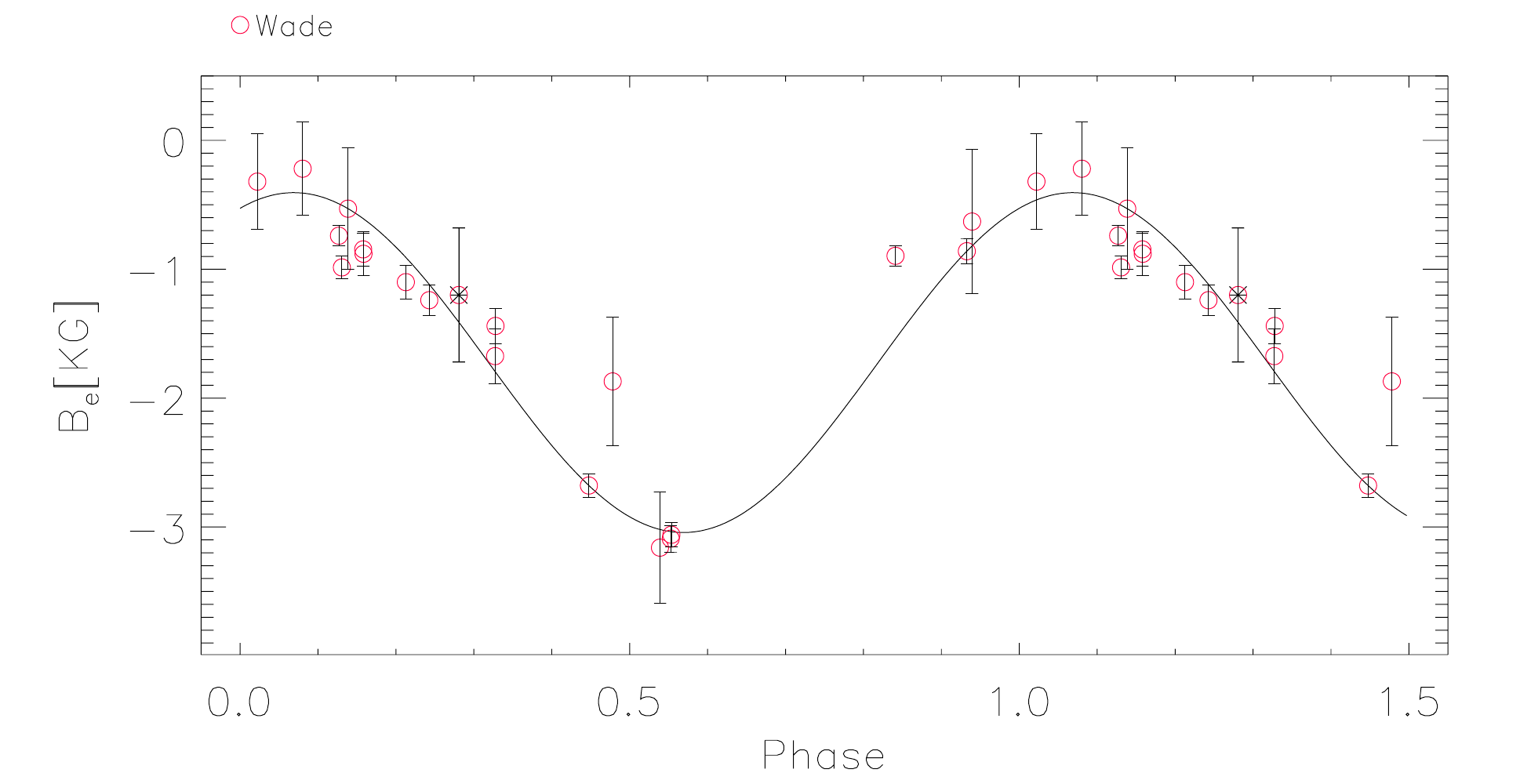}
\includegraphics[trim={0.3cm 0cm 0cm 0cm},width=0.50\textwidth]{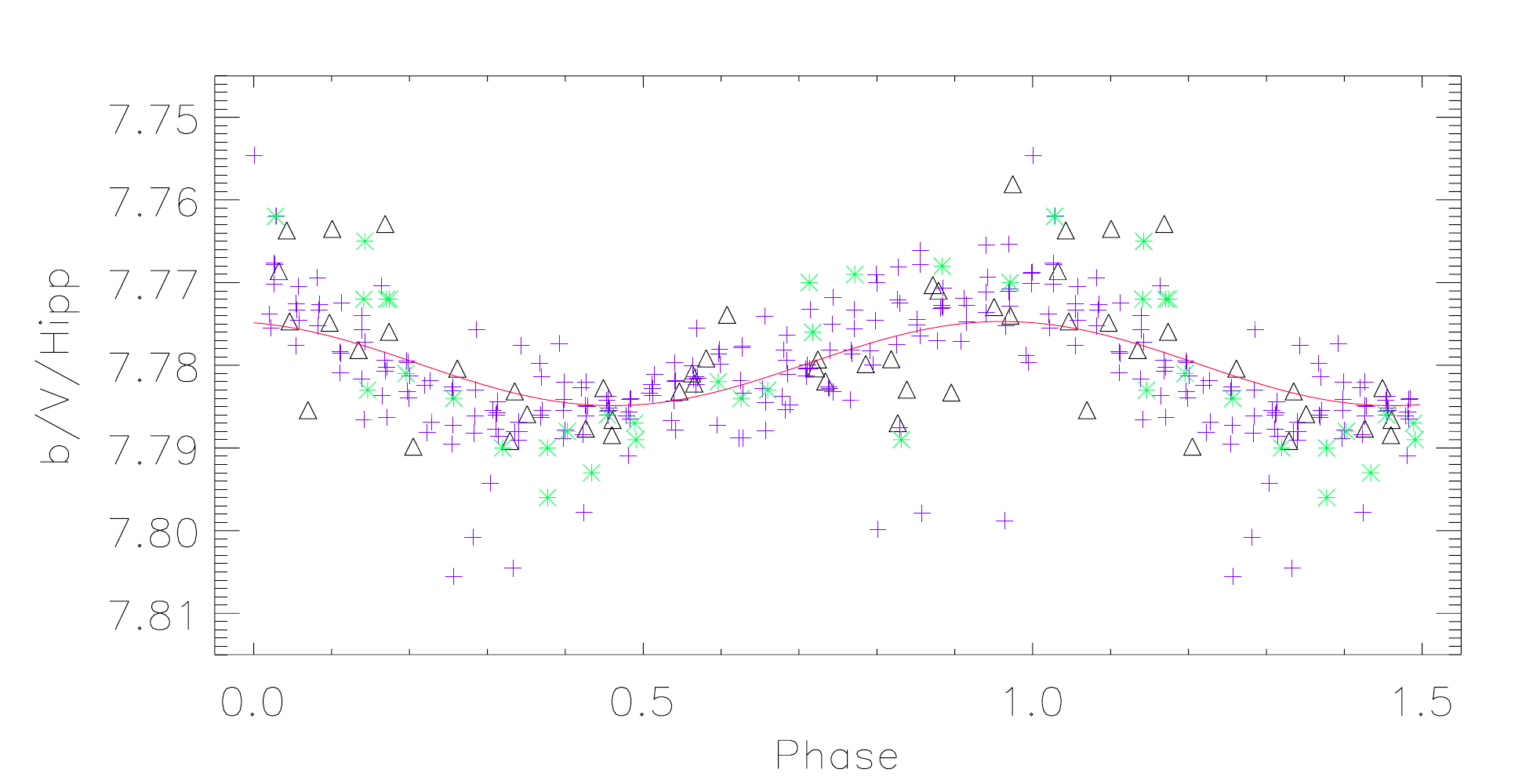}
\caption{HD\,12288. Variability in $B_s$, $B_e$, and photometry -- \citet{Wolff1973}
(\color{green}{$*$}\color{black}), HIPPARCOS ($\triangle$) and MASCARA ($+$) -- folded with the 34.993\,d period.}
\label{Fig_HD12288}
\end{figure}

\subsection{HD\,12288}\label{SecHD12288}
\cite{Bernhard2020} established HD\,12288 as a photometric variable with the ephemeris
JD(V$^{\rm max}$) = 2\,457\,218.6 + 35.73$\pm$ 0.03\,E\,d, presenting a peak-to-peak
difference of 0.02\,mag. According to P17, this star is variable in the Str{\"o}mgren
$u$, $v$ and $b$ filters with a 34.99\,d period, but it presents a constant Str{\"o}mgren
$y$ magnitude. A period of 34.9\,d was adopted by M17 to discuss the magnetic variability
of HD\,12288.

We have observed HD\,12288 eight times and obtained 1 spectrum from the ELODIE archive
and 1 from the CFHT archive. In accord with P17, we found the TESS (600-1000 nm)
magnitudes obtained between 2\,458\,790 and 2\,458\,841 to be constant (7.63688$\pm$0.00006).
A Lomb-Scargle analysis of our measurements (Table\,\ref{Tab_HD12288}), together with
the M97 and M17 $B_s$ data, the \cite{Wade2000} $B_e$ values, and photometry by
\cite{Wolff1973}, MASCARA \citep{Talens2017, Bernhard2020}, and HIPPARCOS
\citep{vanLeeuwen2007} leads to a period of 34.993$\pm$0.003\,d, corresponding to the
highest peak of $LS(B_s, B_e, Mag.)$. The data folded with this period are shown in
Fig.\,\ref{Fig_HD12288}. It would seem that HD\,12288 is presenting some quite
singular behaviour among the MCP stars: $B_s$ remains almost constant (at the maximum
value of 8.5\,kG) during half of the rotation period, at the same time $B_e$ goes from
$-3$\,kG to near zero. The decrease of $B_s$ to about 7.5\,kG is accompanied by a
change in $B_e$ down again towards $-3$\,kG. This star becomes brighter as $B_e$ weakens.

\subsection{HD\,14437}\label{SecHD14437}
\begin{figure}
\includegraphics[trim={0.3cm 0cm 0cm 0cm},width=0.50\textwidth]{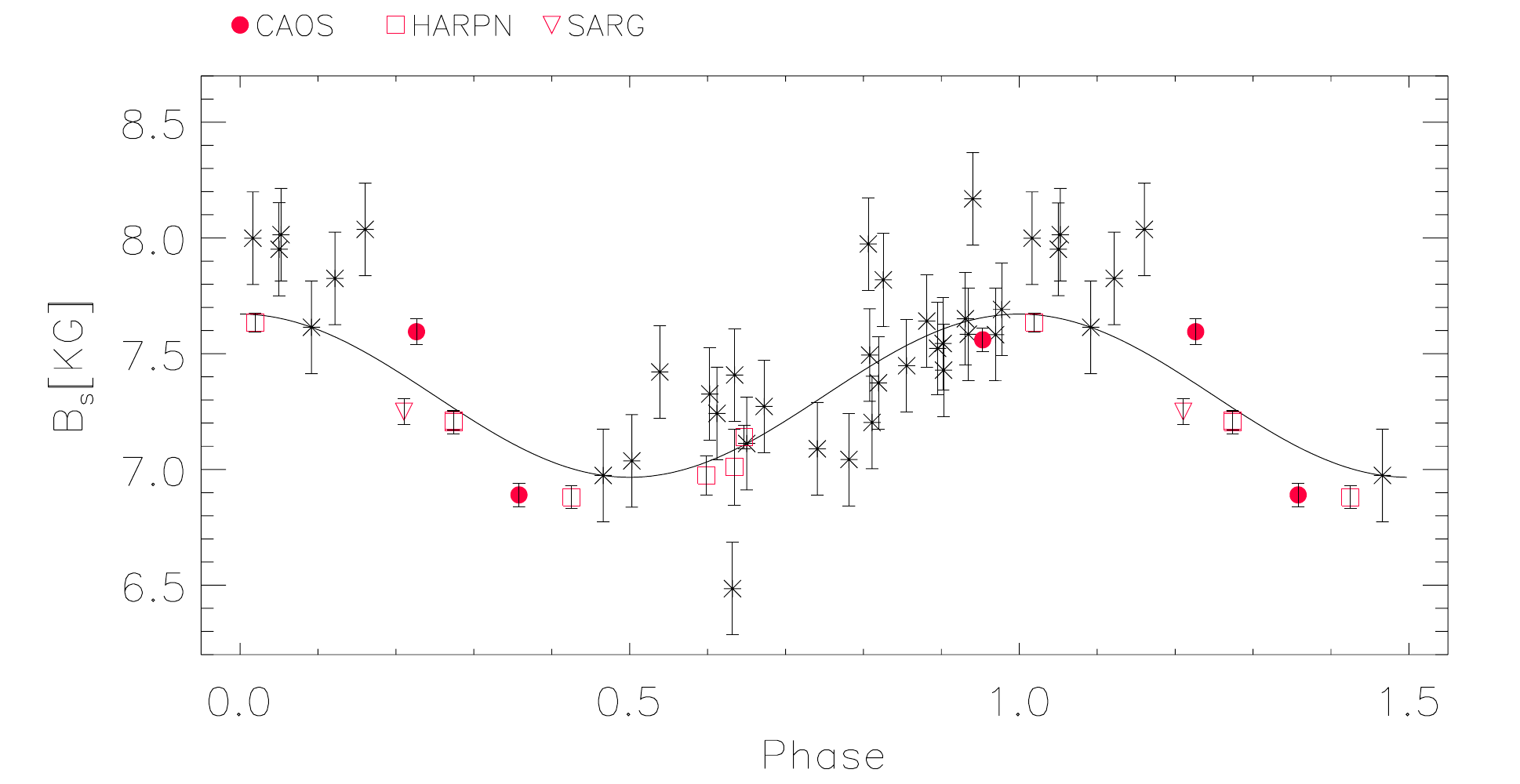}
 \includegraphics[trim={0.3cm 0cm 0cm 0cm},width=0.50\textwidth]{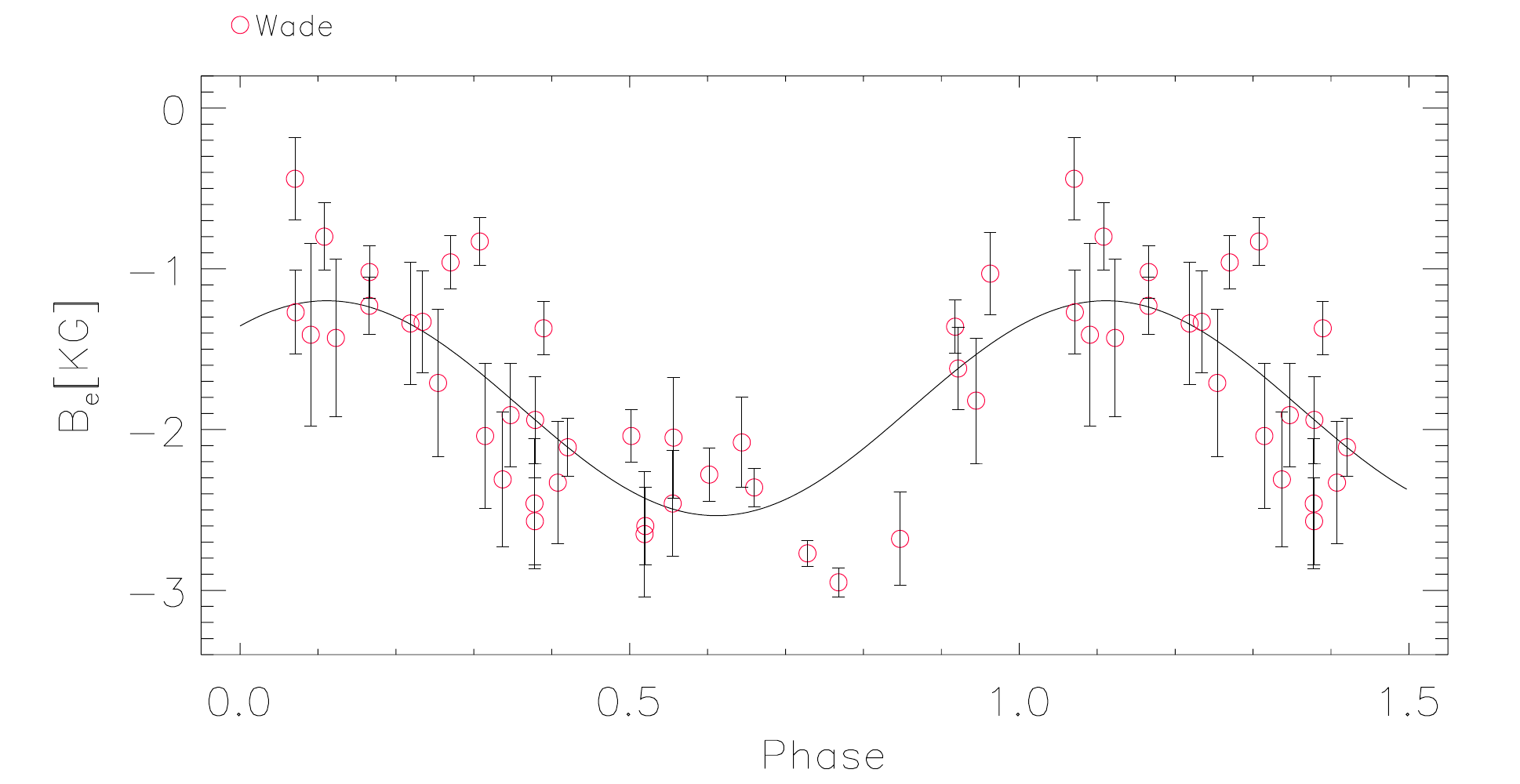}
 \includegraphics[trim={0.3cm 0cm 0cm 0cm},width=0.50\textwidth]{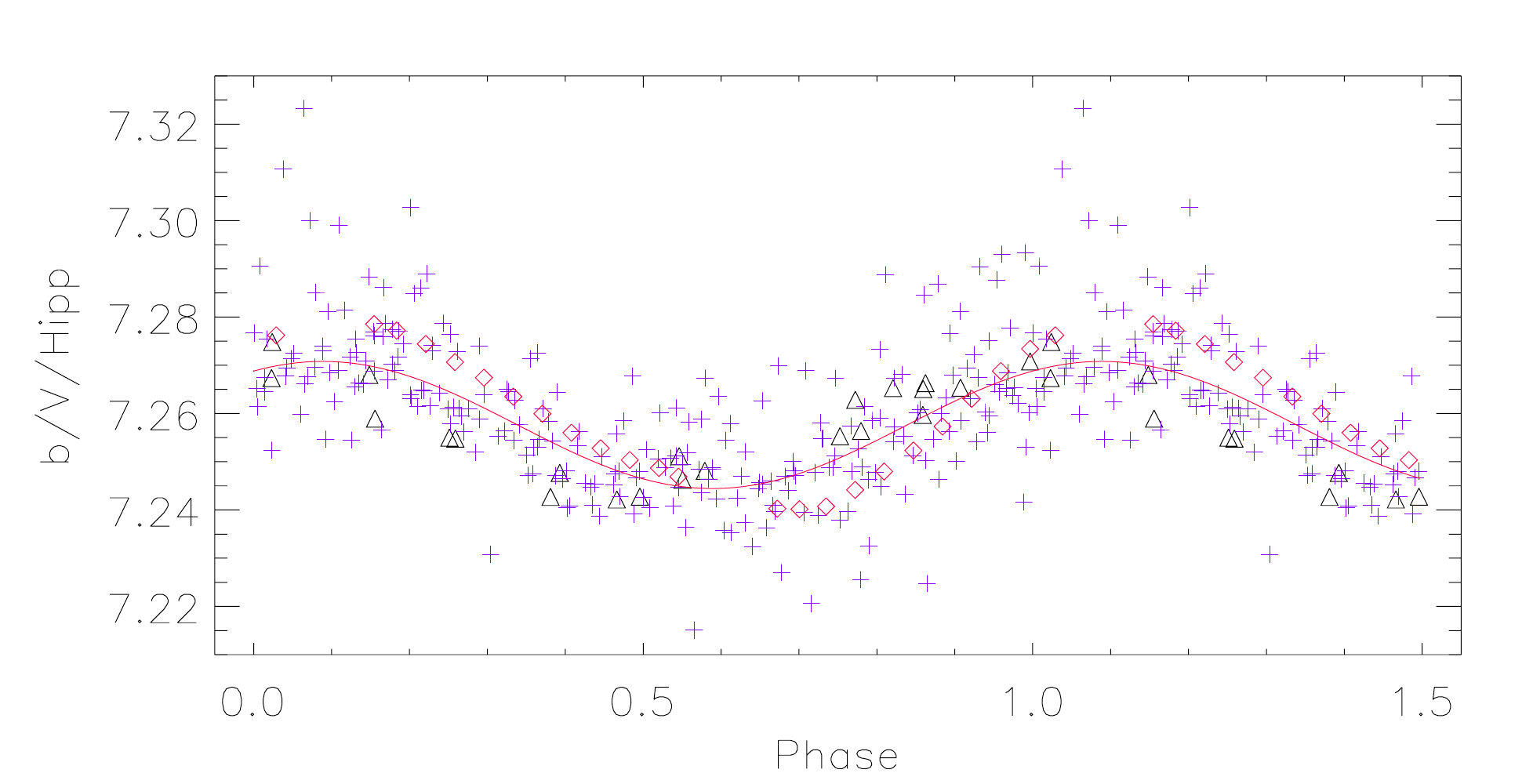}
\caption{HD\,14437. Variability in $B_s$, $B_e$, and photometry -- HIPPARCOS
($\triangle$), TESS  ($\diamond$), MASCARA ($+$) -- folded with the
26.734\,d period.}
\label{Fig_HD14437}
\end{figure}

The variability period of HD\,14437 was determined to 26.87$\pm$0.02\,d by
\cite{Wade2000} from $B_e$ measurements. From MASCARA data, \cite{Bernhard2020} found
a photometric period of 26.78$\pm$0.01\,d. Measurements of $B_s$ by M97 and M17 cover
the years from 1991 to 1997, our 11 spectra were obtained between 2001 and 2021. We
have performed a Lomb-Scargle analysis of our data (Table\,\ref{Tab_HD14437}) together
with the M97 and M17 $B_s$ measurements, the \cite{Wade2000} $B_e$ values, and the
photometry by HIPPARCOS, TESS, and MASCARA. The period that best reproduces all these
variations is 26.734$\pm$0.007\,d (Fig.\,\ref{Fig_HD14437}).

\subsection{HD\,18078}

\begin{figure}
\includegraphics[trim={0.3cm 0cm 0cm 0cm},width=0.50\textwidth]{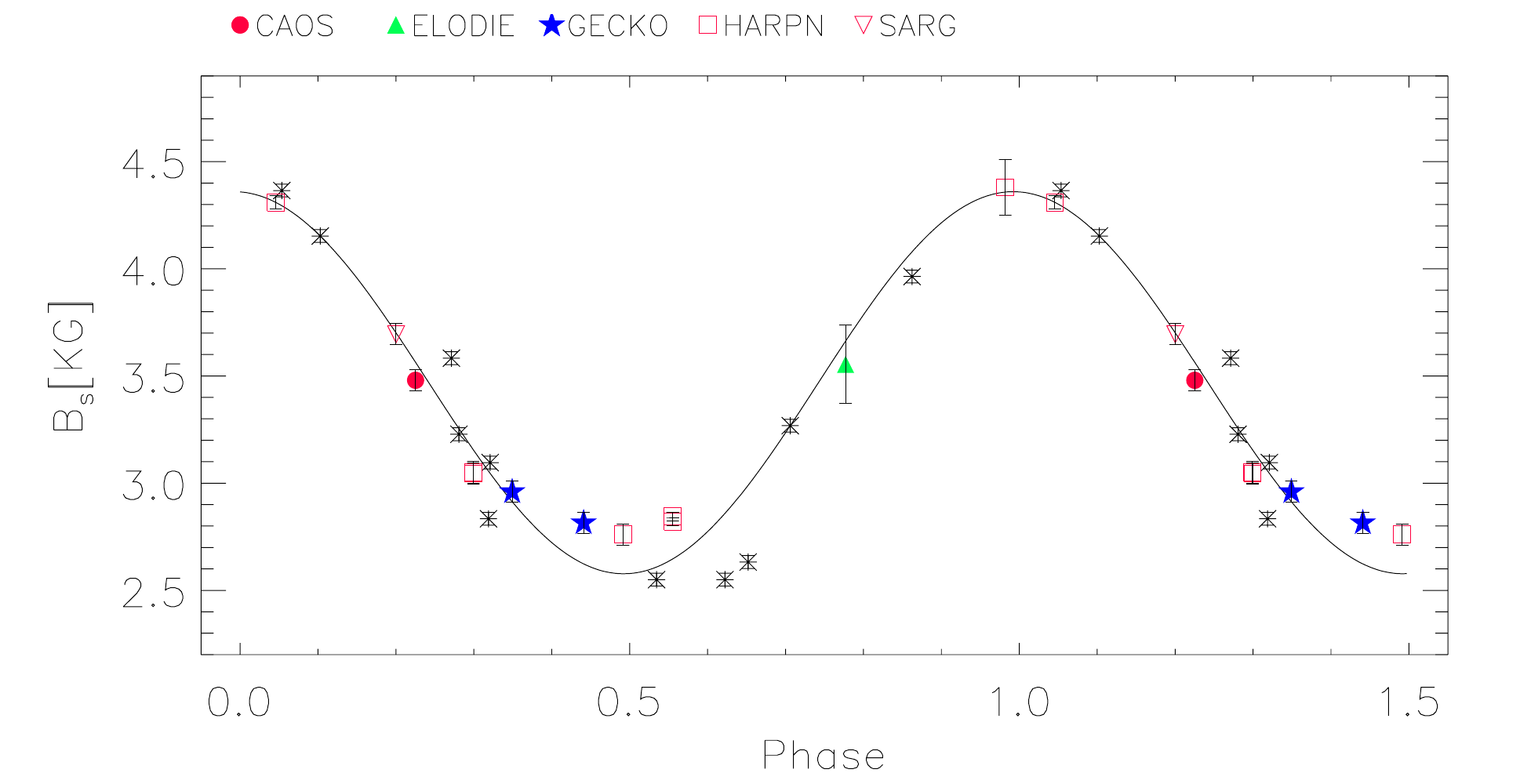}
\includegraphics[trim={0.3cm 0cm 0cm 0cm},width=0.50\textwidth]{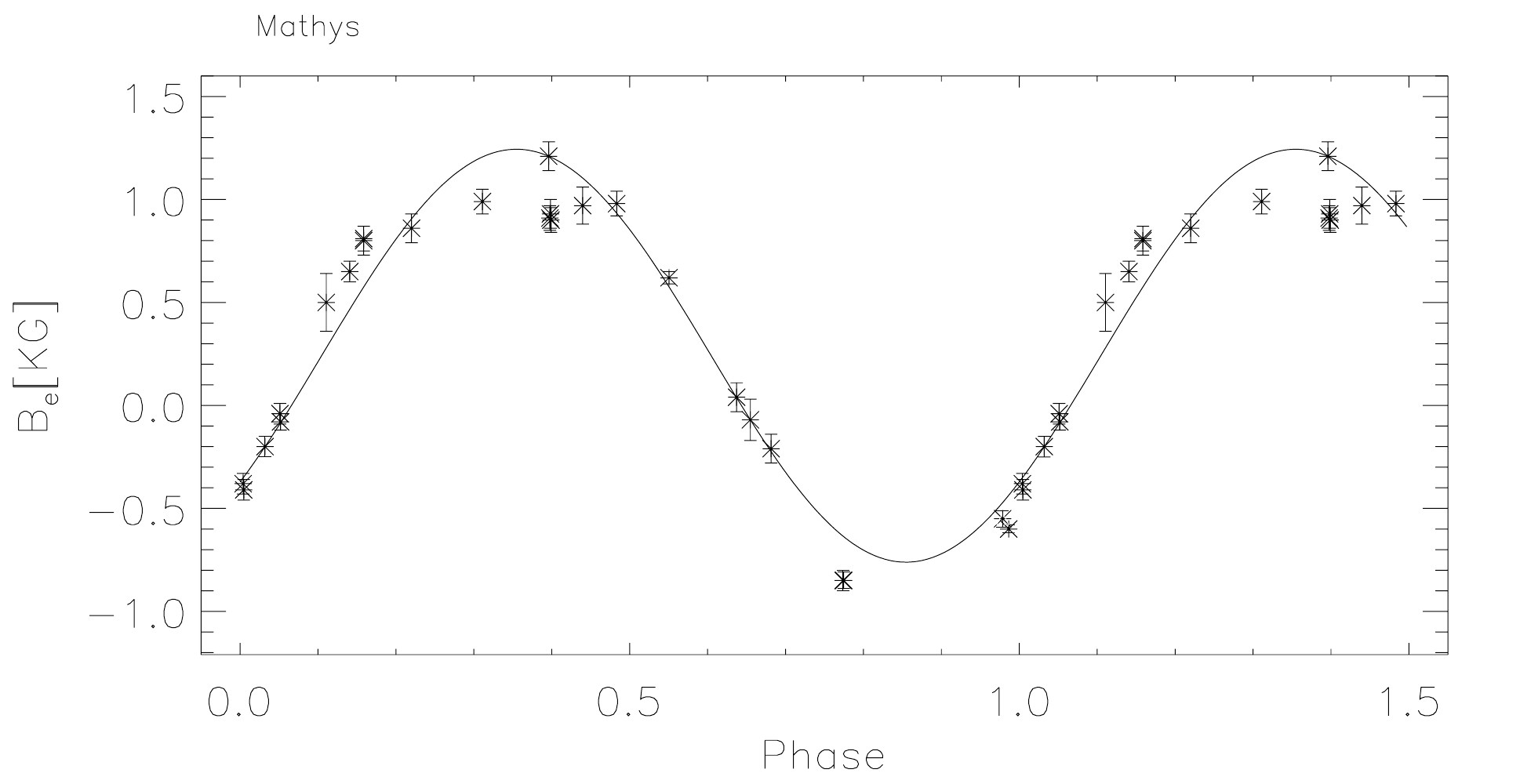}
\includegraphics[trim={0.3cm 0cm 0cm 0cm},width=0.50\textwidth]{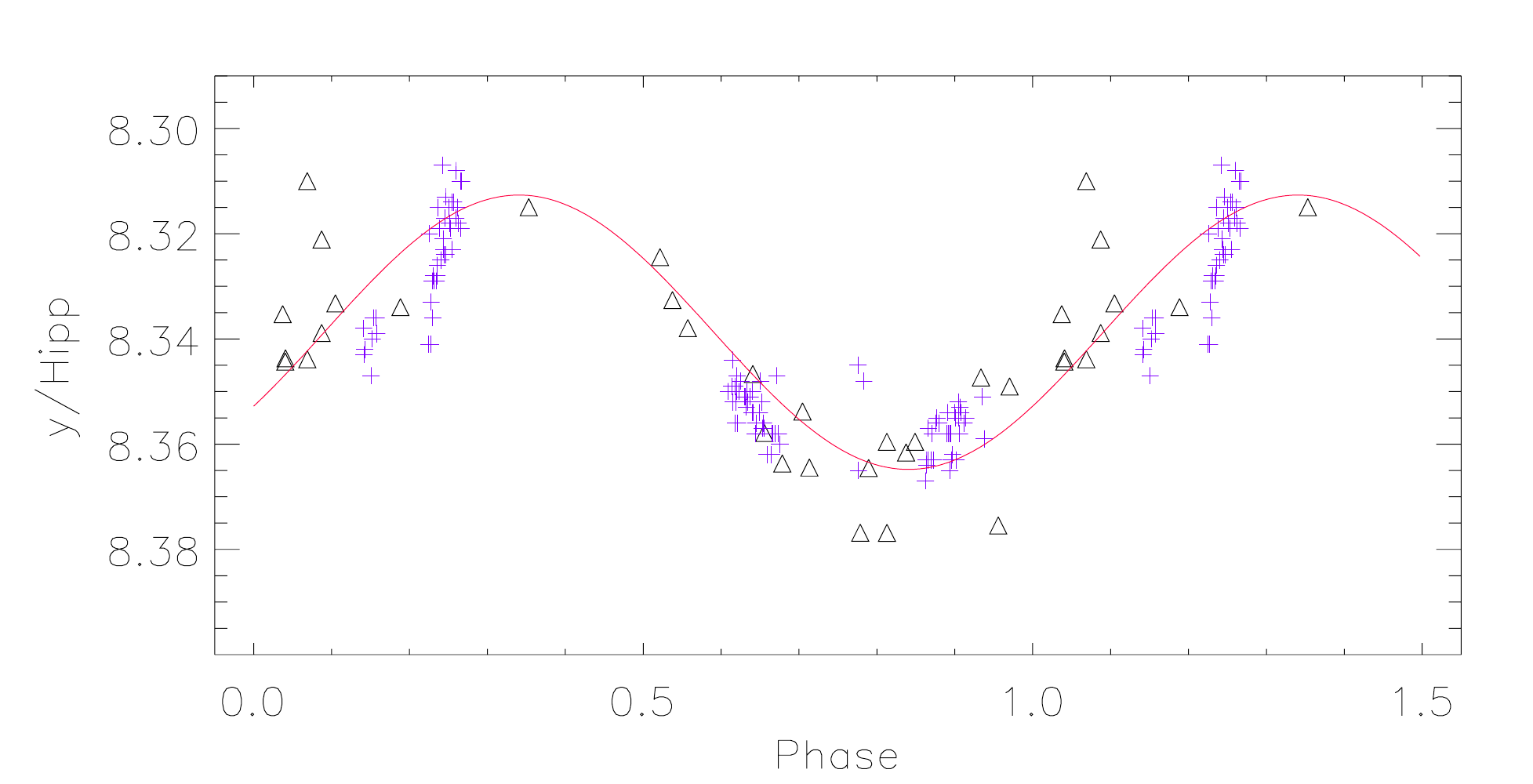}
\caption{HD\,18078. Variability in $B_s$, $B_e$ \citep{Mathys2016_HD18078},
and photometry -- HIPPARCOS ($\triangle$), \citet{Mathys2016_HD18078} ($+$) --
folded with the 1352\,d period. Errorbars for the $B_s$ measurements are barely visible, being often smaller than symbol size.}
\label{Fig_HD18078}
\end{figure}

From $B_s$ and $B_e$ measurements, the variability period of HD\,18078 has been
determined by \cite{Mathys2016_HD18078} to 1358$\pm 12$\,d. We have obtained 10
high-resolution spectra of HD\,18078  from 2003 up to 2021 with SARG, CAOS, and
HARPS-North, and retrieved spectra from the CFHT and Elodie archives. When the period given by \cite{Mathys2016_HD18078} is adopted, our $B_s$
measurements (Table\,\ref{Tab_HD18078}) are slightly in advance. A simultaneous
Lomb-Scargle analysis of our $B_s$ measurements, \cite{Mathys2016_HD18078} $B_s$ and $B_e$ data,
and  HIPPARCOS and Str{\"o}mgren $y$ photometry \citep{Mathys2016_HD18078} 
yields $LS(B_s, B_e, Mag.)$ with the main peak at 1352$\pm$6\,d.
$B_s$ and $B_e$ data, together with HIPPARCOS and Str{\"o}mgren $y$
photometry \citep{Mathys2016_HD18078} are shown in Fig.\,\ref{Fig_HD18078}.

\subsection{HD\,29578}
From spectra collected between JD = 2\,449\,298 and 2\,451\,084 ($\sim$1786 days), M17
found that the surface magnetic field of HD\,29578 changes with a period much longer
than 1800 days. We have obtained one spectrum of HD\,29578 with UCLES at the AAT on
JD = 2\,457\,056.992, but unfortunately the Fe{\sc ii}\,6149.258\,{\AA} line region
was not included. From other lines the surface magnetic field was estimated at
$\sim$2900\,G. We have also obtained 1 UVES and 2 FEROS spectra from the ESO archive.
These $B_s$ measurements (Table\,\ref{Tab_HD29578}), plus the ones by M97, M17 and
\cite{Ryabchikova2004} extend the time-frame of magnetic measurements to 7759 days.

\begin{figure*}
\includegraphics[trim={0.3cm 0cm 0cm 0cm},width=0.47\textwidth]{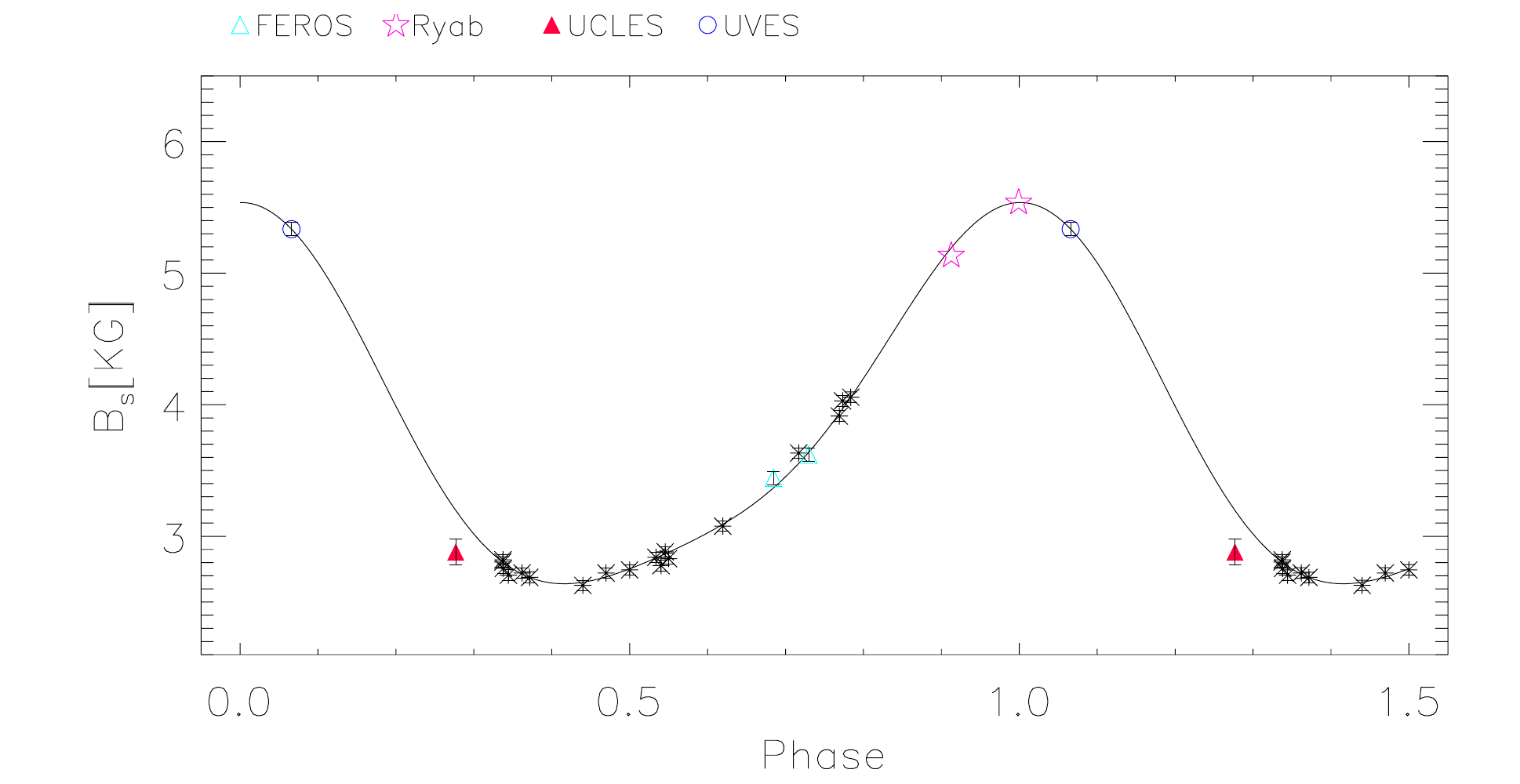}
\includegraphics[trim={0.3cm 0cm 0cm 0cm},width=0.47\textwidth]{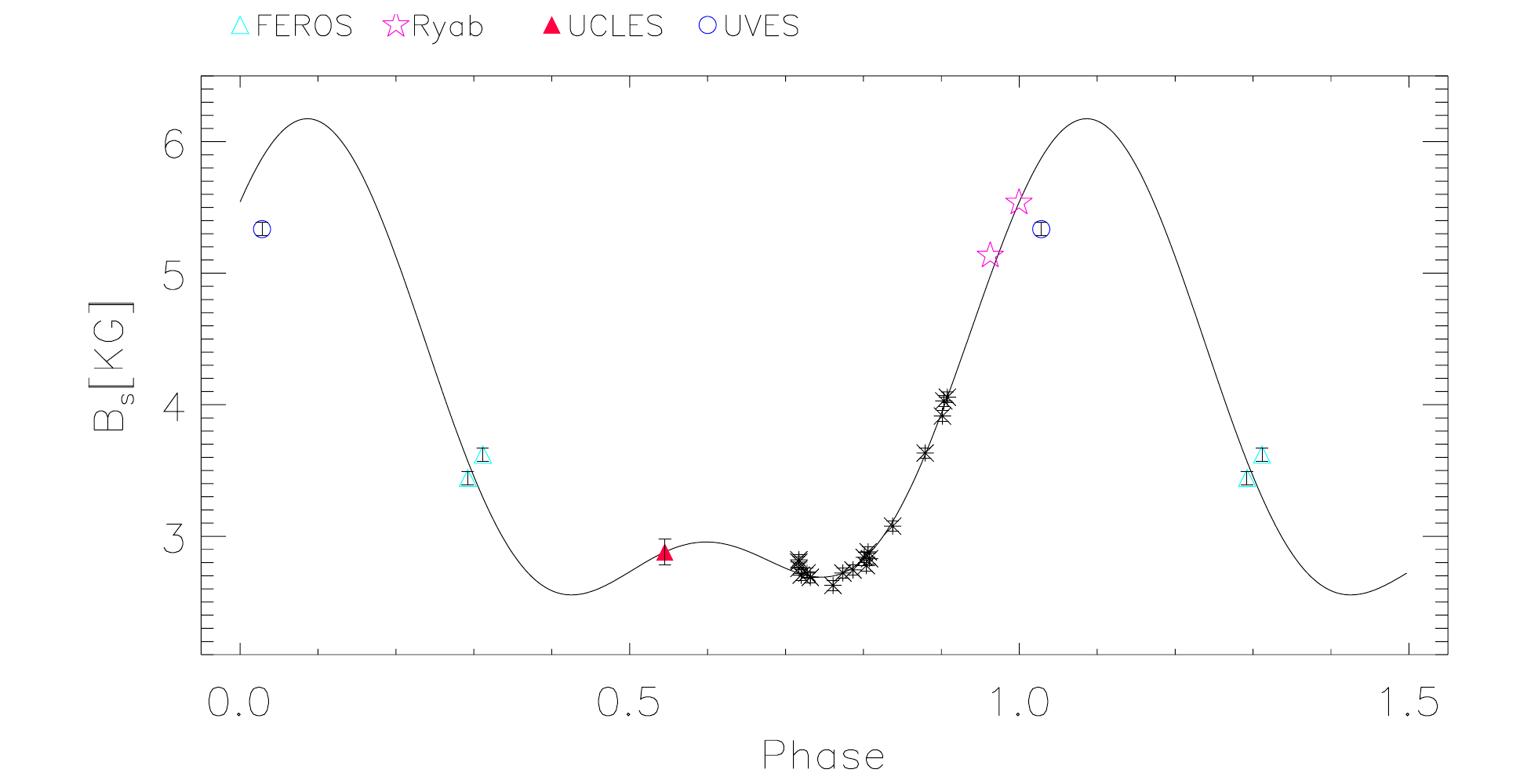}
\includegraphics[trim={0.3cm 0cm 0cm 0cm},width=0.47\textwidth]{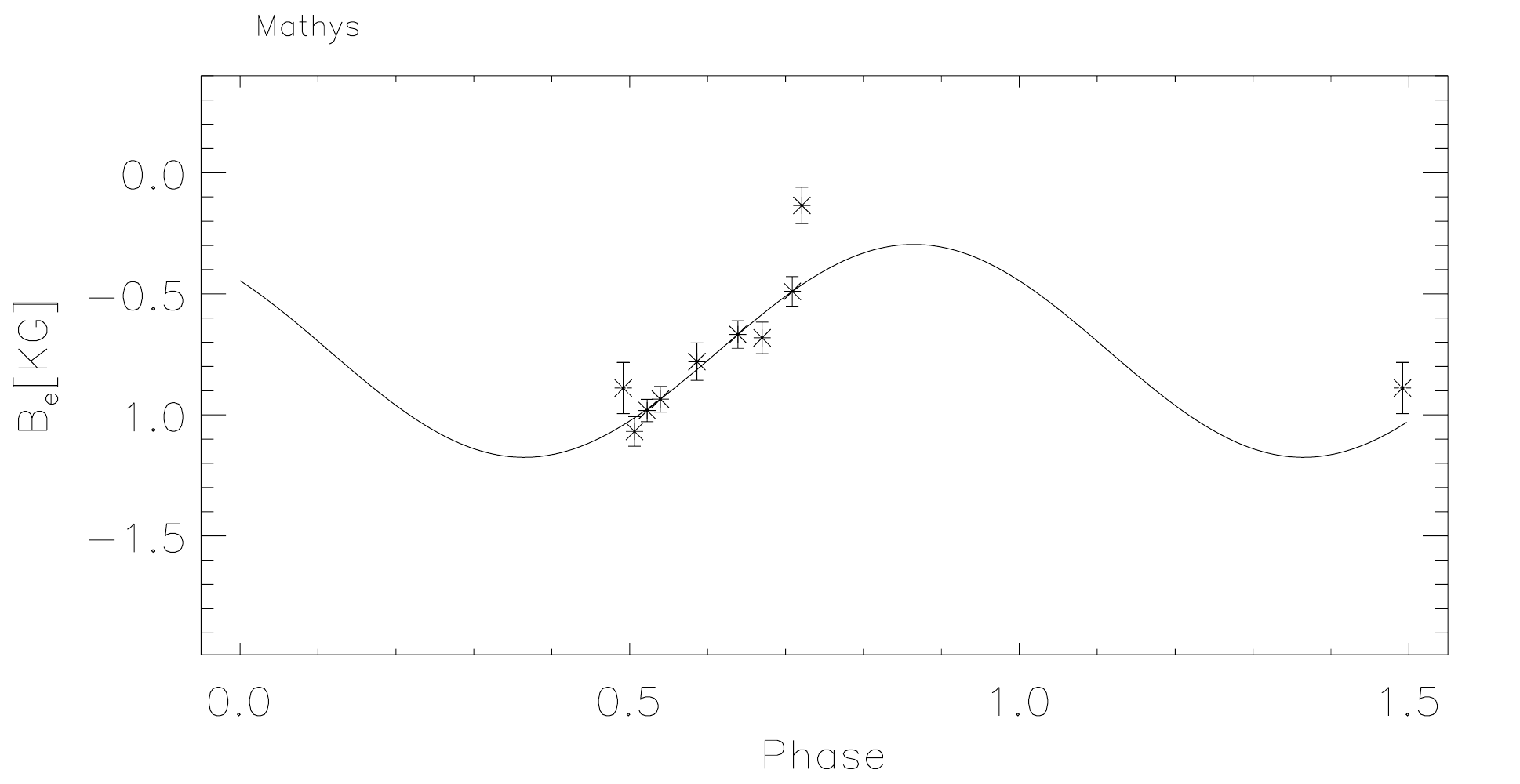}
\includegraphics[trim={0.3cm 0cm 0cm 0cm},width=0.47\textwidth]{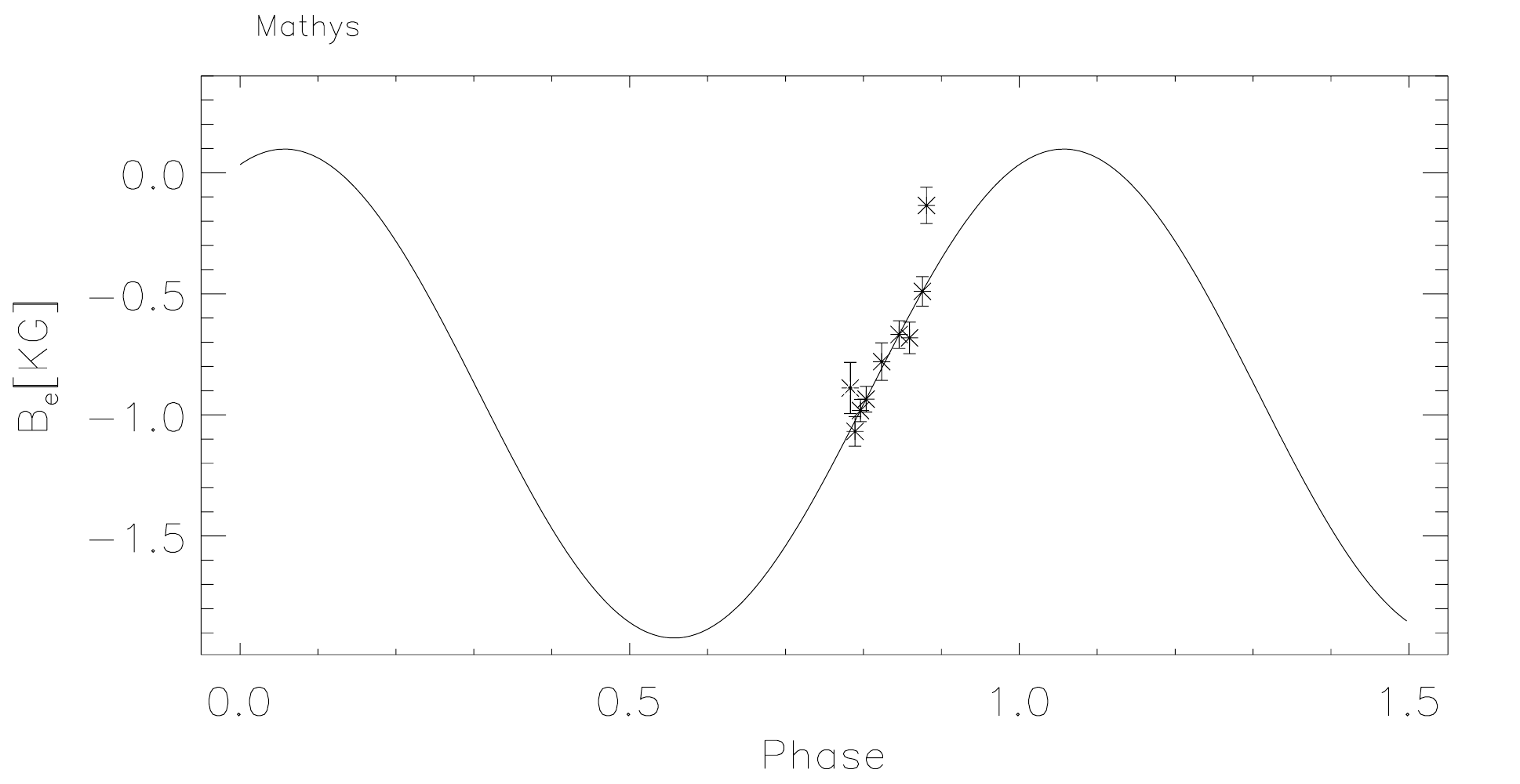}
\caption{HD\,29578. Variability of $B_s$ and $B_e$. In the left panels the data
are folded with the 4000 day period, in the right panels with the 9370 day period.
A sinusoidal fit to the $B_e$ data helps to visualise the possible variability.
Errorbars for the $B_s$ measurements are barely visible, being often smaller than symbol size.}
\label{Fig_HD29578}
\end{figure*}

\begin{figure}\center
\includegraphics[trim={0.5cm 0cm 0.7cm 0cm},width=0.23\textwidth]{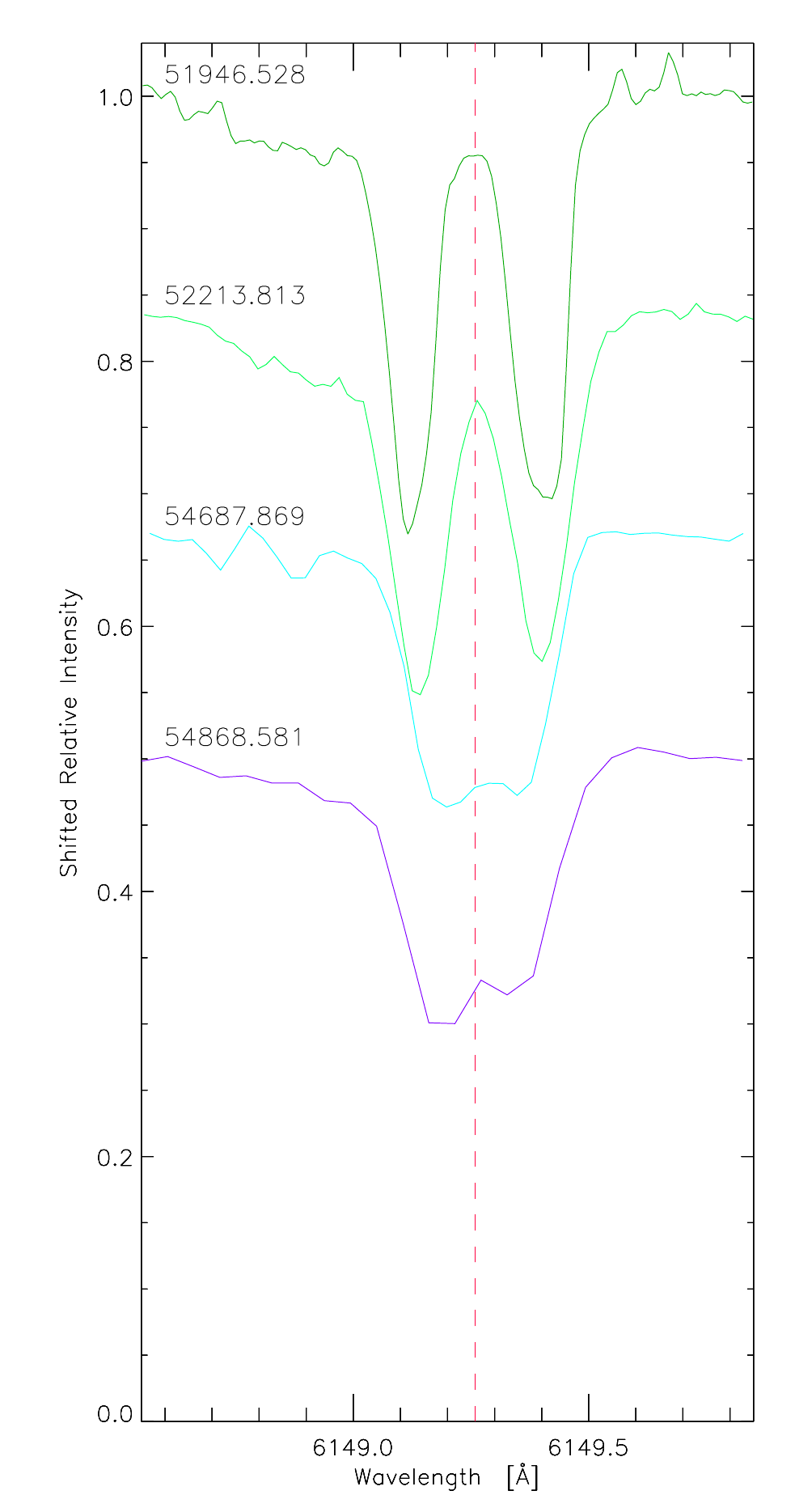}
\includegraphics[trim={0.5cm 0cm 0.7cm 0cm},width=0.23\textwidth]{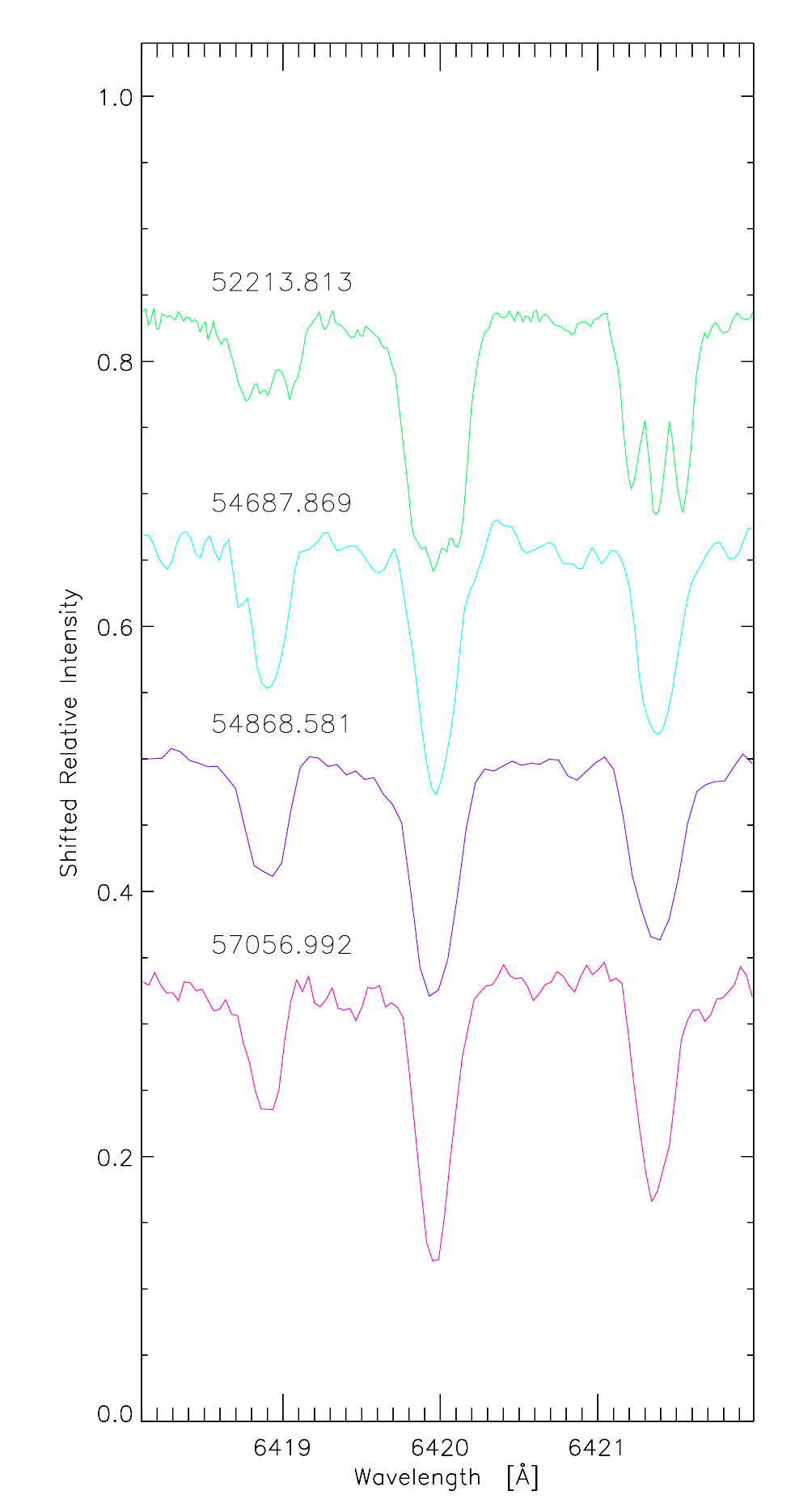}
\caption{HD\,29578. Chunks of spectra ordered with decreasing $B_s$;
x-axis values correspond to JD - 2\,400\,000. The CES@3.6m-ESO spectrum
(JD = 2\,451\,946.528) extends from 6120 to  6150\,{\AA}. The UCLES@AAT spectrum
(JD = 2\,457\,056.992) does not include the Fe{\sc ii}\,6149.258\,{\AA} line;
Zeeman splitting of other metal lines is indicative of a surface magnetic
field of about 2900\,G.}
\label{Fig_HD29578_spectrum}
\end{figure}

We have performed a Lomb-Scargle analysis of all available $B_s$ measurements, finding
two comparable peaks at 4000 and 9370 days in the periodogram (Table\,\ref{Tab_Periods}).
Fig.\,\ref{Fig_HD29578} shows the $B_s$ and the $B_e$ (M17) measurements, folded with
both periods respectively. It appears that new measurements are necessary to reliably
determine the period of HD\,29578. Fig.\,\ref{Fig_HD29578_spectrum} shows the variation
of the HD\,29578 spectrum over time, from JD = 2\,451\,946 (when Zeeman subcomponents
are clearly visible) to 2\,457\,056, when these overlap. NIR high-resolution spectroscopy
is desirable in order to define the $B_s$ variability of HD\,29578.

\subsection{HD\,47103}

\begin{figure}
\includegraphics[trim={0.3cm 0cm 0cm 0cm},width=0.50\textwidth]{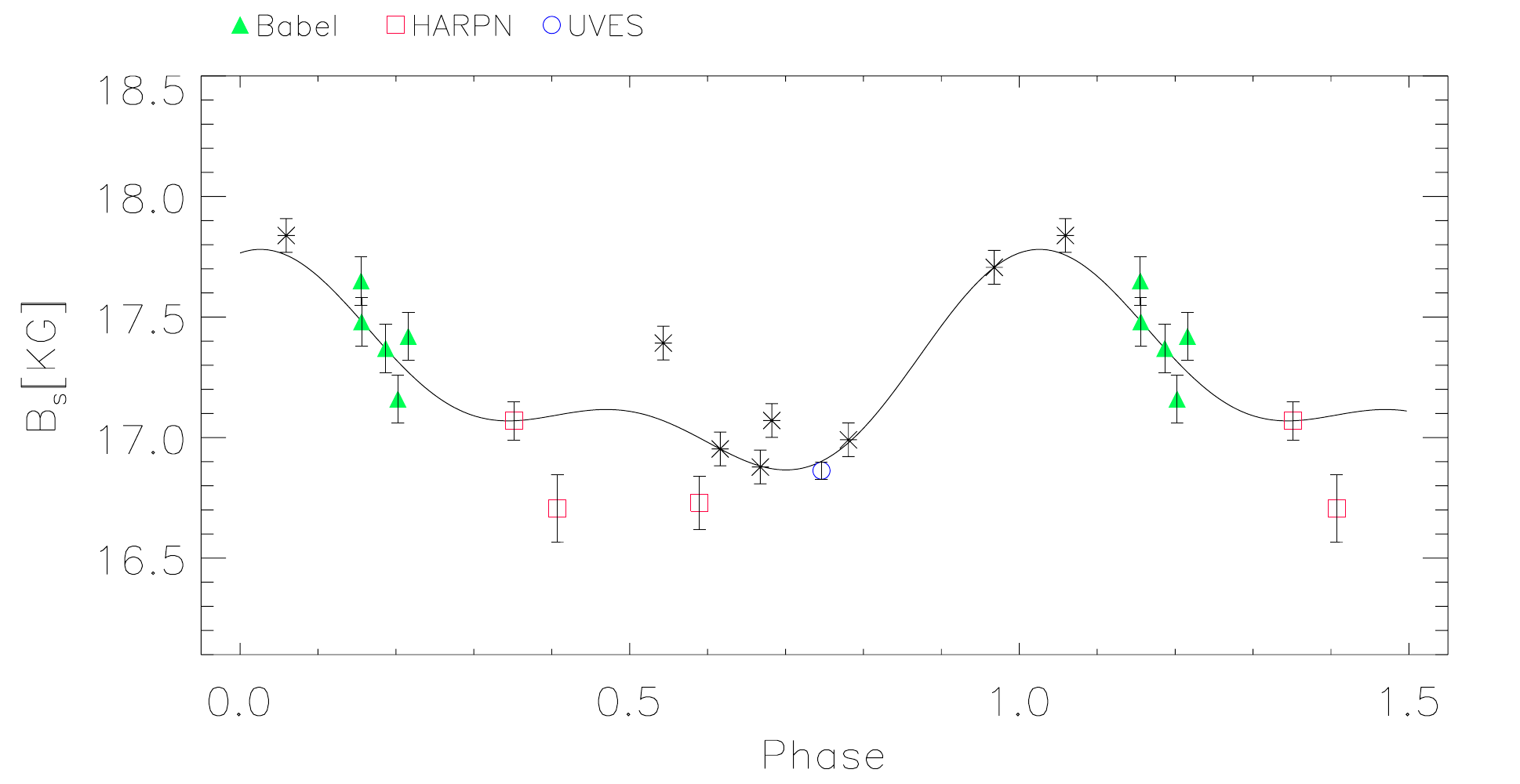}
\includegraphics[trim={0.3cm 0cm 0cm 0cm},width=0.50\textwidth]{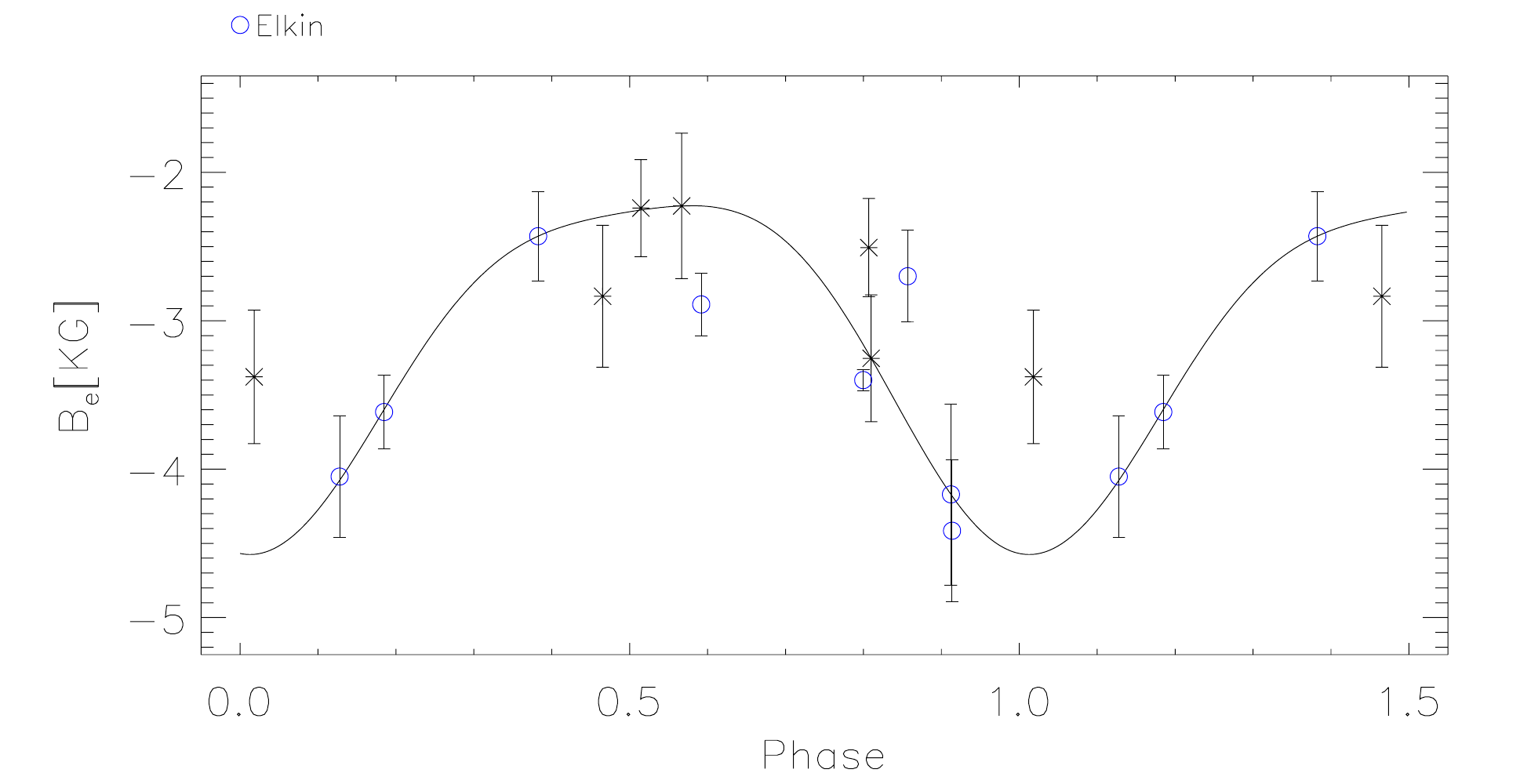}
\caption{HD\,47103. $B_s$ and $B_e$ variability, folded with the 17.683\,d period.
Circles denote data from \citet{Elkin1997}.}
\label{Fig_HD47103}
\end{figure}

From $B_s$ measurements, obtained between JD = 2\,449\,816 and 2\,451\,085, M17 found
HD\,47103 to be variable with an extremely long period. Our measurements of $B_s$
(Table\,\ref{Tab_HD47103}) extend the time baseline to 7600 days. We have computed
the Lomb-Scargle periodogram $LS(B_s)$ of our $B_s$ measurements in combination with
those taken from the literature \citep{Babel1997, Mathys2017}, and the Lomb-Scargle
periodogram of the $B_e$ measurements by \cite{Elkin1997} and M17. $LS(B_s, B_e)$
peaks at 17.683$\pm$0.004\,d. Fig.\,\ref{Fig_HD47103} presents the variability in
$B_s$ and $B_e$.

\subsection{HD\,50169}

\begin{figure}\center
\includegraphics[trim={0.3cm 0cm 0cm 0cm},width=0.50\textwidth]{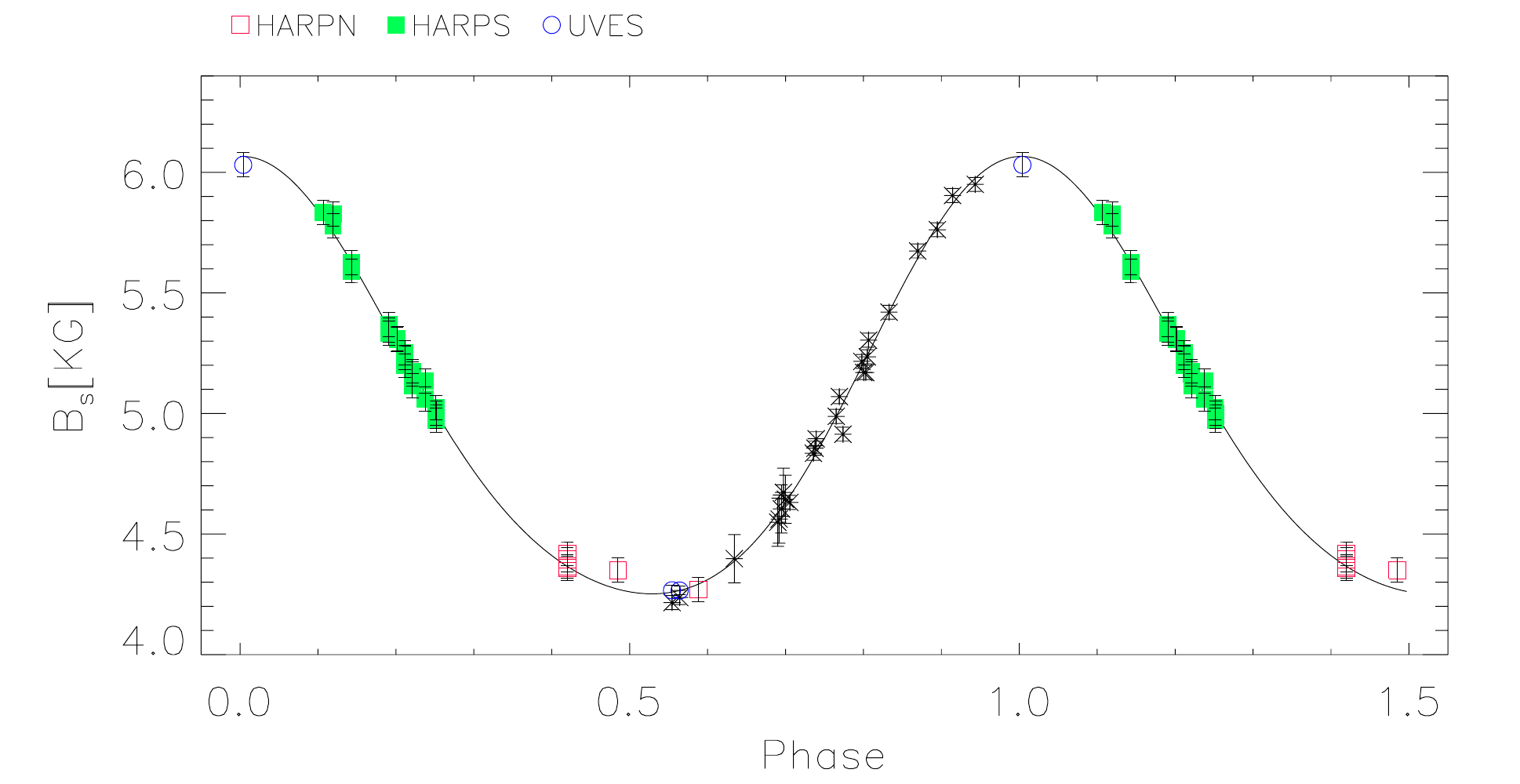}  
\caption{HD\,50169. $B_s$ variability, folded with the 10600\,d period. }
\label{Fig_HD50169}
\end{figure}

\cite{Mathys2019_HD50169} found the $B_e$ variability period of HD\,50169 to be
10600$\pm$300\,d. This value is also representative of their measurements of the
surface magnetic field, which however cover only 9856 days, an interval shorter
than a full rotation cycle. Our $B_s$ measurements (Table\,\ref{Tab_HD50169})
slightly extend the time-frame to 10109 days. These data confirm the period
advanced by \cite{Mathys2019_HD50169} and establish the shape of the minimum of
the variations in $B_s$. The uncertainty in the period of 300 days is due to the
$B_e$ measurements, it cannot be made smaller here. The data are folded in
Fig.\,\ref{Fig_HD50169} with the ephemeris given in Table\,\ref{Tab_Periods}. 

\subsection{HD\,51684}
\begin{figure}\center
\includegraphics[trim={0.3cm 0cm 0cm 0cm},width=0.50\textwidth]{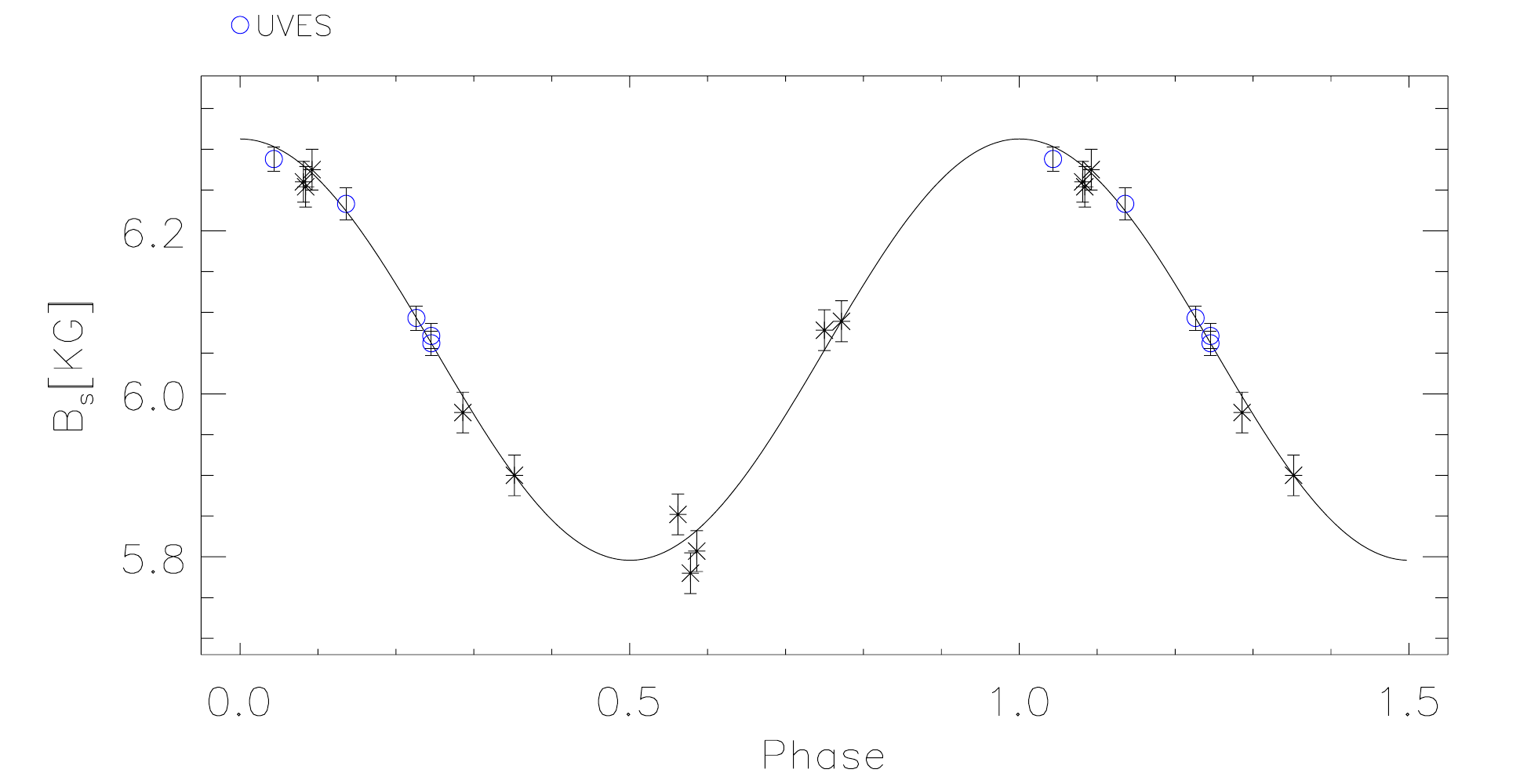}  
\caption{HD\,51684. $B_s$ variability, folded with the 366\,d period.}
\label{Fig_HD51684}
\end{figure}

From 10 measurements of $B_s$ collected between JD = 2\,450\,162 and 2\,451\,086, M17
concluded that the variability period of HD\,51684 is 371$\pm$6\,d. In the ESO archive,
we found 5 additional UVES spectra spread over 330 days. A Lomb-Scargle analysis of 
our measurements (Table\,\ref{Tab_HD51684}) together with those of Mathys gives a
slightly shorter variability period: 366$\pm$1\,d. Data folded with this period are
shown in Fig.\,\ref{Fig_HD51684}. 

\subsection{HD\,55719}
\begin{figure}\center        
\includegraphics[trim={0.3cm 0cm 0cm 0cm},width=0.50\textwidth]{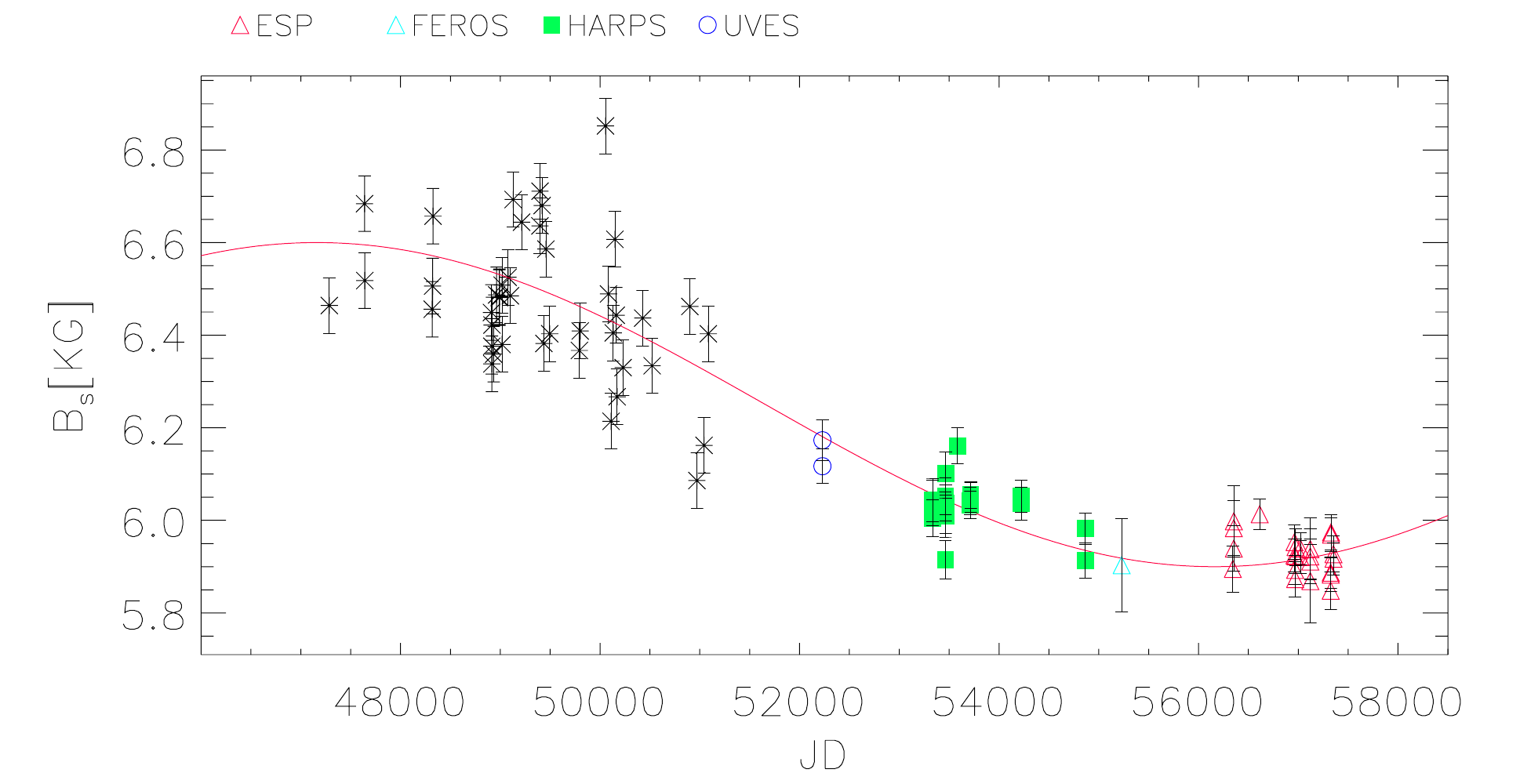}     
\caption{HD\,55719. $B_s$ over time. The sine fit to the data is based on the
shortest possible period of 14000 d.}
\label{Fig_HD55719}
\end{figure}

M17 analysed $B_s$ measurements, collected from JD = 2\,447\,285 to 2\,451\,086,
and found the rotational period of HD\,55719 to be much longer than 10 years. Our
measurements of $B_s$ (Table\,\ref{Tab_HD55719}) and those from the literature cover
more than 10000 days. We note that the surface magnetic field of HD\,55719 has always
been decreasing over the last 27 years, with a resulting rotation period not shorter
than 14000 days (38 years) by assuming a simple sinusoidal variation
(Fig.\,\ref{Fig_HD55719}).

\subsection{HD\,61468}

\begin{figure}\center
\includegraphics[trim={0.3cm 0cm 0cm 0cm},width=0.50\textwidth]{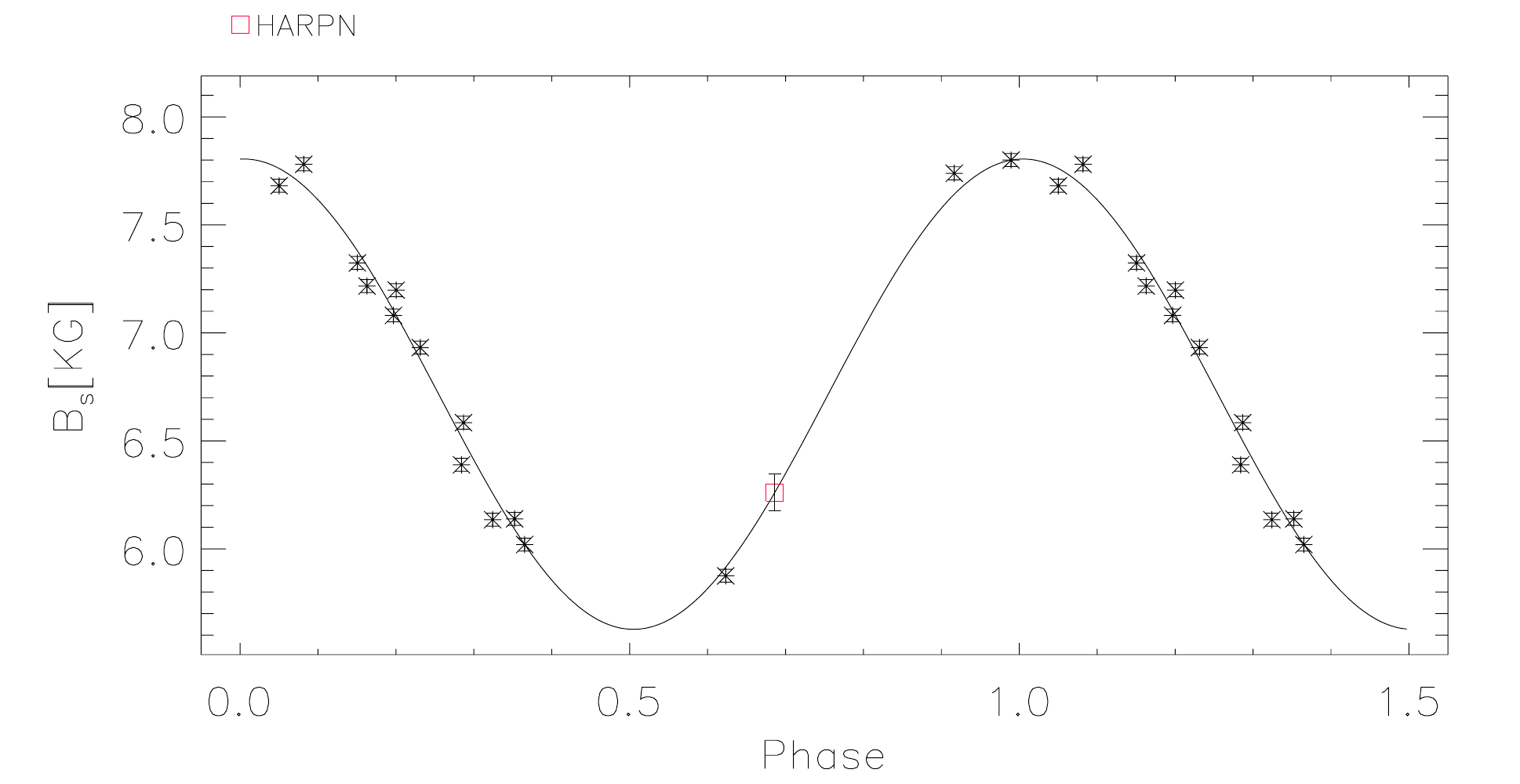}  
\caption{HD\,61468. $B_s$ variability, folded with the 321\,d period. Errorbars (tenths of G) for $B_s$ measurements can be smaller than symbols and
barely visible.}
\label{Fig_HD61468}
\end{figure}

M17 found $B_s$ in HD\,61468 to be variable with a 322$\pm 3$\,d period. We have obtained
a spectrum of this star with HARPS-North on JD = 2\,457\,340.717 and measured a value of
$B_s = 6260\pm 85$\,G. A Lomb-Scargle periodogram of the combined $B_s$ values presents
2 merged peaks centred at 321 and 325.5\,d. A sine fit to the data favours a variability
period equal to 321$\pm$1\,d. Fig.\,\ref{Fig_HD61468} shows the periodic variability of
the surface field of HD\,61468 with the ephemeris given in Table\,\ref{Tab_Periods}.

\subsection{HD\,75445}

M17 found HD\,75445 to be variable in $B_s$ with a small amplitude and a period of
6.291\,d. Our measurements (Table\,\ref{Tab_HD75445}), combined with the surface magnetic
field values from M17 and \cite{Ryabchikova2004}, span the interval from JD = 2\,449\,457
to 2\,454\,205 ($\sim$13 yr) and reveal an average value of $<B_s> = 2936\pm$49\,G. A
Lomb-Scargle analysis results in a large number of peaks of similar importance, with the
highest three at 2.5026, 3.2926 and 6.5500\,d.

\begin{figure}\center        
\includegraphics[trim={0.3cm 0cm 0cm 0cm},width=0.50\textwidth]{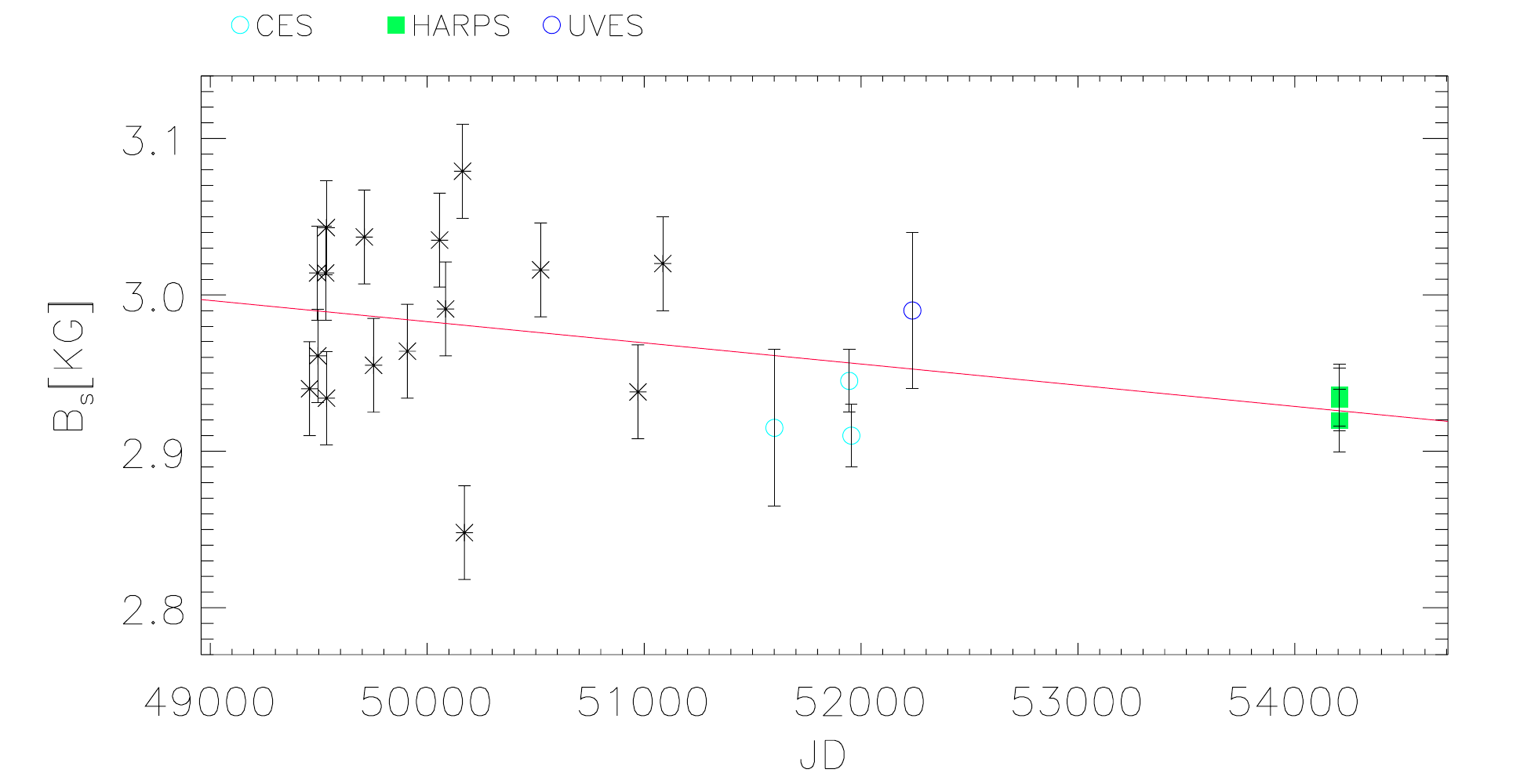}     
\caption{HD\,75445. $B_s$ over time. A straight line equal to the average value of the surface field
is also plotted. }
\label{Fig_HD75445}
\end{figure}

The constant magnitude (6.9000$\pm$0.0003) measured with TESS over 50 days (JD =
2\,458\,517 - 2\,458\,568), together with the $B_s$ r.m.s. error (comparable to the
measurement uncertainty) suggests that the period of variability -- provided there
is any -- of HD\,75445 must be considerably longer than 13 years (Fig.\,\ref{Fig_HD75445}). 

\subsection{HD\,81009}

\begin{figure}\center
\includegraphics[trim={0.3cm 0cm 0cm 0cm},width=0.50\textwidth]{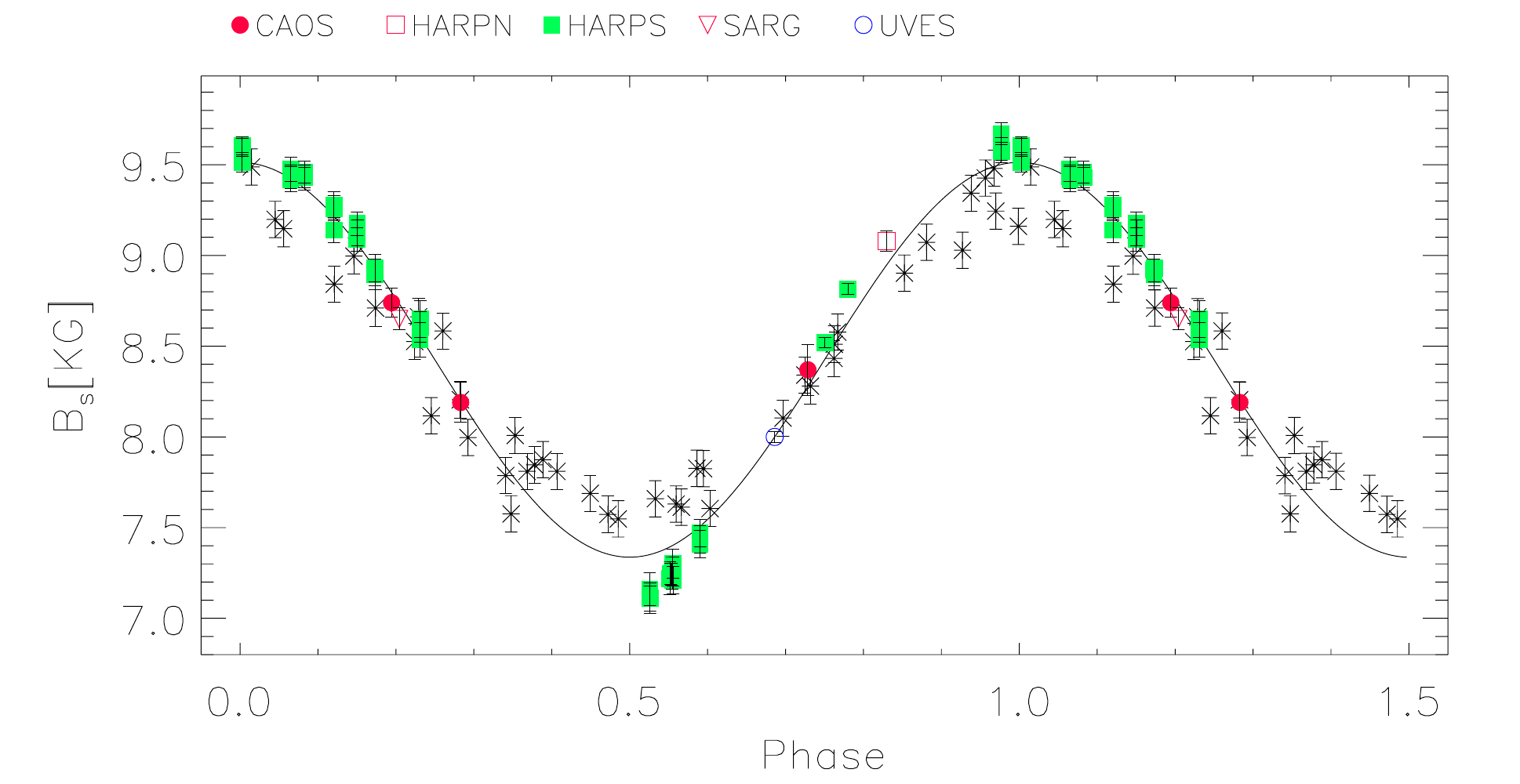}  
\caption{HD\,81009. $B_s$ variability, folded with the 33.987\,d period.}
\label{Fig_HD81009}
\end{figure}

From $B_s$ and $B_e$ measurements, \cite{Wade2000_HD81009} determined the variability
period of HD\,81009 to 33.984$\pm 0.055$\,d. P17 concluded that the photometric
variability period of this star is 33.987$\pm 0.002$\,d. Our $B_s$ measurements
(Table\,\ref{Tab_HD81009}) extend the time coverage of $B_s$ from 1954 to 8210 days.
A Lomb-Scargle analysis of the $B_s$ data confirms the period found by P17.
Fig.\,\ref{Fig_HD81009} shows the  periodic variability of HD\,81009 in $B_s$. 

\subsection{HD\,93507}
\begin{figure}\center
\includegraphics[trim={0.3cm 0cm 0cm 0cm},width=0.50\textwidth]{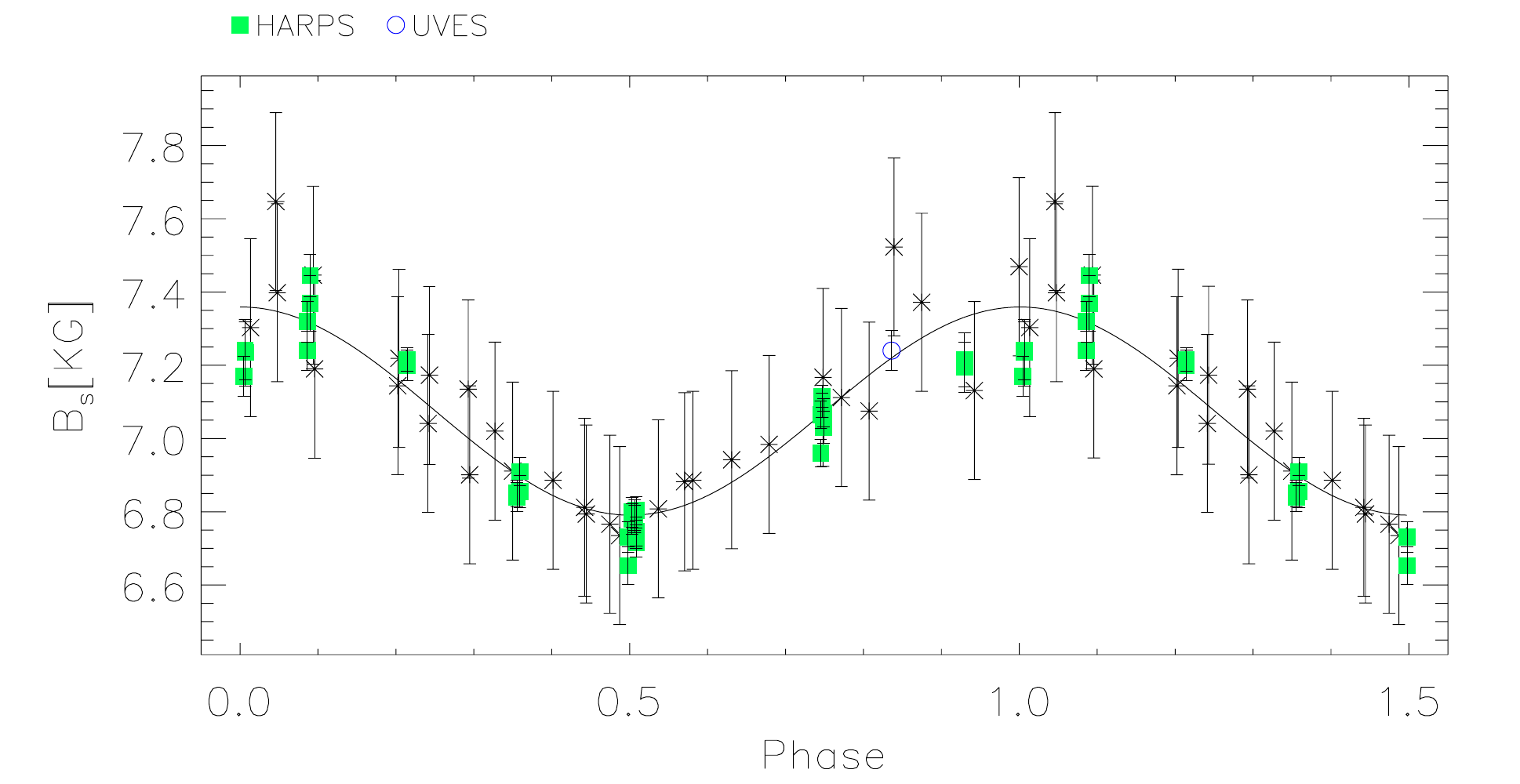}
\includegraphics[trim={0.3cm 0cm 0cm 0cm},width=0.50\textwidth]{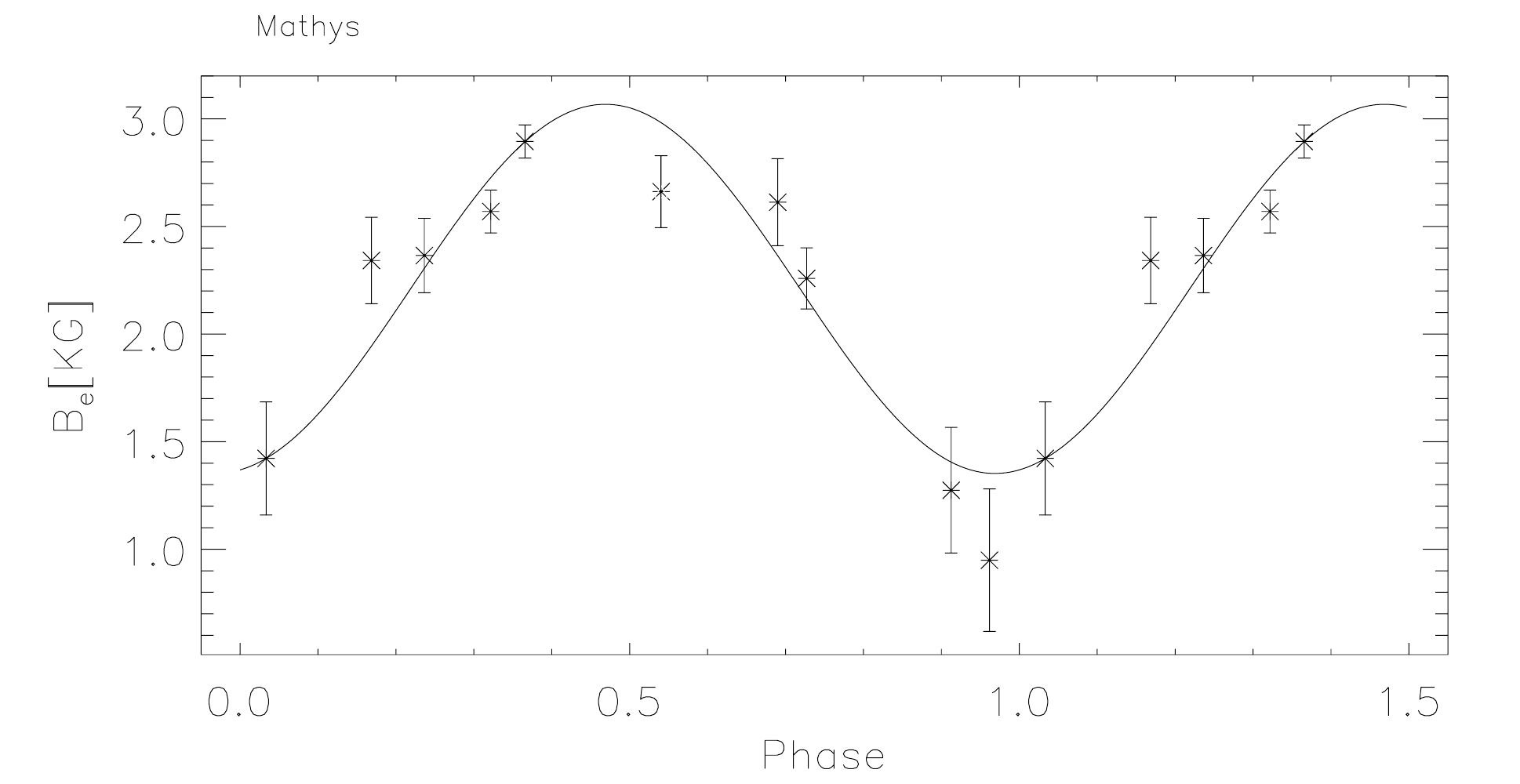}
\caption{HD\,93507. $B_s$ and $B_e$ variability, folded with the 562\,d period.}
\label{Fig_HD93507}
\end{figure}

The period of the variability of HD\,93507 in $B_s$ and $B_e$ is 556$\pm$22\,d (M17).
We have obtained from the ESO archive 1 UVES and 31 HARPS spectra. A Lomb-Scargle
analysis of $B_s$ measurements by M17, our results (Table\,\ref{Tab_HD93507}) and
$B_e$ measurements by \cite{Mathys2017} yields a variability period of 562$\pm$5\,d
(Fig.\,\ref{Fig_HD93507}).

\subsection{HD\,94660}

\begin{figure}\center
\includegraphics[trim={0.3cm 0cm 0cm 0cm},width=0.50\textwidth]{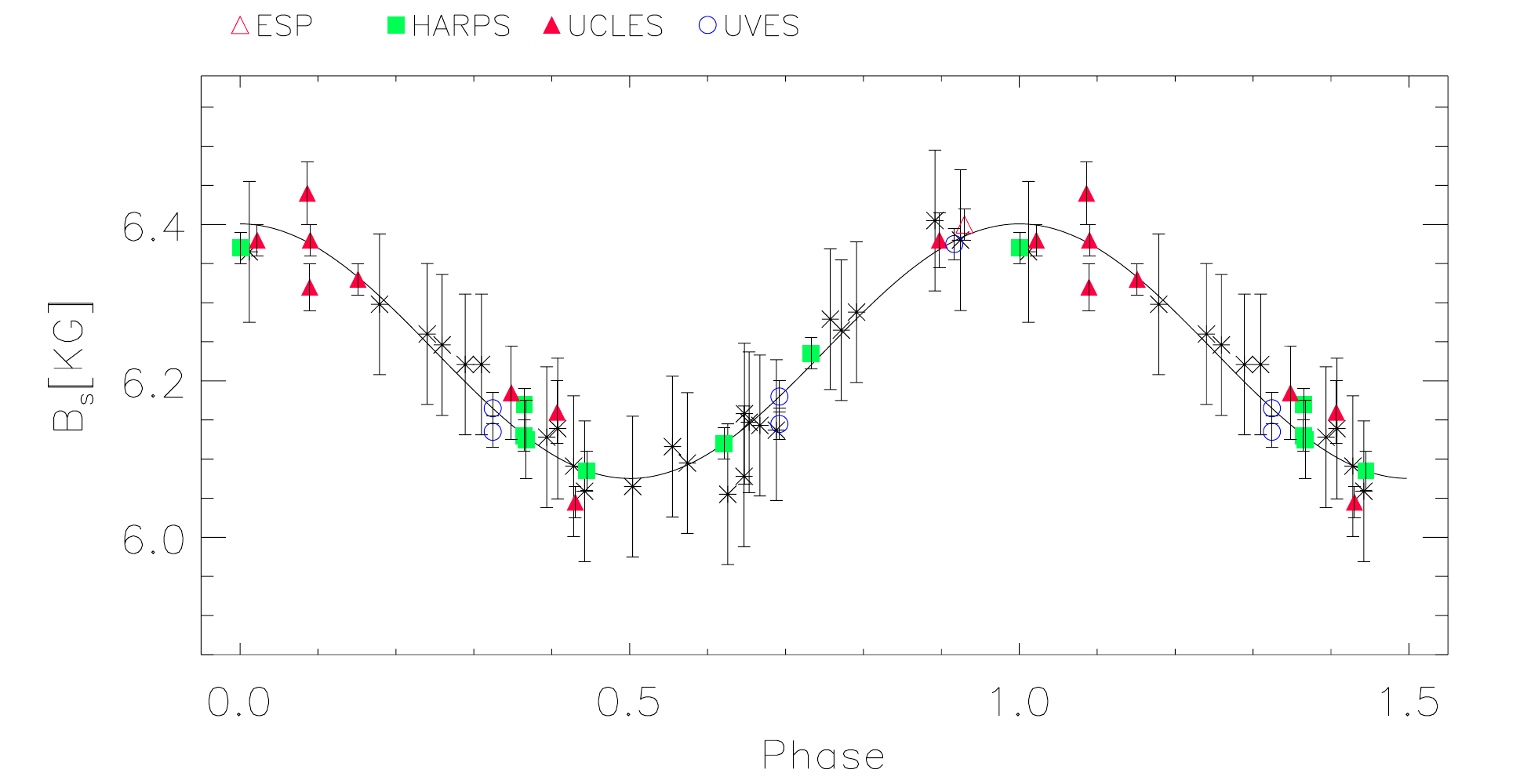}  
\includegraphics[trim={0.3cm 0cm 0cm 0cm},width=0.50\textwidth]{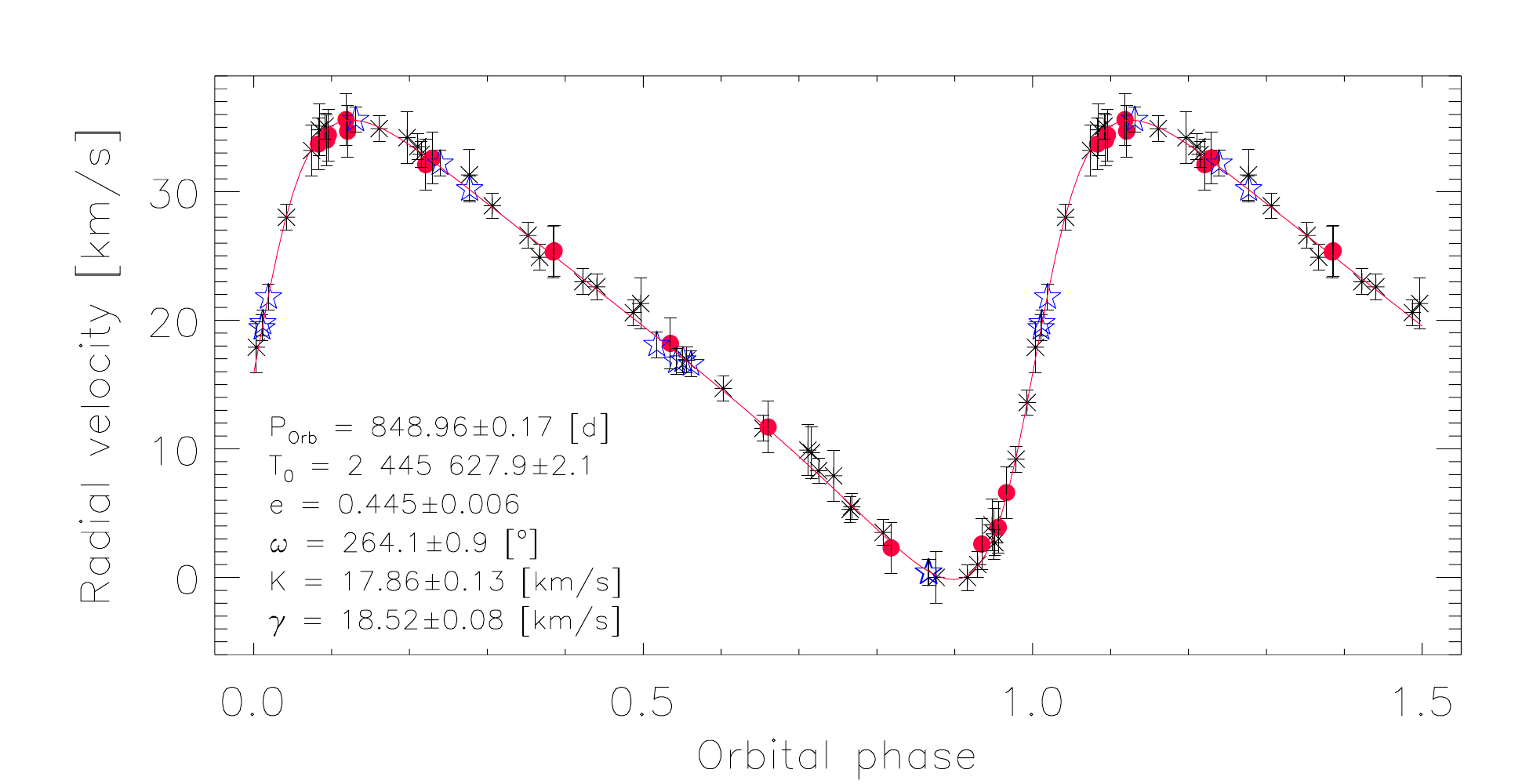}  
\caption{HD\,94660. Top panel, $B_s$ variability folded with the 2830\,d period.
Bottom panel, our radial velocities measured in the UCLES spectra (filled circles) with \citet{Bailey2015} (stars) and M17 (*) values.
Continuous line is the orbital solution fitting at the best the observations.}
\label{Fig_HD94660}
\end{figure}

\cite{Hensberge1993} discovered HD\,94660 to be photometrically variable with a period
close to 2700 days. From spectra acquired on 12 different nights between May 2001
and April 2014, \cite{Bailey2015} found that this star belongs to a binary system
with an orbital period of 840\,d and they adopted the rotational period (P = 2800$\pm$250\,d)
determined by \cite{Landstreet2014}. A similar value (P = 2800$\pm$200 d) was determined independently by M17 from $B_s$ measurements.
We have retrieved the ESO and CFHT archive spectra published by \cite{Bailey2015} and
11 high-resolution spectra were obtained between January 1998 and December 2009 with
UCLES at the AAT. The Lomb-Scargle periodogram of our (Table\,\ref{Tab_HD94660}), M97 and M17 
$B_s$  measurements $LS(B_s)$ peaks at 2830$\pm$140 \,d. Fig.\,\ref{Fig_HD94660}
presents the periodic variability in $B_s$ of HD\,94660.

\begin{figure}\center        
\includegraphics[width=0.50\textwidth]{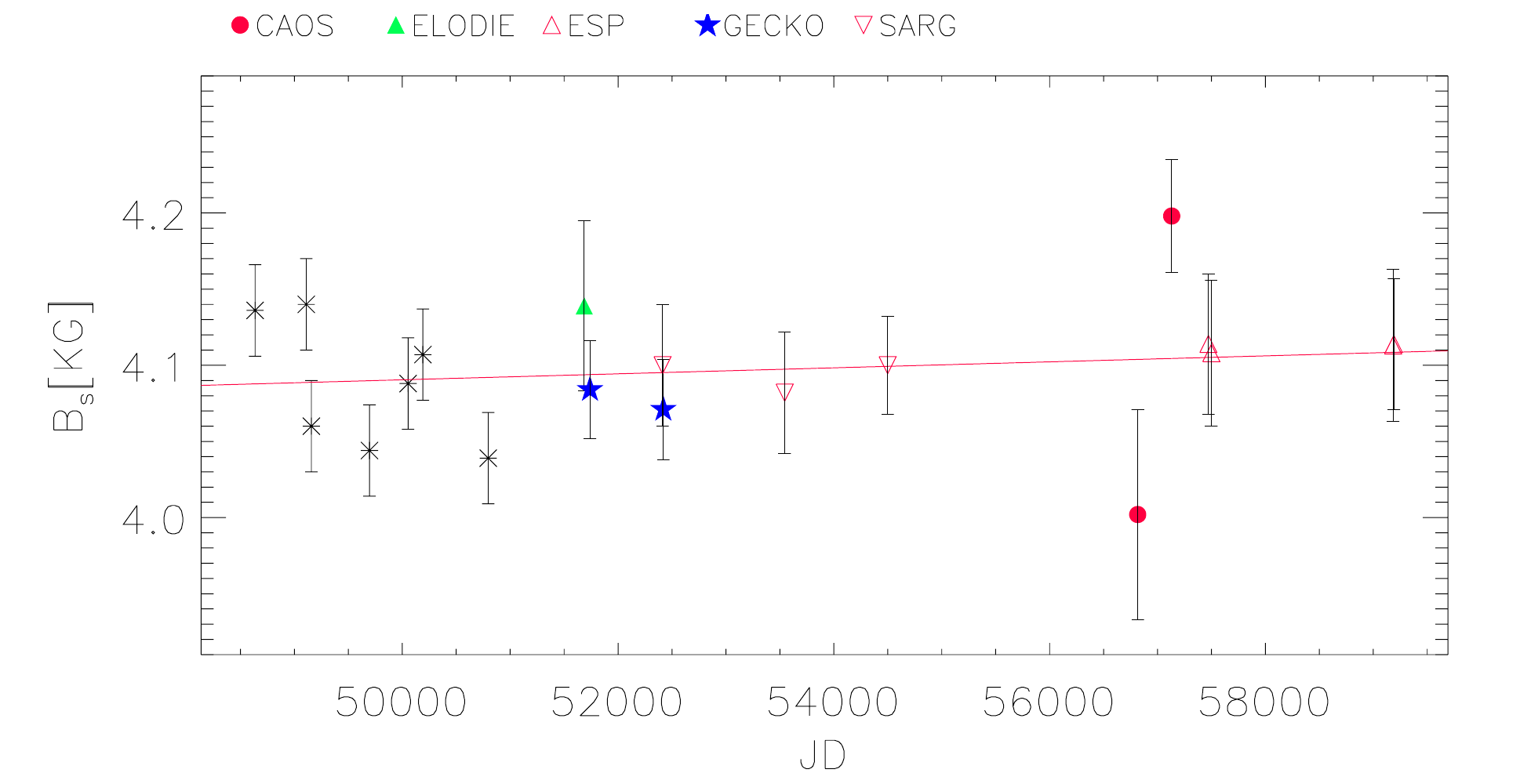}     
\caption{HD\,110066. $B_s$ over time. A linear fit suggests constant behaviour.}
\label{Fig_HD110066}
\end{figure}

Radial velocities measured (Table\,\ref{HD94660_RV}) in the UCLES spectra of HD\,94660 are
combined with the values given by \cite{Bailey2015} and M17 to determine the orbital parameters, following \cite{Catanzaro2016}.
We found the orbital period $P_{Orb.}$ = 848.96$\pm$0.17 d, periastron passage epoch $T_0 = 2\,445\,627.9\pm$2.1,
eccentricity $e$ = 0.445$\pm$0.006, angular anomaly $\omega = 264.1\pm$0.9 degrees, amplitude of radial velocity
variation K =17.86$\pm$0.13 km\,s$^{-1}$ and system velocity $\gamma =18.52\pm$0.08 km\,s$^{-1}$.
Radial velocities of HD\,94660 folded with the here found orbital period are plotted in Fig.\,\ref{Fig_HD94660}.
Errors in the orbital parameters are computed as the difference in the parameter resulting in an increment of the $\chi^2$ equal to 1.
Within errors, these are in agreement with the M17 results. 

\begin{table}
\caption{Radial velocities (RV) of HD\,94660 from archive UCLES spectra. Errors are of 1 km\,s$^{-1}$ order. }
\begin{center}
\begin{tabular}{rrrr}  
\multicolumn{1}{c}{HJD}& \multicolumn{1}{c}{RV} & \multicolumn{1}{c}{HJD} & \multicolumn{1}{c}{RV}\\
 2400000+&km\,s$^{-1}$ & 2400000+ & \multicolumn{1}{c}{km\,s$^{-1}$} \\ \hline
48968.261 &   1.6~~   &52980.254 & 11.7~~   \\    
49350.250 & 25.3~~ &53369.254 &  35.6~~ \\
49351.258 & 25.4~~ &54188.098 &  33.7~~ \\
50060.244 & 32.5~~ &54196.964 &  34.0~~  \\
50824.041 & 35.6~~ &54198.928 &  34.4~~  \\                     
51176.136 & 18.2~~ &54928.960 &   3.9~~   \\
51542.171 &  6.6~~  &55161.232 &  32.6~~ \\
52265.250 &  2.3~~  & &  \\\hline
 \end{tabular}\label{HD94660_RV}
\end{center}
\end{table}

Because of the X-ray emission from the HD\,94660 binary system, the nature of the companion is
still a matter of debate \citep{Oskinova2020, Scholler2020}. 

\subsection{HD\,110066}
\cite{Pyper2017} reported on a 12-year photometric campaign dedicated to HD\,110066,
finding no evidence of variability. M17 noted that the $B_s$ variability of this star
presents an amplitude of 100\,G if the period is as long as 4900 days. \cite{Bychkov2021}
have obtained new $B_e$ measurements and -- discarding the results published by
\cite{Babcock1958} -- established a variability period of 6.4769 days. We have analysed
TESS photometric data and found a constant magnitude (6.346238$\pm$0.000003) between
JD=2\,458\,900.0 and 2\,458\,926.5 (26 days).

We have obtained high-resolution spectra of HD\,110066 with SARG and CAOS and have
retrieved spectra from the Elodie, CFHT and ESO archives.
Our $B_s$ measurements of HD\,110066 are listed in Table\,\ref{Tab_HD110066}. Including
values from M97 and M17, we span a period of 28.8 years without detecting any variability.
All $B_s$ measurements range between 4040 and 4150\,G, with errors not smaller than 50\,G.
On the basis of $B_s$ measurements, we can only conclude that the variability period --
provided there is any -- of HD\,110066 exceeds 3 decades, see Fig.\,\ref{Fig_HD110066}.

\subsection{HD\,116114}

From measurements collected between JD = 2\,448\,732 and 2\,451\,042, M17 found
HD\,116114 to be a magnetic variable with a 27.61$\pm$0.08\,d period. The amplitude
of the $B_s$ variations is 33$\pm 9$\,G, the amplitude of the $B_e$ variations is
84$\pm 33$\,G. M17 ruled out the 4.41156\,d period determined by \cite{Romanyuk2014}
from $B_e$ measurements.   \cite{Wraight2012} concluded that HD\,116114 is a photometric variable with a period of 5.3832 d.

\begin{figure}\center
\includegraphics[trim={0.3cm 0cm 0cm 0cm},width=0.50\textwidth]{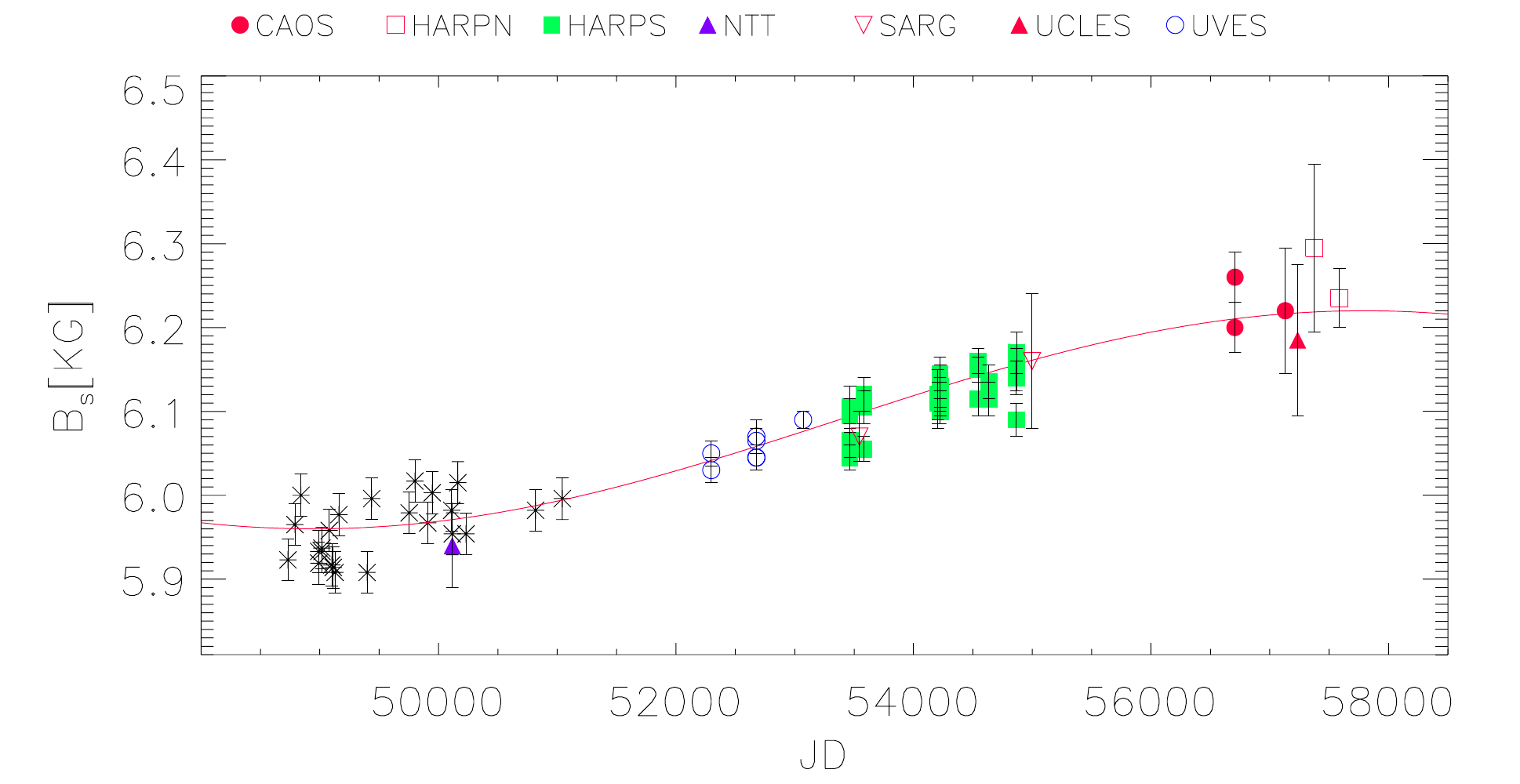}
\includegraphics[trim={0.3cm 0cm 0cm 0cm},width=0.50\textwidth]{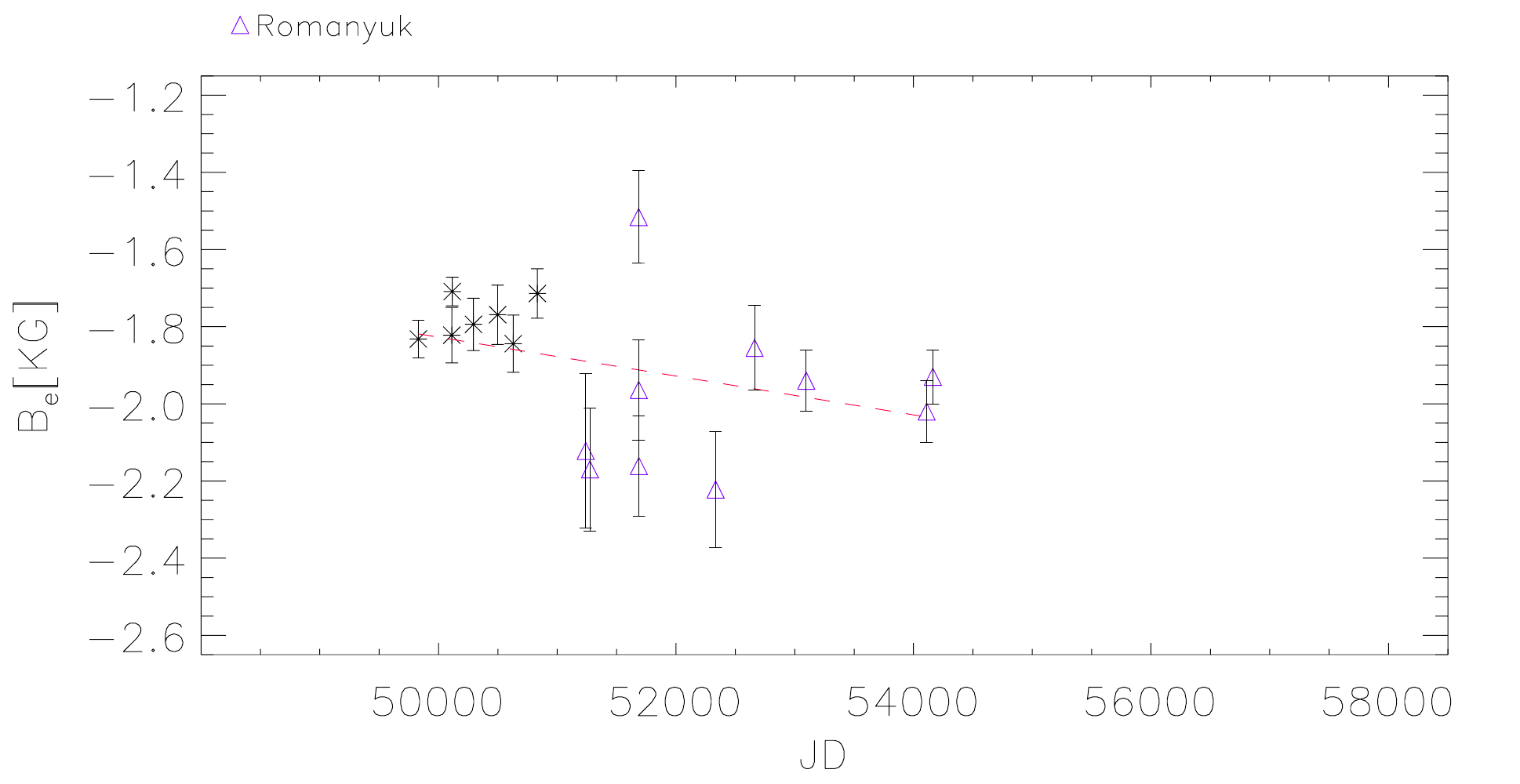}
\caption{HD\,116114. $B_s$ and $B_e$ variations over time. A sine fit with the
shortest possible period of 17700\,d is plotted over the $B_s$ measurements.
The $B_e$ measurements have been linearly fitted.}
\label{Fig_HD116114}
\end{figure}

We have analysed the TESS data of HD\,116114 (between JD = 2\,458\,570 and 2\,458\,595),
finding a constant magnitude (6.76799$\pm 0.00001$) on a time scale of 25 days.
We have observed HD\,116114 with SARG, HARPS, CAOS, UCLES and HARPS-North. In addition,
we have retrieved NTT, UVES and HARPS spectra from ESO archive. Our
$B_s$ measurements are listed in Table\,\ref{Tab_HD116114}. The $B_s$ measurements
cover 8855 days, revealing a steadily increasing field. Assuming sinusoidal variations,
the shortest possible period of HD\,116114 would therefore be of the order of 48
years (Fig.\,\ref{Fig_HD116114}). The $B_e$ measurements by \cite{MathysHubrig1997},
\cite{Romanyuk2014}, and \cite{Mathys2017} on the other hand do not present a clear
trend (Fig.\,\ref{Fig_HD116114}).

\subsection{HD\,126515}

\begin{figure}\center
\includegraphics[trim={0.3cm 0cm 0cm 0cm},width=0.50\textwidth]{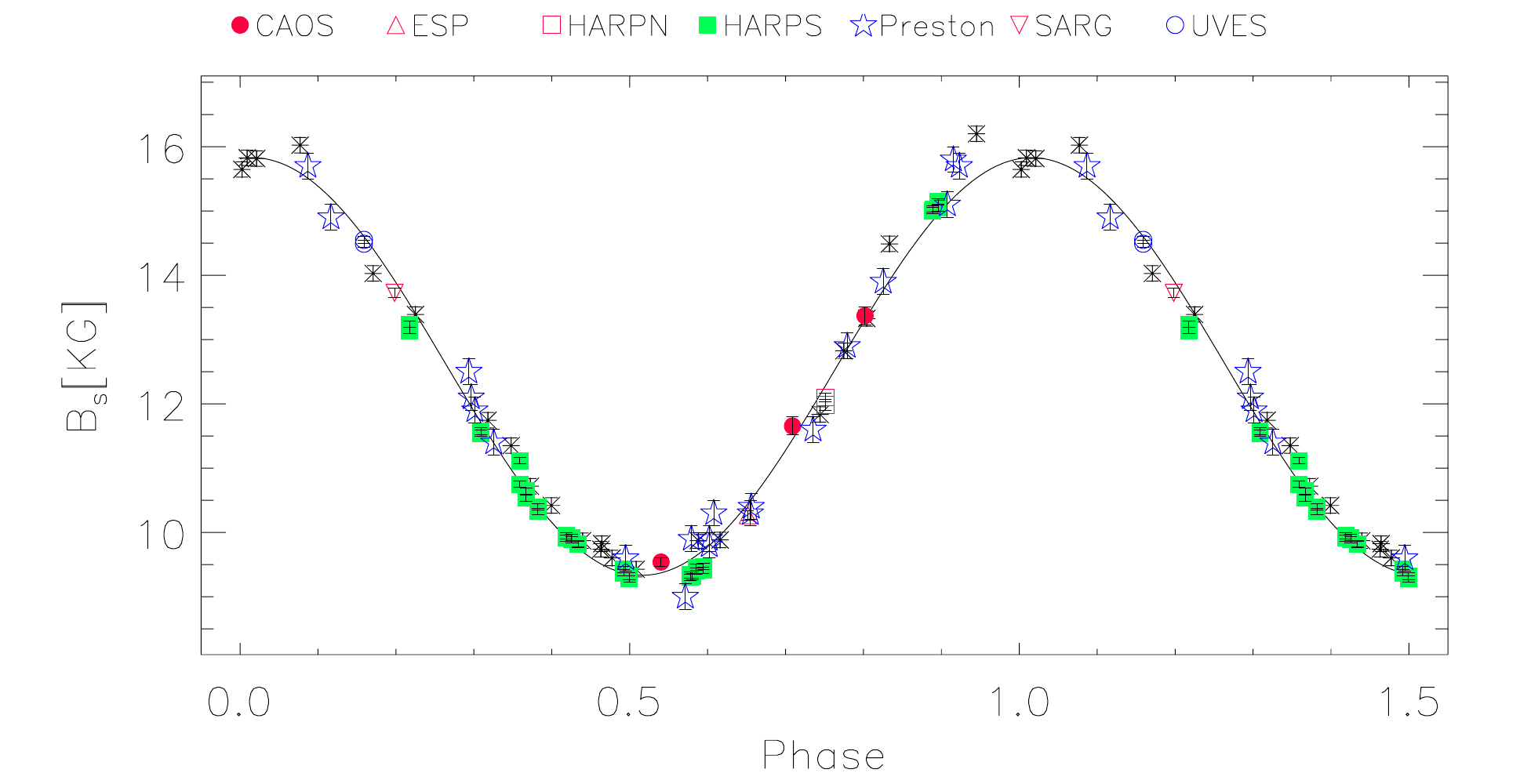}  
\includegraphics[trim={0.3cm 0cm 0cm 0cm},width=0.50\textwidth]{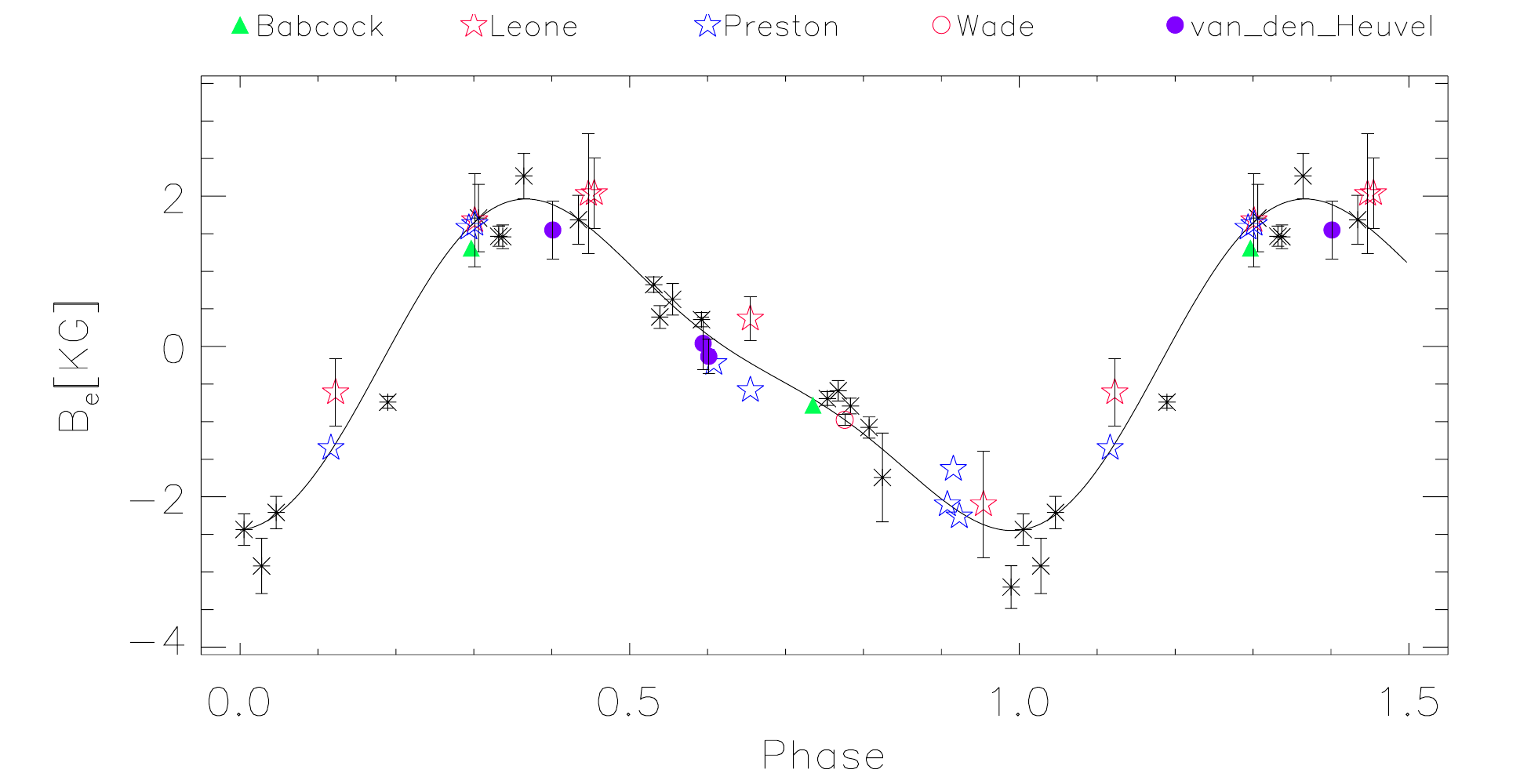}  
\caption{HD\,126515. $B_s$ and $B_e$ variability, folded with the 129.95\,d period.}
\label{Fig_HD126515}
\end{figure}

\cite{Leone2001} noted the coincidence of the extrema of literature light curves of the star
HD\,126515 when a variability period of 129.9474\,d is assumed.
This period turned also out to be representative of the $B_e$ and $B_s$ variations. P17
and M17 established a period of 129.95$\pm$0.02\,d from photometric and magnetic data.
By adding our $B_s$ measurements (Table\,\ref{Tab_HD126515}) to the data published by
\cite{Preston1970_HD126515}, M97 and M17, we span a total of 23250 days. A Lomb-Scargle
analysis of these measurements confirms the validity of the 129.95\,d period.
Fig.\,\ref{Fig_HD126515} shows the respective variations in $B_s$ and in $B_e$, based
on measurements by \cite{Babcock1958}, \cite{vandenHeuvel1971}, \cite{Mathys1997},
\cite{Wade2000_HD126515}, \cite{Leone2001}, and \cite{Mathys2017}.

The $B_s$ maximum coincides with the negative $B_e$ extremum, while the $B_s$ minimum
is found close in phase with zero effective field $B_e$. This constitutes clear evidence
for a non-axisymmetric magnetic field geometry of HD\,126515 \citep{Stift1975, Stift1991}.

\subsection{HD\,137949}

\begin{figure}\center
\includegraphics[trim={0.3cm 0cm 0cm 0cm},width=0.50\textwidth]{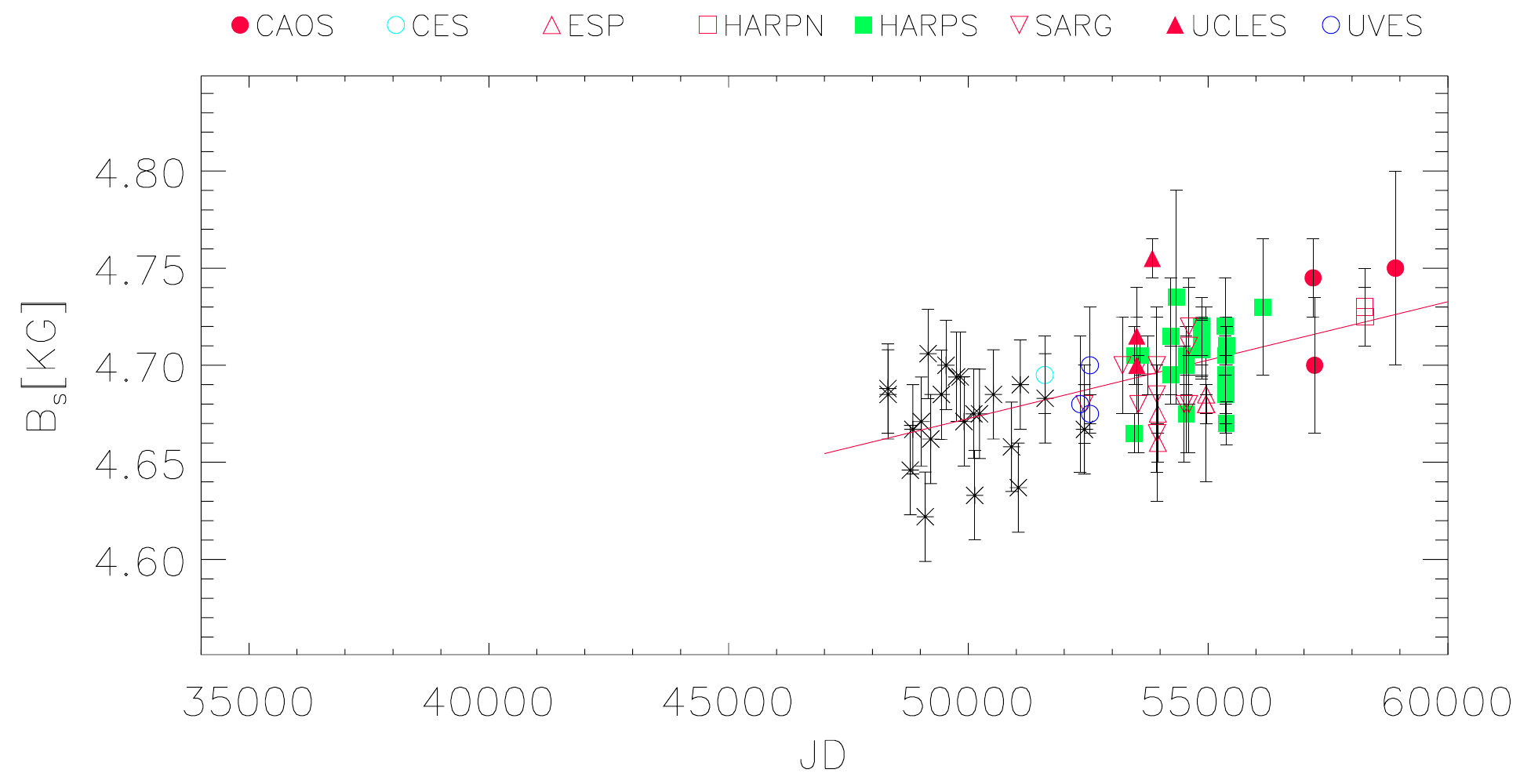}
\includegraphics[trim={0.3cm 0cm 0cm 0cm},width=0.50\textwidth]{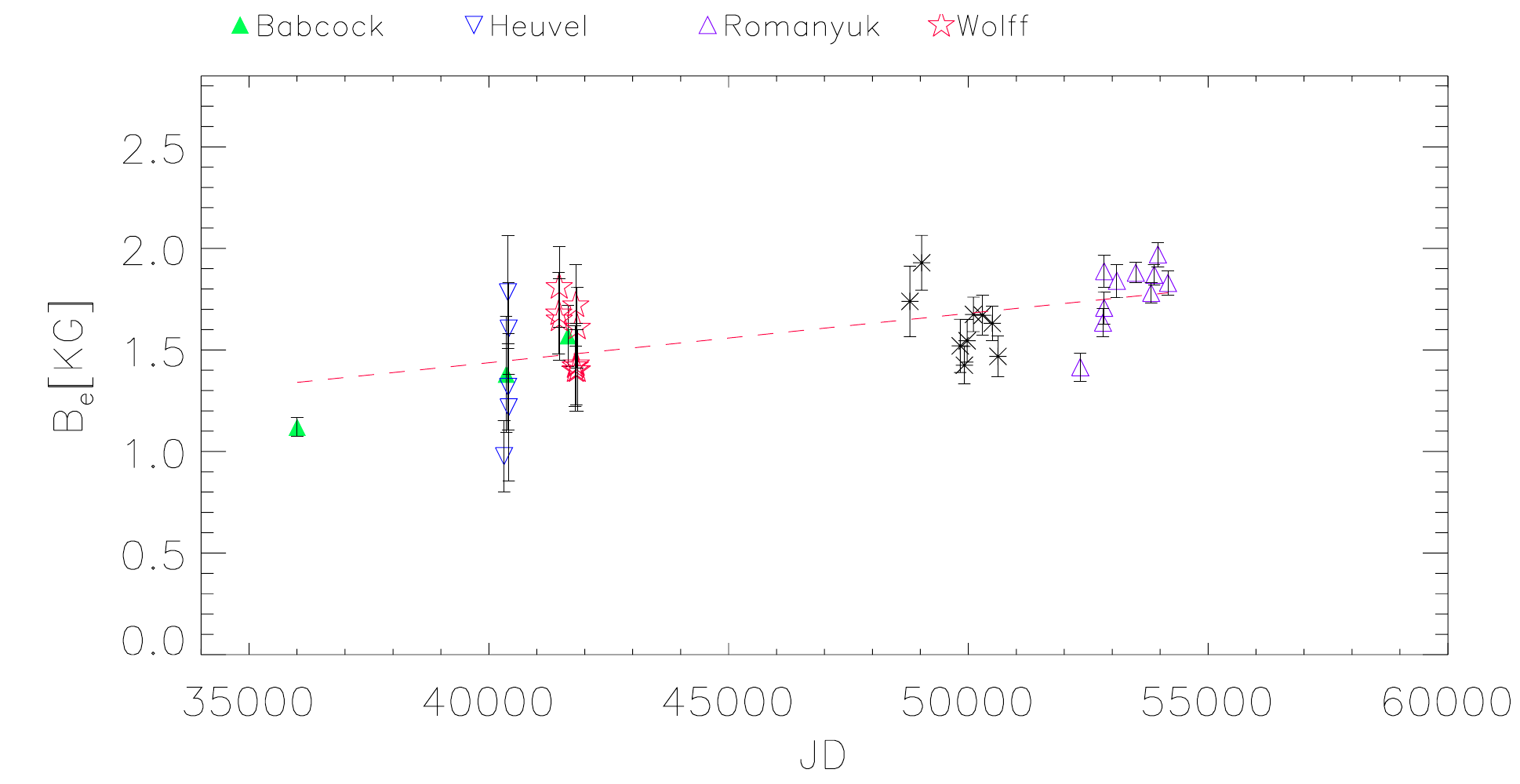}
\caption{HD\,137949. $B_s$ and $B_e$ measurements over time. The straight lines
represent linear fits to the data.}
\label{Fig_HD137949}
\end{figure}

After 14 years of monitoring, no evidence of photometric variability has been detected by
P17. Referring to $B_e$ measurements spanning almost 50 years, \cite{Landstreet2014}
estimated the rotational period of this star to be of the order of one hundred years, M17
concluded on a variability period probably of the order 5195 days ($\sim$14.2\,yr). By combining our
measurements
(Table\,\ref{Tab_HD137949}) with data available in the literature, we cover 10584 days,
corresponding to twice the period put forward by Mathys and coworkers. A Lomb-Scargle
analysis of $B_s$ and $B_e$ data does not show any predominant peak in the periodogram.
Fig.\,\ref{Fig_HD137949} displays $B_s$ and $B_e$ over time. The trends seen in both plots
are not incompatible with a slight increment in field strengths; if real, this would
indicate a rotational period much longer than 27 years.

\begin{figure*}\center
\includegraphics[trim=-0.5cm 0.0cm 0.0cm 0.0cm,clip=true,width=0.53\textwidth]{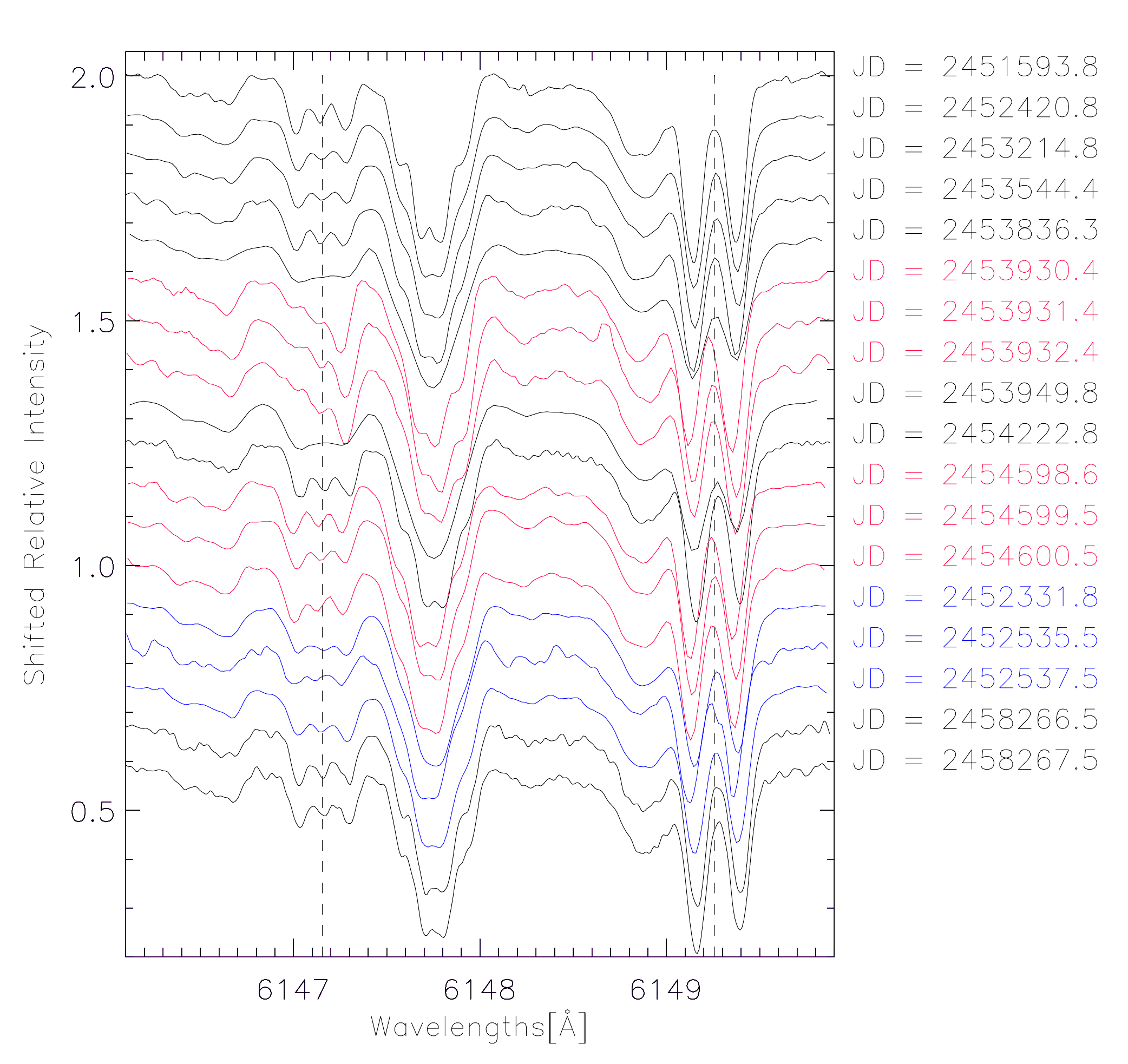}
\includegraphics[trim=-0.5cm 0.0cm 3.0cm 0.0cm,clip=true,width=0.457\textwidth]{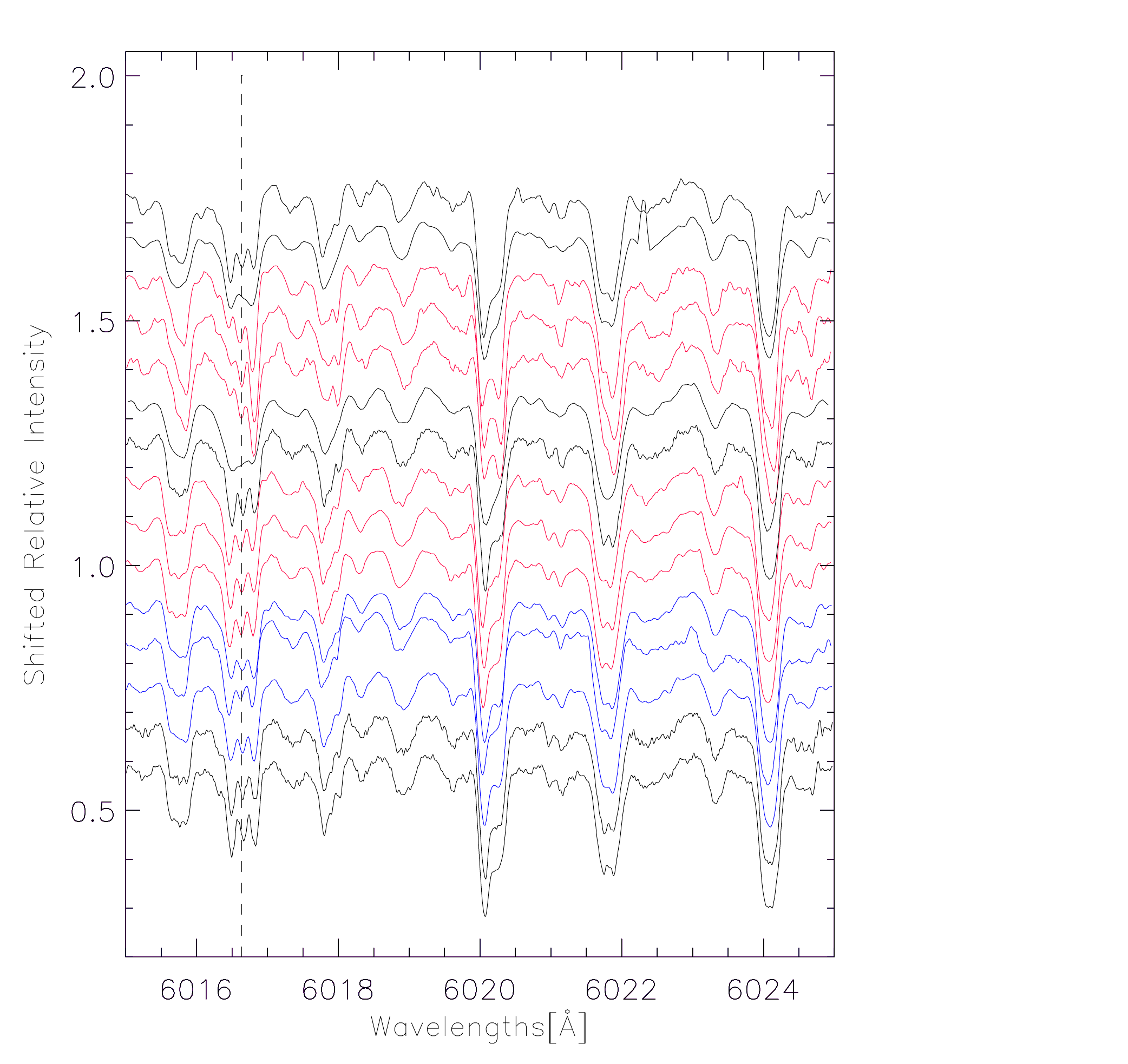}
\caption{HD\,137949. On three consecutive nights, from JD = 2\,453\,930 to 2\,453\,932
(2006 July 13, 14, and 15) SARG spectra presented red $\sigma$ Zeeman subcomponents deeper
than the blue ones, especially in chromium lines. Other series of observations
on consecutive days did not exhibit similar behaviour. ESPaDOns spectra obtained on
JD = 2\,453\,836 and 2\,453\,949 do not show clearly separated Zeeman subcomponents
because of the reduced resolution as compared to other spectrographs, they however reveal
that the phenomenon did not last longer than 110 days.}
\label{Fig_HD137949_Cr}
\end{figure*}

Even if all spectra of HD\,137949 recorded between 2002 and 2018 appear constant, with
equally strong red and blue $\sigma$ components for all species, in the SARG spectra
obtained on 2006 July 13, 14 and 15 the blue $\sigma$ components appear much weaker
than the red ones. This difference is particularly large for chromium
lines, only marginal for iron lines. We have no idea how persistent and recurrent this
phenomenon might turn out; we can only state that it started later than 2006 April 10,
and that it ended before 2006 August 1 (Fig.\,\ref{Fig_HD137949_Cr}). 
Contemporaneous observations of other stars, e.g. HD\,216018 also presented in
this paper, rule out any technical problems that would result in spurious spectral line profiles. 
We are not aware of a similar spectral line change observed before in a magnetic chemically peculiar star.

\subsection{HD\,142070}

From Str{\"o}mgren photometry, \cite{Adelman2001} established HD\,142070 as a photometric
periodic variable with the largest amplitude in the v filter and the ephemeris:
HJD(v$^{min}$) = 2\,450\,837.499 + 3.37189$\pm$0.00007 E. M17 found this star to be also 
magnetically variable with a period of 3.3718$\pm$0.0011\,d. 

\begin{figure}\center
\includegraphics[trim={0.3cm 0cm 0cm 0cm},width=0.50\textwidth]{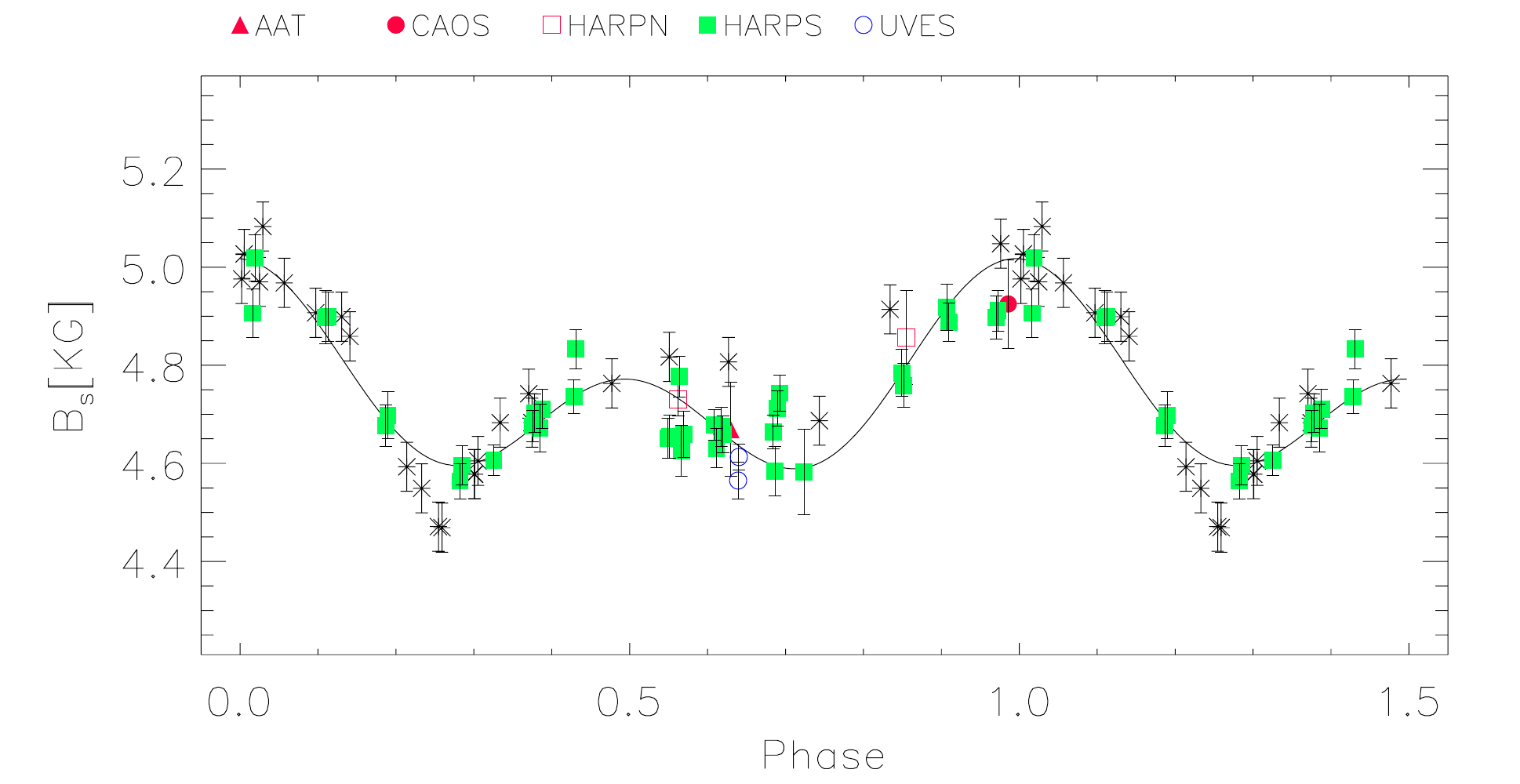}
\includegraphics[trim={0.3cm 0cm 0cm 0cm},width=0.50\textwidth]{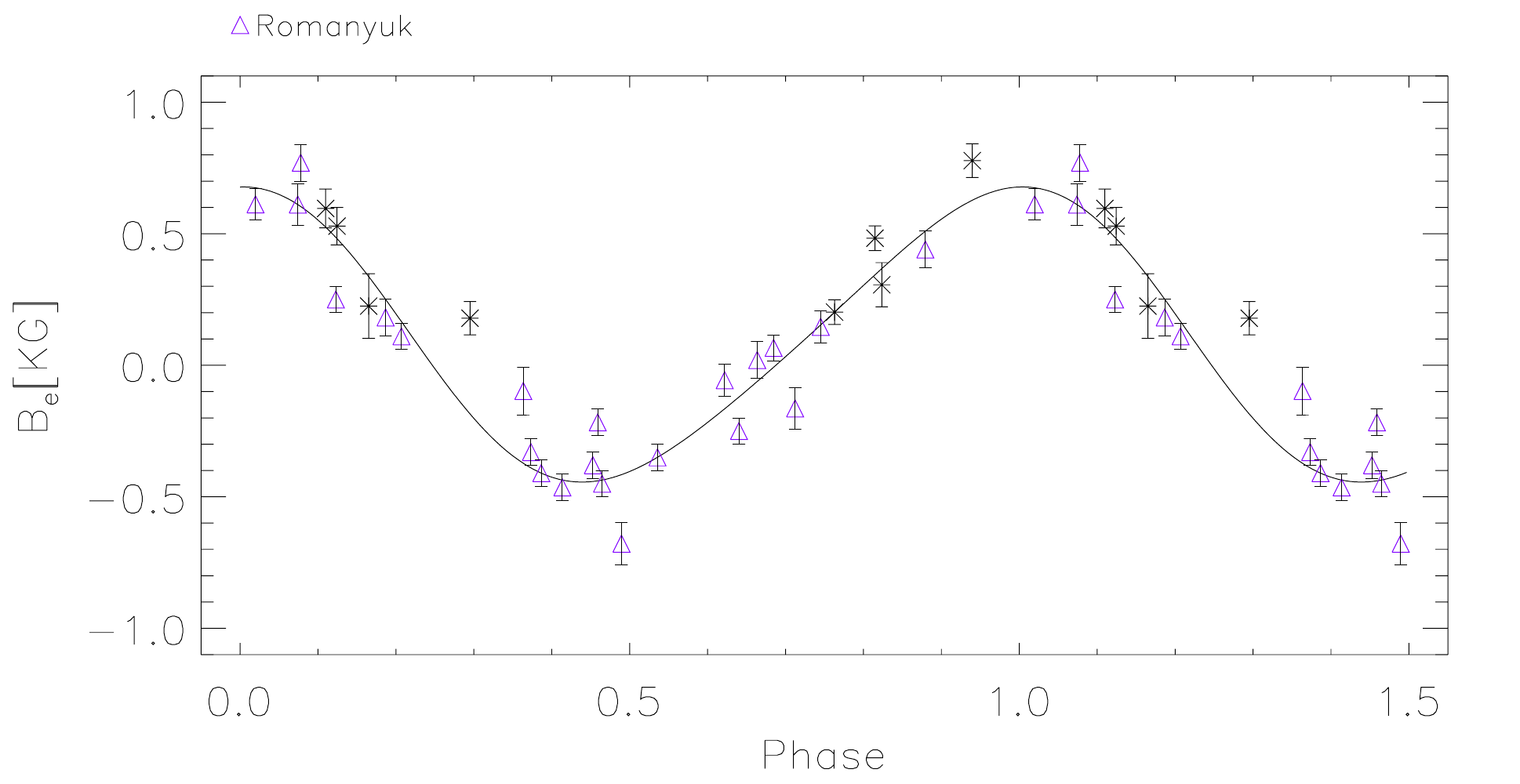}
\includegraphics[trim={0.3cm 0cm 0cm 0cm},width=0.50\textwidth]{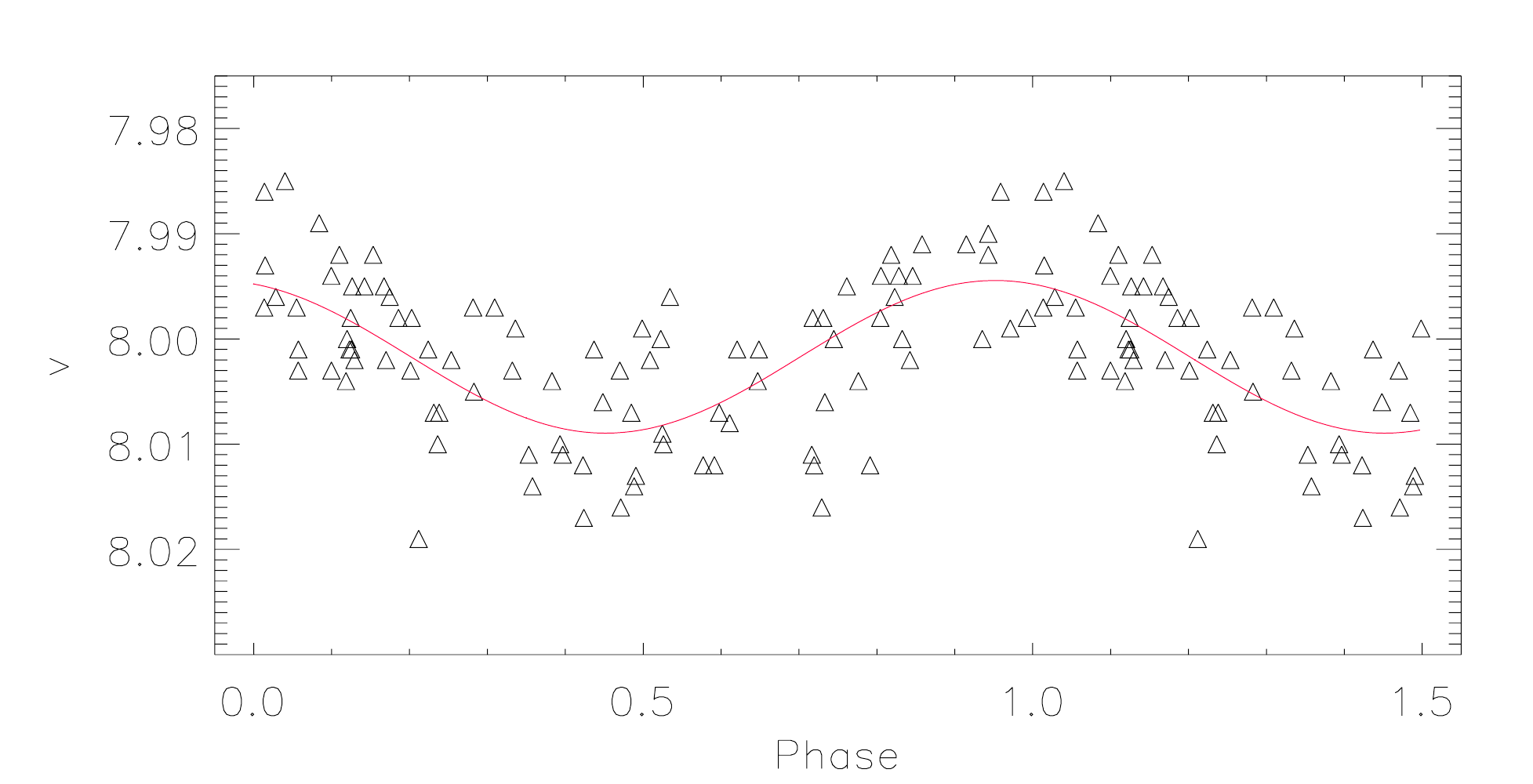}
\caption{HD\,142070. $B_s$, $B_e$ and Str{\"o}mgren $\rm v$ magnitude variations,
folded with the 3.3721\,d period.}
\label{Fig_HD142070}
\end{figure}

We have obtained 4 new spectra of this star with CAOS, UVES and HARPS-North, in addition
we have retrieved 2 UVES and 35 HARPS spectra from ESO archive. Our measurements of $B_s$ are
listed in (Table\,\ref{Tab_HD142070}). The product $LS(B_s, B_e,$v$)$ of the periodograms
of all available $B_s$ measurements, of $B_e$ data from M17 and \cite{Romanyuk2014}, and
of  Str{\"o}mgren $\rm v$ photometry \citep{Adelman2001} peaks at 3.3721$\pm$0.0002\,d.
Fig.\,\ref{Fig_HD142070} shows the double wave variation of $B_s$ whose extrema coincide
in phase with the $B_e$  maxima, based on the ephemeris given in Table\,\ref{Tab_Periods}.

\subsection{HD\,144897}

The variability period of the surface magnetic field of HD\,144897 was determined by M17
to 48.57$\pm 0.15$\,d from data collected over 7.1~years.

\begin{figure}\center
\includegraphics[trim={0.3cm 0cm 0cm 0cm},width=0.50\textwidth]{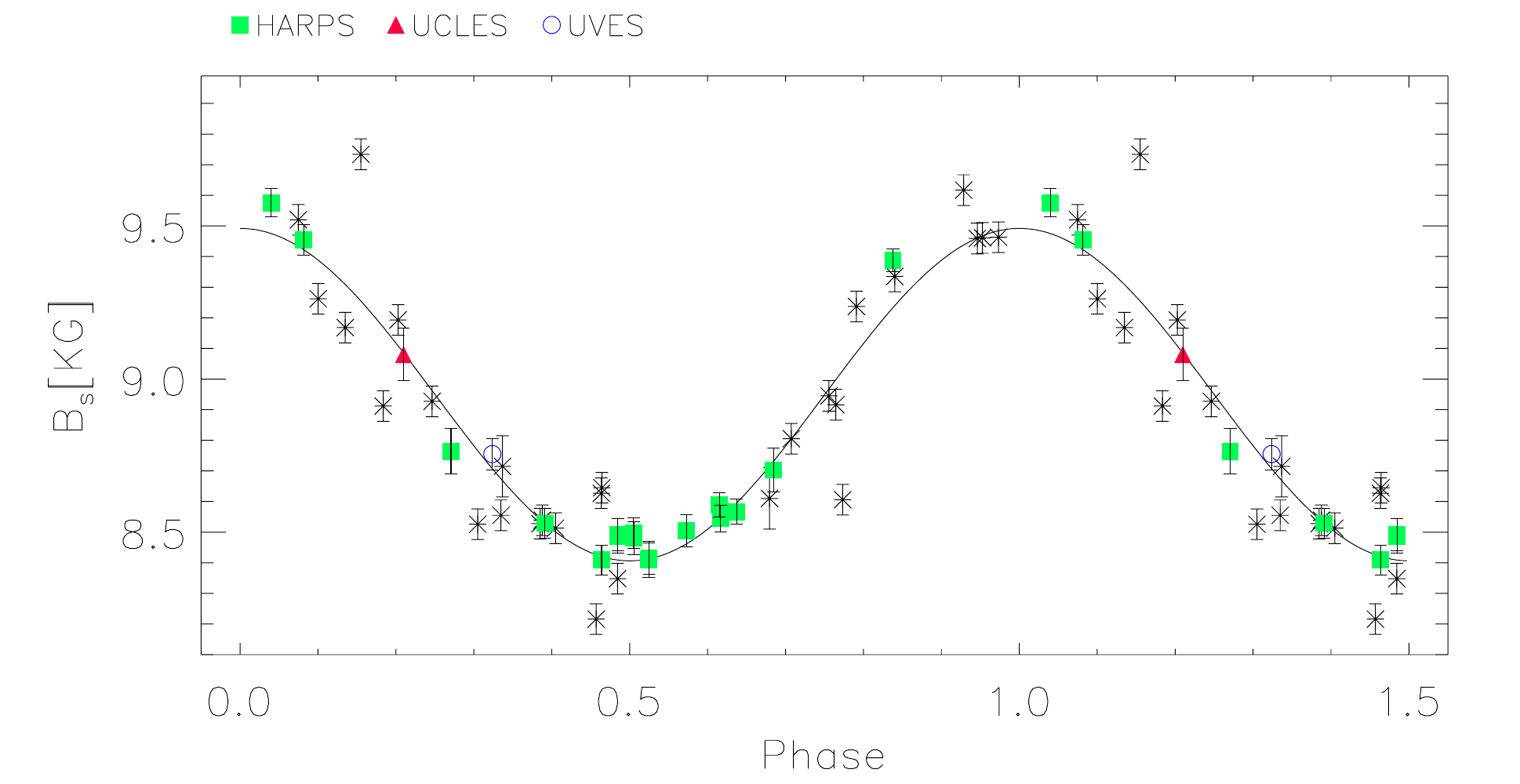}
\includegraphics[trim={0.3cm 0cm 0cm 0cm},width=0.50\textwidth]{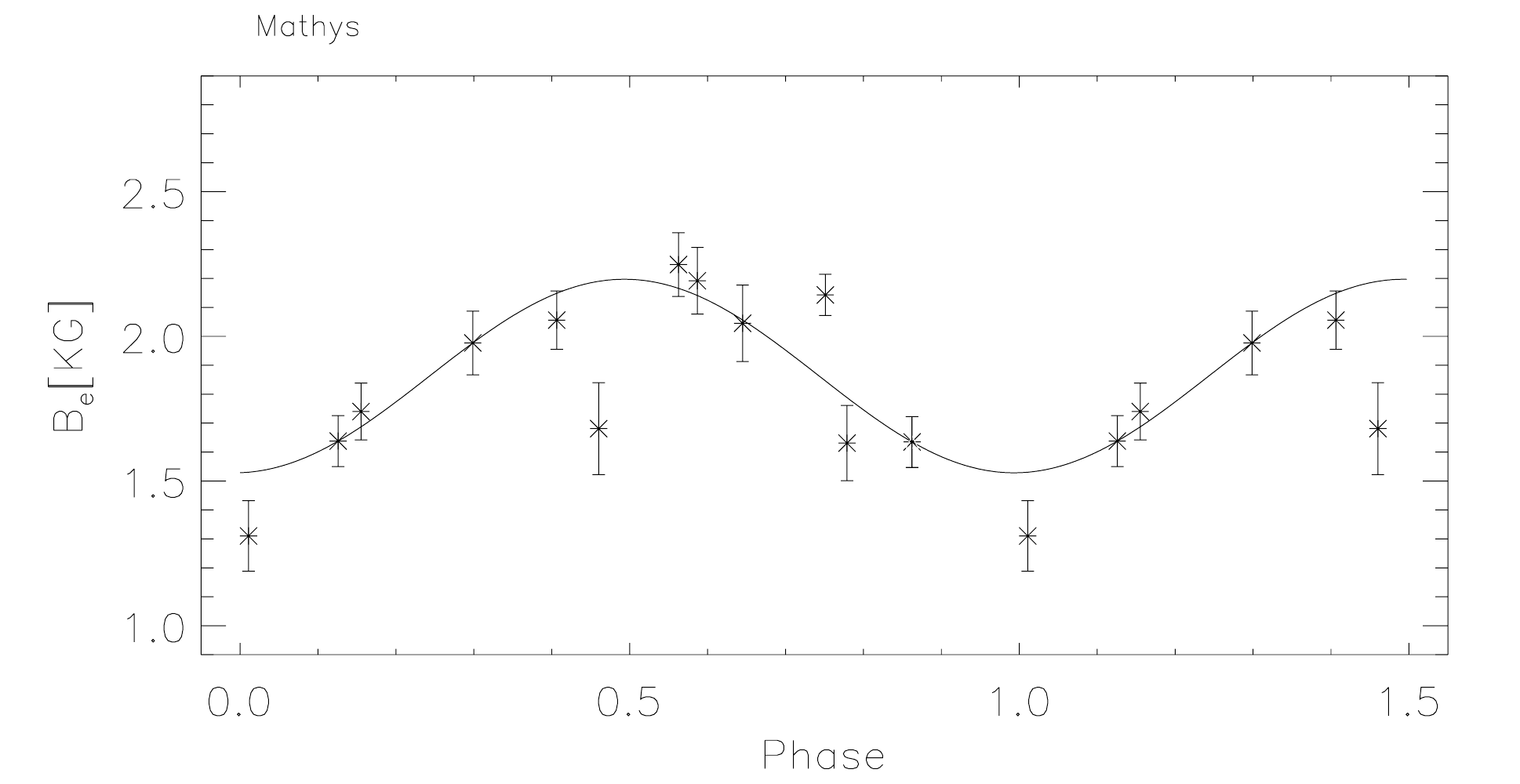}
\caption{HD\,144897. $B_s$ and $B_e$ variability, folded with the 48.60\,d period. }
\label{Fig_HD144897}
\end{figure}

We have obtained 1 spectrum of
HD\,144897 with UCLES and retrieved 1 UVES spectrum and 28 HARPS spectra from the ESO
archive. M17 and our values of $B_s$ (Table\,\ref{Tab_HD144897}) cover a total range of
22.5~years. A Lomb-Scargle analysis gives 48.60$\pm 0.02$\,d. Fig.\ref{Fig_HD144897}
shows all available measurements of $B_s$ and $B_e$ \citep{Mathys2017}, phased with
the ephemeris given in Table\,\ref{Tab_Periods}. Both variations are fairly sinusoidal
and the $B_s$ minimum in coincidence with the $B_e$ maximum is proof of a magnetic
field geometry far from a centred dipole configuration.

\subsection{HD\,150562}

$B_s$ measurements of HD\,150562 have been published by M97 and M17, spanning an interval
of 4.5~years, Mathys and coworkers concluded that the variability period of this star
exceeds 4.5~years. 

\begin{figure}\center
\includegraphics[trim={0.3cm 0cm 0cm 0cm},width=0.50\textwidth]{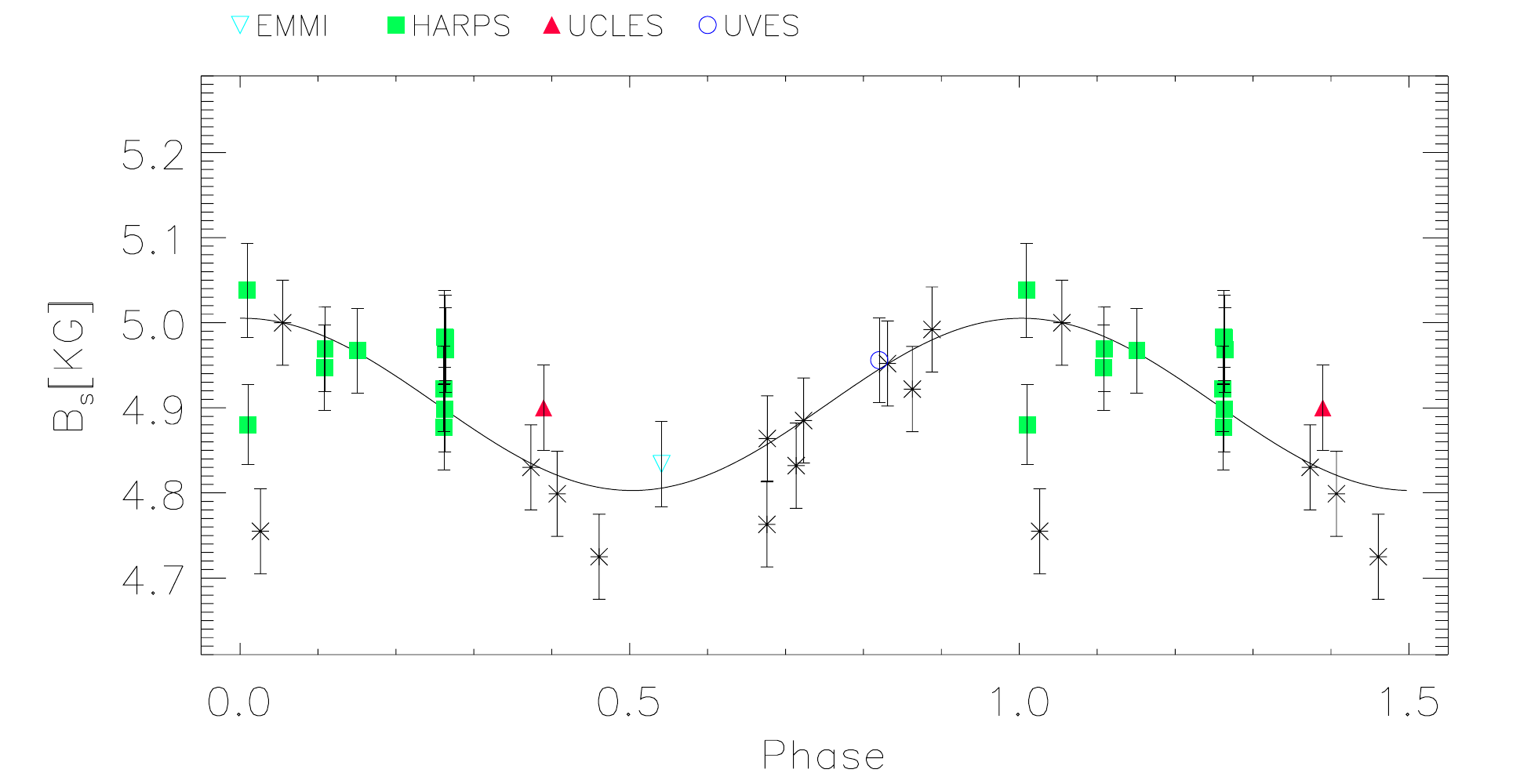}
\includegraphics[trim={0.3cm 0cm 0cm 0cm},width=0.50\textwidth]{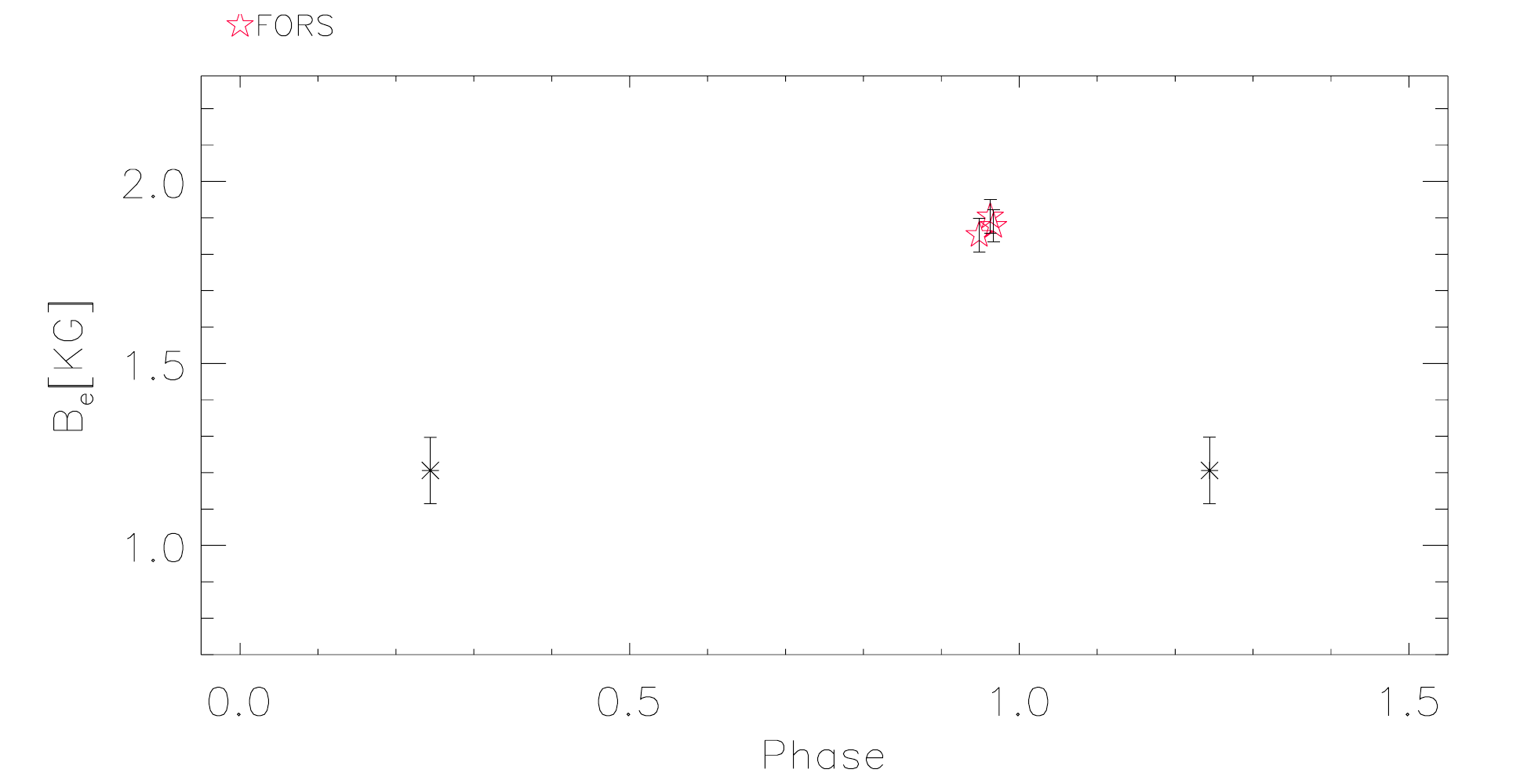}
\caption{HD\,150562. $B_s$ and $B_e$ variability, folded with the 2100\,d period.
$B_e$ measurements with FORS are from \citet{Bagnulo2015}.}
\label{Fig_HD150562}
\end{figure}

We have observed this chemically peculiar star once with UCLES, and we have obtained from ESO archive
1 EMMI, 1 UVES and 11 HARPS spectra.
With our $B_s$ values (Table\,\ref{Tab_HD150562}), the coverage now
extends over 21.3~years. Lomb-Scargle analysis of our and Mathys' measurements produces
a periodogram with the highest peak at 2100~days. Fig.\,\ref{Fig_HD150562} shows the
$B_s$ data folded with this period. Measurements of $B_e$ by \cite{Bagnulo2015} and M17,
plotted in the same figure, are too scanty to draw conclusions as to the variability.

At the phase of the maximum, we find an "unexpected" $B_s$ value obtained on
JD = 2\,450\,171.802 that M17 ascribed to an unidentified cosmic ray. At the same
phase, we find another "unexpected" value measured in a HARPS spectrum acquired on
JD = 2\,454\,338.582. HD\,150562 should be observed in 2024 during the next $B_s$
maximum to check if this is the result of chance or if some physical reason can be
identified (e.g. partial occultation of the visible stellar disk by a secondary star).

\subsection{HD\,154708}

The Zeeman subcomponents of the Fe{\sc ii} 6149.258\,{\AA} line of HD\,154708 are
distant by $\Delta\lambda >  1$\,{\AA}, implying that in the linear Zeeman
approximation the field is about 24000\,G, one of the strongest known fields for a
non degenerate star \citep{Hubrig2005}. 
From $B_e$ measurements, the variability period of
HD\,154708 has been determined by \cite{Hubrig_HD154708} to 5.3666$\pm$0.0007\,d,
whereas \cite{Landstreet2014} arrived at 5.363$\pm 0.003$\,d.  A geometrical model of the magnetic field of HD\,154708 has
been published by \cite{Stift_HD154708}.

\begin{figure}\center
\includegraphics[trim={0.3cm 0cm 0cm 0cm},width=0.50\textwidth]{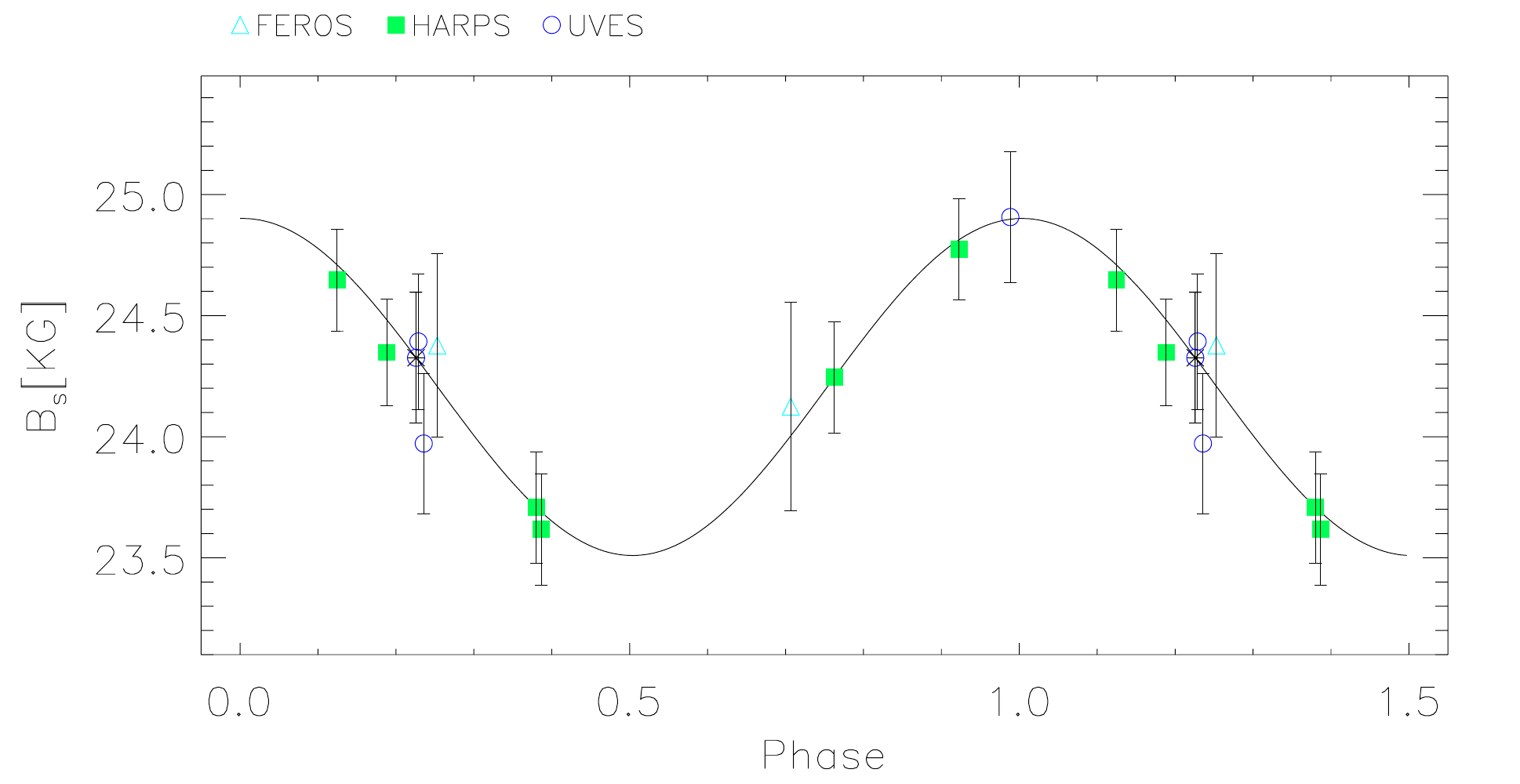}
\includegraphics[trim={0.3cm 0cm 0cm 0cm},width=0.50\textwidth]{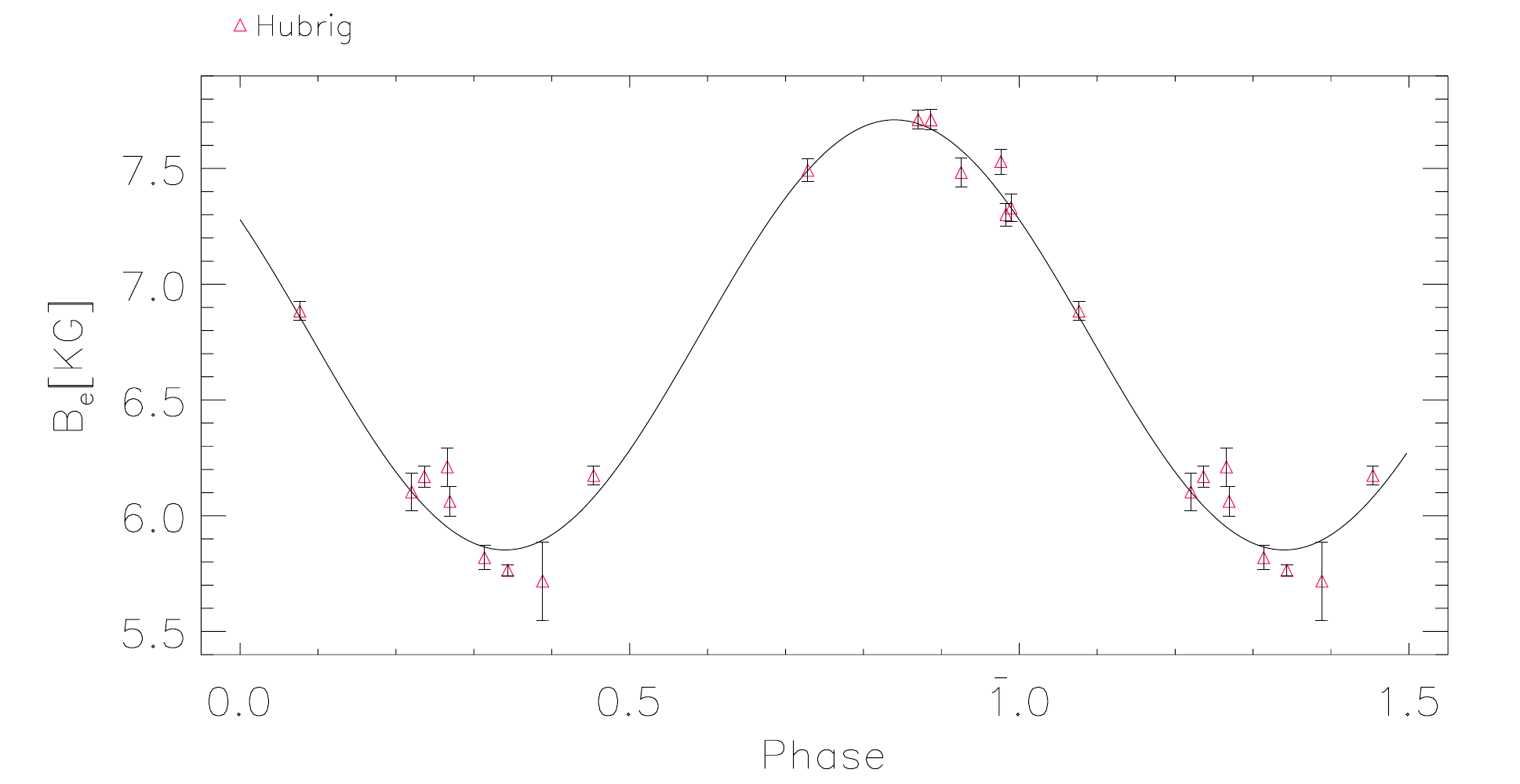}
\includegraphics[trim={0.3cm 0cm 0cm 0cm},width=0.50\textwidth]{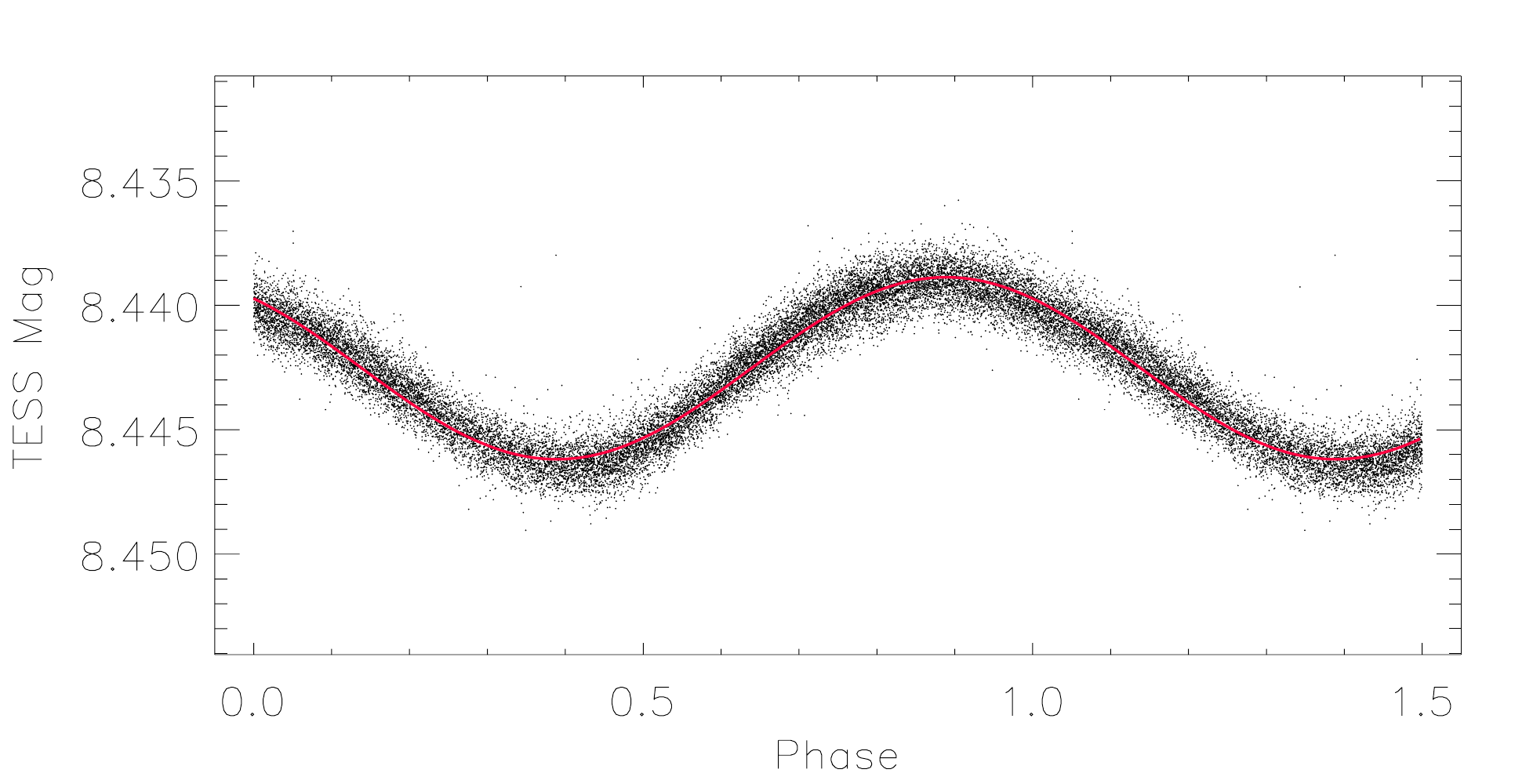}
\caption{HD\,154708. Variability in $B_s$, $B_e$, and TESS photometry, folded with the
5.367\,d period. $B_e$ measurements by \citet{Hubrig_HD154708} are based on  hydrogen
lines.}
\label{Fig_HD154708}
\end{figure}

We have obtained 4 UVES, 6 HARPS and 2 FEROS spectra of HD\,154708 from ESO archive.
Our $\Delta\lambda$ measurements have been converted to $B_s$ values (Table\,\ref{Tab_HD154708}) with
the help of eq.\,\ref{ZWF}. This means that we are neglecting the partial
Paschen-Back effect is expected to play a role on account of the strong magnetic
field \citep[see][]{Stift2008}. The resulting $B_s$ values will thus constitute lower
limits to the actual surface magnetic field, but they remain still representative
of its variability. Including in the Lomb-Scargle analysis the $B_e$ values published by
\cite{Hubrig_HD154708} and TESS magnitudes, the highest peak of the product
$LS(B_s, B_e, TESS)$ occurs at 5.367$\pm$0.001\,d. Fig.\,\ref{Fig_HD154708} shows
the variability in $B_s$, $B_e$, and TESS photometry, based on the ephemeris given
in Table\,\ref{Tab_Periods}. It appears that light variability is driven by the
longitudinal component of the field, while the $B_s$ variations are shifted in
phase by 0.15.

\subsection{HDE\,318107}

\begin{figure}\center        
\includegraphics[trim={0.3cm 0cm 0cm 0cm},width=0.50\textwidth]{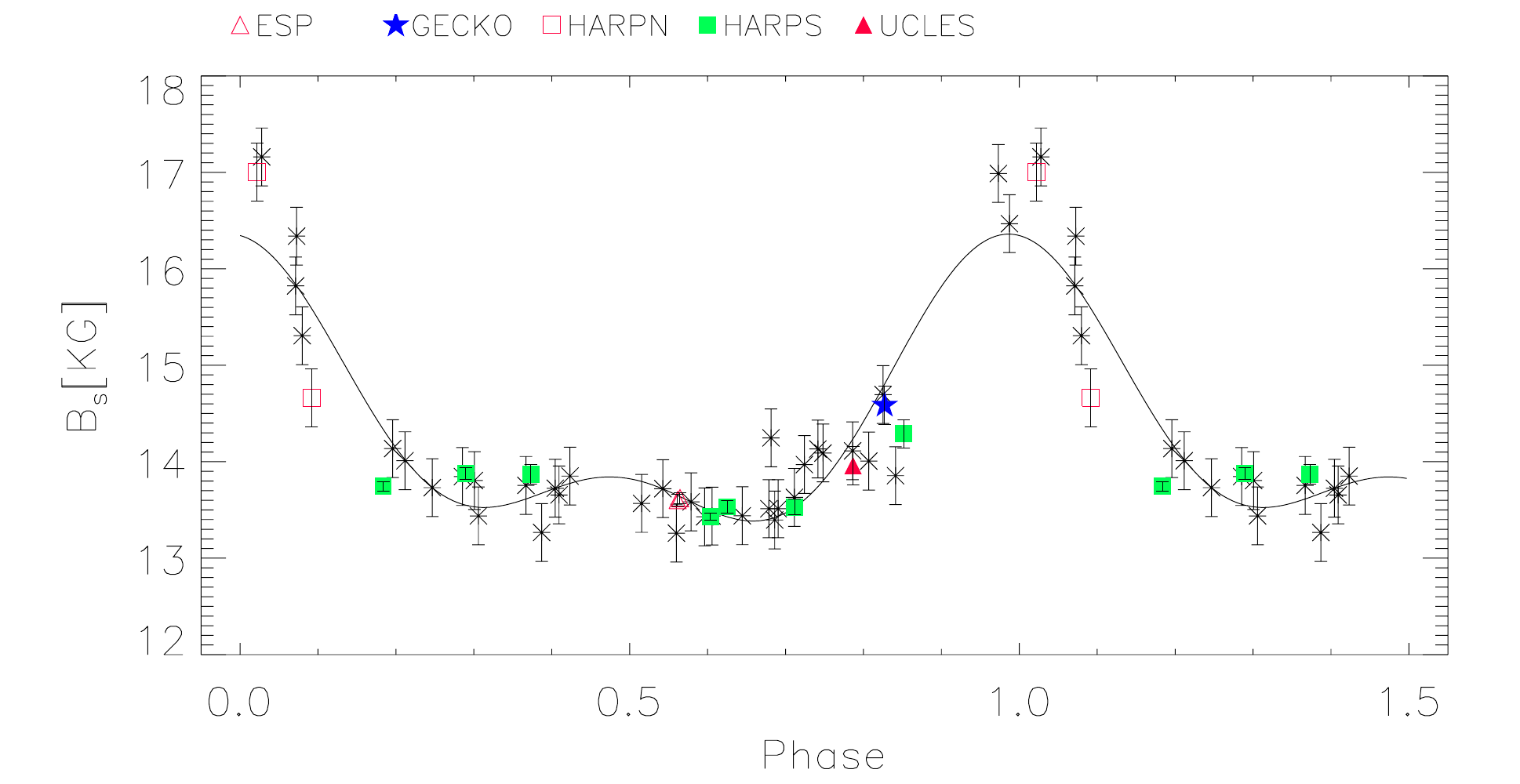}      
\includegraphics[trim={0.3cm 0cm 0cm 0cm},width=0.50\textwidth]{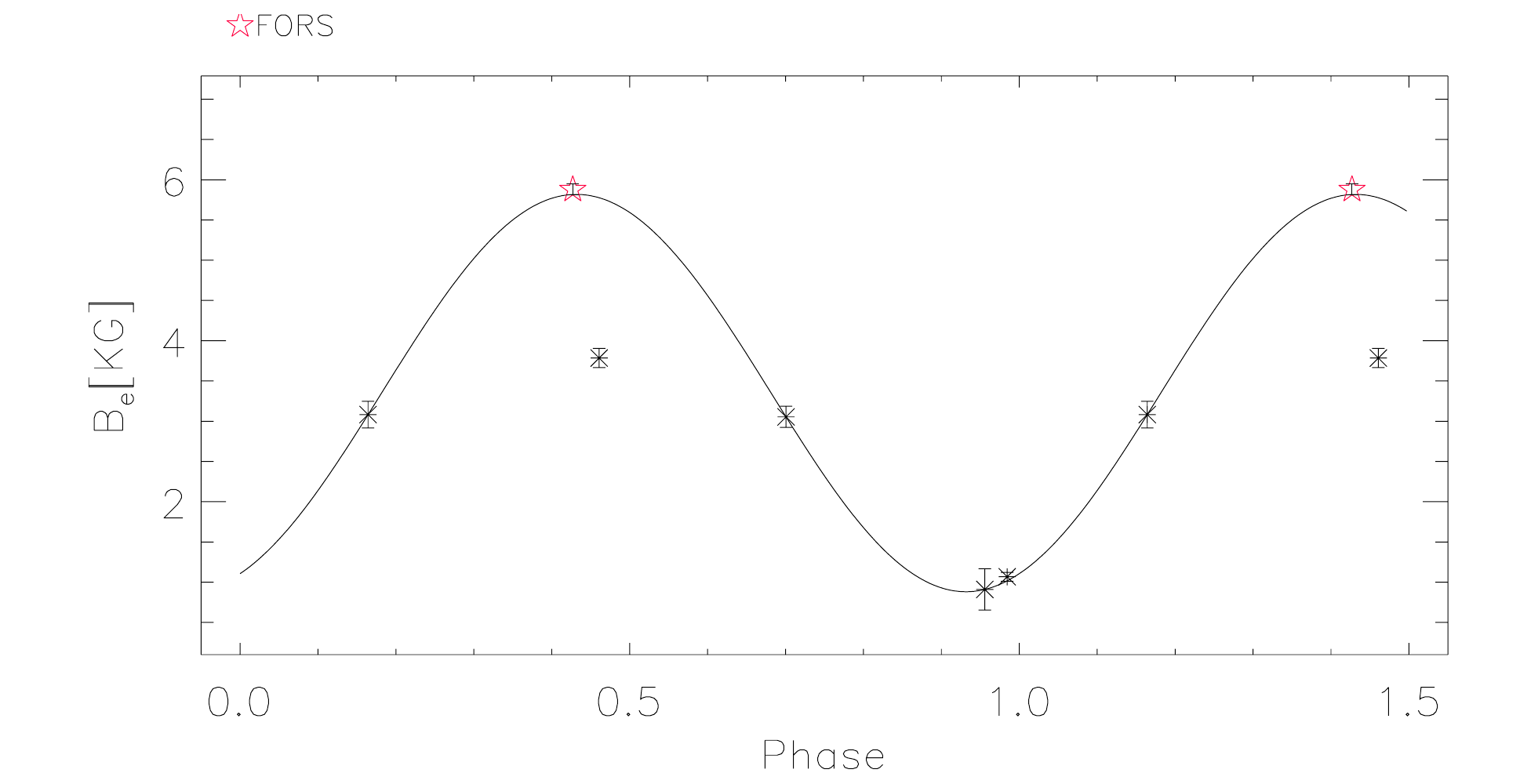}     
\caption{HD\,318107. Variations in $B_s$ and $B_e$ folded with the
         9.7089\,d period.}
\label{Fig_HD318107}
\end{figure}

The $B_s$ and $B_e$ variability period of 9.7088$\pm$0.0007\,d was determined by
\cite{Bailey2011_HDE318107}. We have obtained 7 HARPS spectra from the ESO archive,
1 GECKO spectrum, and 2 ESPaDOns spectra from the CFHT archive. In addition, we
have obtained 1 new spectrum with UCLES and 2 spectra with HARPS-North. Our 13 $B_s$ 
measurements (Table\,\ref{Tab_HD318107}) have been combined with values from M17 for a
Lamb-Scargle analysis that peaks at 9.7089$\pm 0.0002$ days. Fig.\,\ref{Fig_HD318107}
shows the periodic variability  in $B_s$ and $B_e$ (\cite{MathysHubrig1997},
\cite{Bagnulo2015}, M17) of HDE\,318107.

\subsection{HD\,165474}

\begin{figure}\center        
\includegraphics[trim={0.3cm 0cm 0cm 0cm},width=0.50\textwidth]{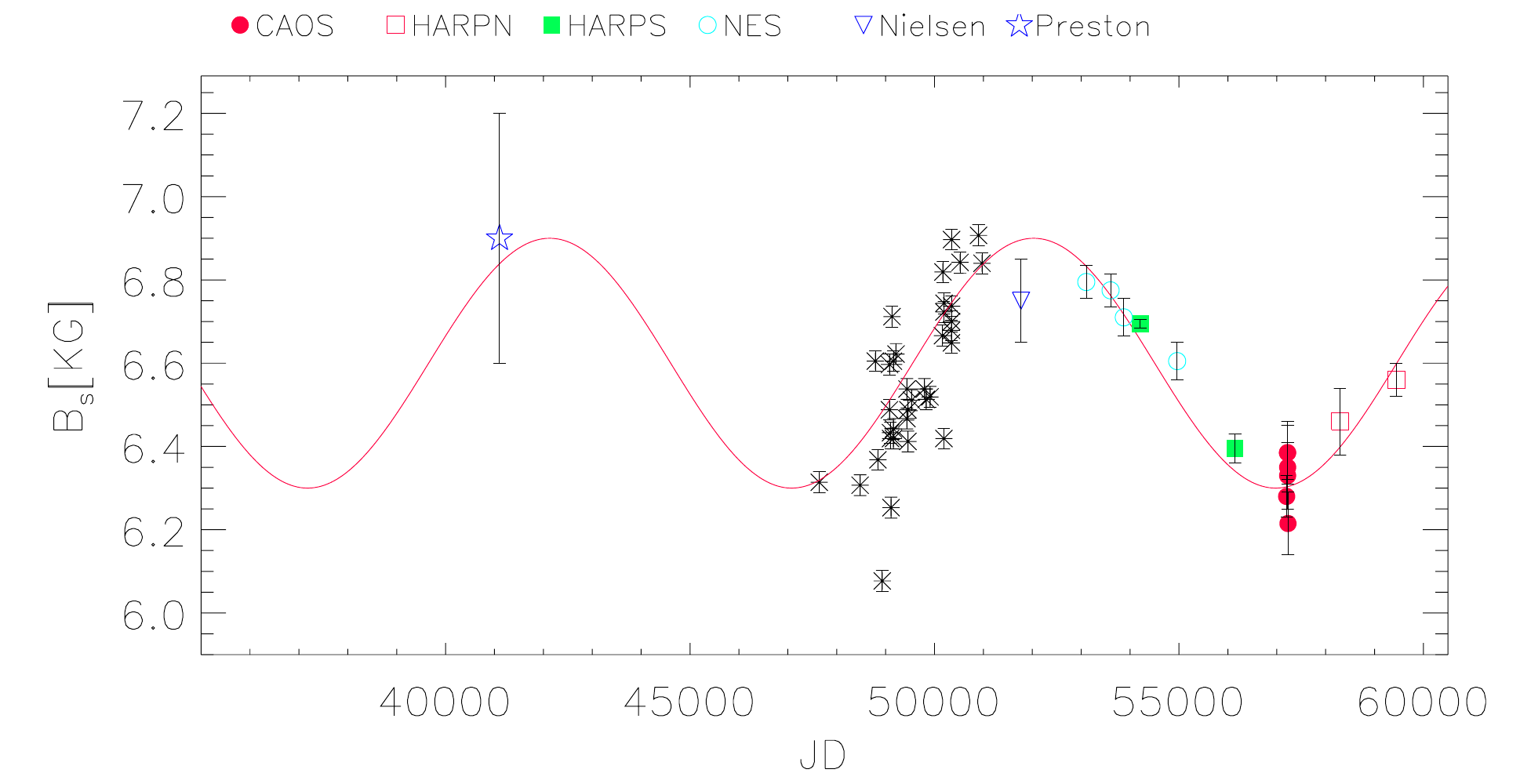}  
\includegraphics[trim={0.3cm 0cm 0cm 0cm},width=0.50\textwidth]{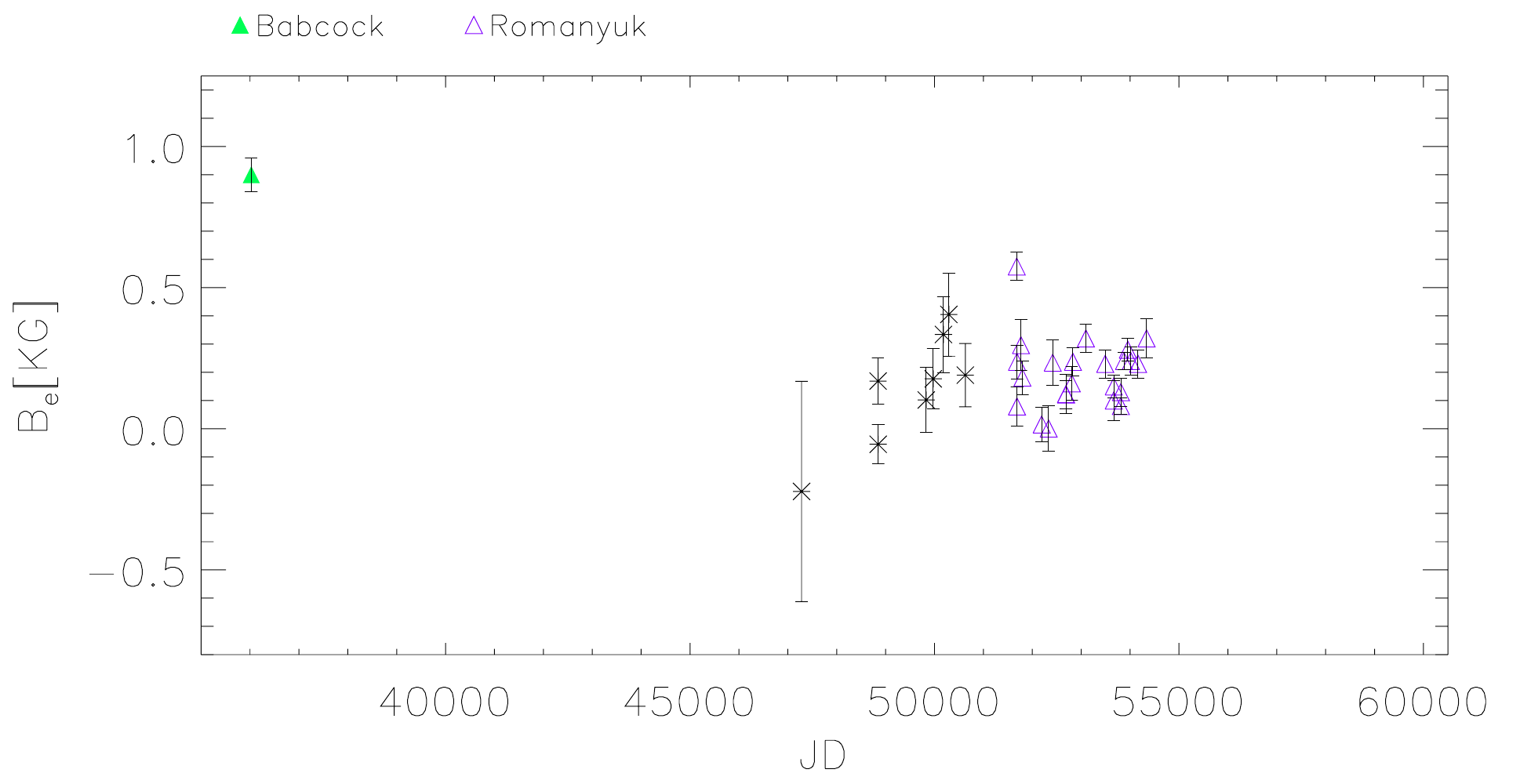}     
\caption{HD\,165474. $B_s$ and $B_e$ measurements over time. It appears that the
shortest possible period for the $B_s$ variability is 9900 days.}
\label{Fig_HD165474}
\end{figure}

M97 and M17 presented $B_s$ measurements of HD\,165474 obtained between 1989
(JD=2\,447\,642) and 1998 (JD=2\,450\,972), leading to the conclusion that the field is
continuously increasing and that the variability period largely exceeds 9 years.
On the other hand, our $B_s$ measurements between 2004 (JD=2\,453\,104) and 2015
(JD=2\,457\,229)  are steadily decreasing, with evidence of a new increase in the
HARPS-North data obtained in 2018 and 2021 (Table\,\ref{Tab_HD165474}). Including
$B_s$ measurements from \cite{Preston1971} and \cite{Nielsen2002}, under the hypothesis
of a single harmonic variation, the period can be no shorter than 9900 days (27.1\,yr):
Fig.\,\ref{Fig_HD165474}. The scarcity of $B_e$ \citep{Babcock1958, Mathys1994, Mathys1997, Romanyuk2014} data does not allow any definite
conclusions.

\subsection{HD\,166473}

\begin{figure}\center        
\includegraphics[trim={0.3cm 0cm 0cm 0cm},width=0.50\textwidth]{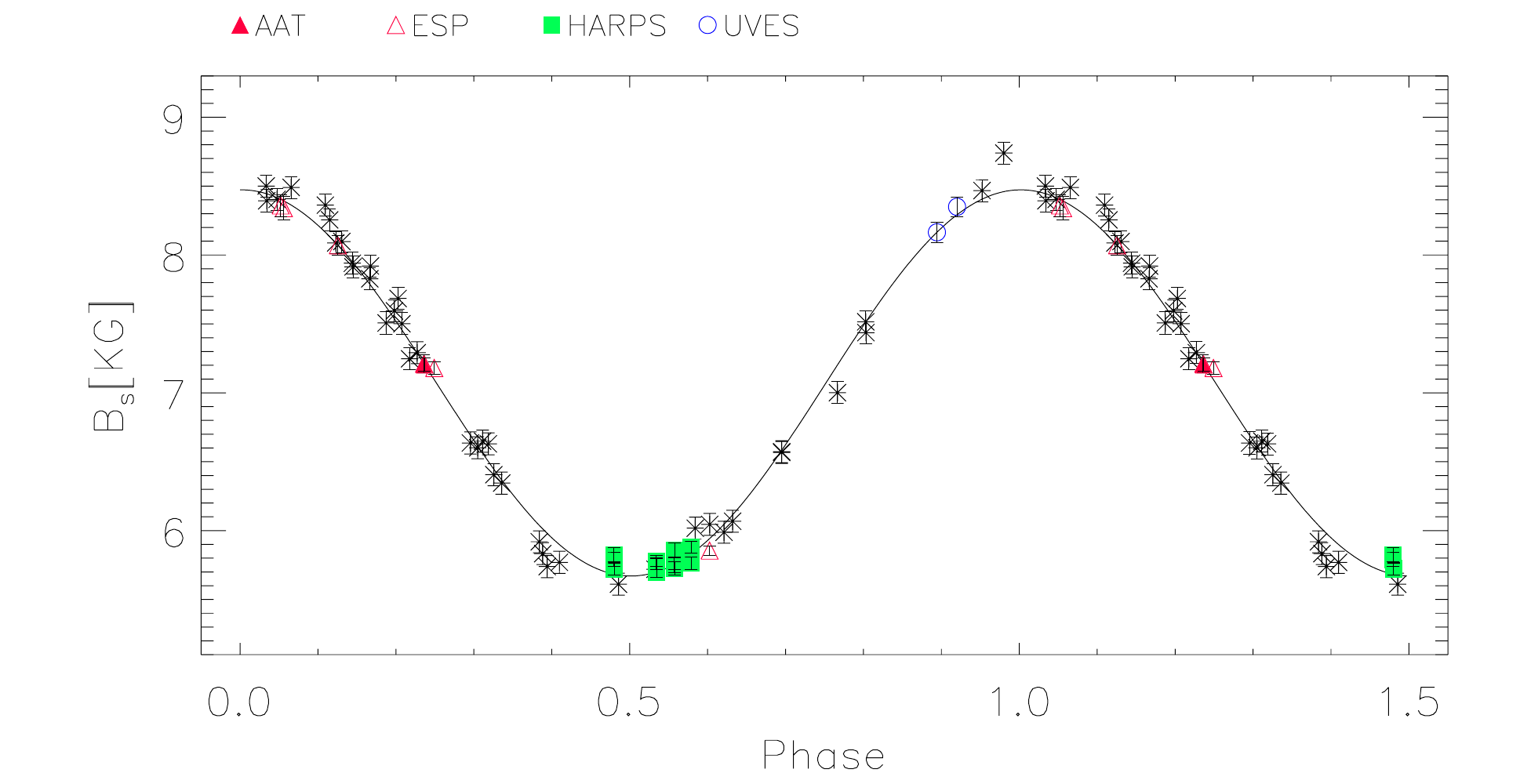}     
\caption{HD\,166473. $B_s$ variability, folded with the 3836\,d period. }
\label{Fig_HD166473}
\end{figure}

The $B_s$ variability of HD\,166473 has very recently been analysed by
\cite{Mathys2020_HD166473}, using spectra collected from ESO and ESPaDOns
archives. These authors determined a period of 3836 days. We too have obtained
HD\,166473 spectra from the ESO and CFHT archives and added 1 new measurement:
$B_s$ = 7210$\pm$65\,G, obtained by us with UCLES at the Anglo Australian
Telescope on JD = 2\,457\,235.008. Our measurements (Table\,\ref{Tab_HD166473})
confirm the published period, and Fig.\,\ref{Fig_HD166473} shows the $B_s$
variability, folded with the ephemeris determined by Mathys and coworkers
(Table\,\ref{Tab_Periods}).

\subsection{HD\,177765}

\begin{figure}\center
\includegraphics[trim={0.3cm 0cm 0cm 0cm},width=0.50\textwidth]{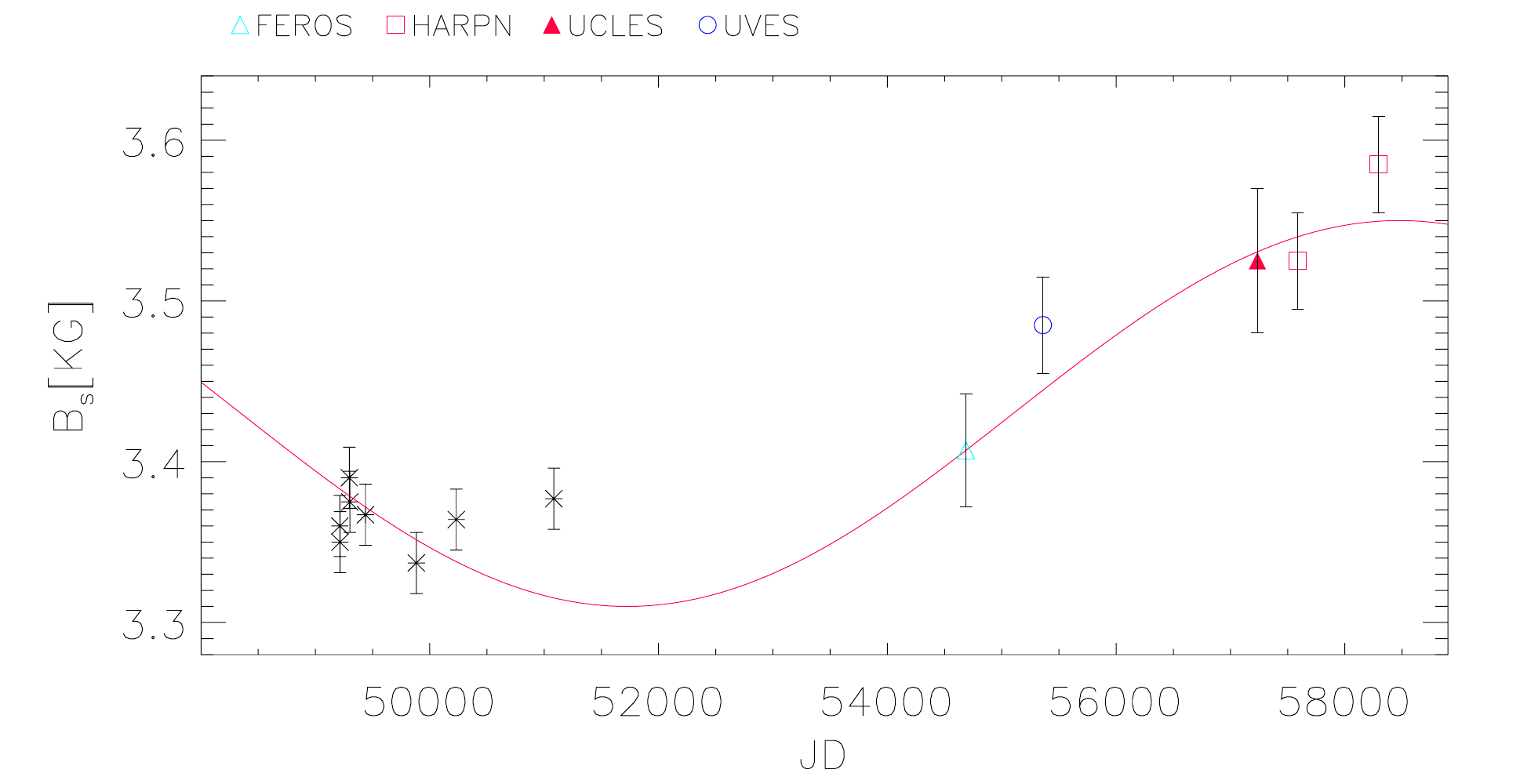}
\caption{HD\,177765. $B_s$ measurements over time. A sine fit to the data with the
shortest possible period of 13500 days is also plotted.}
\label{Fig_HD177765}
\end{figure}

Eight measurements of $B_s$ distributed from 1993 to 1998, plus one measurement taken
from the literature and obtained in 2010, led M17 to the conclusion that the variability
period of HD\,177765 exceeds 17 years. We have recovered 1 FEROS spectrum and 1 UVES
spectrum from the ESO archive and observed HD\,177765 once with UCLES, 2 times with
HARPS-North. Our $B_s$ measurements (Table\,\ref{Tab_HD177765}) extend the temporal
baseline from 6 years to almost 25 years and reveal a continuous increase with time.
Under the assumption of a simple harmonic variation, we estimate the shortest possible
period at 37 years (Fig.\,\ref{Fig_HD177765}).

\subsection{HD\,178892}
From $B_e$ measurements and HIPPARCOS photometry, \cite{Semenko_HD178892} determined
the variability period of HD\,178892 to 8.2549 days. We retrieved 1 HARPS spectrum,
67 UVES spectra obtained between JD = 2\,455\,351.334 and JD = 2\,455\,351.396
(which subsequently we combined into a single spectrum), 1 NES spectrum and 5 ESPaDOns
spectra. In addition, we observed this star twice with HARPS-North. Measured values
of $B_s$ are listed in Table\,\ref{Tab_HD178892}. 

\begin{figure}\center
\includegraphics[trim={0.3cm 0cm 0cm 0cm},width=0.50\textwidth]{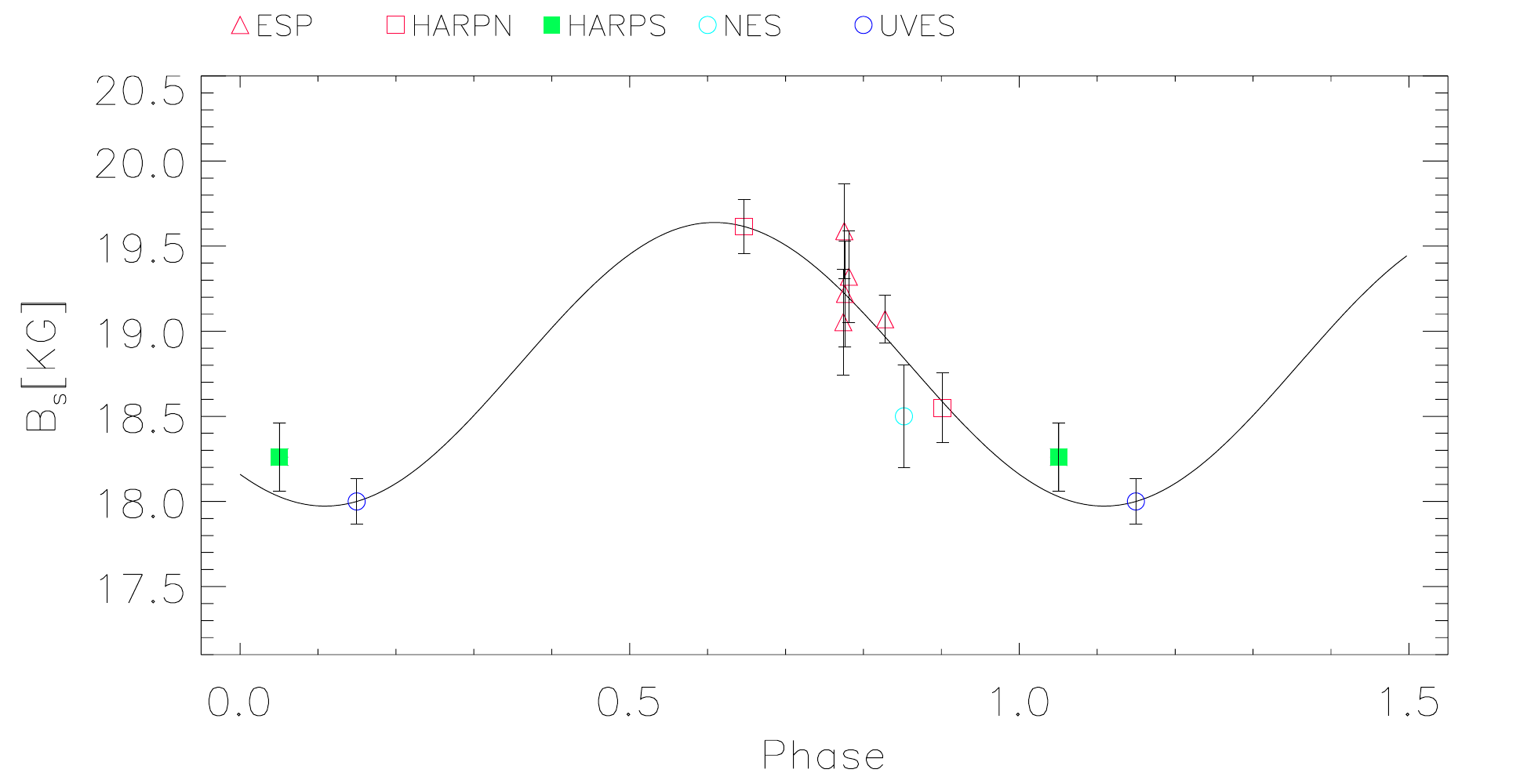}
\includegraphics[trim={0.3cm 0cm 0cm 0cm},width=0.50\textwidth]{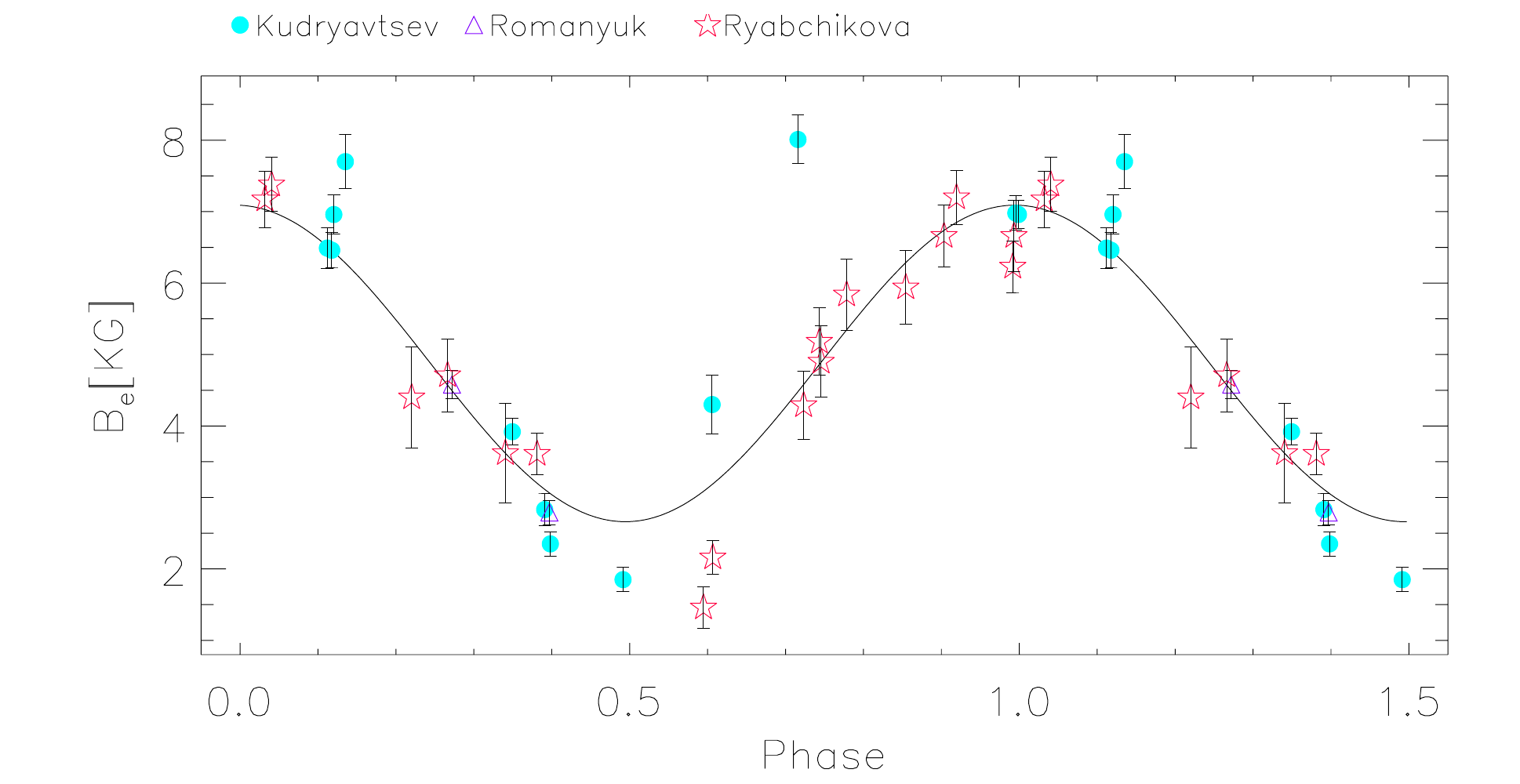}
\caption{HD\,178892: $B_s$ and $B_e$ variability, folded with the 8.2572\,d \,period.}
\label{Fig_HD178892}
\end{figure}

The Lomb-Scargle analysis of our $B_s$ measurements and $B_e$ data from
\cite{Ryabchikova_HD178892}, \cite{Kudryavtsev_HD178892} and \cite{Romanyuk_HD178892}
results in a $LS(B_s, B_e)$ with a peak at 8.2572$\pm$0.0016\,d.
Figure\,\ref{Fig_HD178892} displays the folded $B_s$ and $B_e$ variations.

\subsection{HD\,187474}

\begin{figure}\center
\includegraphics[trim={0.3cm 0cm 0cm 0cm},width=0.50\textwidth]{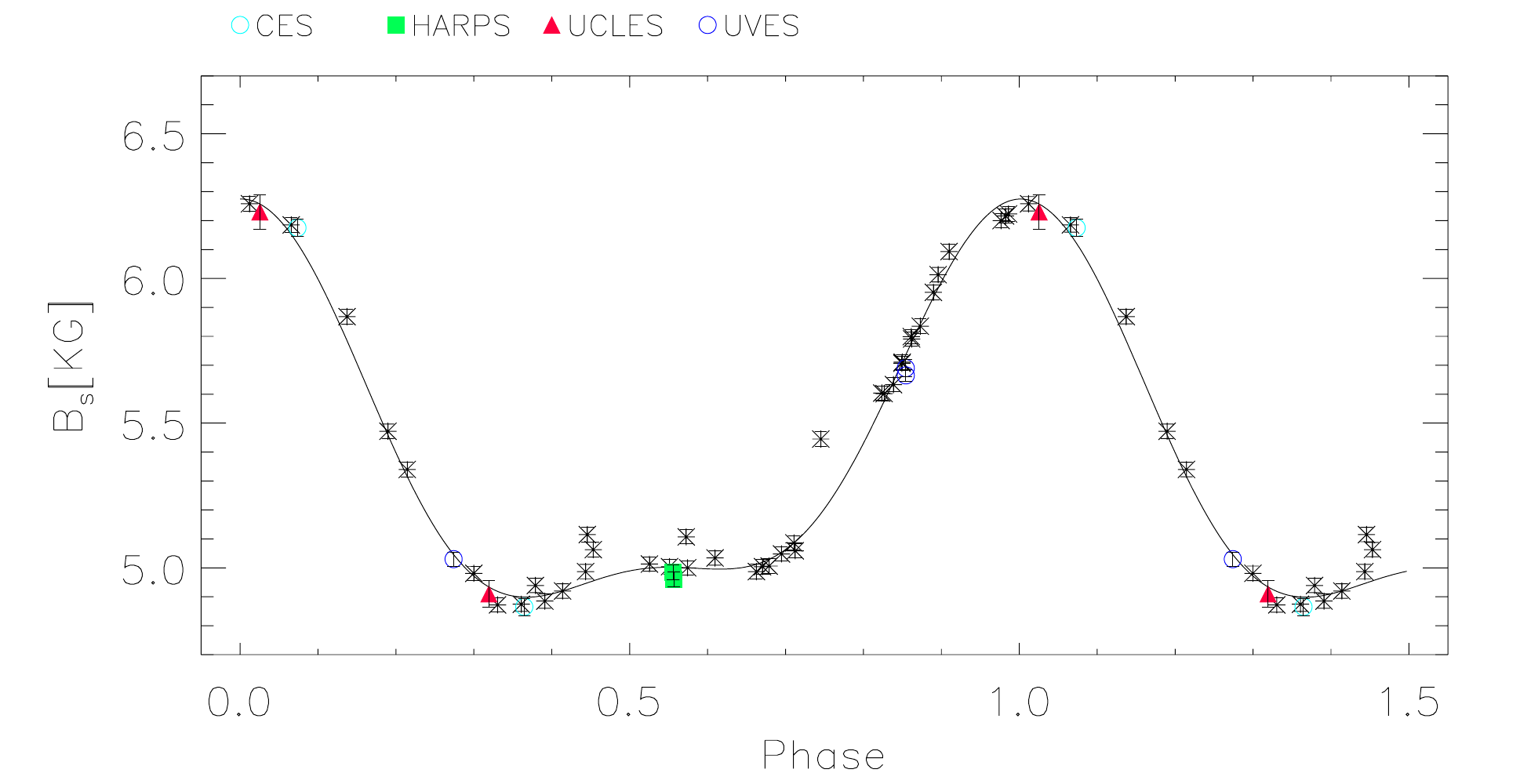}
\includegraphics[trim={0.3cm 0cm 0cm 0cm},width=0.50\textwidth]{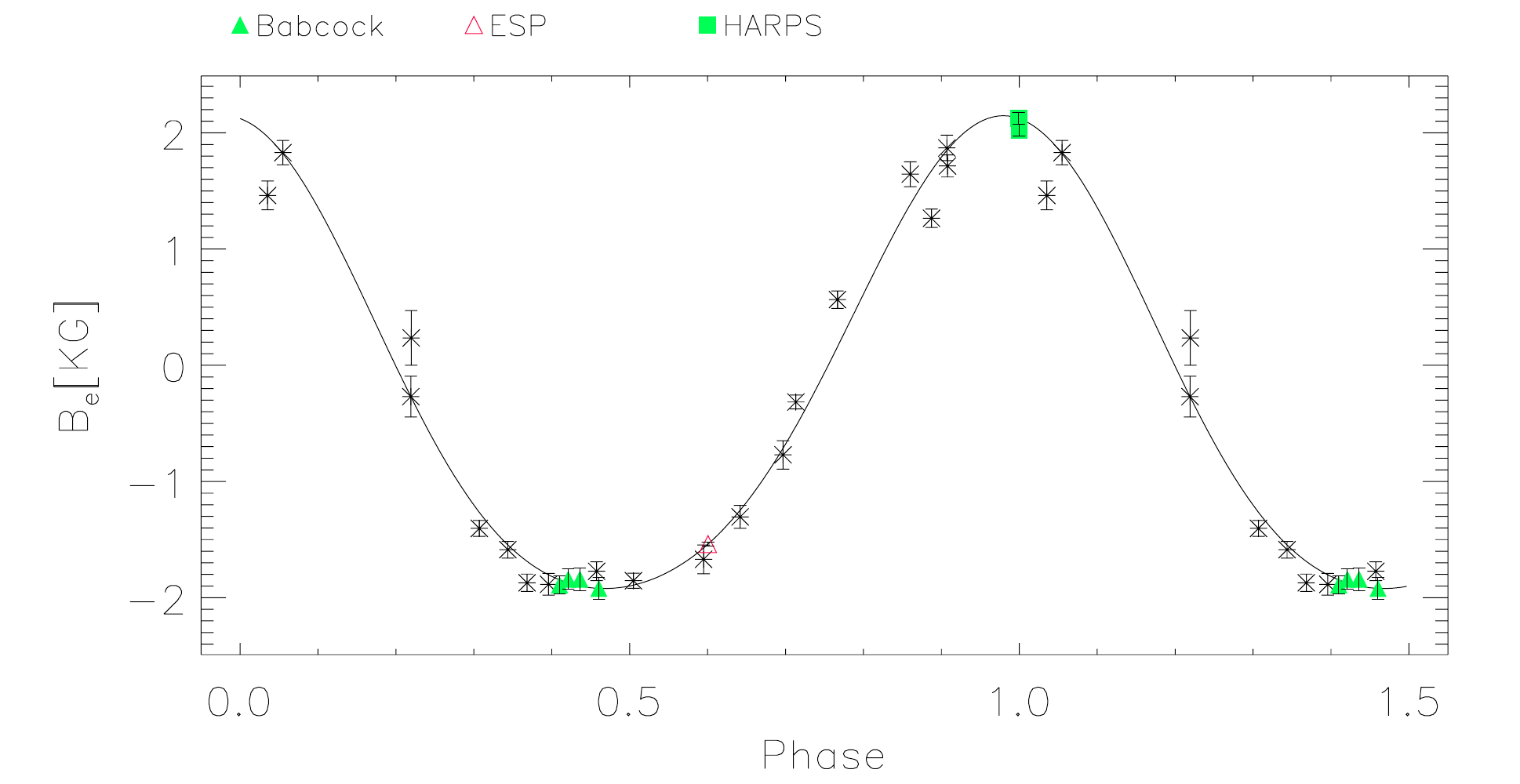}
\caption{HD\,187474. $B_s$ and $B_e$ variability, folded with the 2324\,d period.}
\label{Fig_HD187474}
\end{figure}

\cite{Mathys1991} established a variability period of 2345\,d for HD\,187474
from $B_s$ measurements collected in the interval from JD = 2\,447\,287 to
2\,451\,084 ($\sim$10 yr), and $B_e$ measurements obtained between JD = 2\,436\,002
and 2\,447\,280 ($\sim$31 years). We have acquired high-resolution spectra of
HD\,187474 twice with UCLES and twice with HARPS, extending the $B_s$ time coverage
(Table\,\ref{Tab_HD187474}) by more than 27 years. We have also observed HD\,187474
with the HARPS polarimeter -- see \cite{Gonzalez2014} for details - obtaining 2
measurements of $B_e$:  
\begin{center}HJD\hspace{15mm} $B_e$\\
2\,456\,143.760\hspace{5mm}  2125$\,\pm\,$50\,G\\
2\,456\,145.685\hspace{5mm}  2025$\,\pm\,$50\,G\\
\end{center}
The periodogram of our $B_s$ measurements together with those from M17, and the
periodogram of $B_e$ measurements by \cite{Babcock1958}, \cite{Mathys1991},
\cite{MathysHubrig1997}, M17, \cite{Sikora2019} and by us result in a
$LS(B_s, B_e)$ function with the highest peak at 2324$\pm$40\,d.
Fig.\,\ref{Fig_HD187474} shows the folded $B_s$ and $B_e$ measurements.

An almost constant surface field of some 5\,kG is observed during those phases (from
0.3 to 0.7) where the longitudinal field changes from zero to $+2$\,kG and then back
to zero. When the surface field increases from 5\,kG to reach a maximum at 6.3\,kG,
the longitudinal field turns negative, attaining $-$2\,kG (variability phase from 0.7
to 1.3).

\subsection{HD\,188041}

\begin{figure}\center
\includegraphics[trim={0.3cm 0cm 0cm 0cm},width=0.50\textwidth]{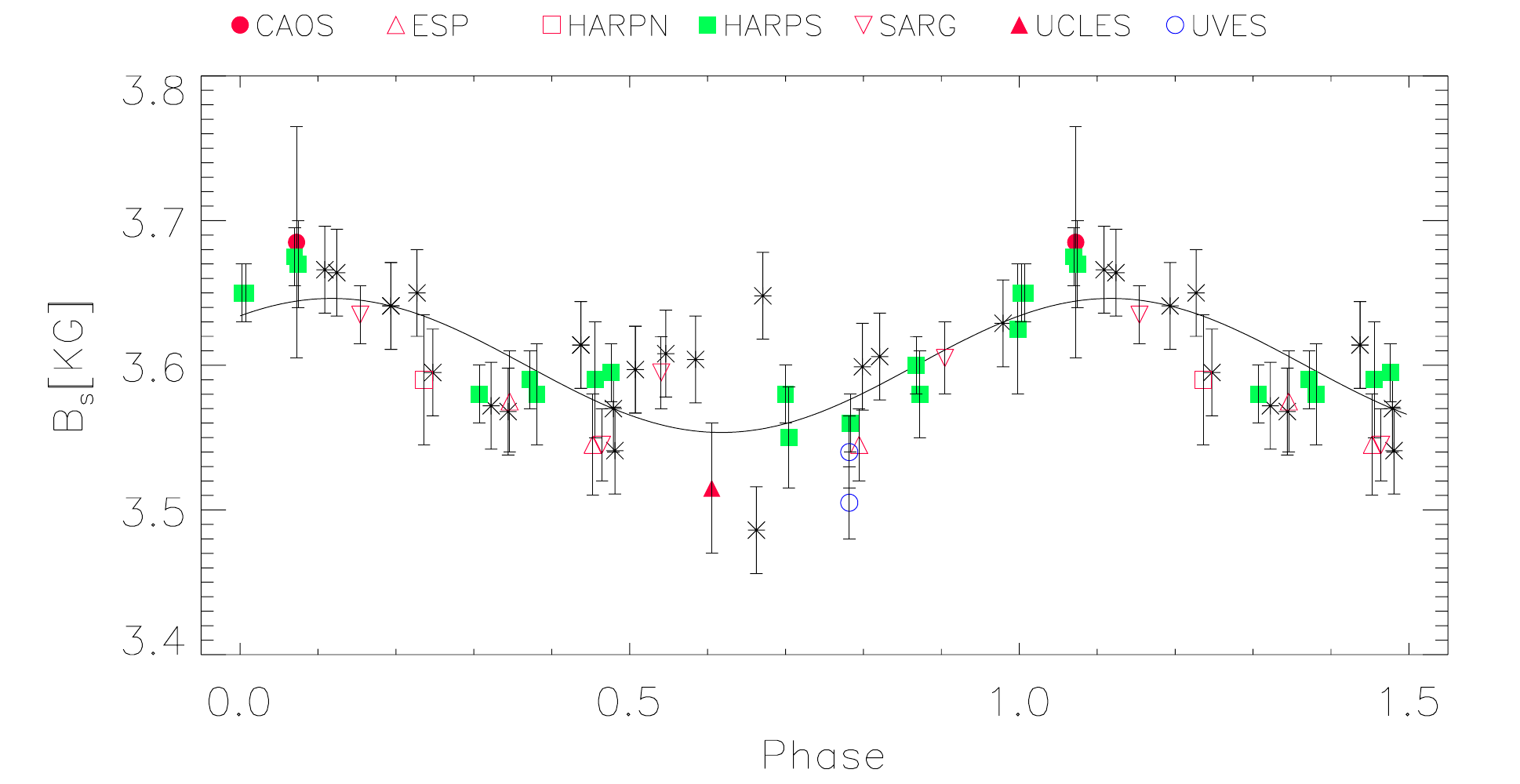}
\includegraphics[trim={0.3cm 0cm 0cm 0cm},width=0.50\textwidth]{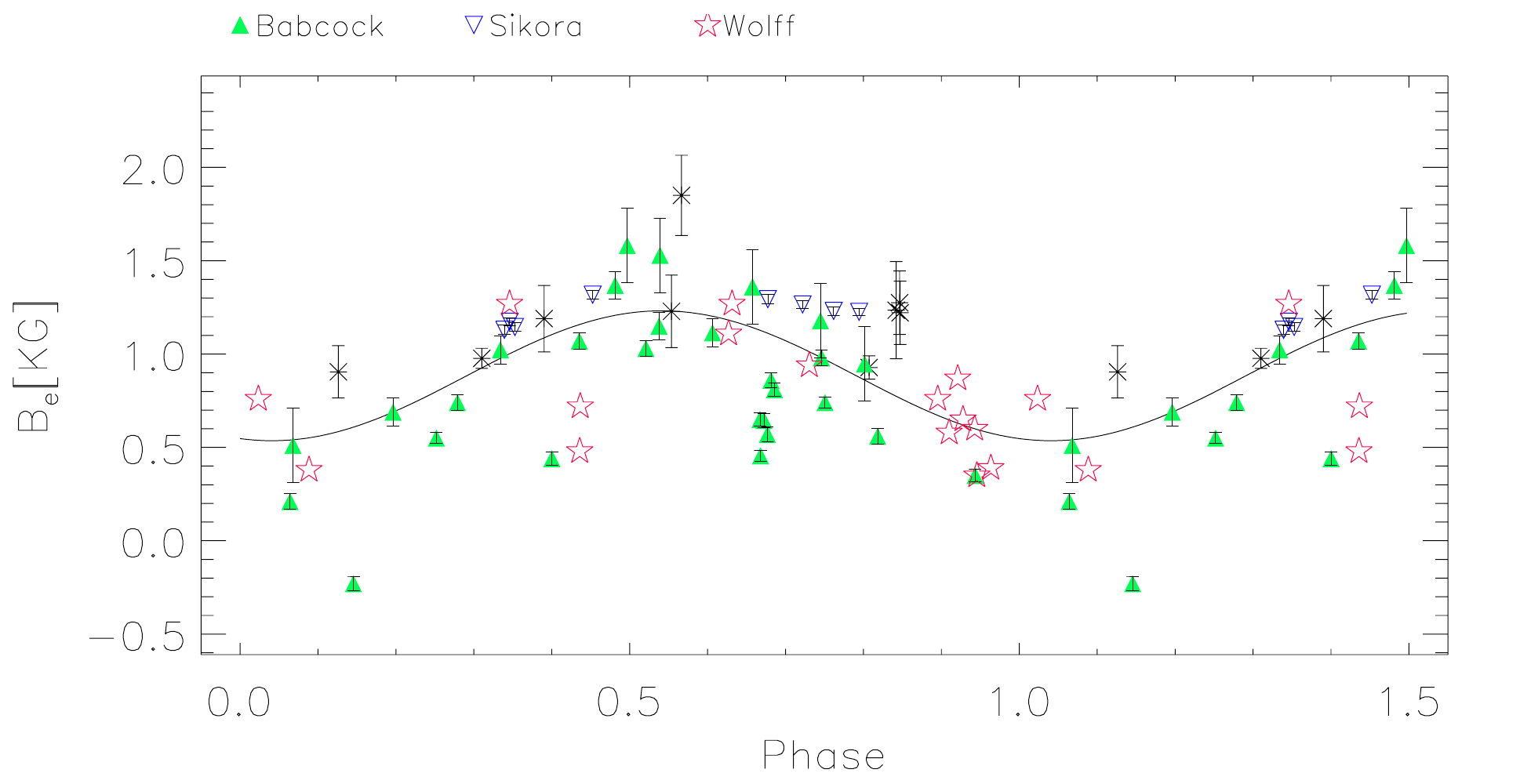}
\caption{HD\,188041. $B_s$ and $B_e$ variability, folded with the 223.826\,d period.}
\label{Fig_HD188041}
\end{figure}

\cite{Mikulasek2003} determined the photometric, magnetic and spectroscopic period
of HD\,188041 to 223.826$\pm$0.040\,d. This period has been adopted by P17 for their
photometric data but from $B_e$ measurements (covering 50 years), M17 concluded on
223.78$\pm$0.10\,d. We have observed this stars since JD = 2\,451\,826.349 with SARG,
UCLES, CAOS and HARPS-North. Our $B_s$ measurements listed in Table\,\ref{Tab_HD187474}
extend the time coverage from 2711 days (7.4\,yr) \citep{Mathys2017} to 10249 days
(28\,yr). The $B_e$ measurements found in the literature come from \cite{Babcock1954,
Babcock1958}, \cite{Wolff1969}, \cite{Mathys1991}, \cite{Mathys1997}, M17, and
\cite{Sikora2019}. The resulting $LS(B_s, B_e)$  presents a peak at $223.82\pm0.32$\,d.
The poor precision of the period determination is a consequence of the very low
amplitude of the $B_s$ variability. In fact, the average value is  $<B_s> = 3620\pm$55\,G
where the r.m.s. is comparable to the typical error in these measurements.
Figure\,\ref{Fig_HD188041} displays the $B_s$ and the $B_e$ data, folded with the
period given by \cite{Mikulasek2003}. It seems that the $B_s$ maximum coincides with
the $B_e$ minimum.

\subsection{HD\,192678}

\cite{Pyper2017} found the period of 6.4193\,d representative of the photometric and
magnetic variability exhibited by HD\,192678. The same period was adopted by M17 to
discuss the $B_s$ variations of this star. However, this author has pointed out
that the $B_e$ measurements by \cite{Wade_HD192678} do not show any variability with
this period. \cite{Bychkov2005} suggested that $B_e$ follows a period equal to
12.91049\,d.

\begin{figure}\center
\includegraphics[trim={0.3cm 0cm 0cm 0cm},width=0.50\textwidth]{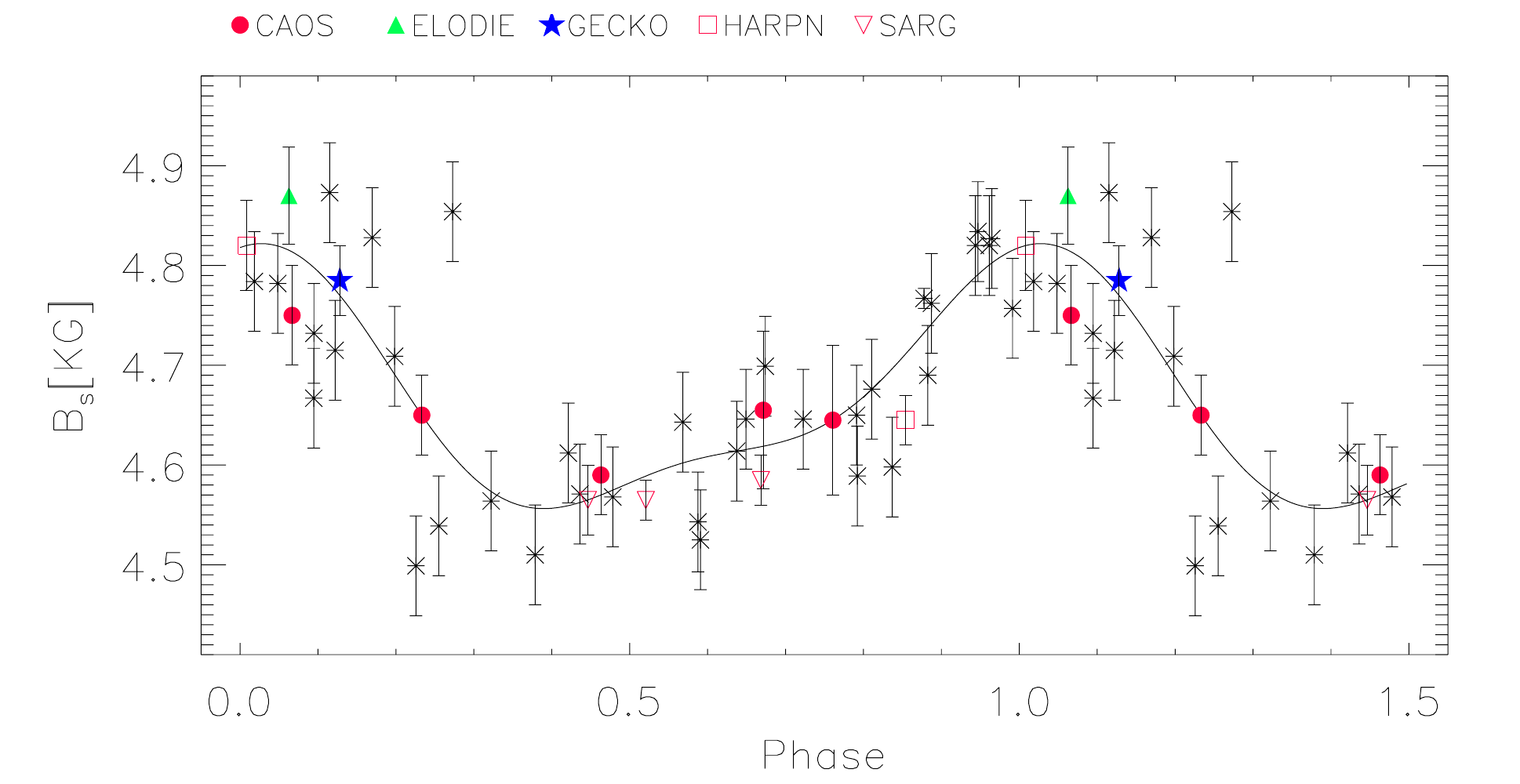}
\includegraphics[trim={0.3cm 0cm 0cm 0cm},width=0.50\textwidth]{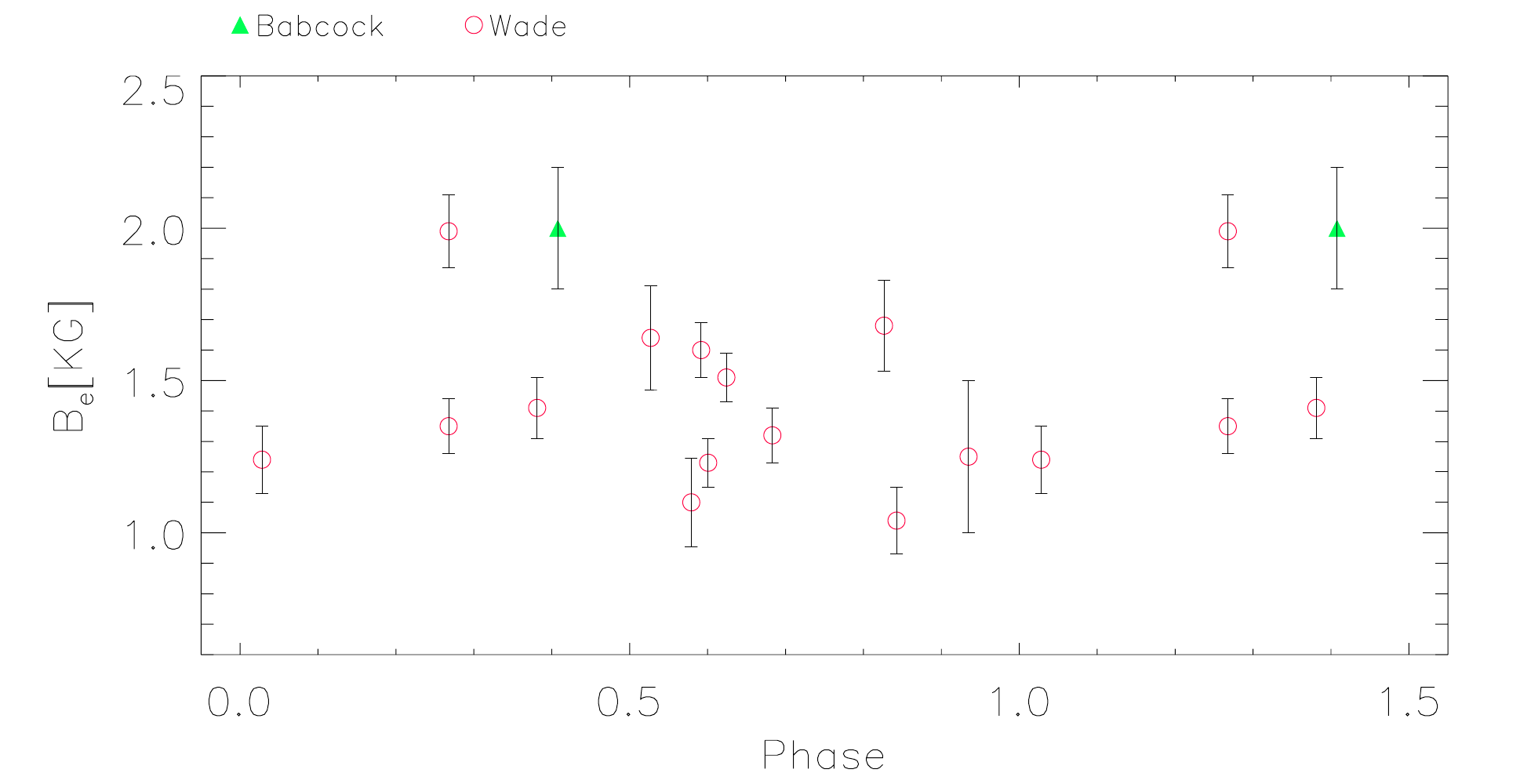}
\includegraphics[trim={0.3cm 0cm 0cm 0cm},width=0.50\textwidth]{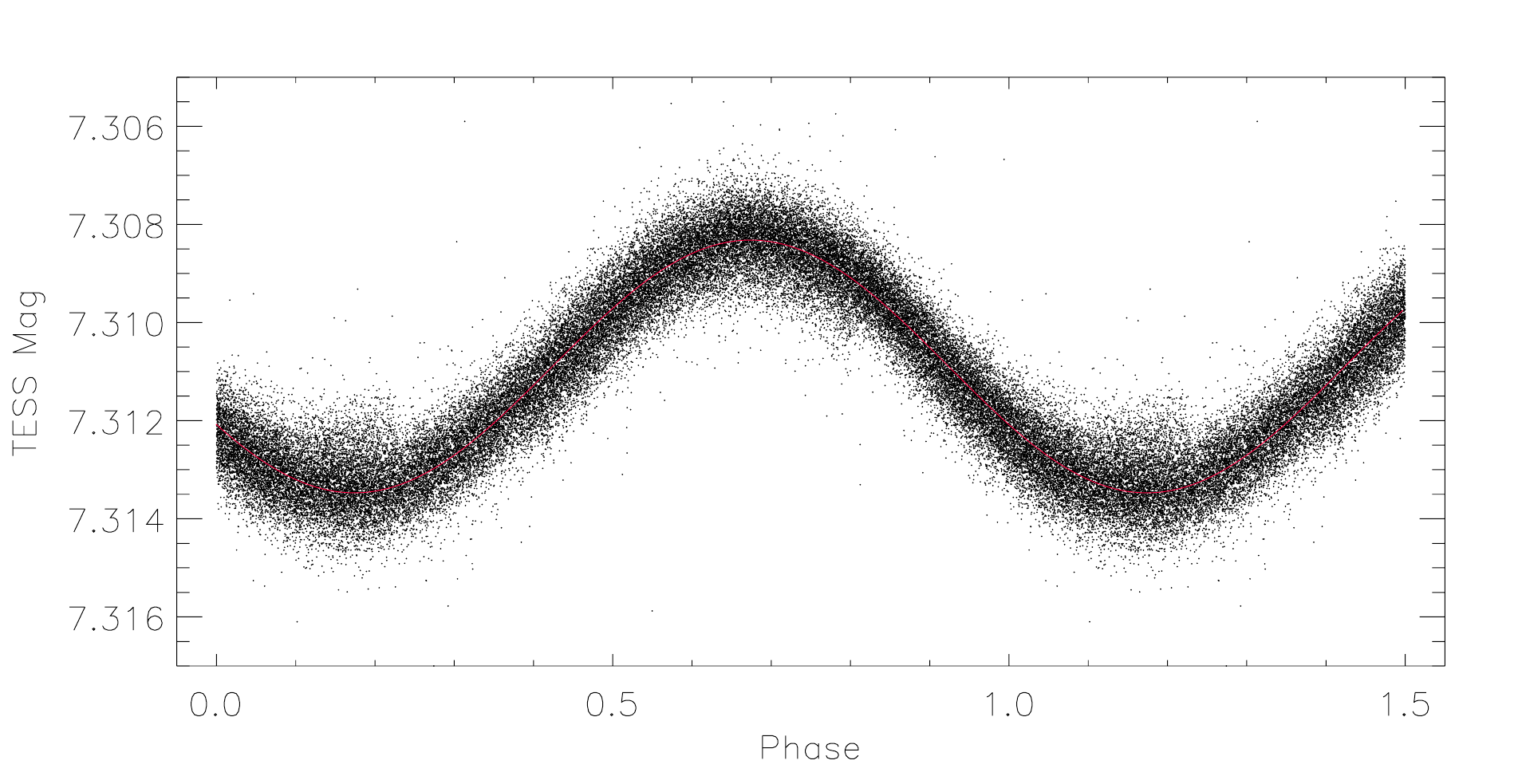}
\caption{HD\,192678. $B_s$ and TESS photometric variability, folded with the 6.4199\,d
 period. $B_e$ measurements by \citet{Babcock1958} and \citet{Wade_HD192678} show no
 clear evidence of variability with this period.}
\label{Fig_HD192678}
\end{figure}

To determine the variability period of HD\,192678, we have measured $B_s$ from our
2 SARG, 5 CAOS and 2 HARPS-North spectra together with one GECKO archive spectra (Table\,\ref{Tab_HD192678}), and we have retrieved TESS photometric data. A Lomb-Scargle analysis of TESS photometry rules out the 12.91049\,d period. We have adopted the maximum of $LS(B_s, TESS)$
at 6.4199$\pm 0.0001$\,d as the variability period (Fig.\,\ref{Fig_HD192678}). In
the same figure we show the folded $B_e$ measurements by \cite{Babcock1958} and
\cite{Wade_HD192678} which however suffer from considerable scatter.

\subsection{HDE\,335238}

\begin{figure}\center
\includegraphics[trim={0.3cm 0cm 0cm 0cm},width=0.50\textwidth]{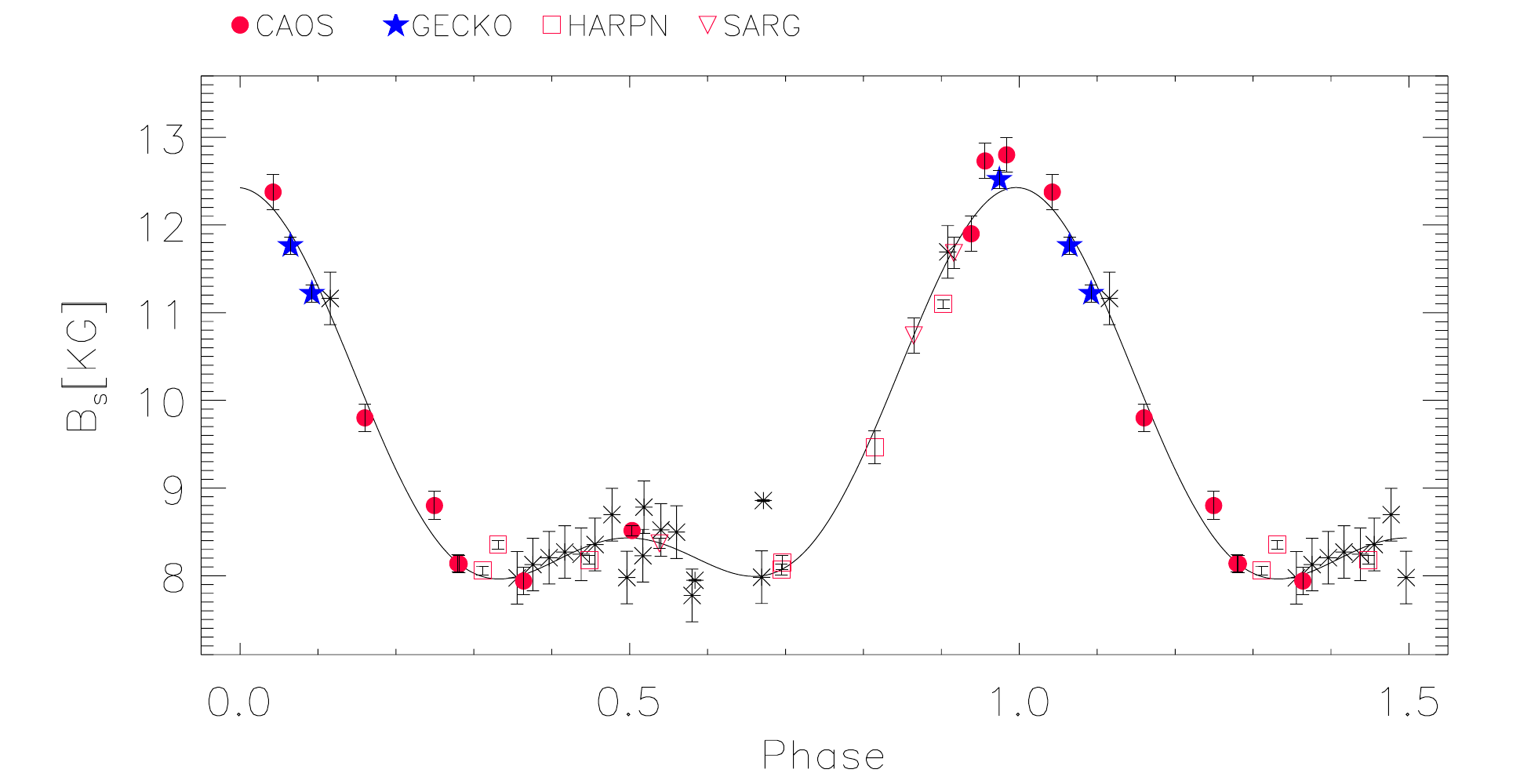}
\caption{HD\,335238. $B_s$  variability, folded with the 48.985\,d period.}
\label{Fig_HD335238}
\end{figure}

M17 established a variability period of 48.7$\pm 0.1$\,d from $B_s$ measurements.
Most of these data were clustered near the minimum value with only two points
shaping the $B_s$ maximum. We have observed this star with SARG (3 spectra),
CAOS (10) and HARPS-North (6). We also have recovered 3 spectra from GECKO. A
Lomb-Scargle analysis of our 22 (Table\,\ref{Tab_HD335238}) data points, together
with those from M97 (16) and M17 (2) produces a periodogram with the highest peak
at 48.985$\pm 0.007$\,d. Fig.\,\ref{Fig_HD335238} shows the $B_s$ measurements,
folded with this period.

\subsection{HD\,201601}

\begin{figure}\center
\includegraphics[trim={0.3cm 0cm 0cm 0cm},width=0.50\textwidth]{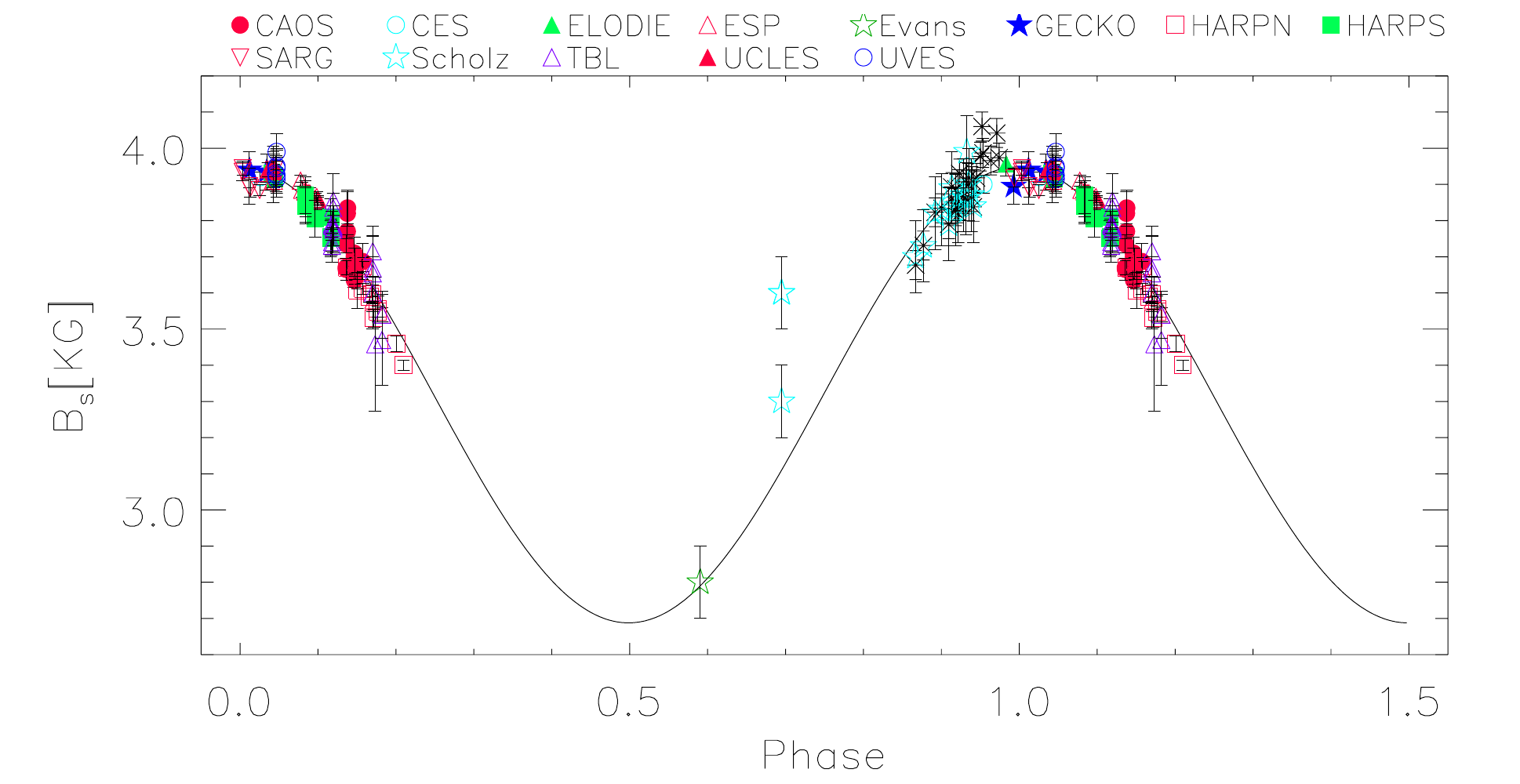}
\includegraphics[trim={0.3cm 0cm 0cm 0cm},width=0.50\textwidth]{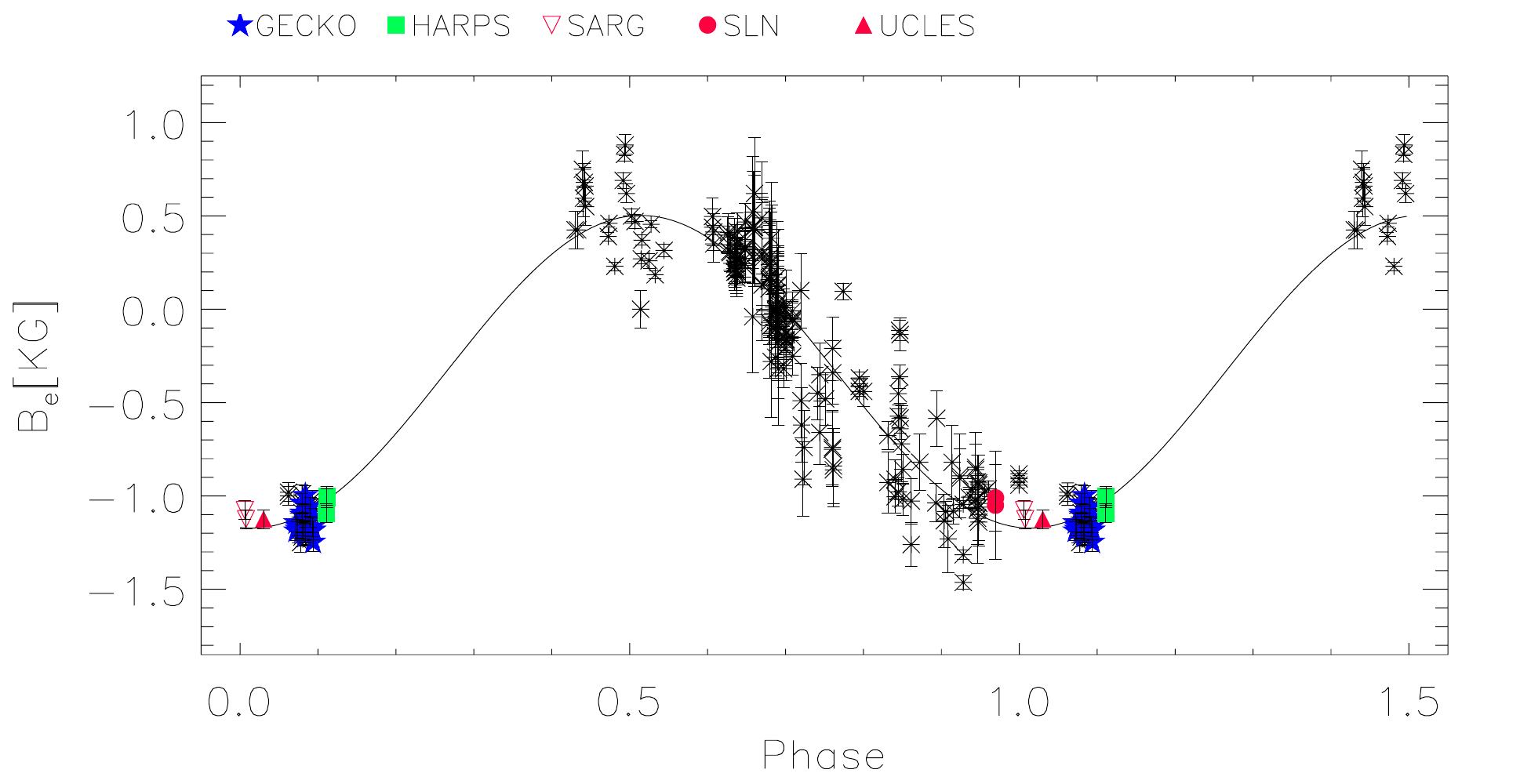}
\caption{HD\,201601. $B_s$ and $B_e$ variability, folded with the 97.16\,yr \,period
determined by \citet{Bychkov_HD201601}.}
\label{Fig_HD201601}
\end{figure}

The most recent estimate, based on the assumption of a simple sine variation of
$B_e$, of the magnetic variability period of HD\,201601 has yielded 97.16 years
\citep{Bychkov_HD201601}. We have acquired high-resolution spectra of HD\,201601
over 24 years with CES (1 spectrum), SARG (8), UCLES (1), HARPS (2), CAOS (14),
HARPS (2), and HARPS-North (10) In addition, 58 spectra have been obtained
from  ESO, CFHT and TBL archives. Measured values of $B_s$ are listed in
Table\,\ref{Tab_HD201601}. In order to extend the time interval of the $B_s$
data as much as possible, we have mined the available literature and found
measurements and estimations, also listed in Table\,\ref{Tab_HD201601} together
with the appropriate references. We have folded these $B_s$ measurements plus
data from \cite{Evans1971} and \cite{Scholz1979} with the period given by
\cite{Bychkov_HD201601}. We establish the $B_s$ maximum at JD = 2\,452\,200, and
observe a peak-to-peak variation of at least 1.3\,kG. From the large number of
$B_e$ measurements reported in the literature, we present a sample well distributed
in time from the first measurements by \cite{Babcock1958} to the present day
(Fig.\,\ref{Fig_HD201601}). Folding the data with the period of \cite{Bychkov_HD201601},
the $B_e$ minimum appears to coincide with the $B_s$ maximum. Just like we found in
the case of HD\,116114, HD\,165474 or HD\,177765, this period is the shortest possible one.
  
\subsection{HD\,208217}

\begin{figure}\center
\includegraphics[trim={0.3cm 0cm 0cm 0cm},width=0.50\textwidth]{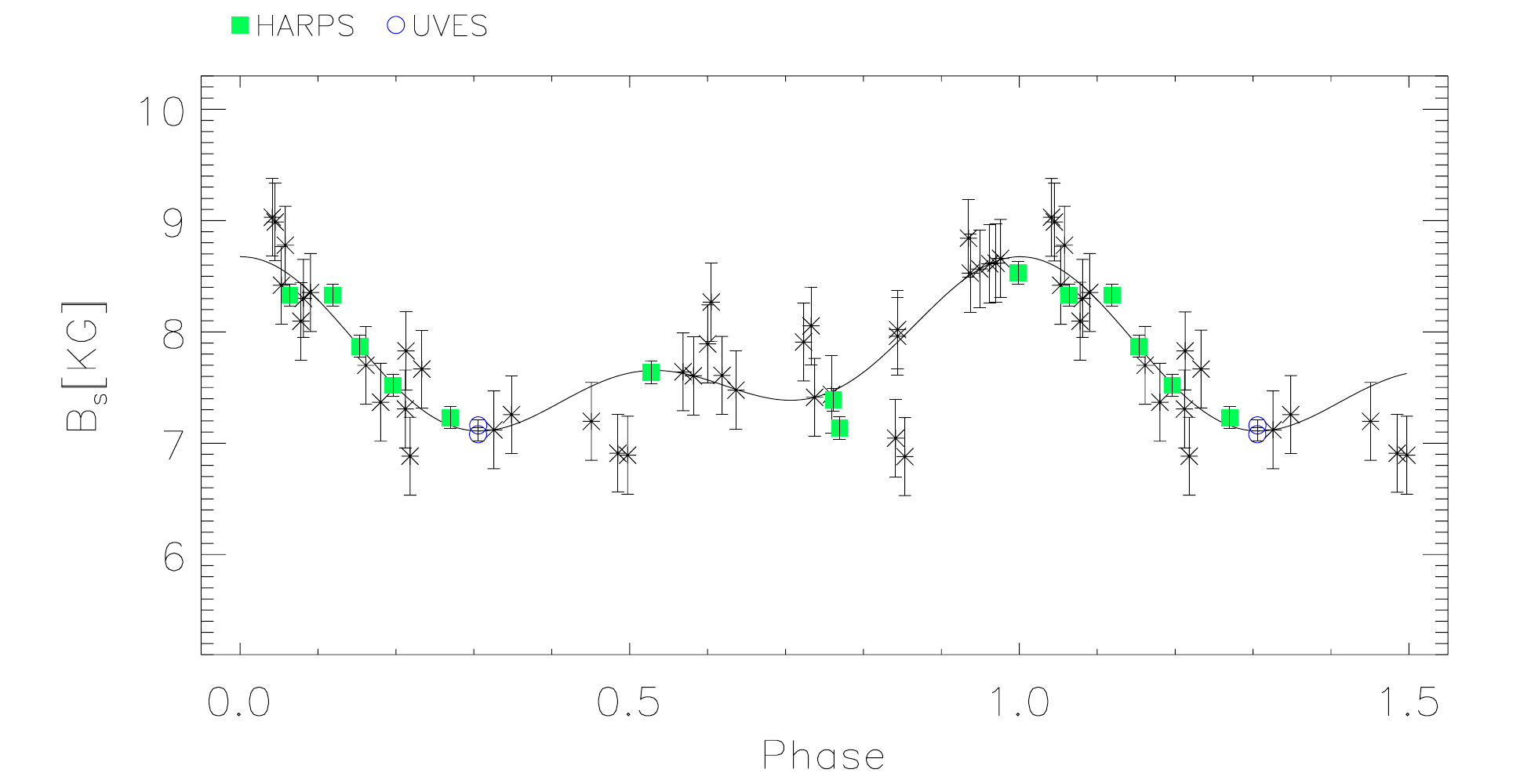}
\includegraphics[trim={0.3cm 0cm 0cm 0cm},width=0.50\textwidth]{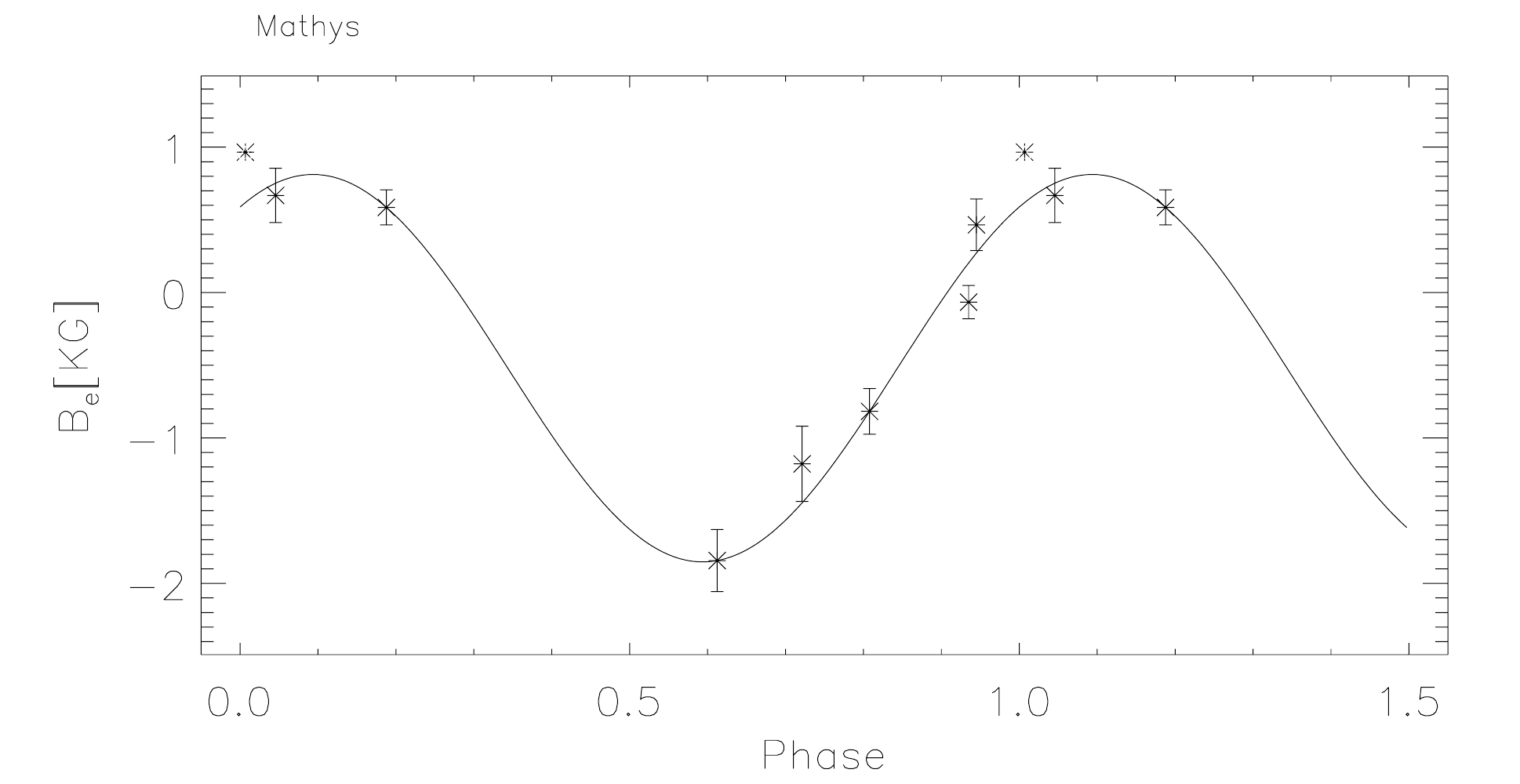}
\includegraphics[trim={0.3cm 0cm 0cm 0cm},width=0.50\textwidth]{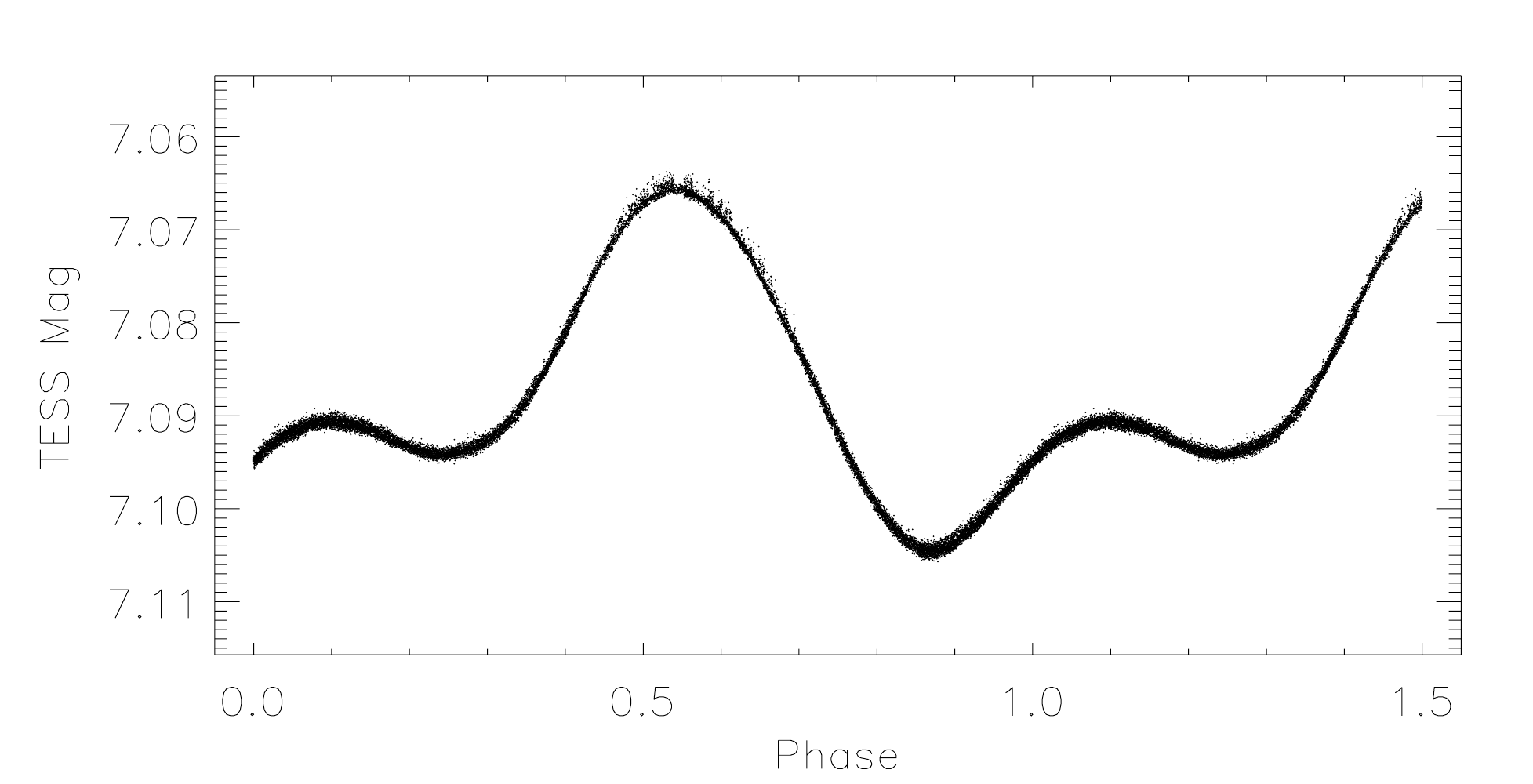}
\caption{HD\,208217. Variations in $B_s$, $B_e$ and TESS photometry, folded with the
         8.445\,d period. }
\label{Fig_HD208217}
\end{figure}

M17 found the $B_s$ measurements of HD\,208217 -- collected between JD = 2\,449\,213 and
2\,451\,085 -- to be variable with the period of 8.44475\,d, photometrically determined
by \cite{Manfroid1997}. From TESS photometry, \cite{David-Uraz2019} determined a period
equal to 8.317$\pm 0.001$\,d. We have obtained 2 UVES spectra and 8 HARPS spectra from
the ESO archive and acquired a new spectrum with HARPS, extending the time coverage to
6932 days. $B_s$ measurements are listed in Table\,\ref{Tab_HD208217}. Additionally,
we have retrieved TESS photometric data. A Lomb-Scargle analysis gives
$LS(B_s, B_e, TESS)$ with a peak at 8.445$\pm$0.005\,d. Figure\,\ref{Fig_HD208217}
shows the variations in $B_s$, $B_e$ (M17) and TESS.

\subsection{HD\,216018}

\begin{figure}\center
\includegraphics[trim={0.3cm 0cm 0cm 0cm},width=0.50\textwidth]{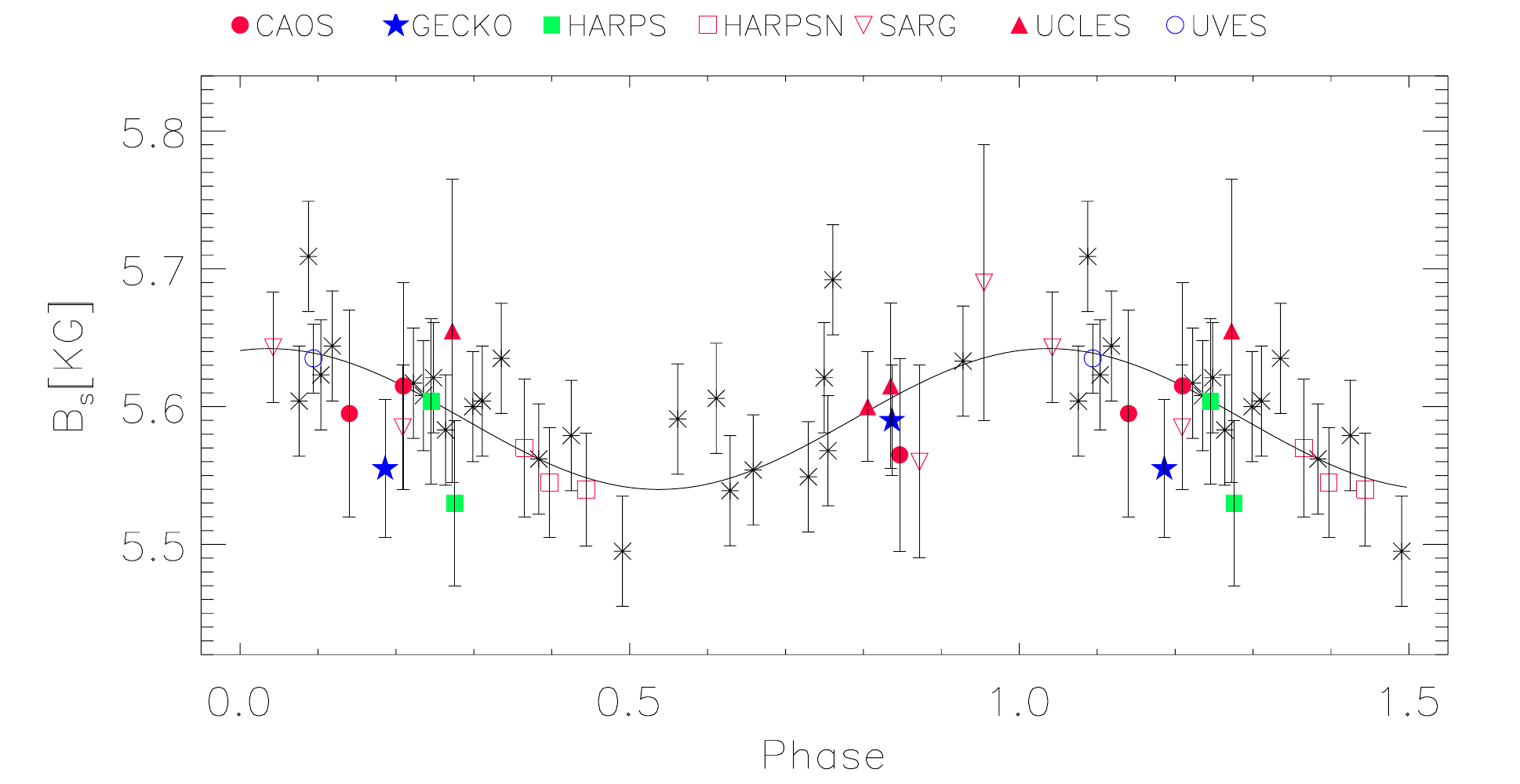}
\includegraphics[trim={0.3cm 0cm 0cm 0cm},width=0.50\textwidth]{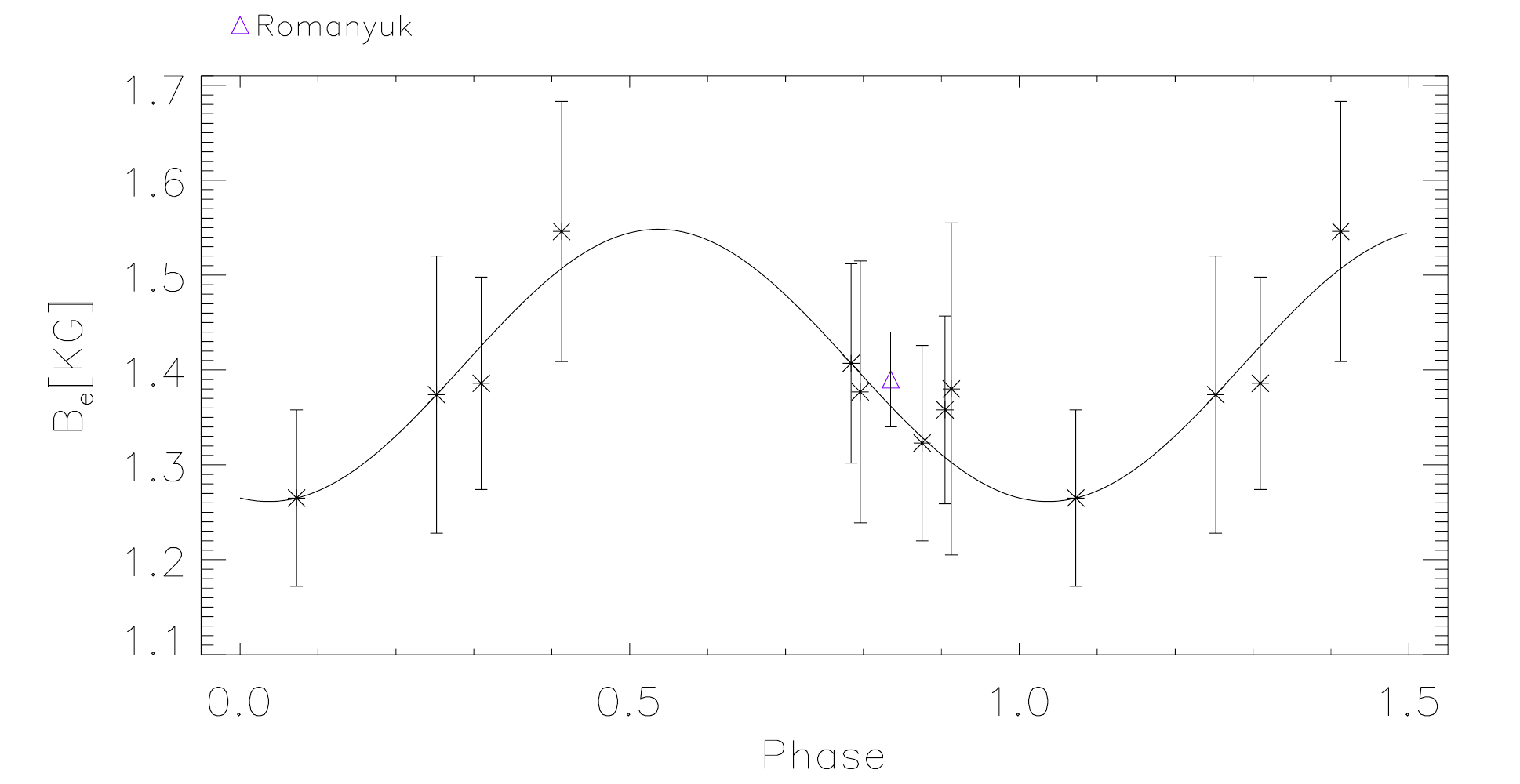}
\caption{HD\,216018. $B_s$  and $B_e$ variability, folded with the 34.044\,d \,period.}
\label{Fig_HD216018}
\end{figure}

From spectra collected between 1992 and 1998, M17 concluded that the period of
the variability of HD\,216018 -- provided there is any periodicity -- exceeds 6 years.
Table\,\ref{Tab_HD216018} lists our measurements of $B_s$ from 
3 UCLES, 2 SARG, 2 HARPS, 3 CAOS, and 3 HARPS-North
spectra collected 
between 2001 and 2018. These plus the M17 measurements result in an average of
$B_s = 5600 \pm 45$\,G. Computing the Lomb-Scargle periodograms of all $B_s$ data
and $B_e$ measurements by M17 and \cite{Romanyuk2016}, we find the main peak of
$LS(B_s, B_e)$ at 34.044$\pm$0.007\,d. Figure\,\ref{Fig_HD216018} displays the
folded $B_s$ and $B_e$ data. Both the $B_s$ and $B_e$ variabilities present rather
modest amplitudes.

\section{Magnetic field strength and rotation period} 

\begin{figure}\center
\includegraphics[width=0.50\textwidth]{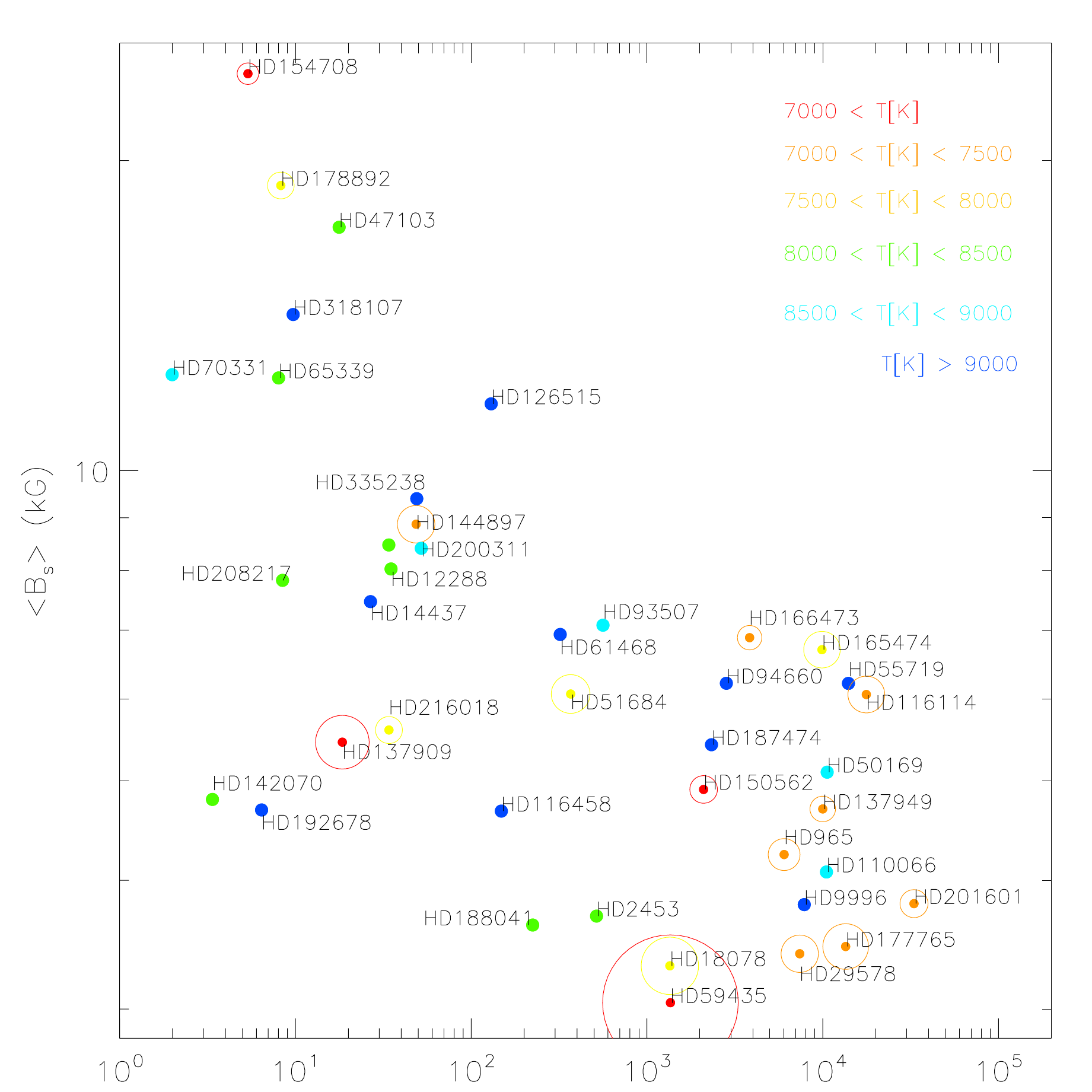}
\center{Rotation period (days)}
\caption{Average surface magnetic field as a function of the rotation period. Colours
of filled circles indicate the stellar temperature. Open circles are plotted with radii
proportional to the stellar radius, taken from \citet{GAIA2018} when available. These
radii range from 1.66\,R$_\odot$ (e.g. HD\,154708) to 10.47\,R$_\odot$ (HD\,59435).}
\label{Fig_BvsP}
\end{figure}

Figure\,\ref{Fig_BvsP} shows the average surface field versus the rotation period for
the stars discussed above, plus HD\,59435, HD\,65339, HD\,70311, HD\,116458 and
HD\,200311, taken from M17. To highlight possible relationships between periods,
field strengths, temperatures and stellar radii (with these two last parameters
taken from \cite{GAIA2018} for homogeneity), colours indicate temperatures and circles
around the position of the star are proportional in size to the stellar radius; radii
range from 1.66\,R$_\odot$ (e.g. HD\,154708) to 10.47\,R$_\odot$ (HD\,59435). Such
unexpectedly large values for stellar radii are probably a consequence of unresolved
components in a binary system or of very small orbital motion perturbing the radius
determination. HD\,59435 for example is a SB2 system \citep{Wade1996}.

It appears there is a general decrease of field strength with the period length of
stellar rotation. Even though it is probably safe to state that the top-right corner
is empty (the result of the Kolmogorov-Smirnov test by M17 has shown that only
stars with rotational periods shorter than 150 days present $B_s$ larger than 7500\,G),
doubts persist as to the left-bottom corner of Figure\,\ref{Fig_BvsP}. It is in fact
not possible to measure weak surface fields from the splitting of the
Fe{\sc ii}\,6149.258\,{\AA} Zeeman subcomponents when these are considerably broadened
by stellar rotation. It would certainly be worthwhile to look at near-infrared lines,
given the $\lambda^2$ dependence of Zeeman splitting,

The spread for a given value of the period cannot be attributed entirely to the random
distribution of angles between line-of-sight, rotation axis and a possible magnetic
symmetry axis. For a centred magnetic dipole, the ratio between the largest and smallest
values of $B_s$ is only 1.2 \citep{Preston1969}. M17 has observationally shown that for
some stars this ratio rises to 2, and here we find that HD\,9996 presents $q \sim 3.3$.

\section{Conclusions}

The surface field of 36 magnetic chemically peculiar stars has been monitored for 20
years. For every star, by adding archive and literature data, we have extended the
time base as much as possible with the aim to determine periods of the order of several
decades. Magnetic fields have been derived via the distance in wavelength of the Zeeman
subcomponents of the Fe{\sc ii}\,6149.258\,{\AA} line under the hypothesis of 
linear Zeeman splitting. For a number of stars, magnetic fields are so strong that the
partial Paschen-Back effect should be taken into account \citep{Stift2008}. The fact that
in these cases we still use the linear approximation does not affect the determination
of the period, but rather only concerns the shape of the variability, with amplitudes
being underestimated.

In summary: we confirm the rotational periods given in the literature for HD\,965,
HD\,50169, HD\,126515 and HD\,166473. We select the correct rotational periods among
possible values quoted in the literature for HD\,2453, HD\,9996, HD\,81009  and
HD\,188041. We revisit the periods of HD\,12288, HD\,14437, HD\,18078, HD\,51684,
HD\,61468, HD\,93507, HD\,94660, HD\,142070, HD\,144897, HD\,154708, HD\,318107,
HD\,178892, HD\,187474, HD\,192678, HD\,335238 and HD\,208217. To our knowledge, the
periods of variability of HD\,29578, HD\,47103, HD\,150562 and HD\,216018 have been
established here for the first time. Lower limits have been estimated or revised for the
variability periods of HD\,55719 ($P > 38$\,yr), HD\,75445 ($P >> 14$\,yr),
HD\,110066  ($P >> 29$\,yr), HD\,116114 ($P >  48$\,yr), HD\,137949 ($P > 27$\,yr),
HD\,165474 ($P > 27$\,yr) and HD\,177765 ($P > 37$\,yr). It has to be noted that
HD\,75445 and HD\,110066 show no evidence of $B_s$ variability at all.

Concerning $\gamma$\,Equ (= HD\,201601), one of the best known magnetic stars, we
have pointed out a clear decrease in the surface field over the last 20 years. Despite
the fact that the 97 yr period comes from the most extended time series for both $B_e$
and $B_s$ for this class of stars, it could still constitute a lower limit to the
rotational period. It appears that the maximum of the surface field has been in phase
coincidence with the negative extremum of the longitudinal field.

The results are summarised in Figure\,\ref{Fig_BvsP}, plotting the average surface
field versus the rotation periods or their lower limits as determined in this study.
As a general rule, it seems that longer periods imply weaker fields.

\section*{Acknowledgements}
Our grateful thanks go to the referee, Dr. Gauthier Mathys, for the careful reading and numberless suggestions.
Based on observations collected at the European Southern Observatory (ESO) and on
data obtained from the ESO Science Archive Facility. "Based on observations made with
the Italian Telescopio Nazionale Galileo (TNG). The TNG is operated on the island of
La Palma by the Fundaci\'on Galileo Galilei of the INAF (Istituto Nazionale di
Astrofisica) at the Spanish Observatorio del Roque de los Muchachos of the Instituto
de Astrofisica de Canarias". This paper includes data collected by the TESS mission
which are publicly available from the Mikulski Archive for Space Telescopes (MAST).
Funding for the TESS mission is provided by the NASA's Science Mission Directorate.
We also acknowledge financial contributions from the agreement ASI-INAF n.2018-16-HH.0
and from the {\it Programma ricerca di Ateneo UNICT 2020-22 linea 2}.

\section*{Data Availability} 

The spectroscopic data underlying this article are available in the AAT, BTA,
CFHT, ESO, TBL and TNG archives at:
\url{https://datacentral.org.au/archives/aat/}\\
\url{https://www.sao.ru/oasis/cgi-bin/fetch?lang=en}\\
\url{https://www.cadc-ccda.hia-iha.nrc-cnrc.gc.ca/en/cfht/}\\
\url{http://archive.eso.org/eso/eso_archive_main.html}
\url{http://polarbase.irap.omp.eu/}\\
\url{http://archives.ia2.inaf.it/tng/}\\
CAOS data will be shared on reasonable request to the corresponding author. \\
TESS photometric data are available in \url{https://archive.stsci.edu/}.

\bibliographystyle{mnras}
\bibliography{long_mnras}

\clearpage

\input{HD965_tab.tex}
\input{HD2453_tab.tex}
\input{HD9996_tab.tex}
\input{HD12288_tab.tex}
\input{HD14437_tab.tex}
\input{HD18078_tab.tex}
\input{HD29578_tab.tex}
\input{HD47103_tab.tex}
\input{HD50169_tab.tex}
\clearpage
\input{HD51684_tab.tex}
\input{HD55719_tab.tex}
\input{HD61468_tab.tex}
\input{HD75445_tab.tex}
\input{HD81009_tab.tex}
\input{HD93507_tab.tex}
\input{HD94660_tab.tex}
\clearpage
\input{HD110066_tab.tex}
\input{HD116114_tab.tex}
\input{HD126515_tab.tex}
\input{HD137949_tab.tex}
\input{HD142070_tab.tex}
\clearpage
\input{HD144897_tab.tex}
\input{HD150562_tab.tex}
\input{HD154708_tab.tex}
\input{HD318107_tab.tex}
\input{HD165474_tab.tex}

\input{HD166473_tab.tex}
\input{HD177765_tab.tex}
\input{HD178892_tab.tex}
\input{HD187474_tab.tex}
\input{HD188041_tab.tex}
\clearpage

\input{HD192678_tab.tex}
\input{HD335238_tab.tex}
\input{HD201601_tab.tex}
\input{HD208217_tab.tex}
\input{HD216018_tab.tex}
\end{document}

%% file: HD965_tab.tex
\begin{table}  
\scriptsize
\caption{Measured $B_s$ of HD965.}
\begin{tabular}{ccc|ccc}
 \hline
{HJD} & {B$_{\rm s}$} & {Sp} & {HJD} & {B$_{\rm s}$} & {Sp} \\
2400000+  & {[G]}                   &       &  {2400000+} & {[G]} &  ~\\\hline
51691.929 &  4190 $\pm$  50 & US &  53715.568 &  4160 $\pm$          50 & HS \\
51740.091 &  4250 $\pm$  60 & GO & 53716.567 &  4140 $\pm$          50 & HS \\ 
51741.038 &  4240 $\pm$  60 & GO & 54336.860 &  4120 $\pm$          50 & HS \\  
52120.670 &  4300 $\pm$  40 & SG & 54338.796 &  4230 $\pm$          50 & HS \\ 
52190.627 &  4280 $\pm$  50 & US & 54406.585 &  4230 $\pm$          40 & HS \\
52420.066 &  4220 $\pm$  40 & GO & 54442.613 &  4190 $\pm$          50 & HS \\  
52535.699 &  4290 $\pm$  50 & US & 54443.560 &  4090 $\pm$          50 & HS \\ 
52535.715 &  4310 $\pm$  50 & US & 54469.515 &  4220 $\pm$          50 & US \\ 
52535.731 &  4300 $\pm$  50 & US & 54469.539 &  4190 $\pm$          50 & US \\ 
52535.748 &  4300 $\pm$  50 & US & 54469.562 &  4190 $\pm$          50 & US \\ 
52889.653 &  4350 $\pm$  40 & SG & 54469.580 &  4200 $\pm$          50 & US \\ 
53270.110 &  4260 $\pm$  50 & UC & 56147.773 &  4200 $\pm$          70 & HS \\ 
53334.505 &  4190 $\pm$  50 & HS &56648.343 &  4190 $\pm$          60 & HN \\ 
53581.741 &  4110 $\pm$  50 & HS &  56892.572 &  4240 $\pm$         110 & CS \\
53582.764 &  4160 $\pm$  50 & HS &  56980.341 &  4260 $\pm$          80 & CS \\ 
53583.867 &  4210 $\pm$  50 & HS &57252.556 &  4300 $\pm$         120 & CS \\
53661.707 &  4110 $\pm$  50 & US & 57634.482 &  4270 $\pm$         150 & CS \\ 
53711.565 &  4160 $\pm$  50 & HS &  58367.515 &  4390 $\pm$          70 & CS \\  
53712.595 &  4150 $\pm$  50 & HS & 58432.484 &  4340 $\pm$          50 & HN \\ 
53713.585 &  4220 $\pm$  50 & HS & 58432.495 &  4300 $\pm$          50 & HN\\ 
53714.583 &  4170 $\pm$  50 & HS & 59441.679 &  4140 $\pm$          50 & HN\\
 \hline
         \end{tabular} \label{Tab_HD965}
         \end{table}

%% file: HD2453_tab.tex
\begin{table} 
\scriptsize
\caption{Measured $B_s$ of HD2453.}
\begin{tabular}{ccc|ccc}
 \hline
 {HJD} & {B$_{\rm s}$} & {Sp} & {HJD} & {B$_{\rm s}$} & {Sp} \\
2400000+  & {[G]}                   &       &  {2400000+} & {[G]} &  ~\\\hline
50349.475 &  3660 $\pm$          60 & EE &  54374.379 &  3590 $\pm$          30 & SG \\
52121.649 &  3690 $\pm$          30 & SG &  58096.318 &  3630 $\pm$          30 & HN \\
52890.562 &  3610 $\pm$          30 & SG &  58135.220 &  3610 $\pm$          40 & CS \\
53215.969 &  3700 $\pm$          20 & GO &  58432.554 &  3680 $\pm$          30 & CS \\
53662.628 &  3680 $\pm$          20 & US  & 59441.749 &  3670 $\pm$ 20 & HN \\
 \hline
         \end{tabular}  \label{Tab_HD2453}
         \end{table}

%% file: HD9996_tab.tex
\begin{table} 
\scriptsize
\caption{Measured $B_s$ of HD9996.}
\begin{tabular}{ccrc|ccrc}
 \hline
{HJD} & {B$_{\rm s}$} & {Sp} & {HJD} & {B$_{\rm s}$} & {Sp} \\
2400000+  & {[G]}                   &       &  {2400000+} & {[G]} &  ~\\\hline
 51740.074 &  1700 $\pm$       20 & GO &  56964.449 &  4710 $\pm$       80 & CS \\ 
 51799.762 &  1530 $\pm$       30 & SG &  57264.629 &  4780 $\pm$      100 & CS \\ 
 52121.660 &  1610 $\pm$       30 & SG &  57340.465 &  4920 $\pm$       70 & HN \\ 
 52212.668 &  1350 $\pm$       40 & SG &  58096.461 &  4190 $\pm$       40 & HN \\ 
 52890.566 &  1520 $\pm$       40 & SG &  58432.605 &  3310 $\pm$       60 & HN \\ 
 53215.113 &  1700 $\pm$       30 & GO &  58828.492 &  2380 $\pm$       50 & HN \\ 
 54374.395 &  1630 $\pm$       20 & SG &  59441.695 &  1680 $\pm$       50 & HN \\ 
 56938.492 &  4750 $\pm$       70 & CS &            &                      &  \\ 
\hline
\end{tabular}\label{Tab_HD9996}
\end{table}

%% file: HD12288_tab.tex
\begin{table}  
\scriptsize
\caption{Measured $B_s$ of HD12288.}
\begin{tabular}{ccc|ccc}
 \hline
 {HJD} & {B$_{\rm s}$} & {Sp} & {HJD} & {B$_{\rm s}$} & {Sp} \\
2400000+  & {[G]}                   &       &  {2400000+} & {[G]} &  ~\\\hline
49614.586 &  8330 $\pm$         110 & EE &56964.478 &  8360 $\pm$          90 & CS \\ 
52121.691 &  7520 $\pm$          50 & SG &57290.555 &  8230 $\pm$         100 & CS \\ 
52212.552 &  8450 $\pm$          90 & SG &57676.762 &  8180 $\pm$         100 & HN \\ 
53216.027 &  8350 $\pm$         100 & GO &58137.269 &  7530 $\pm$          80 & CS \\ 
56938.528 &  8450 $\pm$          90 & CS & 59441.736&  8315 $\pm$          50 & HN \\ 
 \hline
          \end{tabular} \label{Tab_HD12288}
         \end{table}

%% file: HD14437_tab.tex
\begin{table} 
\scriptsize
\caption{Measured $B_s$ of HD14437.}
\begin{tabular}{ccc|ccc}
 \hline
{HJD} & {B$_{\rm s}$} & {Sp} & {HJD} & {B$_{\rm s}$} & {Sp} \\
2400000+  & {[G]}                   &       &  {2400000+} & {[G]} &  ~\\\hline
 52121.719 &  7250 $\pm$        60 & SG    &  56980.426 &  7560 $\pm$         50 & CS \\ 
 56651.438 &  7140 $\pm$        50 & HN &  58137.293 &  7600 $\pm$         60 & CS \\ 
 57340.586 &  6880 $\pm$        50 & HN &  58432.641 &  7210 $\pm$         40 & HN \\ 
 57425.414 &  6980 $\pm$        90 & HN &  58432.633 &  7210 $\pm$         50 & HN \\ 
 57426.383 &  7010 $\pm$       170 & HN &  59441.723 &  7640 $\pm$          40 & HN \\ 
 56964.512 &  6890 $\pm$        50 & CS    &            &           &  \\ 
\hline
\end{tabular} \label{Tab_HD14437}
\end{table}

%% file: HD18078_tab.tex
\begin{table} 
\scriptsize
\caption{Measured $B_s$ of HD18078.}
\begin{tabular}{ccc|ccc}
 \hline
{HJD} & {B$_{\rm s}$} & {Sp} & {HJD} & {B$_{\rm s}$} & {Sp} \\
2400000+  & {[G]}                   &       &  {2400000+} & {[G]} &  ~\\\hline
49614.558 &  3550 $\pm$         180 & EE &57340.499 &  2760 $\pm$          50 & HN \\ 
51740.120 &  2960 $\pm$          50 & GO &57426.397 &  2820 $\pm$          20 & HN \\ 
52890.602 &  3700 $\pm$          50 & SG &57426.408 &  2850 $\pm$          20 & HN \\ 
53216.054 &  2820 $\pm$          50 & GO &58432.653 &  3050 $\pm$          50 & HN \\ 
56651.484 &  4380 $\pm$         130 & HN &58432.664 &  3050 $\pm$          50 & HN \\ 
56980.480 &  3480 $\pm$          50 & CS  &59441.708 & 4310 $\pm$          20 & HN  \\ 
 \hline
         \end{tabular}  \label{Tab_HD18078}
         \end{table}

%% file: HD29578_tab.tex
\begin{table}  
\scriptsize
\caption{Measured $B_s$ of HD29578.}
\begin{tabular}{ccc|ccc}
 \hline
{HJD} & {B$_{\rm s}$} & {Sp} & {HJD} & {B$_{\rm s}$} & {Sp} \\
2400000+  & {[G]}                   &       &  {2400000+} & {[G]} &  ~\\\hline
52213.813 &  5340 $\pm$          50 & US &57056.992 &  2880 $\pm$         100 & UC \\ 
54687.869 &  3440 $\pm$          50 & FS &51600.518 &  5130  & (Ry) \\ 
54868.581 &  3620 $\pm$          50 & FS &51946.528 &  5540  & (Ry) \\ 
 \hline
         \end{tabular} \label{Tab_HD29578}
         \end{table}

%% file: HD47103_tab.tex
\begin{table}  
\scriptsize
\caption{Measured $B_s$ of HD47103.}
\begin{tabular}{ccc|ccc}
 \hline
{HJD} & {B$_{\rm s}$} & {Sp} & {HJD} & {B$_{\rm s}$} & {Sp} \\
2400000+  & {[G]}                   &       &  {2400000+} & {[G]} &  ~\\\hline
49818.318 & 17650 $\pm$ 100 & (Bl) & 52286.703 & 16860 $\pm$    40 & US \\ 
49818.330 & 17480 $\pm$ 100 & (Bl) & 56651.634 & 16730 $\pm$   110 & HN \\
49819.388 & 17420 $\pm$ 100 & (Bl) & 57425.473 & 17070 $\pm$    80 & HN \\  
50119.485 & 17370 $\pm$ 100 & (Bl) & 57426.463 & 16710 $\pm$   140 & HN \\ 
50349.643 & 17160 $\pm$ 100 & (Bl) & &   &  \\
 \hline
          \end{tabular} \label{Tab_HD47103}
         \end{table}

%% file: HD50169_tab.tex
\begin{table} 
\scriptsize
\caption{Measured $B_s$ of HD50169.}
\begin{tabular}{ccc|ccc}
 \hline
{HJD} & {B$_{\rm s}$} & {Sp} & {HJD} & {B$_{\rm s}$} & {Sp} \\
2400000+  & {[G]}                   &       &  {2400000+} & {[G]} &  ~\\\hline
52242.612 &  6030 $\pm$          50 & US &54544.631 &  5180 $\pm$          50 & HS \\ 
53333.857 &  5830 $\pm$          50 & HS &54545.506 &  5120 $\pm$          50 & HS \\ 
53463.496 &  5780 $\pm$          50 & HS &54717.904 &  5060 $\pm$          50 & HS \\ 
53464.494 &  5830 $\pm$          50 & HS &54718.910 &  5140 $\pm$          50 & HS \\ 
53714.661 &  5630 $\pm$          50 & HS &54865.688 &  5020 $\pm$          50 & HS \\ 
53716.633 &  5590 $\pm$          50 & HS &54866.617 &  4970 $\pm$          50 & HS \\ 
54222.497 &  5370 $\pm$          50 & HS &54868.697 &  5000 $\pm$          50 & HS \\ 
54223.509 &  5350 $\pm$          50 & HS &54869.708 &  4990 $\pm$          50 & HS \\ 
54224.523 &  5330 $\pm$          50 & HS &56651.699 &  4390 $\pm$          50 & HN \\ 
54336.917 &  5310 $\pm$          50 & HS &56651.665 &  4420 $\pm$          50 & HN \\ 
54338.914 &  5310 $\pm$          50 & HS &56651.676 &  4360 $\pm$          50 & HN \\ 
54441.674 &  5230 $\pm$          50 & HS &56651.687 &  4370 $\pm$          50 & HN \\ 
54442.731 &  5200 $\pm$          50 & HS &57340.620 &  4350 $\pm$          50 & HN \\ 
54443.689 &  5250 $\pm$          50 & HS &58075.835 &  4270 $\pm$          20 & US \\ 
54443.760 &  5230 $\pm$          50 & HS &58179.551 &  4270 $\pm$          20 & US \\ 
54543.493 &  5160 $\pm$          50 & HS &58432.690 &  4270 $\pm$          50 & HN \\ 
 \hline
         \end{tabular}  \label{Tab_HD50169}
         \end{table}

%% file: HD51684_tab.tex
\begin{table}  
\scriptsize
\caption{Measured $B_s$ of HD51684.}
\begin{tabular}{ccc|ccc}
 \hline
{HJD} & {B$_{\rm s}$} & {Sp} & {HJD} & {B$_{\rm s}$} & {Sp} \\
2400000+  & {[G]}                   &       &  {2400000+} & {[G]} &  ~\\\hline
53333.817 &  6090 $\pm$          20 & US &53632.847 &  6290 $\pm$          20 & US \\ 
53340.802 &  6070 $\pm$          20 & US &53666.765 &  6230 $\pm$          20 & US \\ 
53340.816 &  6060 $\pm$          20 & US &  &   &  \\ 
 \hline
         \end{tabular} \label{Tab_HD51684}
         \end{table}

%% file: HD55719_tab.tex
\begin{table} 
\scriptsize
\caption{Measured $B_s$ of HD55719.}
\begin{tabular}{ccc|ccc}
 \hline
{HJD} & {B$_{\rm s}$} & {Sp} & {HJD} & {B$_{\rm s}$} & {Sp} \\
2400000+  & {[G]}                   &       &  {2400000+} & {[G]} &  ~\\\hline
52229.841 &  6120 $\pm$          40 & US &56353.861 &  6000 $\pm$          80 & ES \\ 
52229.841 &  6170 $\pm$          40 & US &56353.865 &  5980 $\pm$          60 & ES \\ 
53334.680 &  6040 $\pm$          50 & HS &56613.111 &  6010 $\pm$          30 & ES \\ 
53334.681 &  6040 $\pm$          50 & HS &56962.134 &  5920 $\pm$          40 & ES \\ 
53334.682 &  6010 $\pm$          40 & HS &56962.144 &  5950 $\pm$          40 & ES \\ 
53463.535 &  6050 $\pm$          40 & HS &56968.144 &  5890 $\pm$          30 & ES \\ 
53463.536 &  5920 $\pm$          40 & HS &56968.155 &  5870 $\pm$          40 & ES \\ 
53463.537 &  6010 $\pm$          50 & HS &56972.106 &  5940 $\pm$          40 & ES \\ 
53464.536 &  6100 $\pm$          50 & HS &56972.116 &  5920 $\pm$          40 & ES \\ 
53464.538 &  6010 $\pm$          40 & HS &57022.022 &  5940 $\pm$          40 & ES \\ 
53464.539 &  6040 $\pm$          40 & HS &57022.032 &  5920 $\pm$          40 & ES \\ 
53582.949 &  6160 $\pm$          40 & HS &57120.801 &  5920 $\pm$          60 & ES \\ 
53712.674 &  6030 $\pm$          30 & HS &57120.812 &  5940 $\pm$          70 & ES \\ 
53712.678 &  6040 $\pm$          30 & HS &57120.823 &  5910 $\pm$          40 & ES \\ 
53715.631 &  6060 $\pm$          30 & HS &57120.831 &  5870 $\pm$          90 & ES \\ 
53715.636 &  6050 $\pm$          30 & HS &57325.109 &  5850 $\pm$          40 & ES \\ 
54222.581 &  6040 $\pm$          40 & HS &57325.117 &  5880 $\pm$          40 & ES \\ 
54222.585 &  6050 $\pm$          40 & HS &57325.128 &  5890 $\pm$          30 & ES \\ 
54865.672 &  5980 $\pm$          30 & HS &57328.138 &  5970 $\pm$          40 & ES \\ 
54865.677 &  5910 $\pm$          40 & HS &57328.148 &  5970 $\pm$          40 & ES \\ 
55229.615 &  5900 $\pm$         100 & FS &57353.089 &  5920 $\pm$          40 & ES \\ 
56343.815 &  5900 $\pm$          50 & ES &57353.100 &  5930 $\pm$          40 & ES \\ 
56353.857 &  5940 $\pm$          50 & ES &&&\\ 
 \hline
         \end{tabular}  \label{Tab_HD55719}
         \end{table}

%% file: HD61468_tab.tex
\begin{table}  
\scriptsize
\caption{Measured $B_s$ of HD61468.}
\begin{tabular}{ccc|ccc}
 \hline
 {HJD} & {B$_{\rm s}$} & {Sp} & {HJD} & {B$_{\rm s}$} & {Sp} \\
2400000+  & {[G]}                   &       &  {2400000+} & {[G]} &  ~\\\hline
 57340.717 & 6260 $\pm$ 85 &  HN &    & \\
 \hline
         \end{tabular} \label{Tab_HD61468}
         \end{table}

%% file: HD75445_tab.tex
\begin{table}
\scriptsize
\caption{Measured $B_s$ of HD75445.}
\begin{tabular}{ccc|ccc}
 \hline
{HJD} & {B$_{\rm s}$} & {Sp} & {HJD} & {B$_{\rm s}$} & {Sp} \\
2400000+  & {[G]}                   &       &  {2400000+} & {[G]} &  ~\\\hline
51600.623 &  2920 $\pm$          50 & CE & 54205.492 &  2930 $\pm$          20 & HS \\ 
51945.543 &  2950 $\pm$          20 & CE & 54205.546 &  2920 $\pm$          20 & HS \\ 
51955.501 &  2910 $\pm$          20 & CE & 54205.600 &  2935 $\pm$          20 & HS \\
52236.838 &  2990 $\pm$          50 & US & & & \\
 \hline
         \end{tabular}   \label{Tab_HD75445}
         \end{table}

%% file: HD81009_tab.tex
\begin{table} 
\scriptsize
\caption{Measured $B_s$ of HD81009.}
\begin{tabular}{ccc|ccc}
 \hline
{HJD} & {B$_{\rm s}$} & {Sp} & {HJD} & {B$_{\rm s}$} & {Sp} \\
2400000+  & {[G]}                   &       &  {2400000+} & {[G]} &  ~\\\hline
52237.853 &  8000 $\pm$          30 & US & 54635.460 &  8540 $\pm$         100 & HS \\ 
53463.579 &  8520 $\pm$          30 & HS & 54635.464 &  8610 $\pm$          90 & HS \\ 
53464.585 &  8820 $\pm$          30 & HS & 54635.467 &  8650 $\pm$         100 & HS \\ 
54204.530 &  7220 $\pm$          90 & HS & 54864.731 &  9670 $\pm$          60 & HS \\ 
54204.562 &  7250 $\pm$          60 & HS & 54864.734 &  9580 $\pm$          70 & HS \\ 
54204.594 &  7240 $\pm$          60 & HS & 54864.737 &  9580 $\pm$          50 & HS \\ 
54204.627 &  7250 $\pm$          90 & HS & 54865.607 &  9510 $\pm$          60 & HS \\ 
54222.569 &  9460 $\pm$          60 & HS & 54865.610 &  9610 $\pm$          50 & HS \\ 
54222.572 &  9430 $\pm$          70 & HS & 54865.613 &  9600 $\pm$          50 & HS \\ 
54222.575 &  9430 $\pm$          60 & HS & 54867.732 &  9440 $\pm$          70 & HS \\ 
54498.604 &  8650 $\pm$          60 & SG & 54867.735 &  9480 $\pm$          70 & HS \\ 
54543.522 &  7160 $\pm$          90 & HS & 54867.739 &  9420 $\pm$          70 & HS \\ 
54543.525 &  7110 $\pm$          90 & HS & 54869.609 &  9140 $\pm$          70 & HS \\ 
54543.528 &  7110 $\pm$          70 & HS & 54869.612 &  9280 $\pm$          80 & HS \\ 
54544.526 &  7260 $\pm$          80 & HS & 54869.616 &  9260 $\pm$          70 & HS \\ 
54544.529 &  7300 $\pm$          80 & HS & 54870.626 &  9090 $\pm$          70 & HS \\ 
54544.532 &  7210 $\pm$          80 & HS & 54870.630 &  9180 $\pm$          70 & HS \\ 
54545.709 &  7440 $\pm$          80 & HS & 54870.633 &  9130 $\pm$          70 & HS \\ 
54545.712 &  7410 $\pm$          80 & HS & 56707.426 &  8740 $\pm$          80 & CS \\ 
54545.715 &  7470 $\pm$          80 & HS & 56710.441 &  8190 $\pm$         110 & CS \\ 
54633.491 &  8920 $\pm$          90 & HS & 56759.570 &  8370 $\pm$         140 & CS \\ 
54633.494 &  8900 $\pm$          90 & HS & 57340.777 &  9080  $\pm$         60 & CS \\
54633.498 &  8930 $\pm$          80 & HS &&&\\ 
 \hline
         \end{tabular}  \label{Tab_HD81009}
         \end{table}

%% file: HD93507_tab.tex
\begin{table} 
\scriptsize
\caption{Measured $B_s$ of HD93507.}
\begin{tabular}{ccc|ccc}
 \hline
{HJD} & {B$_{\rm s}$} & {Sp} & {HJD} & {B$_{\rm s}$} & {Sp} \\
2400000+  & {[G]}                   &       &  {2400000+} & {[G]} &  ~\\\hline
52244.849 &  7240 $\pm$          60 & US &54545.554 &  7200 $\pm$          70 & HS \\ 
53463.672 &  7170 $\pm$          60 & HS &54545.561 &  7220 $\pm$          70 & HS \\ 
53464.662 &  7240 $\pm$          80 & HS &54633.540 &  7240 $\pm$          50 & HS \\ 
53464.669 &  7240 $\pm$          90 & HS &54633.547 &  7320 $\pm$          60 & HS \\ 
53581.463 &  7220 $\pm$          30 & HS &54635.495 &  7370 $\pm$          80 & HS \\ 
53581.471 &  7200 $\pm$          40 & HS &54635.503 &  7450 $\pm$          60 & HS \\ 
54222.675 &  6840 $\pm$          40 & HS &54864.763 &  6650 $\pm$          50 & HS \\ 
54222.683 &  6850 $\pm$          40 & HS &54864.770 &  6730 $\pm$          40 & HS \\ 
54224.700 &  6860 $\pm$          40 & HS &54867.605 &  6800 $\pm$          40 & HS \\ 
54224.711 &  6910 $\pm$          40 & HS &54867.612 &  6780 $\pm$          40 & HS \\ 
54441.807 &  7070 $\pm$          40 & HS &54869.656 &  6780 $\pm$          40 & HS \\ 
54441.815 &  6960 $\pm$          40 & HS &54869.663 &  6790 $\pm$          40 & HS \\ 
54442.816 &  7090 $\pm$          30 & HS &54870.646 &  6720 $\pm$          40 & HS \\ 
54442.823 &  7120 $\pm$          30 & HS &54870.654 &  6740 $\pm$          40 & HS \\ 
54443.816 &  7070 $\pm$          40 & HS &54870.742 &  6800 $\pm$          40 & HS \\ 
54443.823 &  7030 $\pm$          40 & HS &54870.750 &  6750 $\pm$          40 & HS \\ 
 \hline
         \end{tabular}  \label{Tab_HD93507}
         \end{table}

%% file: HD94660_tab.tex
\begin{table} 
\scriptsize
\caption{Measured $B_s$ of HD94660.}
\begin{tabular}{ccc|ccc}
 \hline
{HJD} & {B$_{\rm s}$} & {Sp} & {HJD} & {B$_{\rm s}$} & {Sp} \\
2400000+  & {[G]}                   &       &  {2400000+} & {[G]} &  ~\\\hline
50824.041 &  6380 $\pm$          40 & UC &54196.964 &  6320 $\pm$          30 & UC \\ 
51176.136 &  6380 $\pm$          20 & UC &54198.928 &  6380 $\pm$          20 & UC \\ 
51542.171 &  6330 $\pm$          20 & UC &54928.959 &  6190 $\pm$          60 & UC \\ 
52031.464 &  6140 $\pm$          20 & US &54975.546 &  6130 $\pm$          20 & HS \\ 
52031.466 &  6170 $\pm$          20 & US &54976.536 &  6170 $\pm$          20 & HS \\ 
52265.250 &  6160 $\pm$          40 & UC &54982.605 &  6130 $\pm$          50 & HS \\ 
53072.664 &  6150 $\pm$          20 & US &55161.232 &  6050 $\pm$          20 & UC \\ 
53072.705 &  6180 $\pm$          20 & US &55202.867 &  6090 $\pm$          30 & HS \\ 
53707.846 &  6380 $\pm$          20 & US &55701.447 &  6120 $\pm$          20 & HS \\ 
53745.174 &  6400 $\pm$          20 & ES &56018.549 &  6240 $\pm$          20 & HS \\ 
54188.098 &  6440 $\pm$          40 & UC &56775.604 &  6370 $\pm$          20 & HS \\ 
 \hline
         \end{tabular}  \label{Tab_HD94660}
         \end{table}

%% file: HD110066_tab.tex
\begin{table} 
\scriptsize
\caption{Measured $B_s$ of HD110066.}
\begin{tabular}{ccc|ccc}
 \hline
{HJD} & {B$_{\rm s}$} & {Sp} & {HJD} & {B$_{\rm s}$} & {Sp} \\
2400000+  & {[G]}                   &       &  {2400000+} & {[G]} &  ~\\\hline
51686.409 &  4140 $\pm$          60 & EE &56816.402 &  4000 $\pm$          70 & CS \\ 
51739.759 &  4080 $\pm$          30 & GO &57131.457 &  4200 $\pm$          40 & CS \\ 
52412.498 &  4100 $\pm$          40 & SG &57471.933 &  4110 $\pm$          50 & ES \\ 
52419.774 &  4070 $\pm$          30 & GO &57498.902 &  4110 $\pm$          50 & ES \\ 
53544.371 &  4080 $\pm$          40 & SG &59183.159 &  4110 $\pm$          50 & ES \\ 
54498.670 &  4100 $\pm$          30 & SG &59191.150 &  4110 $\pm$          40 & ES \\ 
 \hline
         \end{tabular}  \label{Tab_HD110066}
         \end{table}

%% file: HD116114_tab.tex
\begin{table} 
\scriptsize
\caption{Measured $B_s$ of HD116114.}
\begin{tabular}{ccc|ccc}
 \hline
{HJD} & {B$_{\rm s}$} & {Sp} & {HJD} & {B$_{\rm s}$} & {Sp} \\
2400000+  & {[G]}                   &       &  {2400000+} & {[G]} &  ~\\\hline
50115.761 &  5940 $\pm$          50 & EM &54222.786 &  6120 $\pm$          20 & HS \\ 
52296.868 &  6030 $\pm$          20 & US &54223.718 &  6100 $\pm$          20 & HS \\ 
52296.870 &  6050 $\pm$          20 & US &54223.722 &  6110 $\pm$          20 & HS \\ 
52676.855 &  6070 $\pm$          20 & US &54224.827 &  6130 $\pm$          30 & HS \\ 
52676.862 &  6050 $\pm$          20 & US &54224.831 &  6120 $\pm$          20 & HS \\ 
52676.866 &  6070 $\pm$          20 & US &54544.591 &  6120 $\pm$          20 & HS \\ 
52676.869 &  6050 $\pm$          20 & US &54544.595 &  6120 $\pm$          20 & HS \\ 
52676.872 &  6050 $\pm$          20 & US &54545.677 &  6160 $\pm$          20 & HS \\ 
53070.761 &  6090 $\pm$          10 & US &54545.681 &  6150 $\pm$          20 & HS \\ 
53544.389 &  6070 $\pm$          30 & SG &54633.653 &  6120 $\pm$          20 & HS \\ 
53463.725 &  6100 $\pm$          20 & HS &54633.657 &  6140 $\pm$          20 & HS \\ 
53463.729 &  6050 $\pm$          20 & HS &54865.822 &  6140 $\pm$          20 & HS \\ 
53464.703 &  6110 $\pm$          30 & HS &54865.826 &  6090 $\pm$          20 & HS \\ 
53464.707 &  6070 $\pm$          20 & HS &54868.760 &  6170 $\pm$          30 & HS \\ 
53581.520 &  6060 $\pm$          20 & HS &54868.764 &  6150 $\pm$          30 & HS \\ 
53581.524 &  6110 $\pm$          20 & HS &54998.717 &  6160 $\pm$          80 & SG \\ 
53582.544 &  6120 $\pm$          20 & HS &56706.596 &  6200 $\pm$          30 & CS \\ 
53582.548 &  6110 $\pm$          20 & HS &56707.567 &  6260 $\pm$          30 & CS \\ 
54204.668 &  6120 $\pm$          30 & HS &57131.486 &  6220 $\pm$          80 & CS \\ 
54204.724 &  6120 $\pm$          40 & HS &57234.859 &  6190 $\pm$          90 & UC \\ 
54204.789 &  6110 $\pm$          30 & HS &57375.286 &  6300 $\pm$         100 & HN \\ 
54222.782 &  6150 $\pm$          20 & HS &57587.432 &  6240 $\pm$          40 & HN \\ 
 \hline
         \end{tabular}  \label{Tab_HD116114}
         \end{table}

%% file: HD126515_tab.tex
\begin{table} 
\scriptsize
\caption{Measured $B_s$ of HD126515.}
\begin{tabular}{ccc|ccc}
 \hline
{HJD} & {B$_{\rm s}$} & {Sp} & {HJD} & {B$_{\rm s}$} & {Sp} \\
2400000+  & {[G]}                   &       &  {2400000+} & {[G]} &  ~\\\hline
51979.910 & 14490 $\pm$          50 & US &54544.705 & 15050 $\pm$          50 & HS \\ 
51979.913 & 14550 $\pm$          50 & US &54544.710 & 15150 $\pm$          50 & HS \\ 
53463.772 &  9340 $\pm$          50 & HS &54633.662 &  9340 $\pm$          50 & HS \\ 
53463.777 &  9310 $\pm$          50 & HS &54633.667 &  9320 $\pm$          50 & HS \\ 
53464.756 &  9460 $\pm$          50 & HS &54635.481 &  9410 $\pm$          50 & HS \\ 
53464.761 &  9400 $\pm$          50 & HS &54635.486 &  9470 $\pm$          50 & HS \\ 
53544.401 & 13730 $\pm$          70 & SG &54716.476 & 13240 $\pm$          50 & HS \\ 
53582.603 &  9370 $\pm$          50 & HS &54716.481 & 13140 $\pm$          50 & HS \\ 
53582.608 &  9430 $\pm$          50 & HS &54864.815 & 11120 $\pm$          50 & HS \\ 
53583.566 &  9320 $\pm$          50 & HS &54864.820 & 10750 $\pm$          50 & HS \\ 
53583.571 &  9280 $\pm$          50 & HS &54865.832 & 10650 $\pm$          50 & HS \\ 
54222.818 &  9950 $\pm$          50 & HS &54865.837 & 10540 $\pm$          50 & HS \\ 
54222.823 &  9920 $\pm$          50 & HS &54867.823 & 10330 $\pm$          50 & HS \\ 
54223.696 &  9890 $\pm$          50 & HS &54867.828 & 10400 $\pm$          50 & HS \\ 
54223.701 &  9920 $\pm$          50 & HS &56707.654 &  9540 $\pm$          70 & CS \\ 
54224.786 &  9820 $\pm$          50 & HS &56729.579 & 11660 $\pm$         140 & CS \\ 
54224.791 &  9830 $\pm$          50 & HS &57131.517 & 13370 $\pm$         140 & CS \\ 
54338.495 & 11540 $\pm$          50 & HS &58294.454 & 12100 $\pm$          70 & HN \\ 
54338.500 & 11580 $\pm$          50 & HS &58294.447 & 11990 $\pm$          90 & HN \\ 
54543.828 & 15010 $\pm$          50 & HS &59191.162 & 10270 $\pm$          80 & ES \\ 
54543.833 & 15030 $\pm$          50 & HS & &  & \\ 
 \hline
         \end{tabular}  \label{Tab_HD126515}
         \end{table}

%% file: HD137949_tab.tex
\begin{table} 
\scriptsize
\caption{Measured $B_s$ of HD137949.}
\begin{tabular}{ccc|ccc}
 \hline
{HJD} & {B$_{\rm s}$} & {Sp} & {HJD} & {B$_{\rm s}$} & {Sp} \\
2400000+  & {[G]}                   &       &  {2400000+} & {[G]} &  ~\\\hline
51593.835 &  4700 $\pm$          20 & CE &54598.589 &  4680 $\pm$          30 & SG \\ 
52331.803 &  4680 $\pm$          40 & US &54599.521 &  4720 $\pm$          30 & SG \\ 
52420.829 &  4680 $\pm$          20 & SG &54600.478 &  4710 $\pm$          30 & SG \\ 
52535.505 &  4700 $\pm$          30 & US &54867.786 &  4710 $\pm$          20 & HS \\ 
52537.509 &  4680 $\pm$          10 & US &54867.791 &  4710 $\pm$          20 & HS \\ 
53214.796 &  4700 $\pm$          30 & SG &54869.789 &  4720 $\pm$          20 & HS \\ 
53463.817 &  4670 $\pm$          10 & HS &54869.794 &  4720 $\pm$          20 & HS \\ 
53464.798 &  4710 $\pm$          20 & HS &54870.862 &  4710 $\pm$          20 & HS \\ 
53511.082 &  4720 $\pm$          30 & UC &54870.868 &  4720 $\pm$          20 & HS \\ 
53513.126 &  4700 $\pm$          30 & UC &54953.976 &  4680 $\pm$          10 & ES \\ 
53544.424 &  4680 $\pm$          30 & SG &54957.878 &  4690 $\pm$          50 & ES \\ 
53581.602 &  4710 $\pm$          10 & HS &54962.896 &  4680 $\pm$          10 & ES \\ 
53745.190 &  4700 $\pm$          20 & ES &54963.000 &  4680 $\pm$          10 & ES \\ 
53836.299 &  4760 $\pm$          10 & UC &55360.557 &  4720 $\pm$          30 & HS \\ 
53930.384 &  4700 $\pm$          30 & SG &55361.522 &  4710 $\pm$          10 & HS \\ 
53931.389 &  4690 $\pm$          40 & SG &55364.472 &  4690 $\pm$          20 & HS \\ 
53932.382 &  4670 $\pm$          40 & SG &55365.500 &  4690 $\pm$          10 & HS \\ 
53949.814 &  4660 $\pm$          10 & ES &55368.490 &  4690 $\pm$          20 & HS \\ 
53950.739 &  4680 $\pm$          10 & ES &55369.502 &  4700 $\pm$          30 & HS \\ 
53951.738 &  4660 $\pm$          10 & ES &55379.483 &  4670 $\pm$          10 & HS \\ 
54222.804 &  4700 $\pm$          20 & HS &55382.619 &  4710 $\pm$          10 & HS \\ 
54223.702 &  4720 $\pm$          30 & HS &55383.663 &  4710 $\pm$          10 & HS \\ 
54336.599 &  4740 $\pm$          60 & HS &56145.514 &  4730 $\pm$          40 & HS \\ 
54498.750 &  4680 $\pm$          30 & SG &57189.402 &  4750 $\pm$          20 & CS \\ 
54543.859 &  4710 $\pm$          20 & HS &57221.350 &  4700 $\pm$          40 & CS \\ 
54544.719 &  4700 $\pm$          20 & HS &58266.543 &  4730 $\pm$          20 & HN \\ 
54544.724 &  4710 $\pm$          20 & HS &58267.520 &  4730 $\pm$          20 & HN \\ 
54545.764 &  4680 $\pm$          20 & HS &58906.500 &  4750 $\pm$          50 & CS \\ 
54545.770 &  4710 $\pm$          20 & HS &                  &                                    &  \\ 
 \hline
         \end{tabular}  \label{Tab_HD137949}
         \end{table}

%% file: HD142070_tab.tex
\begin{table} 
\scriptsize
\caption{Measured $B_s$ of HD142070.}
\begin{tabular}{ccc|ccc}
 \hline
{HJD} & {B$_{\rm s}$} & {Sp} & {HJD} & {B$_{\rm s}$} & {Sp} \\
2400000+  & {[G]}                   &       &  {2400000+} & {[G]} &  ~\\\hline
52331.872 &  4640 $\pm$          40 & US &54336.496 &  4970 $\pm$          50 & HS \\ 
52331.875 &  4680 $\pm$          30 & US &54543.716 &  4850 $\pm$          40 & HS \\ 
53463.840 &  4680 $\pm$          30 & HS &54543.724 &  4700 $\pm$          50 & HS \\ 
53464.825 &  4750 $\pm$          40 & HS &54543.734 &  4730 $\pm$          50 & HS \\ 
53464.832 &  4730 $\pm$          40 & HS &54543.867 &  4750 $\pm$          30 & HS \\ 
53582.617 &  4720 $\pm$          40 & HS &54543.876 &  4700 $\pm$          40 & HS \\ 
53582.625 &  4720 $\pm$          40 & HS &54544.873 &  4990 $\pm$          50 & HS \\ 
53583.630 &  4850 $\pm$          50 & HS &54544.882 &  4960 $\pm$          40 & HS \\ 
53583.638 &  4830 $\pm$          50 & HS &54545.817 &  4750 $\pm$          40 & HS \\ 
54222.731 &  4750 $\pm$          30 & HS &54545.826 &  4770 $\pm$          50 & HS \\ 
54222.740 &  4770 $\pm$          40 & HS &54866.834 &  4740 $\pm$          50 & HS \\ 
54222.909 &  4810 $\pm$          40 & HS &54866.843 &  4780 $\pm$          40 & HS \\ 
54222.918 &  4900 $\pm$          40 & HS &54867.862 &  4780 $\pm$          40 & HS \\ 
54223.772 &  4730 $\pm$          30 & HS &54867.871 &  4810 $\pm$          40 & HS \\ 
54223.780 &  4650 $\pm$          50 & HS &54869.860 &  4630 $\pm$          40 & HS \\ 
54223.905 &  4650 $\pm$          90 & HS &54869.869 &  4670 $\pm$          40 & HS \\ 
54224.737 &  4970 $\pm$          40 & HS &57229.330 &  5000 $\pm$          90 & CS \\ 
54224.745 &  4980 $\pm$          40 & HS &57234.874 &  4740 $\pm$         100 & UC \\ 
54224.892 &  4980 $\pm$          50 & HS &58266.509 &  4800 $\pm$          30 & HN \\ 
54224.901 &  5090 $\pm$          50 & HS &58267.497 &  4930 $\pm$         100 & HN \\ 
54336.487 &  4970 $\pm$          50 & HS &58267.497 &  4930 $\pm$         100 & HN \\ 
 \hline
         \end{tabular}  \label{Tab_HD142070}
         \end{table}

%% file: HD144897_tab.tex
\begin{table} 
\scriptsize
\caption{Measured $B_s$ of HD144897.}
\begin{tabular}{ccc|ccc}
 \hline
{HJD} & {B$_{\rm s}$} & {Sp} & {HJD} & {B$_{\rm s}$} & {Sp} \\
2400000+  & {[G]}                   &       &  {2400000+} & {[G]} &  ~\\\hline
52331.834 &  8760 $\pm$          50 & US &54543.815 &  9390 $\pm$          40 & HS \\ 
53463.864 &  8550 $\pm$          40 & HS &54543.820 &  9390 $\pm$          40 & HS \\ 
53463.869 &  8550 $\pm$          40 & HS &54633.569 &  8700 $\pm$          70 & HS \\ 
53464.863 &  8570 $\pm$          40 & HS &54633.574 &  8700 $\pm$          70 & HS \\ 
53464.868 &  8570 $\pm$          40 & HS &54716.503 &  8530 $\pm$          50 & HS \\ 
53581.632 &  9580 $\pm$          50 & HS &54716.508 &  8530 $\pm$          50 & HS \\ 
53581.637 &  9580 $\pm$          50 & HS &54865.843 &  8410 $\pm$          50 & HS \\ 
53583.669 &  9460 $\pm$          50 & HS &54865.848 &  8410 $\pm$          50 & HS \\ 
53583.674 &  9460 $\pm$          50 & HS &54866.854 &  8490 $\pm$          50 & HS \\ 
54224.653 &  8760 $\pm$          70 & HS &54866.859 &  8490 $\pm$          60 & HS \\ 
54224.659 &  8760 $\pm$          70 & HS &54867.850 &  8500 $\pm$          50 & HS \\ 
54336.542 &  8510 $\pm$          50 & HS &54867.855 &  8480 $\pm$          50 & HS \\ 
54336.549 &  8510 $\pm$          50 & HS &54868.800 &  8410 $\pm$          60 & HS \\ 
54338.591 &  8590 $\pm$          40 & HS &54868.805 &  8410 $\pm$          50 & HS \\ 
54338.598 &  8590 $\pm$          40 & HS &57234.894 &  9080 $\pm$          90 & UC \\ 
 \hline
         \end{tabular}  \label{Tab_HD144897}
         \end{table}

%% file: HD150562_tab.tex
\begin{table}  
\scriptsize
\caption{Measured $B_s$ of HD150562.}
\begin{tabular}{ccc|ccc}
 \hline
{HJD} & {B$_{\rm s}$} & {Sp} & {HJD} & {B$_{\rm s}$} & {Sp} \\
2400000+  & {[G]}                   &       &  {2400000+} & {[G]} &  ~\\\hline
51252.826 &  4830 $\pm$          50 & EM &54865.885 &  4920 $\pm$          50 & HS \\ 
53939.497 &  4960 $\pm$          50 & US &54866.873 &  4880 $\pm$          50 & HS \\ 
54336.532 &  5040 $\pm$          60 & HS &54867.813 &  4980 $\pm$          60 & HS \\ 
54338.582 &  4880 $\pm$          50 & HS &54868.819 &  4900 $\pm$          50 & HS \\ 
54544.824 &  4950 $\pm$          50 & HS &54869.819 &  4980 $\pm$          50 & HS \\ 
54545.846 &  4970 $\pm$          50 & HS &54870.853 &  4970 $\pm$          50 & HS \\ 
54633.744 &  4970 $\pm$          50 & HS &57234.910 &  4900 $\pm$          50 & UC \\ 
 \hline
         \end{tabular} \label{Tab_HD150562}
         \end{table}

%% file: HD154708_tab.tex
\begin{table} 
\scriptsize
\caption{Measured $B_s$ of HD154708.}
\begin{tabular}{ccc|ccc}
 \hline
 {HJD} & {B$_{\rm s}$} & {Sp} & {HJD} & {B$_{\rm s}$} & {Sp} \\
2400000+  & {[G]}                   &       &  {2400000+} & {[G]} &  ~\\\hline
53631.580 & 24330 $\pm$         270 & US &54633.582 & 24770 $\pm$         210 & HS \\ 
53631.596 & 24390 $\pm$         280 & US &54716.543 & 23710 $\pm$         230 & HS \\ 
53631.633 & 23970 $\pm$         290 & US &54868.871 & 24250 $\pm$         230 & HS \\ 
53662.508 & 24910 $\pm$         270 & US &54870.814 & 24650 $\pm$         210 & HS \\ 
54543.769 & 24350 $\pm$         220 & HS &55021.779 & 24380 $\pm$         380 & FS \\ 
54544.833 & 23620 $\pm$         230 & HS &55029.581 & 24120 $\pm$         430 & FS \\ 
 \hline
         \end{tabular}  \label{Tab_HD154708}
         \end{table}

%% file: HD318107_tab.tex
\begin{table} 
\scriptsize
\caption{Measured $B_s$ of HD318107.}
\begin{tabular}{ccc|ccc}
 \hline
{HJD} & {B$_{\rm s}$} & {Sp} & {HJD} & {B$_{\rm s}$} & {Sp} \\
2400000+  & {[G]}                   &       &  {2400000+} & {[G]} &  ~\\\hline53215.872 & 14590 $\pm$         200 & GO &54633.608 & 14290 $\pm$         150 & HS \\ 
53463.893 & 13870 $\pm$         100 & HS &54869.845 & 13750 $\pm$          50 & HS \\ 
53582.640 & 13430 $\pm$          40 & HS &54870.874 & 13880 $\pm$          60 & HS \\ 
53583.688 & 13530 $\pm$          80 & HS &57234.963 & 13960 $\pm$         200 & UC \\ 
54223.637 & 13530 $\pm$          70 & HS &57587.444 & 14670 $\pm$         300 & HN \\ 
54553.120 & 13600 $\pm$          60 & ES &58295.517 & 17000 $\pm$         300 & HN \\ 
54553.150 & 13620 $\pm$          60 & ES &58295.517 & 17000 $\pm$         300 & HN \\ 
 \hline
         \end{tabular}  \label{Tab_HD318107}
         \end{table}

%% file: HD165474_tab.tex
\begin{table} 
\scriptsize
\caption{Measured $B_s$ of HD165474.}
\begin{tabular}{ccc|ccc}
 \hline
 {HJD} & {B$_{\rm s}$} & {Sp} & {HJD} & {B$_{\rm s}$} & {Sp} \\
2400000+  & {[G]}                   &       &  {2400000+} & {[G]} &  ~\\\hline
53104.452 &  6800 $\pm$          40 & NS &57220.407 &  6390 $\pm$          70 & CS \\ 
53601.345 &  6780 $\pm$          40 & NS &57221.375 &  6390 $\pm$          80 & CS \\ 
53871.451 &  6710 $\pm$          50 & NS &57222.435 &  6330 $\pm$          80 & CS \\ 
54207.848 &  6700 $\pm$          10 & HS &57223.381 &  6350 $\pm$          60 & CS \\ 
54963.440 &  6610 $\pm$          50 & NS &57229.452 &  6220 $\pm$          80 & CS \\ 
56145.616 &  6400 $\pm$          40 & HS &58294.554 &  6460 $\pm$          80 & HN \\ 
57202.477 &  6280 $\pm$          50 & CS &59441.374 &  6560 $\pm$          40 & HN \\ 
 \hline
          \end{tabular}  \label{Tab_HD165474}
         \end{table}

%% file: HD166473_tab.tex
\begin{table}
\scriptsize
\caption{Measured $B_s$ of HD166473.}
\begin{tabular}{ccc|ccc}
 \hline
 \hline
{HJD} & {B$_{\rm s}$} & {Sp} & {HJD} & {B$_{\rm s}$} & {Sp} \\
2400000+  & {[G]}                   &       &  {2400000+} & {[G]} &  ~\\\hline
52090.459 &  8170 $\pm$          80 & US &54634.852 &  5730 $\pm$          50 & HS \\ 
52189.506 &  8350 $\pm$          70 & US &54634.861 &  5860 $\pm$          50 & HS \\ 
54336.612 &  5790 $\pm$          50 & HS &54716.623 &  5770 $\pm$          50 & HS \\ 
54336.622 &  5830 $\pm$          50 & HS &54716.633 &  5880 $\pm$          50 & HS \\ 
54338.630 &  5720 $\pm$          50 & HS &56531.773 &  8360 $\pm$          80 & ES \\ 
54338.639 &  5730 $\pm$          50 & HS &56547.732 &  8340 $\pm$          90 & ES \\ 
54544.782 &  5780 $\pm$          40 & HS &56813.014 &  8070 $\pm$          70 & ES \\ 
54544.791 &  5760 $\pm$          50 & HS &57239.836 &  7210 $\pm$          50 & ES \\ 
54545.866 &  5760 $\pm$          40 & HS &57287.712 &  7180 $\pm$          50 & ES \\ 
54545.872 &  5700 $\pm$          40 & HS &58642.993 &  5860 $\pm$          40 & ES \\ 
54633.767 &  5760 $\pm$          40 & HS &57235.008 &  7210 $\pm$          70 & UC \\ 
54633.777 &  5740 $\pm$          40 & HS   &       &   \\ 
 \hline
 \hline
         \end{tabular}   \label{Tab_HD166473}
         \end{table}

%% file: HD177765_tab.tex
\begin{table}  
\scriptsize
\caption{Measured $B_s$ of HD177765.}
\begin{tabular}{ccc|ccc}
 \hline
{HJD} & {B$_{\rm s}$} & {Sp} & {HJD} & {B$_{\rm s}$} & {Sp} \\
2400000+  & {[G]}                   &       &  {2400000+} & {[G]} &  ~\\\hline
54687.752 &  3410 $\pm$          40 & FS &57586.569 &  3530 $\pm$          30 & HN \\ 
55359.817 &  3490 $\pm$          30 & US &58294.607 &  3590 $\pm$          30 & HN \\ 
57235.029 &  3530 $\pm$          50 & UC &  &    &   \\ 
 \hline
         \end{tabular} \label{Tab_HD177765}
         \end{table}

%% file: HD178892_tab.tex
\begin{table} 
\scriptsize
\caption{Measured $B_s$ of HD178892.}
\begin{tabular}{ccc|ccc}
 \hline
{HJD} & {B$_{\rm s}$} & {Sp} & {HJD} & {B$_{\rm s}$} & {Sp} \\
2400000+  & {[G]}                   &       &  {2400000+} & {[G]} &  ~\\\hline
53871.405 & 18500 $\pm$         300 & NS &53871.405 & 18500 $\pm$         300 & NS \\ 
55351.884 & 18000 $\pm$         140 & US &55351.884 & 18000 $\pm$         140 & US \\ 
56356.149 & 19050 $\pm$         310 & ES &56356.149 & 19050 $\pm$         310 & ES \\ 
56356.160 & 19590 $\pm$         280 & ES &56356.160 & 19590 $\pm$         280 & ES \\ 
56356.166 & 19220 $\pm$         310 & ES &56356.166 & 19220 $\pm$         310 & ES \\ 
56430.906 & 19070 $\pm$         140 & ES &56430.906 & 19070 $\pm$         140 & ES \\ 
56471.808 & 19320 $\pm$         270 & ES &56471.808 & 19320 $\pm$         270 & ES \\ 
57587.510 & 18550 $\pm$         210 & HN &57587.510 & 18550 $\pm$         210 & HN \\ 
58295.517 & 19620 $\pm$         160 & HN &58295.517 & 19620 $\pm$         160 & HN \\ 
53600.561 & 18260 $\pm$         200 & HS &  &   &   \\ 
 \hline
         \end{tabular}  \label{Tab_HD178892}
         \end{table}

%% file: HD187474_tab.tex
\begin{table} 
\scriptsize
\caption{Measured $B_s$ of HD187474.}
\begin{tabular}{ccc|ccc}
 \hline
{HJD} & {B$_{\rm s}$} & {Sp} & {HJD} & {B$_{\rm s}$} & {Sp} \\
2400000+  & {[G]}                   &       &  {2400000+} & {[G]} &  ~\\\hline
50375.159 &  6180 $\pm$          30 & CE &53270.000 &  4910 $\pm$          50 & UC \\ 
51051.667 &  4870 $\pm$          30 & CE &56143.760 &  4990 $\pm$          30 & HS \\ 
53164.788 &  5030 $\pm$          30 & US &56145.685 &  4960 $\pm$          30 & HS \\ 
52189.525 &  5670 $\pm$          20 & US &57235.076 &  6230 $\pm$          60 & UC \\ 
52189.526 &  5690 $\pm$          30 & US &  &    &   \\ 
 \hline
          \end{tabular}  \label{Tab_HD187474}
         \end{table}

%% file: HD188041_tab.tex
\begin{table} 
\scriptsize
\caption{Measured $B_s$ of HD188041.}
\begin{tabular}{ccc|ccc}
 \hline
{HJD} & {B$_{\rm s}$} & {Sp} & {HJD} & {B$_{\rm s}$} & {Sp} \\
2400000+  & {[G]}                   &       &  {2400000+} & {[G]} &  ~\\\hline
51826.349 &  3640 $\pm$          20 & SG &54338.672 &  3580 $\pm$          40 & HS \\ 
52119.549 &  3550 $\pm$          30 & SG &54374.362 &  3600 $\pm$          30 & SG \\ 
52190.600 &  3510 $\pm$          30 & US &54545.899 &  3580 $\pm$          20 & HS \\ 
52190.601 &  3540 $\pm$          30 & US &54633.859 &  3580 $\pm$          20 & HS \\ 
52889.376 &  3610 $\pm$          30 & SG &54634.808 &  3550 $\pm$          40 & HS \\ 
53270.015 &  3520 $\pm$          50 & UC &54716.644 &  3680 $\pm$          20 & HS \\ 
53464.884 &  3600 $\pm$          20 & HS &54717.696 &  3670 $\pm$          30 & HS \\ 
53581.663 &  3630 $\pm$          50 & HS &56145.652 &  3590 $\pm$          40 & HS \\ 
53582.668 &  3650 $\pm$          20 & HS &56507.449 &  3690 $\pm$          80 & CS \\ 
53583.716 &  3650 $\pm$          20 & HS &57239.986 &  3580 $\pm$          40 & ES \\ 
54204.876 &  3560 $\pm$          20 & HS &57263.823 &  3550 $\pm$          40 & ES \\ 
54223.864 &  3600 $\pm$          20 & HS &57564.132 &  3550 $\pm$          30 & ES \\ 
54224.862 &  3580 $\pm$          30 & HS &57886.723 &  3590 $\pm$          50 & HN \\ 
54336.651 &  3590 $\pm$          20 & HS & &   &  \\  
 \hline
         \end{tabular}  \label{Tab_HD188041}
         \end{table}

%% file: HD192678_tab.tex
\begin{table} 
\scriptsize
\caption{Measured $B_s$ of HD192678.}
\begin{tabular}{ccc|ccc}
 \hline
{HJD} & {B$_{\rm s}$} & {Sp} & {HJD} & {B$_{\rm s}$} & {Sp} \\
2400000+  & {[G]}                   &       &  {2400000+} & {[G]} &  ~\\\hline
49620.334 &  4870 $\pm$          50 & EE &57221.522 &  4750 $\pm$          50 & CS \\ 
52120.616 &  4570 $\pm$          20 & SG &57222.591 &  4650 $\pm$          40 & CS \\ 
52121.564 &  4590 $\pm$          30 & SG &57230.486 &  4590 $\pm$          40 & CS \\ 
52890.527 &  4570 $\pm$          40 & SG &57264.497 &  4650 $\pm$          80 & CS \\ 
53215.898 &  4790 $\pm$          40 & GO &58266.601 &  4650 $\pm$          30 & HN \\ 
56904.409 &  4660 $\pm$          80 & CS &58267.591 &  4820 $\pm$          50 & HN \\ 
 \hline
         \end{tabular}  \label{Tab_HD192678}
         \end{table}

%% file: HD335238_tab.tex
\begin{table}  
\scriptsize
\caption{Measured $B_s$ of HD335238.}
\begin{tabular}{ccc|ccc}
 \hline
{HJD} & {B$_{\rm s}$} & {Sp} & {HJD} & {B$_{\rm s}$} & {Sp} \\
2400000+  & {[G]}                   &       &  {2400000+} & {[G]} &  ~\\\hline
51739.542 & 11770 $\pm$         100 & GO &57240.524 &  7940 $\pm$         160 & CS \\ 
51740.896 & 11220 $\pm$         100 & GO &57959.548 & 12380 $\pm$         200 & CS \\ 
52420.916 & 12520 $\pm$         100 & GO &57611.553 & 11900 $\pm$         200 & CS \\ 
52121.622 & 10740 $\pm$         200 & SG &58299.573 & 12800 $\pm$         200 & CS \\ 
52889.396 &  8370 $\pm$          60 & SG &58312.588 &  8800 $\pm$         160 & CS \\ 
53544.714 & 11680 $\pm$         180 & SG &57587.537 &  8180 $\pm$          50 & HN \\ 
56844.590 &  8140 $\pm$         100 & CS &58095.336 &  9470 $\pm$         190 & HN \\ 
56893.490 &  8140 $\pm$         100 & CS &58266.621 &  8060 $\pm$          50 & HN \\ 
56904.447 &  8520 $\pm$          60 & CS &58267.602 &  8360 $\pm$          50 & HN \\ 
57220.548 & 12730 $\pm$         200 & CS &58295.601 & 11100 $\pm$          50 & HN \\ 
57230.548 &  9800 $\pm$         160 & CS &58432.375 &  8070 $\pm$          70 & HN \\ 
 \hline
         \end{tabular} \label{Tab_HD335238}
         \end{table}

%% file: HD201601_tab.tex
\begin{table} 
\scriptsize
\caption{Measured $B_s$ of HD201601.}
\begin{tabular}{ccc|ccc}
 \hline
{HJD} & {B$_{\rm s}$} & {Sp} & {HJD} & {B$_{\rm s}$} & {Sp} \\
2400000+  & {[G]}                   &       &  {2400000+} & {[G]} &  ~\\\hline
37476.500 &  2800 $\pm$         100 & Evans &55168.747 &  3880 $\pm$          20 & ES \\ 
41180.411 &  3600 $\pm$         100 & Scholz &55406.850 &  3860 $\pm$          20 & ES \\ 
41181.410 &  3300 $\pm$         100 & Scholz &55415.600 &  3810 $\pm$          50 & HS \\ 
41182.454 &  3600 $\pm$         100 & Scholz &55490.789 &  3870 $\pm$          20 & ES \\ 
47283.000 &  3700 $\pm$         100 & Scholz &55517.704 &  3840 $\pm$          10 & ES \\ 
47638.000 &  3730 $\pm$         100 & Scholz &55523.729 &  3860 $\pm$          10 & ES \\ 
48169.000 &  3820 $\pm$         100 & Scholz &55524.694 &  3850 $\pm$          20 & ES \\ 
48479.000 &  3830 $\pm$         100 & Scholz &55530.768 &  3860 $\pm$          20 & ES \\ 
48790.000 &  3790 $\pm$         100 & Scholz &56124.801 &  3810 $\pm$          50 & HS \\ 
48925.000 &  3890 $\pm$         100 & Scholz &56126.795 &  3750 $\pm$          50 & HS \\ 
49102.000 &  3830 $\pm$         100 & Scholz &56148.712 &  3760 $\pm$          50 & HS \\ 
49216.000 &  3840 $\pm$         100 & Scholz &56175.495 &  3730 $\pm$          50 & NL \\ 
49457.000 &  3870 $\pm$         100 & Scholz &56176.477 &  3790 $\pm$          40 & NL \\ 
49577.000 &  3860 $\pm$         100 & Scholz &56182.373 &  3740 $\pm$          30 & NL \\ 
49608.000 &  3990 $\pm$         100 & Scholz &56185.408 &  3790 $\pm$          40 & NL \\ 
49696.000 &  3900 $\pm$         100 & Scholz &56234.241 &  3840 $\pm$          40 & NL \\ 
49829.000 &  3870 $\pm$         100 & Scholz &56238.250 &  3860 $\pm$          80 & NL \\ 
49909.000 &  3840 $\pm$         100 & Scholz &56239.237 &  3780 $\pm$          70 & NL \\ 
50374.500 &  3900 $\pm$          30 & CE &56251.237 &  3790 $\pm$          60 & NL \\ 
51404.518 &  3960 $\pm$          30 & EE &56252.224 &  3780 $\pm$          40 & NL \\ 
51740.013 &  3900 $\pm$          50 & GO &56822.587 &  3670 $\pm$          30 & CS \\ 
51773.570 &  3920 $\pm$          30 & SG &56829.600 &  3740 $\pm$          30 & CS \\ 
52117.649 &  3940 $\pm$          30 & SG &56830.592 &  3670 $\pm$          30 & CS \\ 
52118.547 &  3950 $\pm$          20 & SG &56873.544 &  3670 $\pm$          20 & HN \\ 
52412.722 &  3900 $\pm$          50 & SG &56881.466 &  3770 $\pm$          20 & CS \\ 
52420.042 &  3940 $\pm$          50 & GO &56881.474 &  3820 $\pm$          60 & CS \\ 
52463.730 &  3890 $\pm$          30 & SG &56892.403 &  3840 $\pm$          50 & CS \\ 
52889.433 &  3890 $\pm$          20 & SG &57180.587 &  3660 $\pm$          40 & CS \\ 
52920.569 &  3920 $\pm$          20 & CE &57189.576 &  3650 $\pm$          20 & CS \\ 
53214.938 &  3940 $\pm$          50 & GO &57190.580 &  3650 $\pm$          20 & CS \\ 
53270.053 &  3940 $\pm$          20 & UC &57192.517 &  3710 $\pm$          50 & CS \\ 
53464.922 &  3930 $\pm$          10 & HS &57193.515 &  3700 $\pm$          30 & CS \\ 
53508.708 &  3900 $\pm$          50 & SG &57202.542 &  3640 $\pm$          20 & CS \\ 
53512.343 &  3960 $\pm$          50 & UC &57235.083 &  3690 $\pm$          40 & UC \\ 
53513.268 &  3940 $\pm$          50 & UC &57240.045 &  3690 $\pm$          40 & ES \\ 
53515.269 &  3940 $\pm$          50 & UC &57252.503 &  3670 $\pm$          30 & CS \\ 
53541.052 &  3910 $\pm$          20 & ES &57340.357 &  3610 $\pm$          50 & HN \\ 
53544.690 &  3940 $\pm$          20 & SG &57340.362 &  3640 $\pm$          30 & HN \\ 
53581.690 &  3920 $\pm$          20 & HS &57564.136 &  3670 $\pm$          30 & ES \\ 
53582.725 &  3930 $\pm$          10 & HS &57575.590 &  3690 $\pm$          50 & CS \\ 
53583.743 &  3920 $\pm$          20 & HS &57586.595 &  3670 $\pm$          20 & HN \\ 
53631.752 &  3930 $\pm$          50 & US &57587.561 &  3610 $\pm$          20 & HN \\ 
53659.599 &  3990 $\pm$          50 & US &57900.725 &  3590 $\pm$          20 & HN \\ 
53659.600 &  3920 $\pm$          50 & US &58012.494 &  3600 $\pm$          30 & HN \\ 
53659.600 &  3950 $\pm$          50 & US &58036.391 &  3670 $\pm$          90 & NL \\ 
53659.601 &  3930 $\pm$          50 & US &58036.392 &  3600 $\pm$         100 & NL \\ 
53659.602 &  3950 $\pm$          50 & US &58037.396 &  3650 $\pm$         110 & NL \\ 
54753.743 &  3910 $\pm$          20 & ES &58037.397 &  3720 $\pm$          70 & NL \\ 
54818.720 &  3890 $\pm$          20 & ES &58060.696 &  3580 $\pm$          10 & ES \\ 
54963.124 &  3880 $\pm$          20 & ES &58094.806 &  3530 $\pm$          30 & HN \\ 
54975.828 &  3860 $\pm$          50 & HS &58143.240 &  3460 $\pm$         190 & NL \\ 
54975.833 &  3840 $\pm$          50 & HS &58266.684 &  3550 $\pm$          20 & HN \\ 
54976.828 &  3850 $\pm$          50 & HS &58267.637 &  3560 $\pm$          40 & HN \\ 
54976.833 &  3870 $\pm$          50 & HS &58463.222 &  3470 $\pm$         130 & NL \\ 
55028.949 &  3880 $\pm$          20 & ES &58464.225 &  3540 $\pm$          70 & NL \\ 
55029.049 &  3880 $\pm$          10 & ES &58474.715 &  3540 $\pm$          20 & ES \\ 
55110.824 &  3880 $\pm$          20 & ES &59119.381 &  3460 $\pm$          20 & HN \\ 
55160.747 &  3870 $\pm$          20 & ES &59441.366 &  3400 $\pm$          20 & HN \\ 
 \hline
         \end{tabular}  \label{Tab_HD201601}
         \end{table}

%% file: HD208217_tab.tex
\begin{table}
\scriptsize
\caption{Measured $B_s$ of HD208217.}
\begin{tabular}{ccrc|ccrc}
 \hline
{HJD} & {B$_{\rm s}$} & {Sp} & {HJD} & {B$_{\rm s}$} & {Sp} \\
2400000+  & {[G]}                   &       &  {2400000+} & {[G]} &  ~\\\hline
 52189.566 &  7080   $\pm$          60 & US &  53715.523 &  8530  $\pm$         100 & HS \\ 
 52189.570 &  7160   $\pm$          50 & US &  53716.539 &  8330  $\pm$         100 & HS \\ 
 53581.711 &  7870   $\pm$         100 & HS &  54205.883 &  8330  $\pm$         100 & HS \\ 
 53582.691 &  7230   $\pm$         100 & HS &  54223.891 &  7520  $\pm$         100 & HS \\ 
 53711.547 &  7640   $\pm$         100 & HS &  56145.746 &  7140   $\pm$        100 & HS \\ 
 53713.520 &  7390  $\pm$          100 & HS &            &             &  \\ 
\hline
\end{tabular}\label{Tab_HD208217}
\end{table}

%% file: HD216018_tab.tex
\begin{table}  
\scriptsize
\caption{Measured $B_s$ of HD216018.}
\begin{tabular}{ccc|ccc}
 \hline
{HJD} & {B$_{\rm s}$} & {Sp} & {HJD} & {B$_{\rm s}$} & {Sp} \\
2400000+  & {[G]}                   &       &  {2400000+} & {[G]} &  ~\\\hline
52117.670 &  5690 $\pm$         100 & SG &56144.761 &  5600 $\pm$          60 & HS \\ 
52120.654 &  5640 $\pm$          40 & SG &56145.791 &  5530 $\pm$          60 & HS \\ 
52190.510 &  5640 $\pm$          30 & US &56892.507 &  5620 $\pm$          80 & CS \\ 
52420.046 &  5590 $\pm$          40 & GO &57220.589 &  5570 $\pm$          70 & CS \\ 
53214.957 &  5560 $\pm$          50 & GO &57230.588 &  5600 $\pm$          80 & CS \\ 
53270.083 &  5600 $\pm$          40 & UC &57235.088 &  5660 $\pm$         110 & UC \\ 
53271.093 &  5620 $\pm$          60 & UC &57340.376 &  5570 $\pm$          50 & HN \\ 
53544.700 &  5560 $\pm$          70 & SG &58294.694 &  5550 $\pm$          40 & HN \\ 
53930.653 &  5590 $\pm$          50 & SG &58432.465 &  5540 $\pm$          40 & HN \\ 
 \hline
         \end{tabular} \label{Tab_HD216018}
         \end{table}